\documentclass[12pt, a4paper, longbibliography, twoside]{Thesis} 
\usepackage{setspace}
\usepackage{wrapfig}
\usepackage{lscape}
\usepackage{graphicx}
\usepackage{caption}
\usepackage{amsmath}
\usepackage{amssymb}
\usepackage[square, numbers]{natbib} 
\usepackage[most]{tcolorbox}
\usepackage{hyperref}
\usepackage{xcolor}
\usepackage{sectsty}
\usepackage{tikz}
\usepackage{multirow}
\usetikzlibrary{shadows,positioning,shapes}
\usepackage{fancyhdr}

\makeatletter
\definecolor{qtcol}{rgb}{0,0.5,0}
\definecolor{arx}{rgb}{0.2,0.5,0.7}
\definecolor{jnl}{rgb}{0.9,0,0.7}
\definecolor{hcol}{rgb}{0.5,0,0}
\definecolor{chptr}{rgb}{0,0,0.5}
\definecolor{chptrnm}{rgb}{0,0,0.8}
\definecolor{chptrnm1}{rgb}{0,0.4,0.8}
\definecolor{shdw}{rgb}{0.5,0.5,0.5}
\definecolor{sec}{rgb}{0,0,0.5}
\definecolor{ssec}{rgb}{0,0,0.5}
\definecolor{capc}{rgb}{0,0.5,0.5}
\definecolor{citec}{rgb}{0,0.8,0}
\sectionfont{\color{sec}}  
\subsectionfont{\color{ssec}}
\usepackage[labelfont={color=capc,bf},textfont={color=capc}]{caption}
\hypersetup{colorlinks,	citecolor=citec, filecolor=magenta,	linkcolor=blue}

\newenvironment{myfont}{\fontsize{15pt}{15pt}\bf\fontfamily{pzc}\selectfont \color{qtcol}}{\par}

\newcommand{\Autoref}[1]{%
	\begingroup%
	\def\chapterautorefname{Chapter}%
	\def\sectionautorefname{Section}%
	\def\subsectionautorefname{Subsection}%
	\autoref{#1}%
	\endgroup%
}

\renewcommand{\@makechapterhead}[1]{%
	\vspace*{10\p@}%
	{\parindent \z@ \reset@font
		
	\begin{tikzpicture}
		\draw[black, line width=1mm] (-12.50,0) -- (2,0);
		\draw[black, line width=0.3mm] (-12.50,0.1) -- (2,0.1);
		\draw[black, line width=0.3mm] (-12.50,-0.1) -- (2,-0.1);
		\node (1) [ drop shadow={opacity=0.5, shadow xshift=0.1cm}, draw, rounded rectangle, fill=chptr, inner sep=8pt] {\textcolor{white}{\textbf{Chapter  \thechapter}}};
	\end{tikzpicture}
		
	\vskip 10\p@
	\begin{changemargin}{20pt}{20pt}
		\setstretch{1}
		{\fontsize{20}{30} \fontshape{\itdefault}\fontseries{\bfdefault}\selectfont \color{chptrnm1} {#1}\par\nobreak}
	\end{changemargin}
	\par\nobreak
	\vskip 20\p@
	\vskip 100\p@
	{\color{black}\hrule height 0.1ex \hfill}
}}

	\def\position{\centering}
	
	\def\@makeschapterhead#1{%
	\vspace*{10\p@}%
	{\parindent \z@ \position \reset@font
	     \vskip 20\p@
	     {\Huge \bfseries \color{chptrnm} \underline{\textbf{\textit{#1}}}\par\nobreak}
	     \vskip 20\p@
	}}
\usepackage{verbatim}

\newcommand{\mnras}{Mon. Not. Roy. Astron. Soc.}
\newcommand{\jcap}{J. Cosmol. Astropart. Phys.}

\newcommand{\prd}{Phys. Rev. D}
\newcommand{\prc}{Phys. Rev. C}

\newcommand{\ijmp}{Int. J. Mod. Phys.}

\makeatother

\begin{document}
\thispagestyle{empty}
\pagenumbering{roman}
\begin{center}
	\textsc{\bfseries {\Large Compact objects and dark matter dynamics in astroparticle physics}}
	\vskip 0.5cm
	{\large Thesis submitted for the Degree of Doctor of Philosophy (Science)\\
	\vskip 0.3cm
	In {\bf Physics}}
	\vskip 1.5cm
\end{center} 

\begin{figure}[hbt]
	\begin{center}
		\includegraphics[scale=0.5]{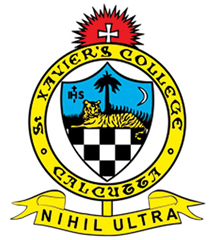}
	\end{center}
\end{figure}

\begin{center}
	\vskip 0.80cm
	{\bf {\em By}} 
	\vskip 0.2cm
	{\bf {\Large Ashadul Halder}}
	\vskip 0.8cm
	{\large Post Graduate  and  Research  Department}
	\vskip 0.1cm
	{\large \emph{of}}
	\vskip 0.1cm
	{\bf {\Large St. Xavier’s College (Autonomous), Kolkata}}
	\vskip 0.1cm
	{\Large Affiliated to the {\bf University of Calcutta}}
	\vskip 0.8cm
	{\bf {\Large 2021}}
	\vfill
\end{center}
\cleardoublepage

\cleardoublepage
\newpage
\cleardoublepage
\thispagestyle{empty}
\includegraphics[trim=60 100 60 50, clip, width=\linewidth]{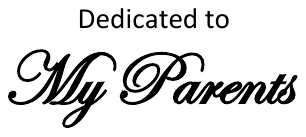}

\cleardoublepage
\vskip 1.0cm
\centerline{{\bf{\large ACKNOWLEDGMENTS}}}
\vspace{1cm}
\begin{changemargin}{40pt}{40pt}
	I would like to use this space to convey my sincere appreciation for all the assistance and support that I have received during the course of this work. First and foremost, I would like to show my gratitude to my parents, to whom this thesis is dedicated.
	
I would like to thank my esteemed supervisor {\bf Dr. Shibaji Banerjee}. Without his assistance, dedicated involvement in every step throughout the process, these papers would have never been accomplished. His insightful suggestions pushed me to sharpen my thinking, not only regarding my research related problems but in several aspects on the way of my life.
 
I am equally grateful to {\bf Prof. Debasish Majumdar}, whose expertise was invaluable in formulating the research works. I feel very fortunate to have the opportunity to have him as a teacher as well as a collaborator. Several inferences expressed in this thesis are the results of long discussions with him.

I extend my heartfelt gratitude to our respected Father Principal of St. Xavier's College (Autonomous), Kolkata {\bf Rev. Fr. (Dr.) Dominic Savio, S.J.} and our former respected Father Principal {\bf Rev. Fr. (Dr.) John Felix Raj, S.J.} for their earnest concerns towards the progress of my thesis work and for providing me with all the necessary facilities at St. Xavier's College (Autonomous), Kolkata for pursuation of my Ph.D. thesis work. I also remain grateful to {\bf Dr. Samrat Roy}, Ph.D. coordinator, for extending me valuable supports.

I am indebted to {\bf University Grants Commission (UGC), Govt. of India}, for the financial support in terms of my fellowship and other grants during the tenure of my Ph.D. 

I would like to convey my sincere gratitude to {\bf Prof. Sibaji Raha} and {\bf Prof. Sanjay K. Ghosh}. Their valuable guidance helps me a lot throughout my studies. I am really glad to have them as my collaborator. 

I am very much fortunate to get associated with our department faculties. Their suggestions and inspirations help me a lot during this period of my academic career. Specially, I want to thank {\bf Prof. Sailendra Narayan Roy Choudhury}, {\bf Dr. Suparna Roy Chowdhury}, {\bf Dr. Indranath Chaudhuri}, {\bf Dr. Tapati Dutta}, {\bf Dr. Soma Ghosh}, {\bf Dr. Sarbari Guha}. 

My thanks extend to my friends and colleagues {\bf Rupa Basu}, {\bf Siddhartha Bhattacharyya} for their precious suggestions about several topics including my works.

	\vskip 1.0cm
	\rightline{Ashadul Halder\hspace{0.9cm}}
\end{changemargin}

\newpage

\normalfont

\tableofcontents
\newpage
\listoffigures
\newpage
\listoftables

\mainmatter
\pagestyle{fancy}

\let\cleardoublepage\clearpage
\setstretch{1.5}
\chapter{Introduction}
	\\ \label{chp:intro}

	
	The recent developments of the theory and instrumental probes have opened up a wider range of investigations, previously unimagined of, and have substantially enhanced the scope of advancements in researchs and refinements of our understandings in several aspects of cosmology and high energy astroparticle physics. From Gravitational Wave to neutrino physics, compact stars to 21-cm cosmology, each probe has enriched our knowledge uniquely. Despite this seamless progress of modern science, the famous quote of Isaac Newton is still valid today.
	\begin{changemargin}{40pt}{40pt}
		\setstretch{1}
		{\myfont ``I do not know what I may appear to the world, but to myself I seem to have been only like a boy playing on the seashore, and diverting myself in now and then finding a smoother pebble or a prettier shell than ordinary, whilst the great ocean of truth lay all undiscovered before me."}
		\vspace{-0.5cm}
		\begin{flushright}
			{\bf \color{qtcol} -Isaac Newton.}
		\end{flushright}
	\end{changemargin}

	The extent of the Universe is essentially beyond our imagination. Whatever has been explored about our Universe, is indeed negligible in comparison to the ocean of the hidden mysteries of the Universe. The current cosmological paradigm tells that a tiny percentage of the entire energy budget of our Universe belongs to the ordinary baryonic matter ($\approx 4.9\%$) \cite{planck15}. The remaining unseen part is essentially distributed among the two dark sector components, namely Dark Energy ($\approx 68.5\%$) and Dark Matter ($\approx 26.6\%$). The characteristics of these dark components are still unknown.
	
	\section{Basics of Cosmology and Dark Energy}
		\begin{changemargin}{40pt}{40pt}
			\setstretch{1}
			{\myfont ``It happened five billion years age. That was when the Universe stopped slowing down and began to accelerate, experiencing a cosmic jerk."}
			\vspace{-0.1cm}
			\begin{flushright}
				{\bf \color{qtcol} -Adam Riess.}
			\end{flushright}
		\end{changemargin}
		
		\begin{figure}[b!]
			\centering{}
			\includegraphics[width=0.9\linewidth]{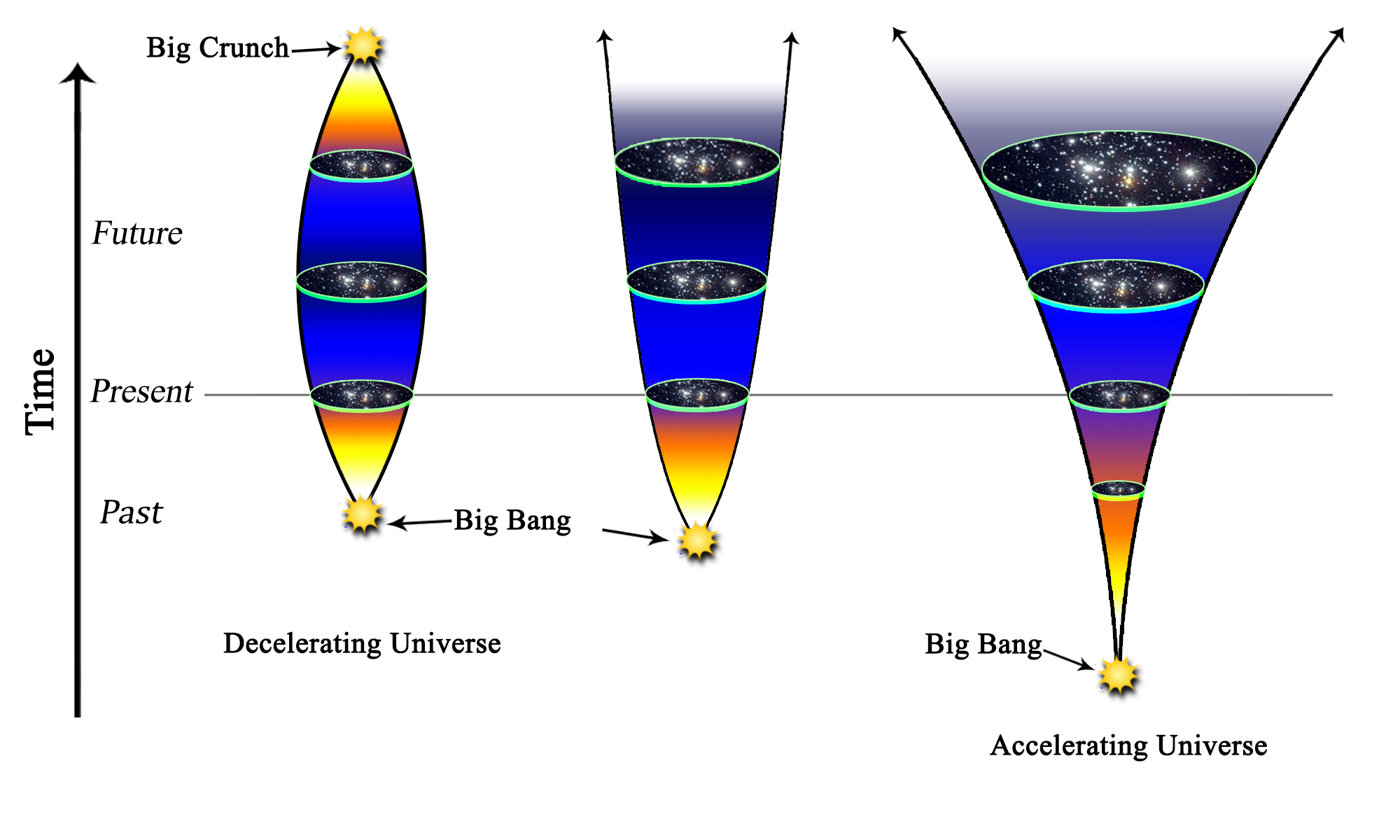}
			\caption{\label{fig:xpnsn} Different models of the expanding Universe.}
		\end{figure}
	
		The nature of dark energy is certainly one of the most interesting open questions in the field of astrophysics and cosmology. Even in the 90s decade, it was believed that the expansion rate of the Universe is gradually decreasing over time due to gravity. Consequently, the Universe may recollapse in the future if there exists a substantial amount of mass in the Universe or at least keeps expanding at a certain rate (first two models of Figure~\ref{fig:xpnsn}). But, at the end of the 20$^{\rm th}$ century (1998-1999), two groups of astronomers obtained a completely unexpected result, that is, {\bf\fontfamily{pzc}\selectfont \color{qtcol} \large``our Universe is not only expanding, but the expansion rate is continuously increasing over time''} \cite{Riess:1998cb,Perlmutter:1998np}. These two supernova-based observations were the first evidence of dark energy. Later, the detailed measurement of CMB (Cosmic Microwave Background) by Planck satellite-borne experiments \cite{planck15,planck18} in accordance with WMAP (Wilkinson Microwave Anisotropy Probe) \cite{wmap} and COBE (Cosmic Background Explorer) provides more precise information about this dark sector component. 
		
		In order to estimate the expansion rate as well as the evolution of the Universe theoretically, first we need to scale up the extent of the Universe by a dimensionless quantity, the cosmic scale factor, given by $a(t)=d(t)/d_0$. Here, the quantity $d_0$ is denoting a certain distance at the current time ($t_0$) while $d(t)$ representing same at the time $t$. According to the broadly accepted theory of Big Bang, everything begins after the very first cosmic event, the Big Bang. Around $\sim10^{-33}$ second after the Big Bang, everything suffered a very intense expansion during a very small period of time ($10^{-33}\sim 10^{-32}$ second). This mysterious cosmic phenomenon is known as Cosmological Inflation \cite{infl}. Following this rapid expansion, the Universe did not stop expanding but continues at a slower rate. Since, during this epoch, the radiation was the dominated component of the Universe (till the radiation-matter equality, at the age of $\sim 47000$ years of the Universe), the Universe evolved more likely a radiation-dominated single component Universe (characterized by the equation of state parameter $\omega=1/3$) \cite{Kolb:1990vq,liddle2015introduction}. Therefore, from the Friedmann equations, the scale factor $a(t)$ is related to the time $t$ as,
		\begin{equation}
		a(t)\propto t^{1/2}.
		\end{equation}
		After the radiation-matter equality, matter started dominating over radiation, because the matter density ($\rho_{\rm Mat}$) changes with scale factor $a(t)$ as $\rho_{\rm Mat}\propto a^{-3}$, while in the case of radiation dominated era, the density ($\rho_{\rm Rad}\propto a^{-4}$) \cite{liddle2015introduction}. During this epoch, the scale factor varies as, 
		\begin{equation}
		a(t)\propto t^{2/3}.
		\end{equation}
		We already discussed that the matter density is related with the scale factor as $\rho_{\rm Mat}\propto a^{-3}$, but in the case of dark energy, its density remains constant over time, as it is the property of space. As a consequence, dark energy has turned out to be the most abundant constituent of the Universe around $9.8\times 10^9$ years after the Big Bang. Therefore, the Universe starts expanding with acceleration and is continuing till the present epoch. Solving the Friedmann equation by incorporating the FRLW metric, the evolution of the scale factor for dark energy dominated Universe can be estimated as
		\begin{equation}
		a(t)\propto \exp \left(H_0 t\right).
		\label{eq:aDE}
		\end{equation}
		In the above expression, $H_0$ represents the Hubble constant at the current epoch ($H_0\approx 70$ km/s/Mpc). From Eq.~\ref{eq:aDE}, one can observe that the Universe is expanding exponentially, which means it not only expanding, the rate of the expansion is also increasing over time. In the word of famous astronomer Prof.~Chris Blake,
		\begin{changemargin}{40pt}{40pt}
			\setstretch{1}
			{\myfont ``The action of dark energy is as if you threw a ball up in the air, and it kept speeding upward into the sky faster and faster"}
			\vspace{-0.5cm}
			\begin{flushright}
				{\bf \color{qtcol} -Chris Blake}
			\end{flushright}
		\end{changemargin}
		
		\begin{figure}[h!]
			\centering{}
			\includegraphics[width=0.7\linewidth]{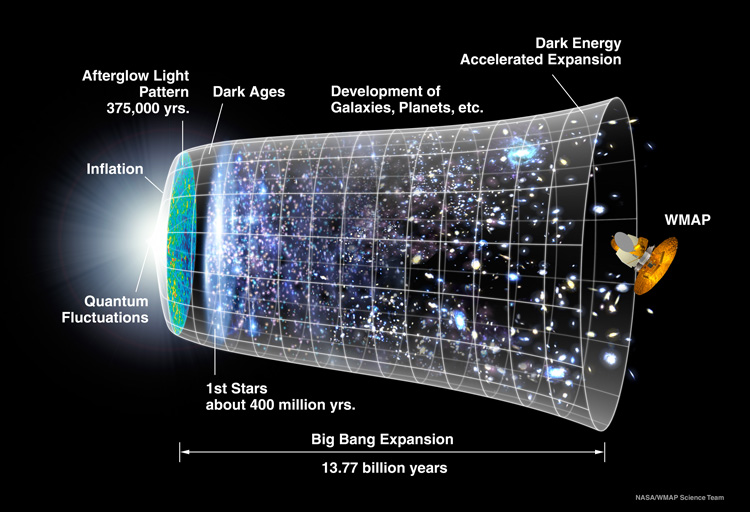}
			\caption{\label{fig:bb} The schematic diagram of the evolution and expansion of the Universe. (Photo credit: NASA / WMAP Science Team.)}
		\end{figure}
		
		At the current epoch, the Universe is expanding with acceleration, so one can conclude that the dominating component of our Universe is gravitationally repulsive, which is considered as Dark energy. So, the pressure ($p$) of the dark energy must be $< -\rho /3$, where $\rho$ denotes the dark energy density. The cosmological constant ($\Lambda$) is the simplest form of dark energy, which was introduced in the context of Einstein's theory of relativity in 1917 \cite{cosmo_const}. In this particular case, the density $\rho_{\Lambda}$ is considered to be a constant and is related to the pressure ($p_{\Lambda}$) as, $\rho_{\Lambda}=-p_{\Lambda}$. So, the equation of state parameter becomes $\omega=p_{\Lambda}/\rho_{\Lambda}=-1$. Quintessence \cite{qntsence} is another explanation of dark energy, which is basically a hypothetical scalar field \footnote{Sometimes it is considered as the fifth fundamental force.}. In this case, the equation of state parameter ($\omega_q$) is not static but depends on the scalar field $Q$ and the corresponding potential $V(Q)$, given by
		\begin{equation}
		\omega_q=\dfrac{p}{\rho}=\dfrac{\dot{Q}^2-2 V(Q)}{\dot{Q}^2+2 V(Q)}.
		\end{equation}
		In the above expression, $\dot{Q}$ represents the variation of the scalar field with time.
		
		Rather introducing any new form of energy, the gravitational field can be precisely modified to explain the accelerated cosmic expansion \cite{Jain_2010,RevModPhys.93.015003}. But, till the present time, there is no modified gravity theory that can explain the observed cosmic acceleration properly. At the end of the day, we do not have sufficient information about this major part of the Universe. Either they are merely the property of space or a mysterious dynamical fluid or a completely new theory of gravity.
		
		\begin{changemargin}{40pt}{40pt}
			\setstretch{1}
			{\myfont ``Dark energy is not only terribly important for astronomy, it's the central problem for physics. It's been the bone in our throat for a long time."}
			\vspace{-0.5cm}
			\begin{flushright}
				{\bf \color{qtcol} -Steven Weinberg}
			\end{flushright}
		\end{changemargin}
	
	\section{Dark Matter}
		Dark matter is another mystery of the Universe, which occupies $\sim 84.34\%$ ($\sim26.57\%$ of the total content of the Universe) of the total matter budget according to the recent observational result \cite{planck15,planck18}. The remaining $\sim 15.66\%$ exists in the form of visible baryonic matter. Whatever we can see in the sky i.e. stars, planets, nebula, etc. belong to this tiny part ($\sim 15\%$) of the matter, while the major part (dark matter) is still unknown. 
		
		In the year 1884, famous British mathematical physicist Lord Kelvin commented in a conference talk \cite{kelvin1904baltimore},
		\begin{changemargin}{40pt}{40pt}
			\setstretch{1}
			{\myfont ``It is nevertheless probable that there may be as many as $10^9$ stars [within a sphere of radius $3.09\:\cdot\:10^{16}$ kilometres] but many of them may be extinct and dark, and nine-tenths of them though not all dark may be not bright enough to be seen by us at their actual distances. [...] Many of our stars, perhaps a great majority of them, may be dark bodies."}
			\vspace{-0.5cm}
			\begin{flushright}
				{\bf \color{qtcol} -Lord Kelvin.}
			\end{flushright}
		\end{changemargin}
		That was the first powerful theoretical investigation for the existence of dark matter in the Milky Way galaxy. Later, Henri Poincar\'{e} termed those dark bodies as ``dark matter" or ``mati\`{e}re obscure" (in french) in his book ``The Milky Way and Theory of Gases" (1906) \cite{hpoincare}. After that, several observational evidences have been detected, that justify the conjecture of dark matter \cite{1932BAN.....6..249O,1933AcHPh...6..110Z}.
		\subsection{Evidences for Dark Matter}
			The idea of dark matter is well motivated in the framework of cosmology and astrophysics. Several astrophysical observations favour the existence of dark matter, which are essentially based on the gravitational effects. In this section, we summarize such few most compelling pieces of evidence of dark matter. 
		
		\subsection*{$\bullet$ Rotation of Spiral Galaxies}
			\begin{figure}[h!]
				\centering{}
				\includegraphics[width=0.6\linewidth]{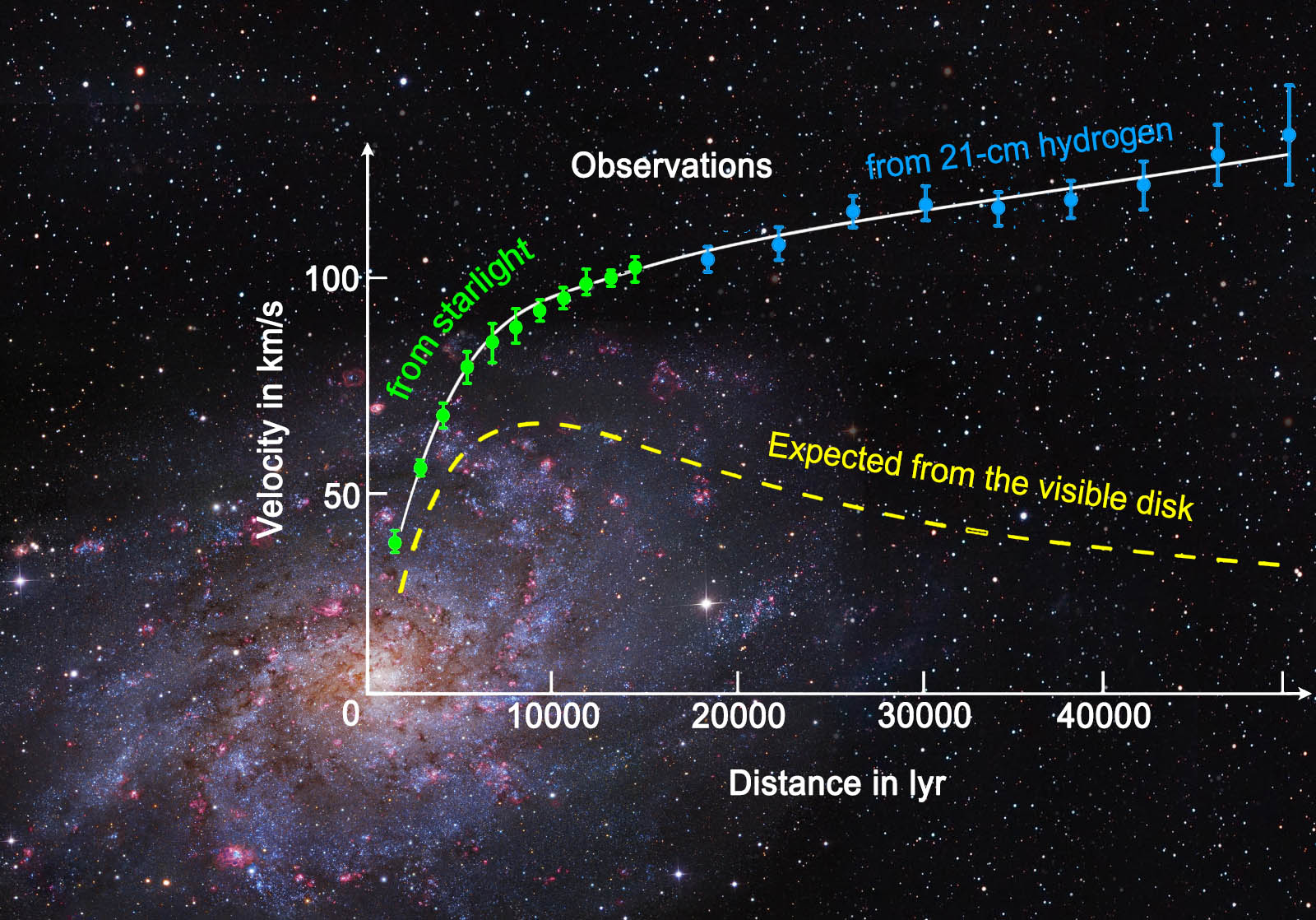}
				\caption{\label{fig:g_rot}Galaxy rotation curve of Messier 33. The yellow dashed line is the expected line from the visible disk, while the solid line represents the fitted line of observed points \cite{g_rot_plt}.}
			\end{figure}
			Flattening of the galactic rotation curve of the spiral galaxies is one of the most promising evidence for the existence of galactic dark matter. In the year 1975, famous American astronomer Vera Florence Cooper Rubin brought into light this sensational fact with other two astronomers W. Kent Ford Jr. and Norbert Thonnard \cite{rubin}. They investigated the rotational curve of several spiral galaxies and obtained remarkable discrepancies with the calculated curves as estimated by incorporating only the visible part of the galaxies. After their extensive studies, the availability of galactic dark matter is universally accepted. Although, in 1959, Louise Volders also pointed out identical discrepancies beforehand in the case of M33 galaxy (see Figure~\ref{fig:g_rot}).
			
\begin{figure}
				\centering{}
				\begin{tabular}{ccc}
					\includegraphics[height=5.5cm, width=6cm]{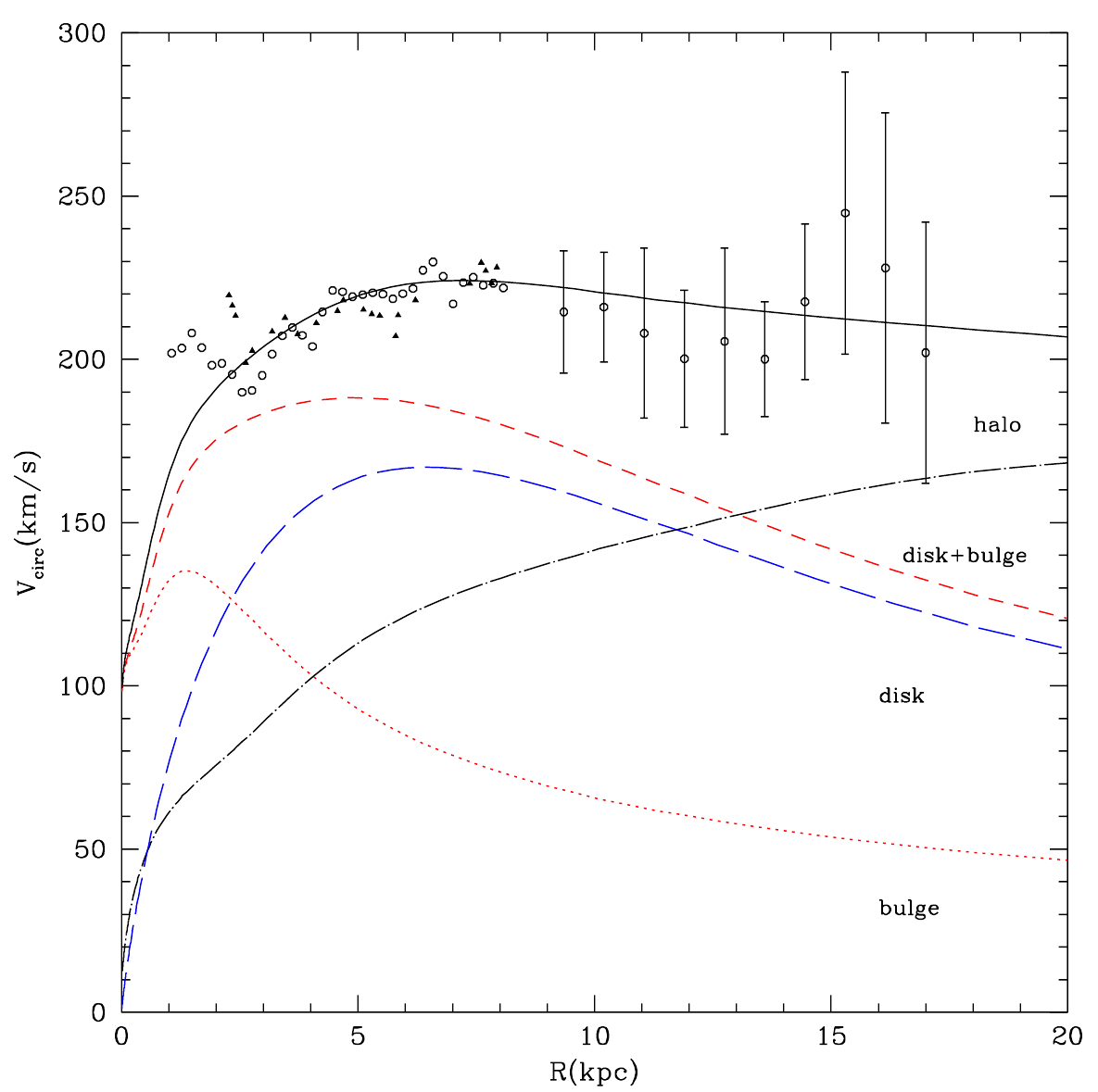}&&
					\includegraphics[height=5.5cm, width=6cm]{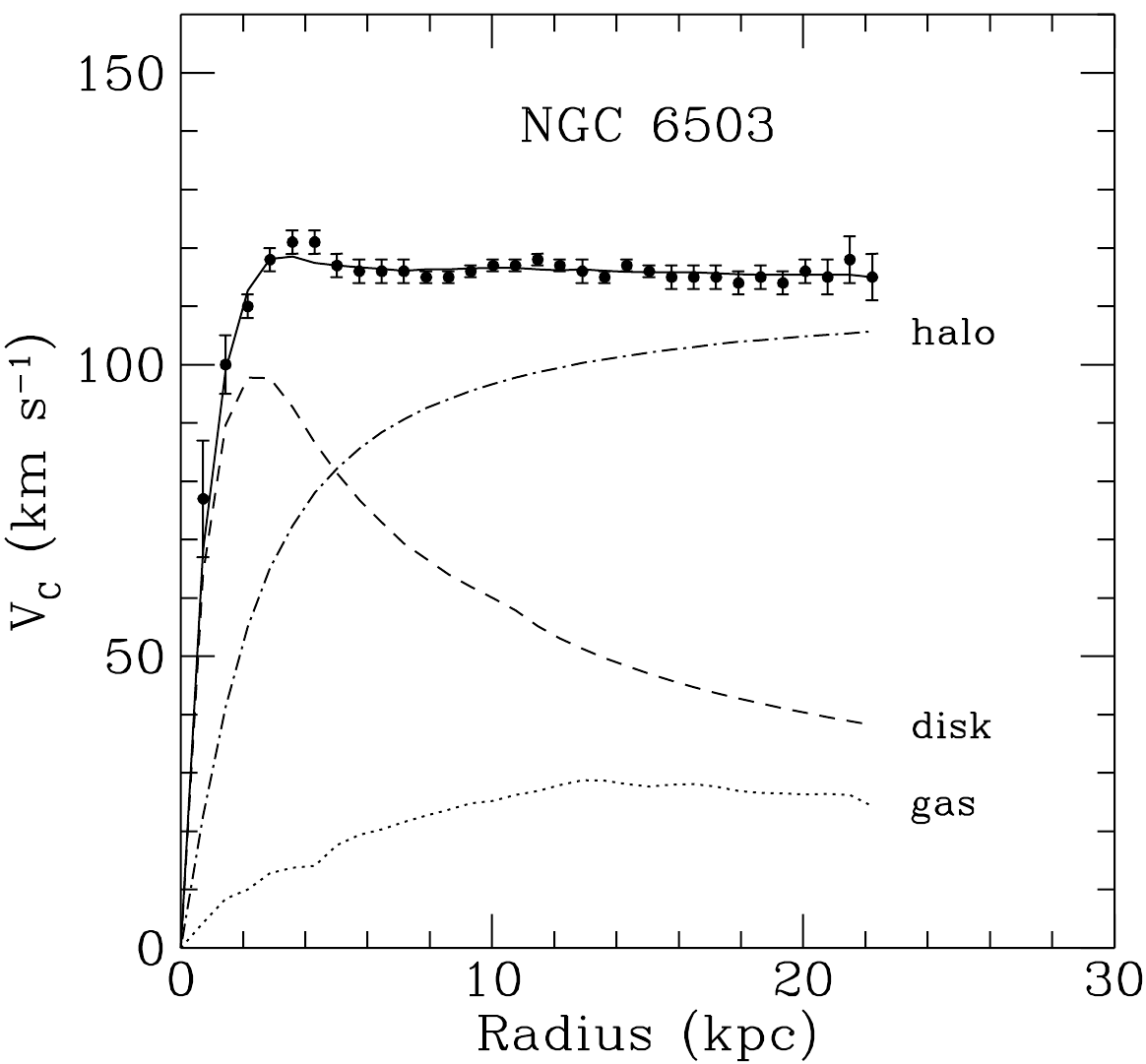}\\
					(a)&&(b)\\
				\end{tabular}
				\caption{\label{fig:g_rot_2} Galactic rotation curve of (a) Milky Way galaxy (Figure from Ref.~\cite{Klypin}), (b) NGC 6503 galaxy (Figure from Ref.~\cite{Kamionkowski, Begeman}.)}
			\end{figure}

			The rotation curve describes the variation of the circular velocity of stars and Interstellar Medium (ISM) with radial distance from the center of the spiral galaxy. The velocity can be estimated by measuring the Doppler shift of the hydrogen spectra. Tracking the motion of stars around the galactic centre (GC) is another significant method in measuring the circular velocity at a particular radial distance from GC. According to the classical theory of gravity, the stars revolve around the galactic centre, keeping the centrifugal force equals to the gravitation force given by
			\begin{eqnarray}
			&\dfrac{m_{*} v(r)^2}{r}=G\dfrac{M(r) m_{*}}{r^2} \nonumber\\
			or,&v(r)=\sqrt{\dfrac{G M(r)}{r}}.
			\label{eq:gal_rot_newt}
			\end{eqnarray}
			In the above equations, $m_{*}$ denotes the mass of the star, orbiting the galactic centre with radius $r$ and $M(r)$ is the mass containing the spherical region of radius $r$ around the galactic centre given by $M(r)=\displaystyle \int_{\rm sphere}\rho {\rm d}V$, where $V$ is the volume and $\rho$ is the average density of the central bulge of the galaxy. From Eq.~\ref{eq:gal_rot_newt}, it can be estimated that, the circular velocity $v(r)\propto r$ within the central bulge. On the other hand, outside the central bulge of the galaxy, velocity varies with $r$ as $v(r)\propto 1/\sqrt{r}$ (the yellow dashed line in Figure~\ref{fig:g_rot}) as the almost entire mass of a spiral galaxy contains within the central bulge. However, the observational evidence indicates that $v(r)\propto r^0$ for higher values of $r$ (see Figure~\ref{fig:g_rot_2}a \cite{Klypin}, Figure~\ref{fig:g_rot_2}b \cite{Kamionkowski, Begeman} and Figure~\ref{fig:g_rot} \cite{g_rot_plt}). As a consequence, there must exist an additional invisible matter distribution, which induces this nature of the rotation curve. This invisible (dark) matter is believed to exist in the form of ``dark matter halo". The density profile of galactic dark matter halo can be estimated by accommodating an approximately flat rotational curve, as obtained from observational evidence. There are several popular galactic DM halo density profiles those are discussed later in the chapter (see \Autoref{tab:profile}).
		
		\subsection*{$\bullet$ Cosmic Microwave Background}
			\begin{figure}[h!]			
				\centering
				\includegraphics[width=0.7\linewidth]{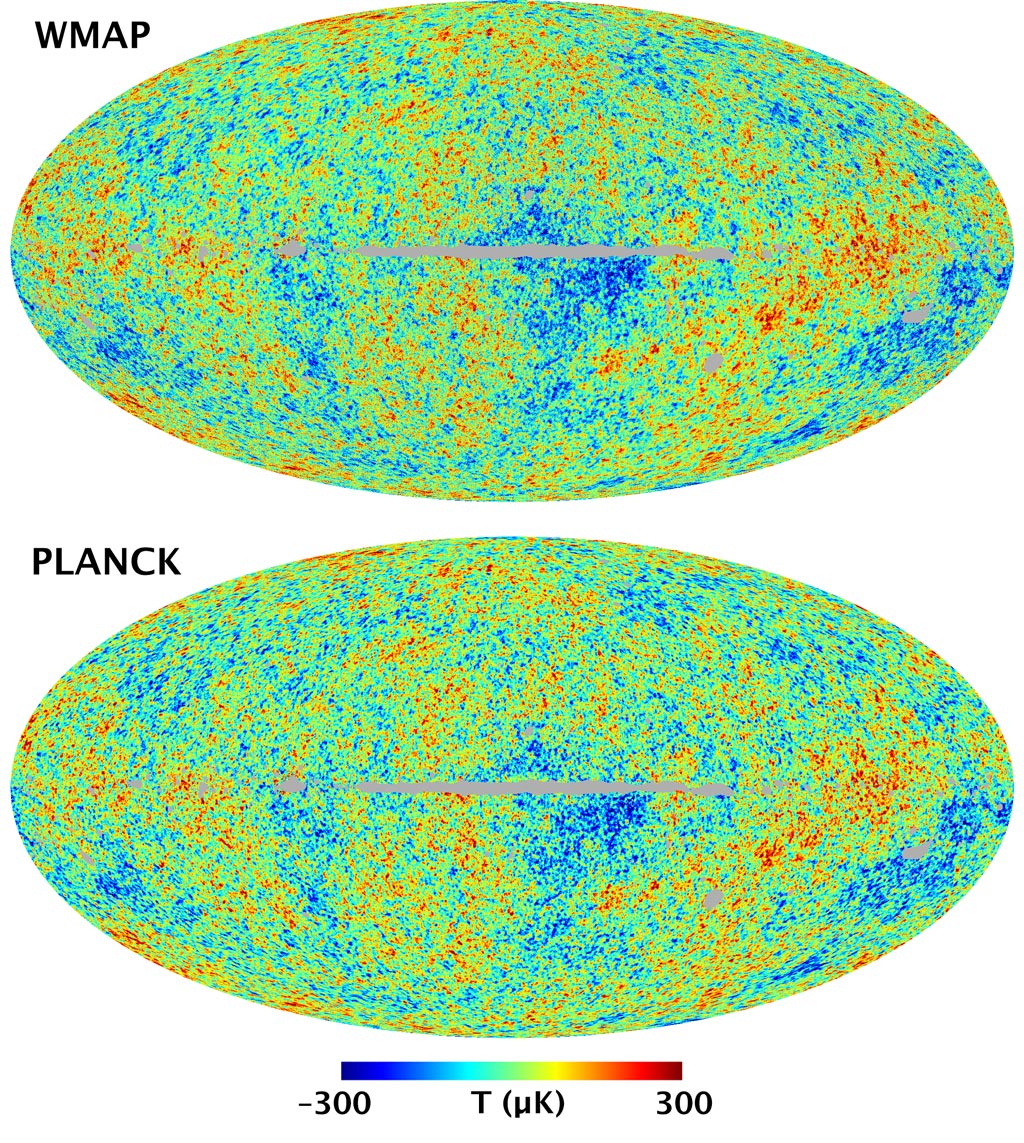}
				\caption{\label{fig:CMB}The all-sky cosmic microwave background radiation map from WMAP and Planck data. Both images show a temperature fluctuations of $\pm300\,\,\mu$K around the average temperature 2.73 K. The top image represents the WMAP W-band CMB map while the bottom image is the Planck SMICA CMB map. (Photo credit: NASA / WMAP Science Team.)}
			\end{figure}
			The cosmic microwave background radiation (CMB, CMBR) is the most ancient cosmic glow, which contains several important information regarding the early epoch of the Universe. This ancient radiation provides the most precise measurement of the abundance of baryonic matter and dark matter. American radio astronomers Arno Penzias and Robert Wilson discovered this cosmic radiation in 1965 \cite{cmbdis} and achieved the Nobel prize as a consequence of their discovery (1978). However, Gamow predicted the same beforehand in 1948 \cite{gmw1,gmw2}.
					
			\begin{figure}[h!]			
				\centering
				\includegraphics[width=0.7\linewidth]{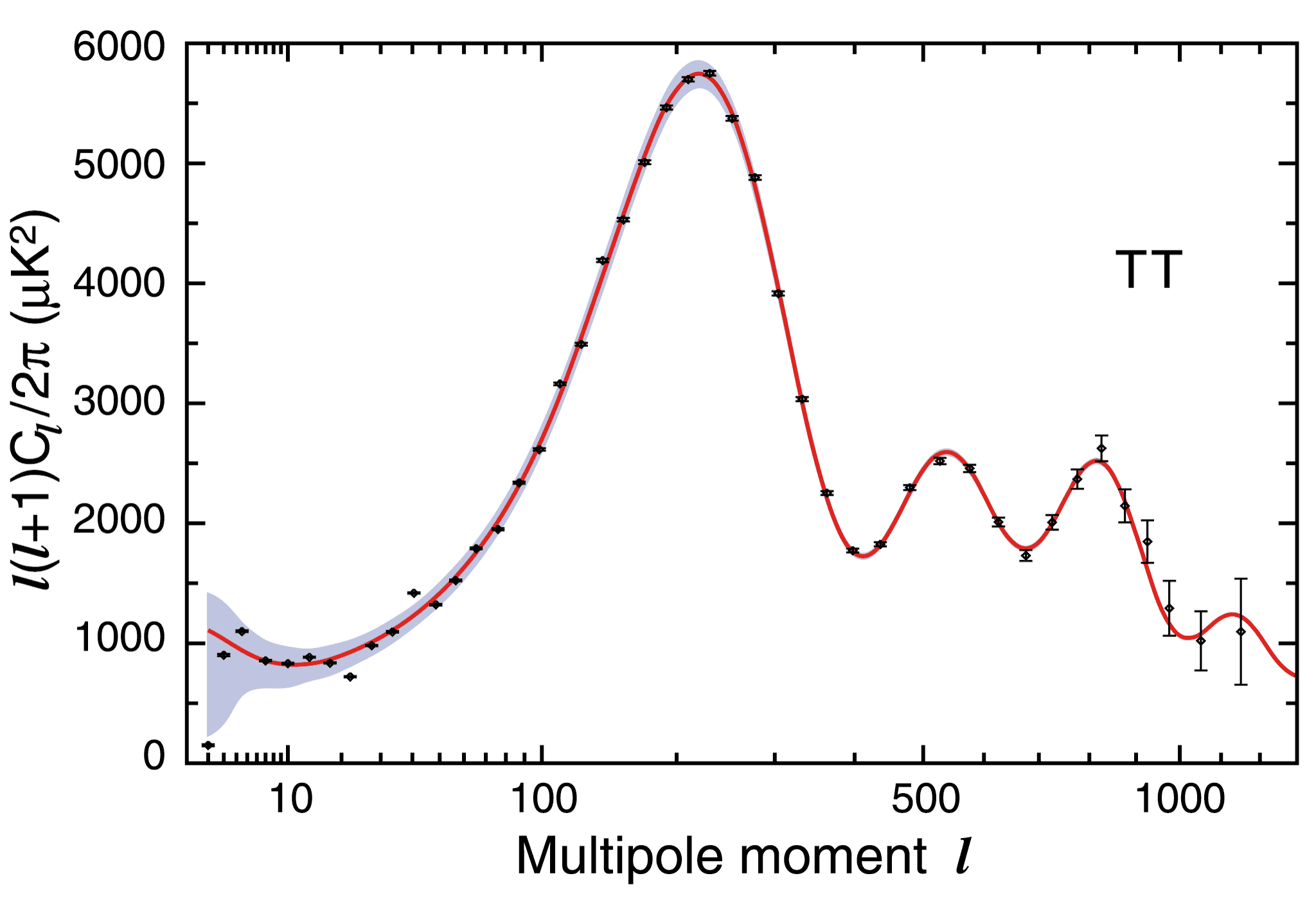}
				\caption{\label{fig:cmbans}The TT (temperature-temperature) angular power spectrum from nine-year WMAP data. The WMAP data are represented by black points and corresponding error bars. The red line describing the best fit model. The smoothed binned spectral function is shown using the gray region. (Photo credit: NASA / WMAP Science Team. \cite{wmap})}
			\end{figure}
		
			The CMB is basically the relic of electromagnetic radiation when the radiation was decoupled from matter ($\sim 380000$ years after the Big Bang). At the time, CMB was mostly visible, but due to the expansion of the Universe, it has been redshifted toward a longer wavelength over time and eventually, it appears in the microwave band at the current epoch (peak at $\approx2$ mm). At present, the CMB spectrum almost perfectly fits with the black body radiation of 2.73 K.
			 
			The faint glow of the CMB is distributed almost uniformly in the entire sky. However, a very tiny fluctuation can be observed on both sides of the peak value of the spectrum (see Figure~\ref{fig:CMB}). The observational results from the satellite-borne experiments (Planck \cite{planck15,planck18} and WMAP (Wilkinson Microwave Anisotropy Probe) \cite{wmap}) measure the spectral anisotropies of the CMB. These anisotropies can be expressed in terms of spherical coordinates ($\theta$, $\phi$) given by,
			\begin{equation}
				\dfrac{\delta T}{T}(\theta, \phi) = \sum_{l=2}^{\infty}\sum_{m=-l}^{l} a_{lm}Y_{lm}(\theta, \phi).
			\end{equation}
			The variance of the term $a_{lm}Y_{lm}(\theta, \phi)$ can be estimated as
			\begin{equation}
				C_l\equiv\langle \left|a_{lm}^2\right|\rangle\equiv\dfrac{1}{1l+1}\sum_{m=-l}^{l}\left|a_{lm}\right|^2.
			\end{equation}
			
			In Figure~\ref{fig:cmbans}, the variation of $C_l$ (in the form of $l(l+1)C_l/2\pi$) with multipole moment $l$ is shown, where the red line denotes the best-fitted spectra and the black error bars are representing the nine-year WMAP data points \cite{wmap}. The shape of the power spectrum depends on the oscillations of the primordial hot gas, while the amplitude and frequency are determined by its composition. As a consequence, several significant information can be explored from this demonstrative graph (Figure~\ref{fig:cmbans}), regarding the constituents of the Universe. For example, the height ratio of the first and second peak measures the abundance of baryon (not the baryonic dark matter), while the position of the first peak tells about the curvature of the Universe. On the other hand, the third peak corresponds to several significant aspects of dark matter.
		
		\subsection*{$\bullet$ Large Scale Structure of the Universe}
			
			\begin{changemargin}{45pt}{45pt}
				\setstretch{1}
				{\myfont ``I sometimes think that the universe is a machine designed for the perpetual astonishment of astronomers."}
				\vspace{-0.5cm}
				\begin{flushright}
					{\bf \color{qtcol} -Arthur C. Clarke.}
				\end{flushright}
			\end{changemargin}
			
			\begin{figure}[h!]			
				\centering
				\includegraphics[width=0.6\linewidth]{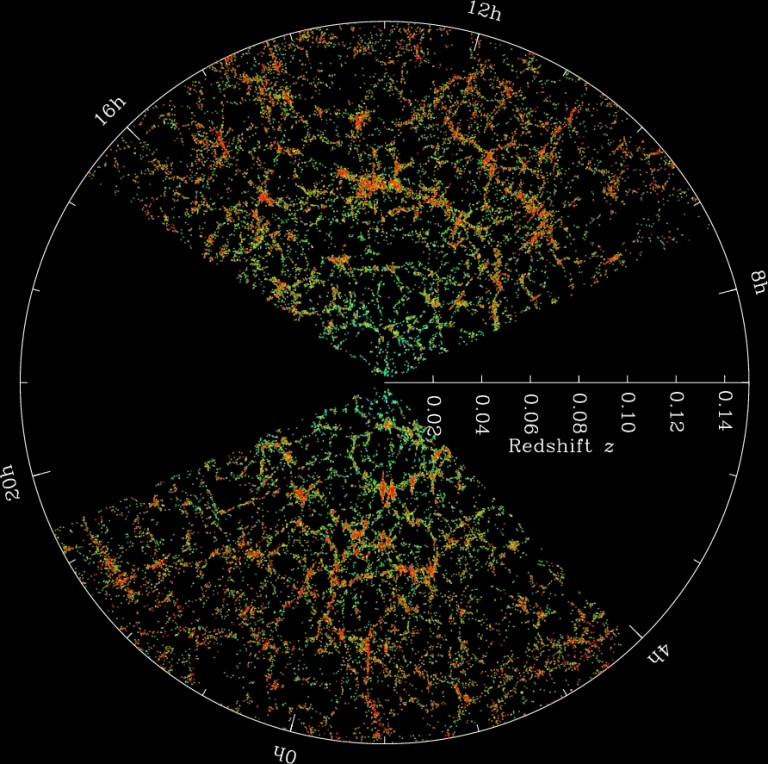}
				\caption{\label{fig:sdss} The map of galaxies discovered by Sloan Digital Sky Survey (SDSS) showing the large-scale structure of the Universe. The radial distance from the centre of the circle representing the redshifts of the corresponding galaxies. (Photo credit: M. Blanton and SDSS)}
			\end{figure}
		
			Although the cosmological principle tells that, {\bf\fontfamily{pzc}\selectfont \color{qtcol} ``the Universe is homogeneous and isotropic''}, the Universe is full of patterns of stars, galaxies, and clusters. In addition, those galaxies and clusters are not scattered uniformly in the Universe, but most of them are found in groups and form large-scale sheets, voids and galactic filaments (see Figure~\ref{fig:sdss}). Galactic filaments are considered to be the largest known structures in the Universe. The large-scale structures are assumed to be evolved from an almost homogeneous matter distribution of matters with small primordial density fluctuations. Later those small fluctuations are amplified under the influence of the gravitational field. As a consequence, these large-scale patterns of the galactic filaments may carry the footprints of gravitational clustering on the Universe. Cosmological N-body simulations (Millennium \cite{msim}, Aquarius \cite{NBsim} etc.) suggests that the large-scale structure could be formed only if a significant amount of dark matter is present in the Universe.
			
			The pattern and amount of fluctuation of the structure indicate the total mass containing the structure as well as the nature of matter. From recent investigations, it is estimated that the density parameter \footnote{Density parameter $\Omega=\rho/\rho_c$, where $\rho_c$ is the critical density of the Universe.} of matter (both visible and dark matter) is $\Omega_m \approx0.29$. If all the dark matters containing the Universe belong to the hot dark matters category, the galactic filaments would be smoother, as the relativistic hot dark matters erase the small-scale structures of matter as an outcome of baryon-dark matter collisions. On the other hand, cold dark matters are unable to reduce small structures due to low velocity. The cold dark matter influences the formation of small clumps of matter, which subsequently grow into larger structures (for example, galaxies, local groups, clusters etc.). As a consequence, one can conclude from the observational data that, both hot dark matter and cold dark matter contribute significantly to the structure formation of the Universe. However, it is also predicted that a large fraction of dark matter containing the Universe must belong to the cold dark matter category, in order to form such large-scale structures as observed by SDSS (the Sloan Digital Sky Survey) \cite{sdss} and 2dFGRS (the 2-degree Field Galaxy Redshift Survey) \cite{2dFGRS,2dFGRS1}.

		\subsection*{$\bullet$ Big Bang Nucleosynthesis}
			The theory of Big Bang cosmology is the prevailing dynamical model of our expanding Universe. According to this cosmological theory, initially, everything in the Universe was extremely dense and hot. But just few minutes after the Big Bang, the temperature drops to $\sim 100$ KeV. During this time (temperature $\sim100$ KeV), the protons ($p$) and neutrons ($n$) are fused together and form nuclei of few light elements. Helium ($^4$He) is the most abundant element among them (mass fraction $\sim1/4$). Moreover, a small amount of Deuterium ($d$), Tritium ($^3$H), $^6$Li, $^7$Li, $^7$Be nuclei are also produced during this process.
			\begin{center}
				\begin{tabular}{ c c c }
					$n+p$ & $\rightarrow$ & $d+\gamma$ \\ 
					$p+d$ & $\rightarrow$ & $^3$He + $\gamma$ \\  
					$d+d$ & $\rightarrow$ & $^3$He + $n$ \\ 
					$n+^3$He & $\rightarrow$ & $^4$He + $\gamma$ \\ 
					$d+^3$He & $\rightarrow$ & $^4$He + $p$   
				\end{tabular}
			\end{center}

			This mechanism of nuclei synthesis is termed ``Big Bang Nucleosynthesis'' (BBN). The baryon-to-photon ratio ($\eta$) is the only free parameter in the BBN model, which is proportional to the baryon abundance of the Universe ($\Omega_b$). \footnote{$\eta\equiv\Omega_b h^2$, where $h$ is the Hubble parameter given by, $h=H_0/(100\,{\rm km\,s^{-1}\,Mpc{-1}})$} 
			
			\begin{figure}[h!]			
				\centering
				\includegraphics[width=0.7\linewidth]{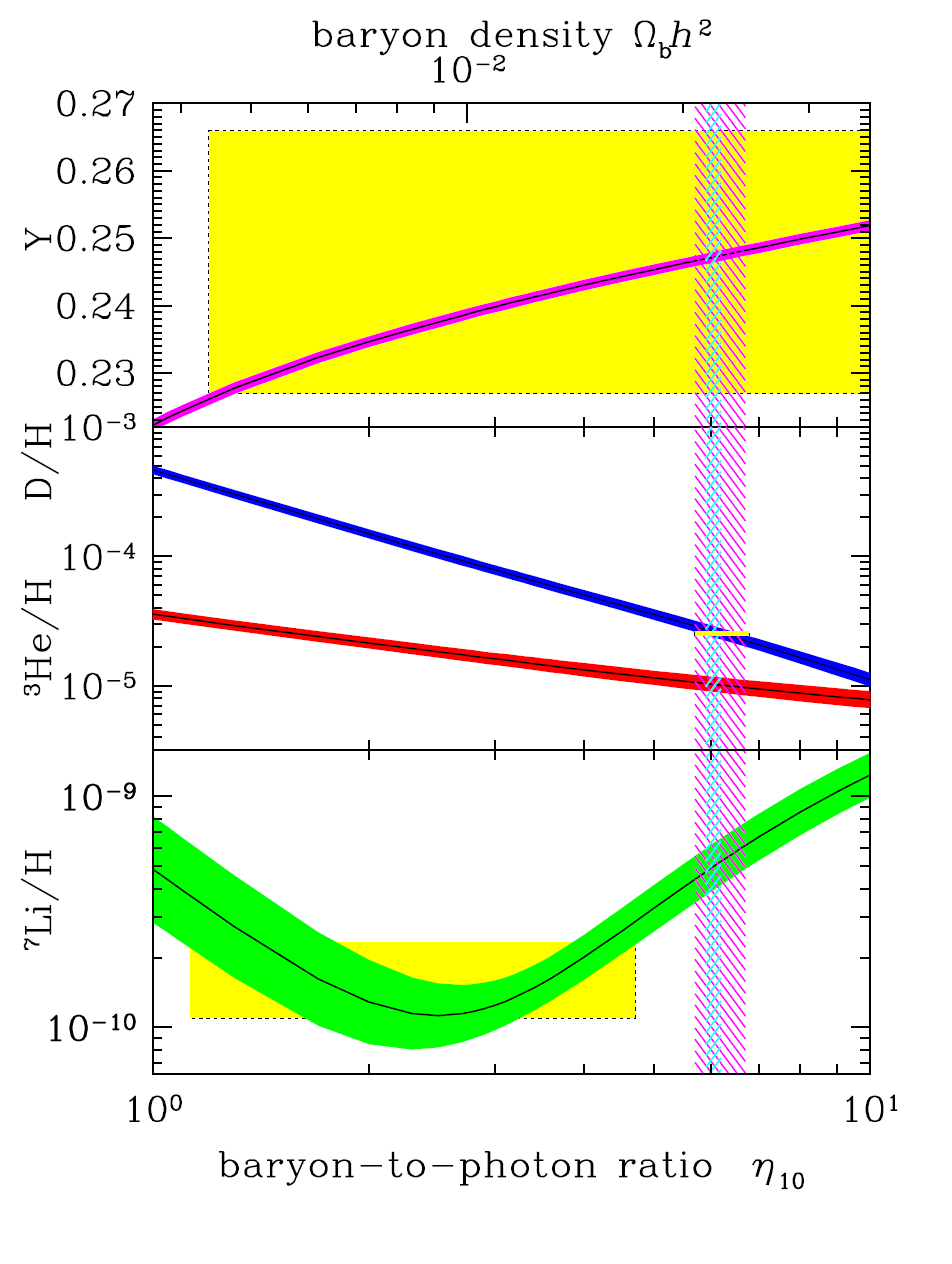}
				\caption{\label{fig:bbn} The predicted abundances of $^2$H, $^3$He, $^4$He, $^7$Li according to the standard model of Big Bang Nucleosynthesis (BBN). The bands denote the range of 95\% confidence level (CL) for different values of ($\eta$). The yellow boxes in this plot represent the range of the observed abundances for those light elements. The pink (wide) vertical band denotes the BBN concordance range, while the blue (narrow) band indicates the CMB measure of the cosmic baryon density (From Ref.~\cite{bbn}).}
			\end{figure}
			
			As the Universe expands, the rate of nucleosynthesis slows down and eventually stops within few minutes after the big bang, as free neutrons are highly unstable (lifetime $\sim$ 15 minute). So, the universal abundances of $^2$H, $^3$He, $^4$He, $^7$li saturates during this time period. However, after the cosmic dark age, the abundance of those light elements gets perturbed as an outcome of stellar nucleosynthesis. In addition, several heavy elements (such as C, N, O, Fe) are also generated in this stellar process. Consequently, the astrophysical sites with low metallicities are ideal to investigate the primordial abundances of such light elements.
			
			Several theoretical and experimental investigations indicates the primordial abundances of different baryonic elements as a function of $\eta$ (see Figure~\ref{fig:bbn}) \cite{Coc_2004,Iocco_2009}. The predicted data of nucleosynthesis suggest that, for $\eta\approx6\times10^{-10}$, the calculated baryonic density is 0.04 times the critical density of the Universe \cite{Dunkley_2009}. But according to the evidence of large-scale structure formation, the matter density parameter is remarkably higher ($\Omega_m\approx 0.29$) than that. Therefore, the remaining part of matter must exist in the form of dark matter. Moreover, from this estimation, one can conclude that the major part of the dark matter is non-baryonic, which does not interact with the electromagnetic field. However, there may exist a small amount of baryonic dark matter in the form of low luminous astrophysical bodies (discussed in section~\ref{subsec:dm_nat}).

		\subsection*{$\bullet$ Motion of the Galaxies (Coma and Virgo Cluster)}
		
			The relation between the gravitational potential and kinetic energy plays an important role in the investigation of the existence of dark matter in the scale of galaxies and galaxy clusters. The Swiss astronomer Fritz Zwicky tried to estimate the dynamical masses of galaxies, by measuring the velocity dispersion of those in the clusters using their Doppler shifts \cite{1933AcHPh...6..110Z,1937ApJ....86..217Z}. In this analysis, he applied the virial theorem for the system in the Coma cluster. The virial theorem provides a simple relationship between the gravitational potential energy and kinetic energy of the gravitationally bounded system. If the  Hamiltonian of a non-relativistic and interacting system can be written as
			
			\begin{equation}
				H=\sum_i^n \dfrac{p_i^2}{2 m_i}+V(r_i),
			\end{equation}
			then the according to the virial theorem for `$i^{\rm th}$' particle, we obtain,
			\begin{equation}
				\left< \dfrac{p_i^2}{2 m_i} \right> = \left< \dfrac{\delta V(r_i)}{\delta r_i}. r_i \right>,
				\label{eq:vir_1}
			\end{equation}
			where, $r_i$, $m_i$, $v_i$, $V_i$ and $p_i$ are the position, mass, velocity, potential energy and momentum of the $i^{\rm th}$ particle respectively, and the $\left<.....\right>$ denotes the average over time. So, from the above equation (Eq.~\ref{eq:vir_1}) one can write,
			\begin{equation}
				2T+U=0,
				\label{eq:vir}
			\end{equation}
			where, $T$ and $U$ are the kinetic energy and potential energy respectively.
			
			Assuming spherically symmetric distribution of galaxies, the total gravitational potential of the entire galaxy cluster (which is considered to be a self gravitating system) of mass $M$ and radius $R$ (that containing the galaxies) is given by
			\begin{equation}
				U = -\dfrac{3}{5} \dfrac{G M^2}{R}.
				\label{eq:int_U}
			\end{equation}
			Now each galaxies of the cluster may have certain velocities. So, the total kinetic energy of the galaxy cluster can be written as, 
			\begin{equation}
				T=\dfrac{1}{2}M \left<v^2\right>,
				\label{eq:int_T}
			\end{equation}
			where, $\left<v^2\right>$ is the average velocity of the galaxies inside the cluster. Now from Eq.~\ref{eq:int_U} and Eq.~\ref{eq:int_T}, one can estimate the entire mass of the galaxy cluster, which is essentially depends on the radius $R$ and the average velocity $\sqrt{\left<v^2\right>}$. In the case of Coma Cluster, Zwicky estimated the average velocity from the seven galaxies using Doppler shift \cite{1933AcHPh...6..110Z,1937ApJ....86..217Z}, which provide the mass of the Coma Cluster $M\approx 1.9\times 10^{13}$ M$_{\odot}$. However, Zwicky obtained the mass $M\approx8.0 \times 10^{11}$ M$_{\odot}$ for the Coma Cluster, from the assumption of the stellar populations in the galaxy, which is $\sim400$ times smaller that the previous estimation. Although, from recent observation and methods, one can predicts the mass of the Come Cluster more precisely, which gives $\approxeq 1.6 \times 10^{14}$ M$_{\odot}$ for hubble parameter $h=0.673\pm0.012$, which is more massive that the previous estimation. 
			
			In 1936, Sinclair Smith performed a similar investigation of the mass of Virgo cluster \cite{1936ApJ....83...23S}, which contains several elliptical and lenticular galaxies. But, Smith also came to the similar conclusions as addressed by Zwicky \cite{1933AcHPh...6..110Z,1937ApJ....86..217Z}. As a consequence, one can inferred that these exists a huge amount of invisible mass distribution in those galaxy cluster (Coma Cluster and Virgo Cluster), which are possibly dark matter candidates.
			
			\begin{figure}
				\centering{}
				\begin{tabular}{c}
					\includegraphics[width=0.6\linewidth]{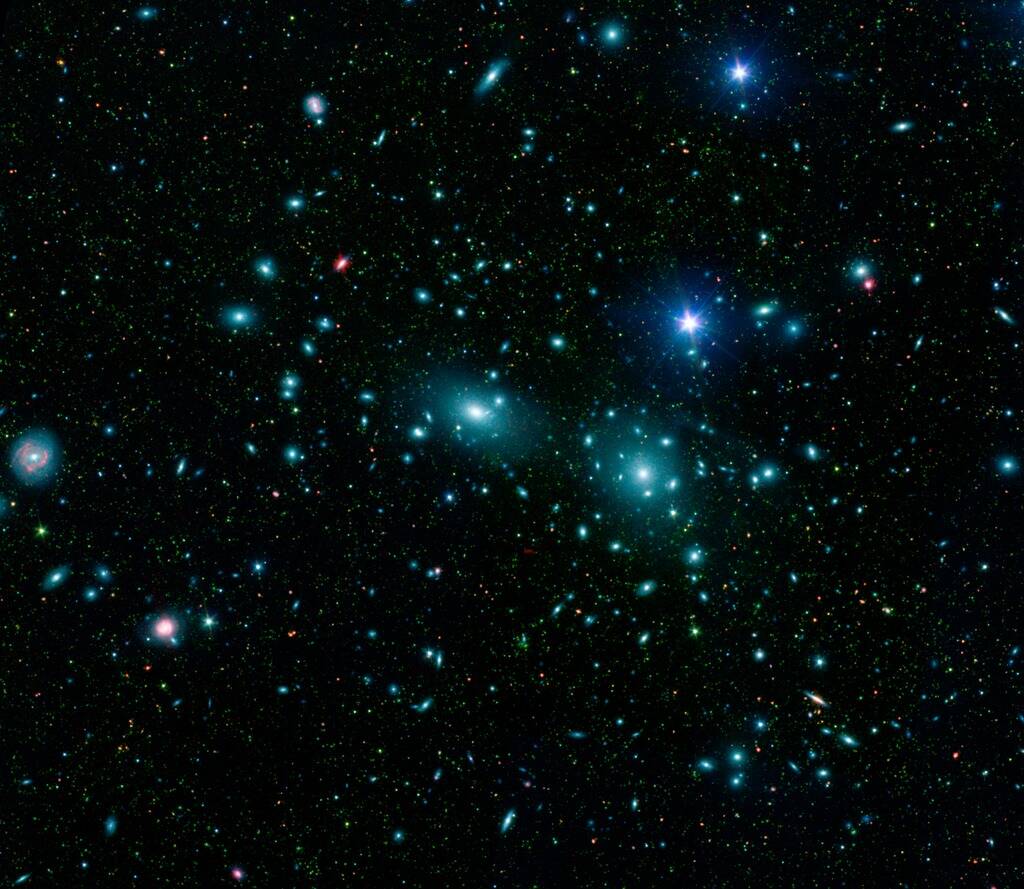}\\
					(a)\\
					\\
					\\
					\includegraphics[width=0.6\linewidth]{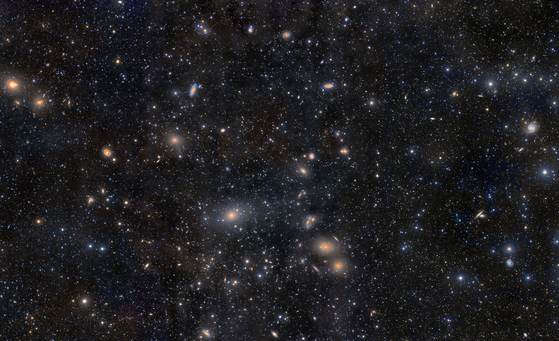}\\
					(b)\\
				\end{tabular}
				\caption{\label{fig:coma_virgo} (a) Galaxies inside the Coma Cluster (Photo credit: Credit: NASA/JPL-Caltech/L. Jenkins (GSFC)). \\(b) Virgo Cluster (Photo credit: and Copyright: Rogelio Bernal Andreo,\\ deepskycolors.com).}
			\end{figure}
		
		\subsection*{$\bullet$ Gravitational Lensing}
			\begin{figure}[h!]			
				\centering
				\includegraphics[width=0.7\linewidth]{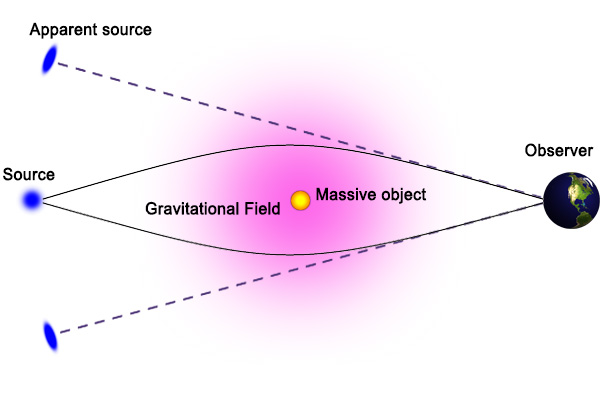}
				\caption{\label{fig:g_lens}Bending of incoming light due to the gravitational field of a massive object. The dashed lines show the apparent position of the light source.}
			\end{figure}
		
			
			The notion of the general theory of relativity tells that, the light follows curved trajectories as described by the geometry of space in the vicinity of massive astrophysical objects (see Figure~\ref{fig:g_lens}). As a result, the observer may see the distorted and multiple images of the light source, which is actually situated beyond the massive lensing astrophysical body. This phenomenon is known as the gravitational lensing of light. The concept of lensing by a massive astrophysical object was first proposed by Henry Cavendish (1784) in his unpublished article. Later German physicist, astronomer Johann Georg von Soldner addressed the same in the framework of Newtonian gravity (1804) \cite{gl_2nd}. Eventually, in 1915, Albert Einstein successfully calculated this phenomenon in the context of his famous theory of relativity. On May 29$\rm{^{th}}$ May 1919, A. Eddington and F. Dyson confirmed Einstein's prediction using their observational consequences. It was the first observational evidence of gravitational lensing.
			
\begin{figure}
				\centering{}
				\begin{tabular}{ccc}
					\includegraphics[height=6cm, width=6cm]{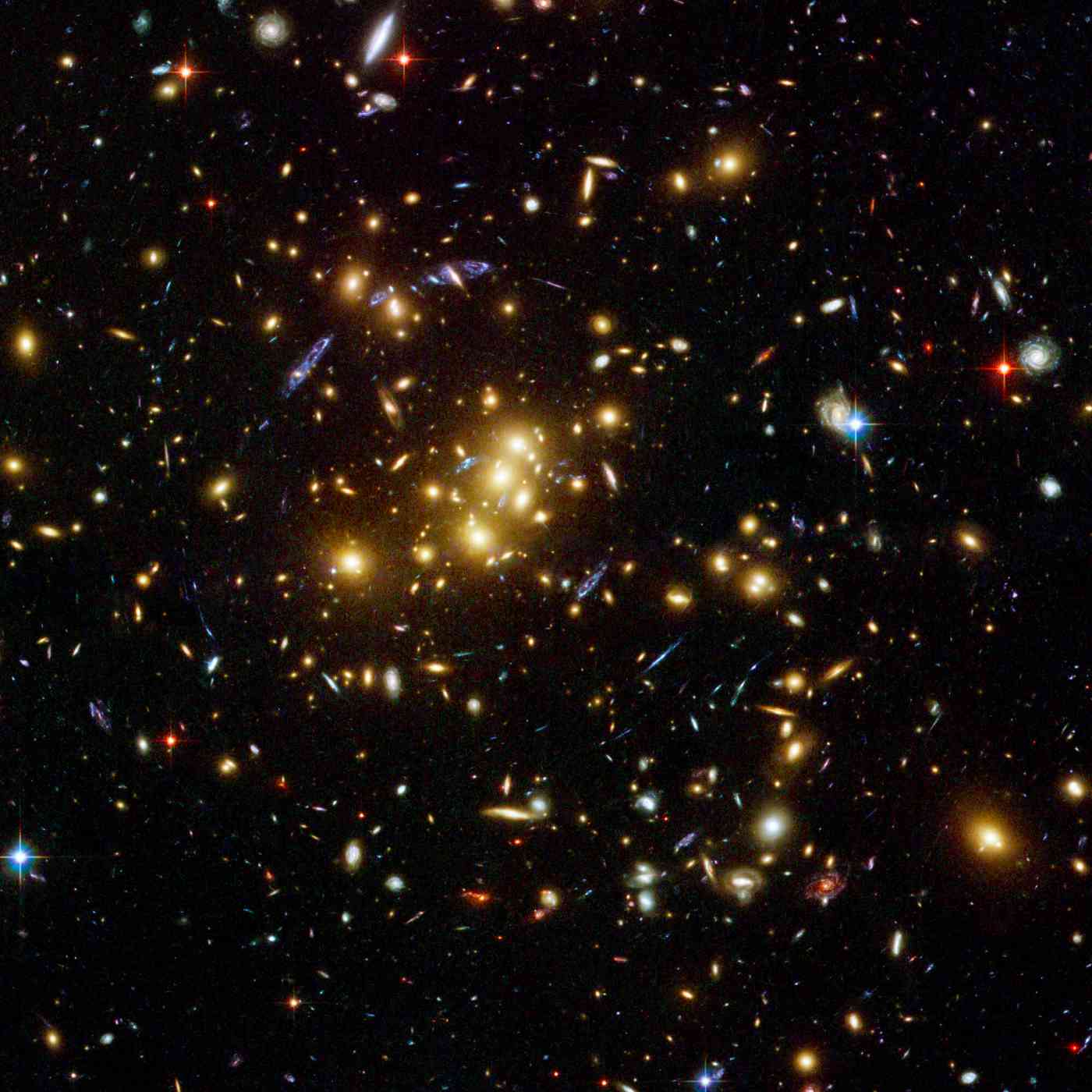}&&
					\includegraphics[height=6cm, width=6cm]{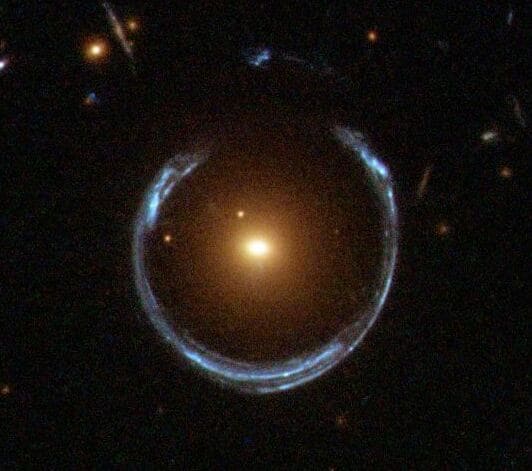}\\
					(a)&&(b)\\
				\end{tabular}
				\caption{\label{fig:g_lens_exmpl} (a) The galaxy cluster CL0024+1654 produces multiple images of a distant blue galaxy (Photo credit: NASA, ESA, H. Lee \& H. Ford (Johns Hopkins U.)\cite{Jee}). \\(b) Horseshoe Einstein Ring from Hubble. The gravitational field of the red galaxy LRG 3-757 lenses the incoming light from a distant blue galaxy (Photo credit: ESA/Hubble, NASA).}
			\end{figure}

			The effect of gravitational lensing can be classified into three categories namely, microlensing, weak lensing and strong lensing. In the case of microlensing, only the intensity of background stars increases apparently, while the weak gravitational lensing manifests a slightly deformed image of the background sources. On the other hand, the effect of strong lensing is easily observable in the forms of Einstein's ring or multiple images (see Figure~\ref{fig:g_lens_exmpl}a and \ref{fig:g_lens_exmpl}b). In general, the strong lensing generates multiple images of the distant object. However, in some special cases, the lensing object and the observed are perfectly aligned to the distant object. In this particular case, a ring-shaped lensed image manifested with angular separation (i.e. Einstein radius),
			\begin{equation}
			\theta_{\rm EC}=\sqrt{\dfrac{4GM}{c^2} \dfrac{d_{\rm LS}}{d_{\rm L} d_{\rm S}}}
			\end{equation}
			where $G$, $c$, $M$ are the universal gravitational constant, the velocity of light in space and the mass of the lensing body respectively. In the above equation, $d_{\rm L}$ and $d_{\rm S}$ are the distance of the lensing body and the background source respectively, while $d_{\rm LS}=d_{\rm S}-d_{\rm L}$.
			
		\subsection*{$\bullet$ Bullet Cluster}
			The Bullet Cluster is another convincing example of dark matter against Modified Newtonian dynamics (MOND). 
			Almost $4\times 10^{9}$ lightyears away from us, a striking piece of evidence for dark matter has been observed at the bullet cluster (1E 0657-56, 1E 0657-558). The Bullet Cluster is basically two colliding sub-cluster located at the Carina constellation. However, in several cases, only the smaller sub-cluster is considered to be the bullet cluster.
			
\begin{figure}				
				\centering
				\includegraphics[width=0.6\linewidth]{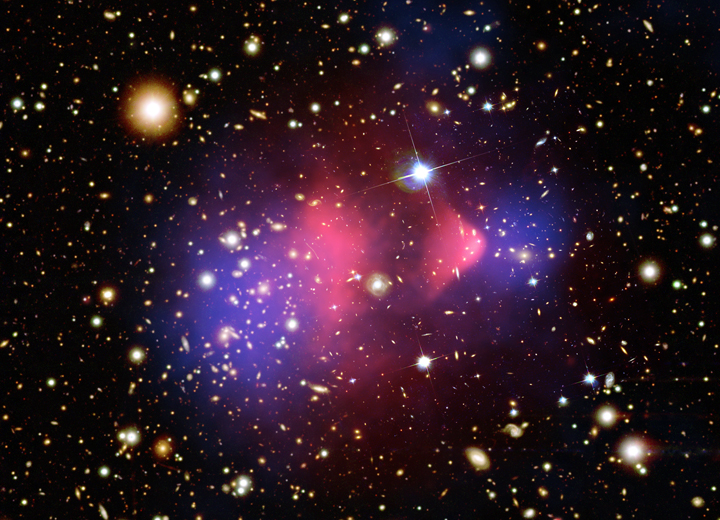}
				\caption{\label{fig:blt_clstr} The Bullet Cluster\\ (\footnotesize{Photo credit: X-ray: NASA/CXC/CfA/M.Markevitch et al.; Optical: NASA/STScI; Magellan/U.Arizona/D.Clowe et al.; Lensing Map: NASA/STScI; ESO WFI; Magellan/U.Arizona/D.Clowe et al.})}
			\end{figure}

			Multi-wavelength observations of the Bullet Cluster provided strong observational evidence of dark matter. The observations describe that, during the high-velocity collision (4500 km/s \cite{bullet_v_1}) between the two clusters, the shape of the baryonic (visible) mass distribution perturbed significantly (pink part in Figure~\ref{fig:blt_clstr}), while the dark matter halo part passed through each other undistorted. This shows that while the visible part is affected by the collision of the two clusters the invisible dark matter part suffers no distortion. It is to be noted that, the invisible dark matter halos for the two clusters are identified by the method of gravitational lensing. This shows that the dark matter particles do not have any interaction. They have of course gravitational interactions between them but that had little effect on the dark matter halos of the two colliding clusters since the halos are too large (and hence the dark matter density is low inside the halos) to affect the shape of the dark matter halos gravitationally. Therefore, one can conclude that a significant amount of dark matter present in this blue zone, which is responsible for the gravitation lensing. This is strong evidence for non-interacting dark matter. An identical phenomenon is also seen at the MACS J0025.4-1222 cluster \cite{bullet_other}.
		
		\subsection{\label{subsec:dm_nat}Nature of Dark Matter}
			The nature of dark matter is still unknown. Depending on different aspects such as production mechanism, particle nature etc., dark matter can be classified into different classes. Depending on the velocity during the structure formation, the dark matter candidates are categorized into three groups.
			\begin{itemize}
				\item {\bf Hot Dark Matter}
				
				At the time of structure formation, some dark matter particles moved with relativistic velocities. Those relativistic DM candidates are termed as {\bf Hot Dark Matter}. Hot dark matters are characterized as $x_f=\frac{m_{\chi}}{T_f}\lesssim 3$ where, $m_{\chi}$ is the mass of the DM particle and $T_f$ is the freeze-out temperature of that DM candidate. Consequently, hot dark matter candidates are comparatively lower in mass. Massive neutrinos belong to this category of dark matter. Observations of Lyman$\alpha$ forest, Planck, WMAP impose significant constraints on hot dark matter. Massive neutrinos, axions etc. low-mass DM particles belong to this category of dark matter. 
				
				\item {\bf Cold Dark Matter}
				
				The dark matters belong to this category were non-relativistic during the structure formation. The {\bf Cold Dark Matters} are characterized as $x_f\gtrsim 3$. Consequently, they are comparatively cool. Such candidates of dark matter are highly contextual in cosmological large structure formation. One of the most popular candidates of this class is Weakly Interacting Massive Particle (WIMP), which is studied extensively in the context of beyond standard model particle physics. However, few discrepancies are observed between theory and observational results namely, `missing satellite problem', 'too-big-to-fail problem', `core vs. cusp problem' etc., which cannot be explained using these two categories of DM. As a consequence, a third category is introduced, i.e. Warm Dark Matter.
				
				\item {\bf Warm Dark Matter}
				
				The {\bf Warm Dark Matters} are the intermediate dark matter candidate, which is less relativistic than hot dark matters but more than cold dark matters. The factor $x_f$ is $\sim 3$ in the case of warm dark matters \cite{Kolb:1990vq}. Such candidates of DM are assumed to decouple at the temperature $T \gg T_{\rm QCD}$. So, the mass of wark dark matter particles is $\sim \mathcal{O}(10)$ times more massive than hot dark matter particles. 
				
			\end{itemize}
			Besides pure hot dark matter, cold dark matter or warm dark matter, an alternative dark matter system may exist, i.e. {\bf Mixed Dark Matter}. Such type of DM is just a combination of hot dark matter and cold dark matter. Moreover, primordial black holes, topological modifications of Newtonian gravity on large scales are also proposed as alternative candidates of DM candidates.
		
			Dark matters also can be categorized into two different classes based on the nature of their constituent.
			\begin{itemize}
				\item {\bf Baryonic Dark Matter}
				
				Being baryonic in nature, MACHOs (Massive Astrophysical Compact Halo Objects), black holes, neutron stars are considered to be the candidates of {\bf Baryonic Dark Matter}. The nature of the galactic rotation curves indicates that galactic halos are enriched with such massive DM candidates. The detection of several microlensing events at LMC (Large Magellanic Cloud) is possibly another significant evidence of baryonic DM in the Milky Way galaxy \cite{Alcock1993,Alcock2000}. As the fuel of the white dwarf runs out, it cannot radiate and hence cannot be observed directly anymore. So, an exhausted white dwarf star also can be considered as a baryonic DM candidate. Moreover, primordial black holes, brown dwarfs etc. are included in this category of dark matter.
				
				The density of the visible matter ($\rho_{\rm vis}$) can be obtained from the mass to luminosity ratio $\Upsilon_{\rm vis}=M_{\rm vir}/L$ ($M_{\rm vir}$ is the mass and $L$ is the luminosity) \cite{Persic1992} \footnote{Visible matter density $\rho_{\rm vis} = \sum \int\phi(L)\Upsilon_{\rm vis} {\rm d}L$, where $\phi(L)$ is the luminosity function of different systems \cite{Persic1992}.}. The satellite-borne Planck experiment \cite{planck15,planck18} in accordance with COBE and WMAP \cite{wmap} estimates the baryon density (including X-ray emitting gas, luminous stars etc.) more precisely, which is roughly the same as obtained from the mass-to-luminosity ratio. The bounds on the baryon density, calculated from the BBN (Big Bang Nucleosynthesis) theory also agree with the previous estimations. So, there is not much room for such DM candidates except intergalactic gas and non-luminous stars.
				
				\item {\bf Non-baryonic Dark Matter}
				
				According to the recent Planck result \cite{planck15,planck18} and CMB spectrum, the total amount of dark matter containing the Universe is several times larger than that of the total baryonic matter. As a consequence, there is only a very small percentage of DM exists in the form of baryonic matters, while almost the entire DM population is {\bf Non-baryonic}. Such kind of dark matter interacts very weakly with SM (Standard Model) particles. As a result, direct detection of such DM candidates is extremely difficult. Most of the candidates of non-baryonic dark matters are extended standard model particles. Supersymmetric particles, Kaluza-Klein dark matter (inspired by the extra dimensions theories) \cite{Cheng:2002ej,servant_tait,Hooper:2007gi,Majumdar:2003dj}, Q-balls \cite{qball1,qball2}, WIMPZillas \cite{wimpzilla}, Superheavy or Heavy Dark Matters (HDM) \cite{kuz,PhysRevD.59.023501}, axions \cite{axion_1,axion_2} etc. are some plausible candidates of non-baryonic dark matter.
			\end{itemize}
		
			According to the production mechanism, dark matter can also be classified into two more different classes i.e. Thermal Dark Matter and Non-thermal Dark Matter.
			
			\begin{itemize}
				\item {\bf Thermal Dark Matter}
				
				Initially, the {\bf Thermal Dark Matter} candidates were assumed to be in thermal equilibrium in the early epoch of the Universe. Later, as the interaction rate decreases as an outcome of the expansion of the Universe, they started decoupling from the cosmic plasma. WIMPs are one of the notable candidates of such dark matters, which were assumed to be in the thermal equilibrium when the Universal temperature was sufficiently high. After being decoupled, the created pairs of particle-antiparticle can also undergo annihilation and hence produce SM particles. Initially, these two processes were in equilibrium. The number density of such candidate at the temperature $T$ can be expressed using the Boltzmann distribution function,
				\begin{equation}
				\Delta n_{\chi}=n_{\chi}-n_{\bar{\chi}}\sim \left(\dfrac{m_{\chi}T}{2 \pi}\right)e^{-m_{\chi}/T},
				\end{equation}
				where $n_{\chi}$ ($n_{\bar{\chi}}$) is the number density of the DM particles (antiparticles). As the expansion rate of the Universe dominates over the particle-antiparticle annihilation (and pair production) rate, the number density in comoving volume freezes. The density of dark matter after freeze-out is known as relic density and the corresponding temperature is termed as the freeze-out temperature of that DM species. The relic density ($\Omega_{\chi}$) of such thermally decoupled DM candidate depends on its annihilation cross-section $\langle \sigma v\rangle$ as $\Omega_{\chi}\propto1/\langle \sigma v\rangle$, where $\Omega_{\chi}$ is the DM density parameter \cite{dm_book}. The relic density and the maximum allowed cross-section of WIMP-type thermal DM provide a maximum limit of DM particle mass i.e. $\sim 10^5$ GeV.
				
				\item {\bf Non-thermal Dark Matter}
				
				On the other hand, the {\bf Non-thermal} candidates of Dark Matter are thought to be produced from bosonic coherent motion of oscillating pseudo-scalar fields or produced gravitationally. As the pre-BBN cosmic history is still unknown,  it is possible that some heavy dark matter ($\sim 10^5 \leq$ mass $\leq 10^{16}$ GeV) candidates can be produced gravitationally in the post-inflationary epoch. WIMPZilla is such a candidate of non-thermal dark matter ($\leq 10^{13}$ GeV) which are possibly formed during the reheating or post inflationary phase transition epoch \cite{PhysRevD.64.043503}. In the case of such superheavy candidates of dark matter, the abundance is suppressed as the power of the temperature-to-mass ratio \cite{PhysRevD.60.063504}.  
				
				Axions and axion-like particles (APLs) are other examples of non-thermal dark matter.  Although at a higher temperature, axions are almost massless, at the QCD scale, the mass of the axion hovers near the minima of the axion potential. Such kinds of DMs are generally of lower masses ($\lesssim 10^{-3}$(eV)).
			\end{itemize}
			
		\subsection{Masses of Different Dark Matter Candidates}

\begin{figure}				
				\centering
				\includegraphics[width=\linewidth]{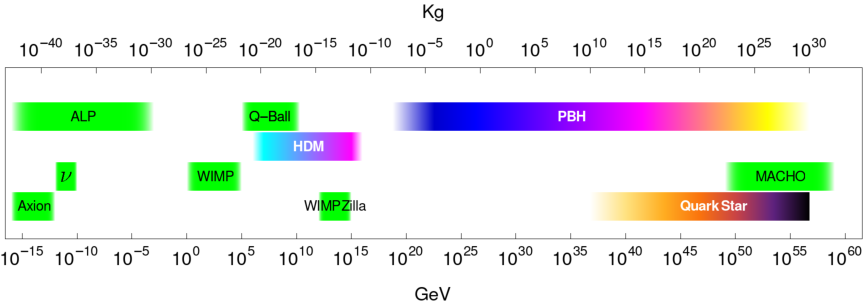}
				\caption{\label{fig:dm_scl}Mass range for different dark matter candidates. Here $\nu$ represents the heavy right-handed neutrino rather than the neutrino in a general form.}
			\end{figure}

			Dark matters cannot be observed directly as they do not interact with the electromagnetic force, unlike visible baryonic elements. As a consequence, besides particle dark matters (namely WIMP, axion etc.), several kinds of non-luminous (or low-luminous) astrophysical objects (e.g. massive compact halo object or MACHO, brown dwarf, pure quark star etc.) are also considered as dark matter candidates. According to observational evidences and theoretical models, there are various kind of dark matters exists in our Universe. Mass ranges of different types of dark matter are addressed graphically in Figure~\ref{fig:dm_scl}. The Weakly Interacting Massive Particle (WIMP) is the leading particle dark matter candidate, as it satisfies all cosmological and astrophysical constraints. WIMPs are assumed to be in thermal equilibrium at the epoch of the early Universe. The annihilation of WIMP-type DM particles produces highly energetic standard-model particles, which are studied extensively in the indirect detection of dark matter. The masses of WIMP type dark matter lies in the range from $\sim 2$ GeV (Lee-Weinberg bound) \cite{Kolb:1990vq} to $\sim 10^5$ TeV (unitarity bound \cite{unitary_bound}). 
			
			WIMPZilla \cite{wimpzilla} is a comparatively heavier dark matter candidate ($\geq 10^{13}$ GeV), which is probably originated at the epoch of preheating and the post-inflation reheating. The WIMPZillas are considered as a part of a wider group i.e., superheavy or heavy dark matter. Superheavy dark matters (HDM) ($10^6\sim10^{16}$ GeV) \cite{PhysRevD.59.023501} are generally assumed to be originated gravitationally in the early epoch of the Universe \cite{kuz,Kuzmin1998} or due to spontaneous symmetry breaking in the Grand Unified scale \cite{PhysRevD.59.023501,PhysRevD.64.043503,PhysRevD.60.063504,Kuzmin1998,PhysRevLett.81.4048,gelmini2010dm,bertone}. In another possibility, phase transition during the inflationary state of the Universe can be a possible source of several such non-thermal heavy dark matter candidates \cite{PhysRevD.64.043503}.	
			
			Moreover, there are few alternative models of dark matter indicating the evidence of right-handed neutrinos, axions and axion-like particles (APLs) etc. as dark matter candidates. These DM candidates are comparatively lighter than other candidates of dark matter ($\lesssim 1$ MeV). The concept of axion ($\lesssim10^{-3}$ eV) was put forwarded by Peccei–Quinn \cite{axion_1,axion_2} in 1977 in the context of strong CP problem in QCD. On the other hand, primordial black holes (PBHs), MACHOs, pure quark stars etc. astrophysical objects also behave as dark matter candidates due to their extremely low luminosity. The maximum masses of such astrophysical bodies are in the order of stellar mass. However, the mass of PBH may decrease significantly by evaporating in the form of hawking radiation \cite{hawking}.

		\subsection{\label{ssec:detection_of_DM} Detection of Dark Matter}
\begin{figure}		
				\centering
				\includegraphics[width=0.5\linewidth]{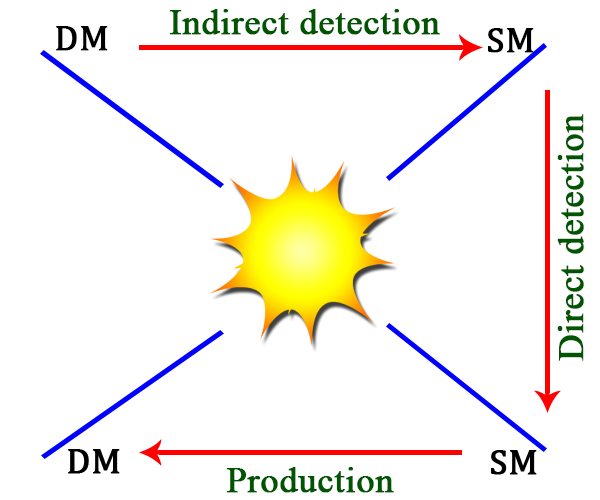}
				\caption{\label{fig:dm_detection} Schematic diagram of the possible channels of dark matter detection.}
			\end{figure}

			Detection of dark matter has turned out to be a worldwide endeavor in the current century. As dark matter does not interact with EM force, it neither absorbs nor reflects or emits light. Consequently, it is extremely difficult to spot them out. Besides the gravitational evidences (discussed earlier), there are three methods to confirm the existence of dark matter namely, direct detection, indirect detection and the production of artificial dark matter at the colliders. All these three possible detection techniques are schematically described in Figure~\ref{fig:dm_detection}. 
			
			In the case of direct detection, a tiny recoiled energy is measured, which is generated as an outcome of dark matter collision with the nucleus of the detector material. On the other hand, the indirect detection of the dark matter dealt with the production of SM particles by annihilation/decay of DM. Dark matters can also be generated at the lab by colliding SM particles at extreme energy. 
			
			\subsection*{$\bullet$ Direct Detection}
			The incoming WIMP-type dark matter can be detected by ground-based detectors from low-energy signals, which are produced as an outcome of the collision between incoming dark matter particles and the SM nuclei of the detector. Such interactions are extremely weak (recoiled energy is only a few keV) and assumed to be elastic in nature. The recoiled low-energy signals from the detector can be tracked out by adopting several optimistic methods namely, scintillator light, phonon excitation, ionization, bolometric current etc. In such detectors, Ge, Si, NaI are used as the detecting materials. DAMA (NaI) \cite{dama1,dama2}, CDMS($^{73}$Ge) \cite{cdms1,cdms2} are some working experiments the direct detection method is used in order to detect dark matter. Time Projection Chamber (TPC) is another effective detector, which allows complete localization of the event in 3D space. In this particular case, the drifting ions produced by recoiled nuclei are traced, where noble liquids (i.e. Xe, Ar, Ne) are used as the detecting material. The TPC method is used in the experiments XENON \cite{xenon1,xenon2,xenon3,xenon4,xenon5}, COUPP \cite{coupp}, LUX (Xe) \cite{lux}, CLEAN (Ar, Ne), DEAP (Ar) \cite{deap} etc.
			
			The direct detection method can be classified into two classes, namely spin-independent and spin-dependent. In the case of spin-independent interaction, scalar type interaction is taken place. XENON10 \cite{xenon1, xenon2}, XENON100 \cite{xenon3, xenon4,xenon5}, DAMA, CDMS, SuperCDMS \cite{scdms}, Zeplin \cite{ziplin1,ziplin2}, KIMS \cite{kims1,kims2,kims3} are the experiments, which are searching for the spin-independent interaction. On the other hand NAIAD \cite{naiad}, SIMPLE, Tokyo/NaF \cite{naf}, PICASSO \cite{picasso} etc. experiments use light nuclei as detecting material in order to observe spin-dependent signals.
			
			\subsection*{$\bullet$ Indirect Detection}
			Unlike direct detection, the indirect process detects the final SM products, which are produced by the decay or annihilation of WIMP or other heavy dark matter candidates. Such dark matter particles undergo pair annihilation and hence produce lepton-antilepton and quark-antiquark pairs. Those elementary particle pairs give rise to neutrino, photons etc. by undergoing subsequent decay. In some cases, dark matters directly decay into neutrino-antineutrino, electron-positron and photons \cite{kuz,Ando_2015,mpandey}.   
			
			\begin{itemize}
				\item {\bf Neutrinos}
				
				Neutrino is a promising probe in investigating several aspects of dark matter and stellar evolution. Besides the stellar process, the annihilation and decay of dark matter particles produce a significant amount of neutrinos. Neutrinos do not interact with the celestial magnetic field and interact weakly with matters. As a result, they can pass through astrophysical objects and remains almost unscathed. But this nature of the neutrino makes the detection process highly challenging. ANTARES, IceCube, SUPER-Kamiokande etc. are some renowned detectors, where incoming neutrinos are detected using indirect methods. In those experiments, high-energy neutrinos generate charged particles (for example muons) during their propagation through ice, water or rock. Then the newly generated and highly energetic charged particles emit Cherenkov radiation while passing through the detectors. Now analyzing the emitted radiation, the energy and direction of the incoming parent neutrinos can be estimated. On the other hand, tracking calorimeters are highly efficient neutrino detectors in measuring the direction of highly energetic incoming neutrinos. In this case, several layers of detector materials are used in order to track the direction of the neutrino propagation. Iron is the most popular detector material as it is a cheap and dense ferromagnetic material. However, liquid or plastic scintillators are often used in order to track the incoming neutrinos. NUTEV, MINOS, ICAL etc. are some notable detectors where this method is used.
				
				\item {\bf Antimatters}
				
				Annihilation or decay of dark matters also produces electron-positron, proton-antiproton pairs. Although there are several sources of electron and proton in the Universe, only a few numbers of sources exist for their antiparticles. Therefore, the excess amount of antimatter may indicate the sources of dark matter. Unlike neutrinos and photons, these particles are affected by the magnetic field of celestial bodies, as well as the earth's magnetic field. As a consequence, only a diffused spectrum can be obtained by the ground-based detectors. In order to minimize the background flux, the detectors can be operated at higher altitudes using satellite-based or balloon-based experiments. PAMELA \cite{pamela}, AMS etc. are notable example of such detectors. On the other hand, HESS, MAGIC, HAWC, CTA observe the Cherenkov radiations, which are produced by those charged particles during their propagation through the earth's atmosphere.
				
				\item {\bf Photons}
				
				Although photons do not get affected significantly by the magnetic field, they get attenuated over a large distance. Although annihilation and decay of WIMP-like dark matter particles produce a large number of high-energy photons, it is difficult to observe the entire band of the spectrum by a ground-based observer due to atmospheric opacity. Therefore space-based telescopes for high-energy photons are much effective. EGRET and \emph{Fermi}-LAT are two satellite-borne $\gamma$-ray telescopes, which are operational within a wide range of energy (20 MeV-300 GeV). On the other hand, HESS, MACE, MAGIC, CTA etc. ground-based $\gamma$-ray telescopes are capable to detect $\gamma$-rays events indirectly (Cherenkov radiation).
			\end{itemize}
			
			\begin{figure}				
				\centering
				\includegraphics[width=\linewidth]{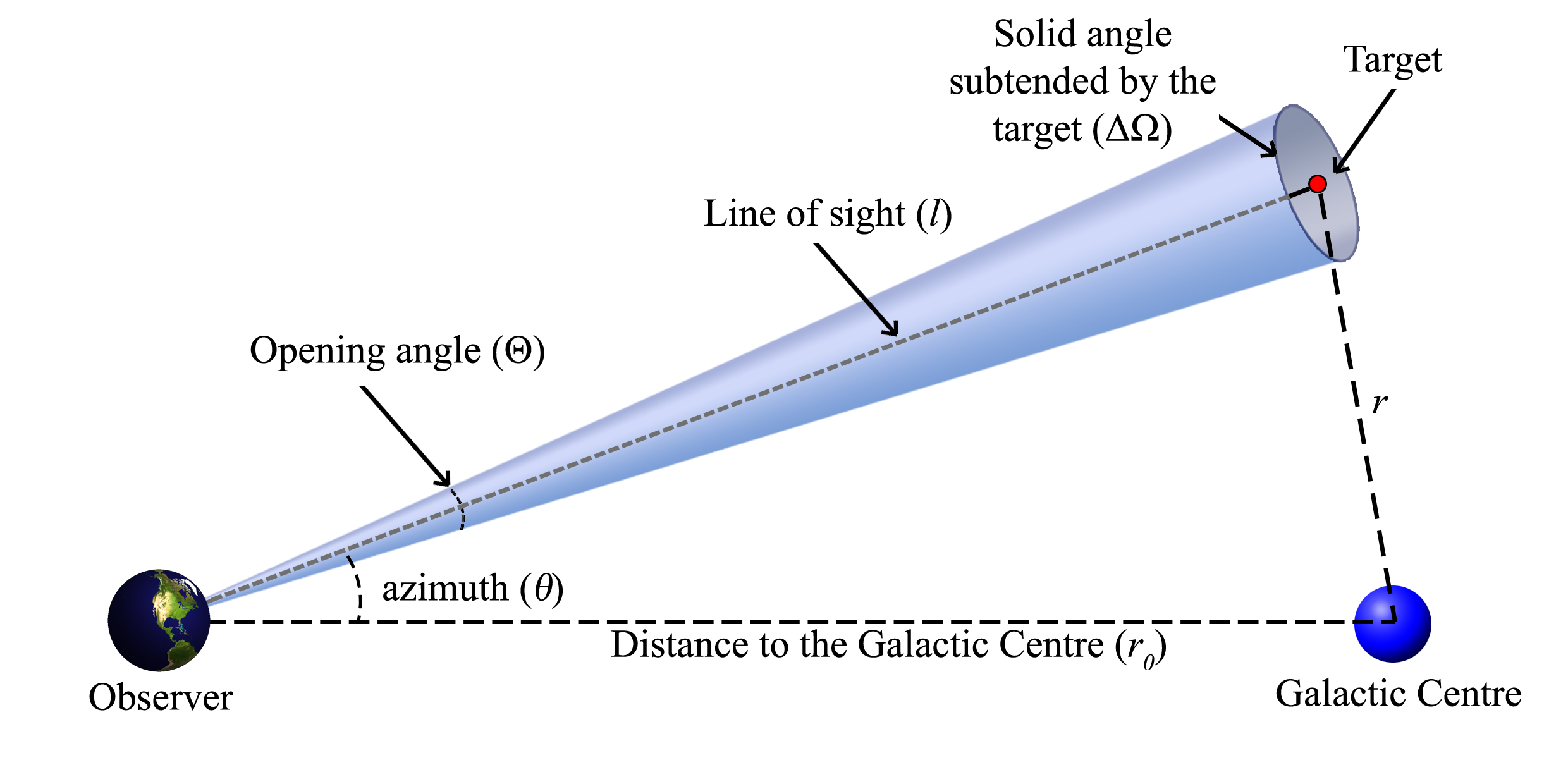}
				\caption{\label{fig:indirect} The schematic diagram of the indirect detection of galactic dark matter halo.}
			\end{figure}
			
			The observed flux from DM annihilation or decay depends on the total amount of dark matter contained within the solid angle $\Delta \Omega$. This quantity is known as $\mathcal{J}$-factor. The $\mathcal{J}$-factor essentially depends on the production mechanism of the emitted flux. In the case of DM annihilation, the expression of $\mathcal{J}$-factor takes the form,
			\begin{equation}
				\mathcal{J}_{\rm ann}=\int_{l.o.s} \rho(r)^2 {\rm d}l
				\label{eq:j_ann}
			\end{equation}
			while, in the case of decay process,
			\begin{equation}
				\mathcal{J}_{\rm dec}=\int_{l.o.s} \rho(r) {\rm d}l.
				\label{eq:j_dec}
			\end{equation}
			The $\mathcal{J}$-factor for the DM decay process ($\mathcal{J}_{\rm dec}$) is often termed as $D$-factor. In the above two equations (Eqs.~\ref{eq:j_ann} and \ref{eq:j_dec}), $\rho(r)$ is the density of DM halo at a radial distance $r$ from the galactic centre (assuming spherical symmetry in the DM halo distribution). The distance $r$ from the galactic centre can be expressed in terms of the coordinate of the target object ($r$, $\theta$) and the line of sight $l$ as,
			\begin{equation}
				r=\sqrt{l^2 + r_{\odot}^2 - 2 l r_{\odot} \cos \theta},
			\end{equation}
			where $r_{\odot}$ represents the distance between the galactic centre and the observer (or roughly Sun).
			Often the $\mathcal{J}$-factor for both cases are expressed in terms of dimensionless quantity given by,
			\begin{equation}
				J_{\rm ann}=\int_{l.o.s} \dfrac{1}{r_{\odot}} \left(\dfrac{\rho(r)}{\rho_{\odot}}\right)^2 {\rm d}l
				\label{eq:jdl_ann}
			\end{equation}
			\begin{equation}
				J_{\rm dec}=\int_{l.o.s} \dfrac{1}{r_{\odot}} \dfrac{\rho(r)}{\rho_{\odot}} {\rm d}l.
				\label{eq:jdl_dec}
			\end{equation}
			In the above equation, $\rho_{\odot}=0.3 {\rm GeV/cm^3}$ is the average halo density in the vicinity of the Sun.
			
			\begin{figure}				
				\centering				\includegraphics[width=0.7\linewidth]{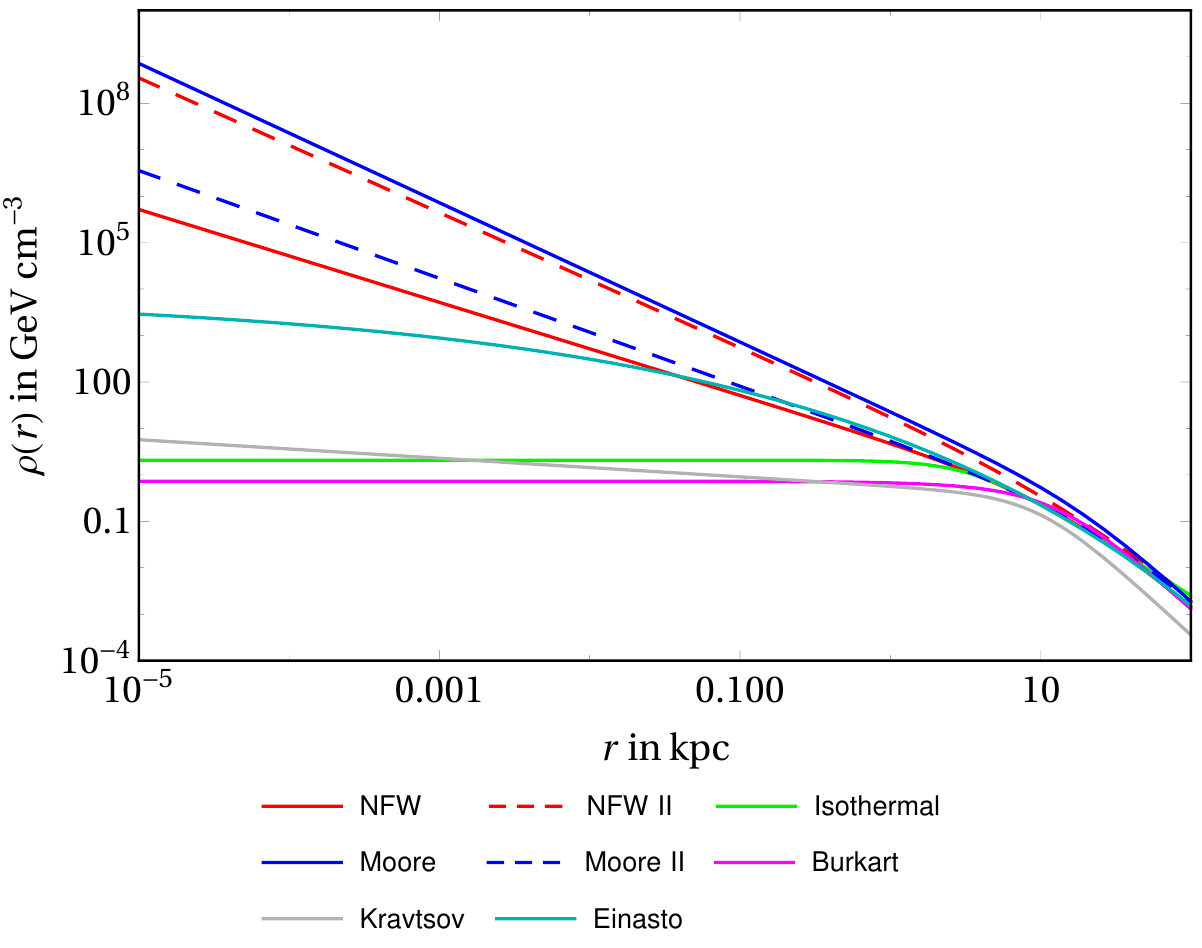}
				\caption{\label{fig:profile} Different dark matter halo profiles (see Table~\ref{tab:profile}).}
			\end{figure}
			The dark matter halo is not uniformly distributed throughout the Universe. As the dark matter halos are essentially distributed around the galaxies, the density of the dark matter increases as the radial distance from the centre of the corresponding galaxies decreases. Moreover, the numerical simulation indicates that, in the outer region of the galaxy, the density of the DM halo falls faster (density $\propto$ distance$^{-3}$) in comparison to the inner region (density $\propto$ distance$^{-1}$) \cite{Shen_2010}. There are several popular spherically symmetric DM halo density profiles available, which provide an approximate distribution of dark matter around our Milky Way galaxy. Some popular scaled halo density models are tabulated in the Table~\ref{tab:profile}. Among these halo models, the first seven profiles can be expressed in the general form,
			\begin{equation}
				\rho(r) = \dfrac{\rho_s}{\left(\kappa+\left(\dfrac{r}{r_s}\right)^{\gamma}\right)\left(1+\left(\dfrac{r}{r_s}\right)^{\alpha}\right)^{\frac{\beta-\gamma}{\alpha}}},
				\label{gen_prof}
			\end{equation}
			where $\alpha$, $\beta$, $\gamma$, $\kappa$ are model parameters and $r_s$ and $\rho_s$ are the scale distance and scale density respectively. This general form of the density profile is known as Hernquist profile \cite{gen_prfl}. 
			
			\begin{table*}
				\centering
				\begin{tabular}{llcc}
					\hline\hline
					&&&\\
					Model&$\rho(r)$&$r_s$ (kpc)&$\rho_s$ (GeV/cm$^3$)\\
					&&&\\
					\hline
					NFW \cite{Navarro:1995iw,Navarro:1996gj} & $\rho_s\dfrac{r_s}{r}\left(1+\dfrac{r}{r_s}\right)^{-2}$&20&0.259\\
					
					NFW II \cite{Navarro:1996gj} & $\rho_s\left(\dfrac{r_s}{r}\right)^{1.45}\left(1+\left(\dfrac{r}{r_s}\right)^{0.8}\right)^{-1.5625}$&20&0.257\\
					
					Isothermal \cite{isotherm} & $\dfrac{\rho_s}{1+\left(r/r_s\right)^2}$&3.5&2.069\\
					
					Moore \cite{moore} & $\rho_s\left(\dfrac{r_s}{r}\right)^{1.5}\left(1+\left(\dfrac{r}{r_s}\right)^{1.5}\right)^{-1}$&20&0.256\\
					
					Moore II \cite{moore2} & $\rho_s\left(\dfrac{r_s}{r}\right)^{1.16}\left(1+\dfrac{r}{r_s}\right)^{-1.84}$&30.28&0.108\\
					
					Burkert \cite{burkert1,burkert2} & $\dfrac{\rho_s}{\left(1+r/r_s\right)\left(1+\left(r/r_s\right)^2\right)}$&12.67&0.729\\
					
					Kravtsov \cite{kravt} & $\rho_s\left(\dfrac{r_s}{r}\right)^{0.2}\left(1+\left(\dfrac{r}{r_s}\right)^{2}\right)^{-1.4}$&10&0.361\\	
									
					Einasto \cite{einasto} & $\rho_s\exp{\left[-\dfrac{2}{\alpha}\left\{\left(\dfrac{r}{r_s}\right)^{\alpha}-1\right\}\right]}$&20&0.061\\
					\hline\hline
				\end{tabular}
				\caption{\label{tab:profile}Density profiles of dark matter halo and corresponding parameters.}
			\end{table*}
						
			The NFW profile is the most popular density profile, motivated by the result of cosmological N-body simulation. But, this profile is very steep near the galactic centre. The NFW II \cite{Navarro:1996gj}, Moore \cite{moore} and Moore II \cite{moore2} profiles are even steeper than the NFW profile in the vicinity of the galactic centre. On the other hand, observational evidence of the galactic centre shows that the dark matter halo density profile is flat near the core part of the galaxy. The density profiles namely Isothermal \cite{isotherm}, Burkert \cite{burkert1,burkert2} and Kravtsov \cite{kravt} agree with that observational result. All these seven profiles converge to different power laws at the core part of the galactic halo. But the Aquarius simulations \cite{NBsim} suggest that the density profile at the central region of the galaxy is not very cuspy \cite{navarro_mnras} (we do not obtain such results from these seven profiles). Such trend can be observed in the Einasto profile \cite{einasto}, which fits better to the result of the Aquarius simulations. The expression of the Einasto profile is given in Table~\ref{tab:profile}, where the values of the model parameters are $r_s=20$ kpc, $\rho_s=0.061$ (GeV/cm$^3$) and $\alpha=0.17$ (shape parameter) for the Milky Way galaxy.
			
	\section{Dark Age of The Universe}
		Besides dark matter and dark energy, the evolution of the Universe in the dark age is another mystery of astrophysics and cosmology. The dark age started subsequently after the end of recombination in the early Universe ($z\approxeq 1100$). At that epoch, the Universe became substantially cool that, the electrons were decoupled from the cosmic background radiation. Therefore, the Universe became transparent to the electromagnetic radiations and simultaneously filled with almost uniform reddish electromagnetic glow. Although after the recombination, the Universe was transparent, there was no large-scale structure that can ignite any significant light source. The only source of light was the cosmic background radiation, but due to the cosmological redshift, the reddish (orange) glow of CMB turned into inferred as the redshift decreases in this epoch. As a consequence, from the end of recombination to the time when the first star was born, the entire Universe was dark and hence this period is termed as Cosmic Dark Age. As there was no new source of light in the dark age, the dynamic of the Universe at that period is difficult to explore. In this particular case, the redshifted signal of the 21-cm hydrogen absorption spectrum is turned out to be the only promising probe. The 21-cm spectrum appears as a result of the transition between two hyperfine spin-state of neutral hydrogen atoms (the most abundant baryonic component of the Universe, which occupies $\sim 75\%$ of the entire baryonic mass), namely a singlet spin 0 state and a triplet spin 1 state. In 1944, dutch astronomer {\bf H. C. Hulst} first predicted the application of the 21-cm spectrum in cosmology and astrophysics. Later, {\bf H. Ewen} and {\bf E. M. Purcell} observe the global 21-cm signal in 1951 at Harvard. That was the first observation of the cosmic 21-cm signal. A detail discussion on 21-cm signal is described in the \Autoref{chp:21_feb}, \Autoref{chp:21_jan} and \Autoref{chp:21_mar}.
		

	\section{Summary}
		
		We addressed in this thesis some significant aspects of cosmology and astroparticle physics. In this introductory chapter, an account of the elementary idea of dark energy and dark matter is given. The rest of the Chapters in the thesis are organized as follows. In \Autoref{chp:compact_obj}, the introduction is extended for the compact astrophysical objects. We proposed a simple model of rotating quark stars in \Autoref{chp:chandra} and hence investigate its maximum possible mass. \Autoref{chp:21_feb} dealt with 21-cm cosmology in presence of primordial black holes (PBHs). In this chapter, the bounds on the initial mass fraction of PBHs and the dark matter mass are estimated by using the observational result of 21-cm brightness temperature by the EDGES experiment. In \Autoref{chp:21_jan} we extend our studies of \Autoref{chp:21_feb}, where the effects of dark matter - dark energy interaction are also included. In this chapter, we investigated the bounds on the dark matter dark energy coupling parameter and their variations with other model parameters (such as PBH mass and initial mass fraction, DM mass etc.) in the framework of 21-cm cosmology. We explore the possible multimessenger signals from the decays of primordial superheavy or heavy dark matter candidates in  \Autoref{chp:21_mar}. One of the multimessenger signals is related to the heating of baryon temperature and its consequent effects in the global 21-cm signature, while the other signal is considered to be the ultra high energy neutrino signal at the IceCube experiment. In \Autoref{chp:IC}, the possible baryon asymmetry that could be generated in case of the decay of heavy dark matter candidates, is addressed. Finally, in \Autoref{chp:conclu}, some concluding remark is given with the summary and the future outlooks.
		
\chapter{Compact Astrophysical Objects} \label{chp:compact_obj}
	\\
	Compact astrophysical objects are considered as the end stages of stellar evolution. Besides neutron stars, white dwarf and black holes, a few exotic stars namely quark star, preon star, boson star etc. lie in this group. Their extreme compactness gives rise to several extreme phenomena which are highly significant in highenergy astrophysics and cosmology. In many cases, such compact objects are non-luminous (low-luminous) for example black holes, MACHOs (Massive Astrophysical Compact Halo Objects), black dwarf (exhausted white dwarf) etc. Those non-luminous (low-luminous) astrophysical bodies are often considered as the candidate of baryonic dark matter.
	
	\section{White Dwarfs}
		\begin{changemargin}{40pt}{40pt}
			\setstretch{1}
			{\myfont ``we have a star of mass about equal to the sun and of radius much less than Uranus."}
			\vspace{-0.5cm}
			\begin{flushright}
				{\bf \color{qtcol} -A. Eddington}
			\end{flushright}
		\end{changemargin}
		Famous English astronomer and physicist Sir Arthur Stanley Eddington has concluded this, about the first ever observed white dwarf (i.e. Sirius B, companion of a binary star system) in his book ``The Internal Constitution of the Stars'' \cite{eddington} (1926). However, the first white dwarf was discovered in 1914 by W. S. Adam. Later, in 1931, S. Chandrasekhar put forward a realistic model of white dwarfs by incorporating the special relativistic effect of degenerate electrons. Moreover, Chandrasekhar estimated the maximum possible mass of white dwarfs (i.e. 1.4 M$_{\odot}$) in his analysis \cite{chandrasekhar}. This limiting mass is named as {\bf ``Chandrasekhar limit"} in honor of the discoverer.
		
		\begin{changemargin}{40pt}{40pt}
			\setstretch{1}
			{\myfont ``The life history of a star of small mass must be essentially different from the life history of a star of large mass. For a star of small mass the natural white-dwarf stage is an initial step towards complete extinction. A star of large mass cannot pass into the white-dwarf stage and one is left speculating on other possibilities."}
			\vspace{-0.5cm}
			\begin{flushright}
				{\bf \color{qtcol} -S. Chandrasekhar}
			\end{flushright}
		\end{changemargin} 
		
		White dwarfs are the first discovered compact stars. Above 95\% of the stars will undergo this phase at the end of their lifetime. According to the recent models, our Milky Way galaxy contains $\sim 10^{10}$ white dwarfs at the present time \cite{2009JPhCS.172a2004N}. It is assumed that a main-sequence star having masses within the range 0.07 M$_{\odot}$ to 10 M$_{\odot}$ \cite{Heger:2002by,2001PASP..113..409F} transforms into a white dwarf at the end of its lifetime. As the hydrogen at the stellar core is completely fused, it starts fusing into heavier elements and sheds its outer layers. Eventually, a major part of the star spreads out into space in form of planetary nebula and leaving a compact core as a white dwarf. The electron degeneracy pressure resists the star in further collapse, which essentially depends on the core density of the star. But if the mass lies beyond the Chandrasekhar limit, the electron degeneracy pressure of the star is unable to withstand the gravitation collapse and hence the star collapse into a neutron star. Moreover, if a white dwarf exists as a companion of any binary system, it may accreate a significant amount of mass from the companion stars and transform into a neutron star as its mass reaches the Chandrasekhar limit. Since white dwarfs are highly compact, their surface gravity reaches a few hundred thousand times that of the earth. A typical white dwarf shines over several tens of billions of years. But after that, there does not remain any energy to radiate and hence turns into a Black Dwarf. We know that the age of the Universe is only $13.6 \times 10^9$ years, so, no observational evidence of black dwarf can be recorded yet.
		
		\section{Neutron Stars}
		\begin{changemargin}{40pt}{40pt}
			\setstretch{1}
			{\myfont ``These are the brightest and most frequent stellar eruptions in the galaxy, and they're often visible to the naked eye"}
			\vspace{-0.5cm}
			\begin{flushright}
				{\bf \color{qtcol} -Przemek Mr\'{o}z}
			\end{flushright}
		\end{changemargin}
		
		Neutron stars are believed to be the most energetic celestial object in the Universe after the black holes. A neutron star is so compact that a 1 M$_{\odot}$ star can fit within a sphere of few kilometers (for non-rotating cases, rotating neutron stars are not perfectly spherical). The existence of such an extreme star was first predicted by two European astronomers Walter Baade and Fritz Zwicky in 1934 \cite{ns}, within two years of Chadwick's famous discovery of the neutron. They also suggested that neutron stars are originated as an outcome of supernova explosions. Within five years of that, Oppenheimer and Volkoff put forward a realistic model for neutron stars \cite{tov}. Being extremely compact, such stars are very small in size. As a consequence, direct observation of this special kind of star turned out to be very difficult at that time. But X-ray observation of Iosif Shklovsky (1967) \cite{nsX} and the discovery of radio pulsar by Jocelyn Bell Burnell and Antony Hewish presented an efficient way to observe neutron stars.
		
		We already mentioned that a comparatively lower mass main sequence stars generally transforms into white dwarfs. However, in the case of more massive stars ($\gtrapprox10$ M$_{\odot}$), as the fuel of the star gets exhausted, the star undergoes extreme gravitational collapse, leading to a more compact and hotter core than typical white dwarfs. As a result, the constituents of the stars (Hydrogen, Helium, carbon etc.) further take part in nuclear fusion and ends up in Iron. After that, due to tremendous gravity, electrons and protons fuse into neutrons and the density reaches near the nuclear density. Along with is process, a shock is also generated, that explodes the outer layers of the star in the form of a Supernova. It is to be mentioned that, Supernova is the most energetic astronomical phenomenon, we have ever seen. The remaining highly compact core is left as a neutron star.
		
		The neutron stars are the remarkable paradigm of the extreme gravitational field. Such stars are so compact that their surface gravity may reach up to 200 billion times the gravitational field at the Earth's surface. Not only the extreme compactness makes the neutron star unique, as a star collapses into a neutron star, it gains tremendous magnetic field and angular velocity in order to conserve the angular momentum. The spinning Neutron stars radiate their energy in the form of radio wave beams. Such spinning Neutron stars are known as Pulsar (a white dwarf also can be a pulsar). They emit EM waves out of the magnetic poles of the spinning star. Alongside the Pulsers, another category of Neutron stars can be observed, namely Magnetar. In the case of Magnetars, the surface magnetic field reaches several 1000 times higher than a typical Neutron star. Till now, only $\sim30$ Magneters are discovered so far, out of which, 6 stars can be categorized in both Pulsars and Magneters. Moreover, the binary Neutron star merger may produce another significant astrophysical phenomenon caller Kilonova. 
		
	\section{Quark Stars}
		Our Universe seems to be full of objects which elude common interpretation. The quark star \cite{Witten} is such an exotic star of our Universe, which was first predicted by two Soviet physicists D. D. Ivanenko and D. F. Kurdgelaidze in 1965, just five years after the discovery of quark. The quark star is one of the most bizarre celestial objects. Quark stars are generally obtained at the core part of massive neutron stars. Moreover, primordial strange stars are another possible source of quark stars. As quarks are the fundamental constituents of matter, quark stars are more compact than a typical neutron star. An analytical study by Banerjee \emph{et~al.} \cite{SBa} reveals that at the extreme case, a quark star can be so compact that its Schwarzschild radius can be up to 3/8 times of the star radius.
		
		Usually, a neutron star is believed to be the end product of the stellar evolution of massive stars. However, if the extreme gravitational pull of the star becomes sufficiently high to overcome the degeneracy pressure, a neutron star further collapses and the constituent neutrons are forced to transform into the ultra-dense quark phase. In another possibility, a quark star can be generated due to the accumulation of ambient quark matter in presence of sufficient strange quarks \cite{Witten,u_lim_mass1}. Further details about quark stars as well as strange stars are discussed in \Autoref{chp:chandra}.

	\section{Massive Compact Halo Objects}

		In the early twentieth century, lensing (microlensing) based observations \cite{Alcock1993,Aubourg1993} suggest the existence of some non-luminous astrophysical bodies in the Milky Way halo. Such non-radiating, stellar-mass objects are termed as Massive Compact Halo Objects (MACHOs). MACHOs may come with a broad range of masses ($6\times 10^{-8}\sim 15$ M$_{\odot}$ \cite{macho_mass}). However, in most of the case the mass lies within the range $0.5^{+0.3}_{-0.2}$M$_{\odot}$ \cite{Alcock2000,sutherland_1999}. 
		
		In the previous section (\Autoref{chp:intro}), we already mentioned that MACHOs are considered as the candidate of baryonic DM. It has been conjectured that the MACHOs are evolved from primordial strange quark nuggets (mass $\sim 10^{44}$ GeV \cite{macho_mass}) via coalescence process \cite{Quark_galaxy}. Quark nuggets are generated via first-order (quark-hadron) phase transition when Universe cools down significantly (temperature $\sim100$ MeV) \cite{Witten,Quark_galaxy}. Such transition took place just after $\sim10^{-5}$ sec of the big bang. 
		
		\section{\label{sec:bhs} Black Holes}
		
		\begin{changemargin}{40pt}{40pt}
			\setstretch{1}
			{\myfont ``If there should really exist in nature any bodies, whose density is not less than that of the sun, and whose diameters are more than 500 times the diameter of the sun, since their light could not arrive at us; or if there should exist any other bodies of a somewhat smaller size, which are not naturally luminous; of the existence of bodies under either of these circumstances, we could have no information from sight; yet, if any other luminous bodies should happen to revolve about them we might still perhaps from the motions of these revolving bodies infer the existence of the central ones with some degree of probability, as this might afford a clue to some of the apparent irregularities of the revolving bodies, which would not be easily explicable on any other hypothesis; but as the consequences of such a supposition are very obvious, and the consideration of them somewhat beside my present purpose, I shall not prosecute them any further."}
			\vspace{-0.5cm}
			\begin{flushright}
				{\bf \color{qtcol} -John Michell.}
			\end{flushright}
		\end{changemargin}
		
		British philosopher John Michell first proposed the idea of ``Dark Star'' (or Black Hole) in 1783 in the framework of Newtonian gravity. About 200 years later (in 1965), Penrose mathematically described this interesting phenomenon using Einstein's theory of relativity. Black holes (BHs) \cite{chandra} are one of the most mysterious phenomena of the Universe. they are so compact that, even light cannot get rid of their extreme gravitational pull. Such compact astrophysical bodies are assumed to be generated from the gravitationally collapsed massive stars. When the fuel of a massive star gets exhausted, the core region of the star collapses as an outcome of extreme gravity. Now in some cases, the star becomes substantially high that the escape velocity on its surface equals the speed of light. As a consequence, the star can no longer radiate, nor reflect any electromagnetic wave. So, such astrophysical objects turned to be dark or black, i.e. Black Hole. 
				
		\begin{figure}
			\centering{}
			\includegraphics[width=0.6\linewidth]{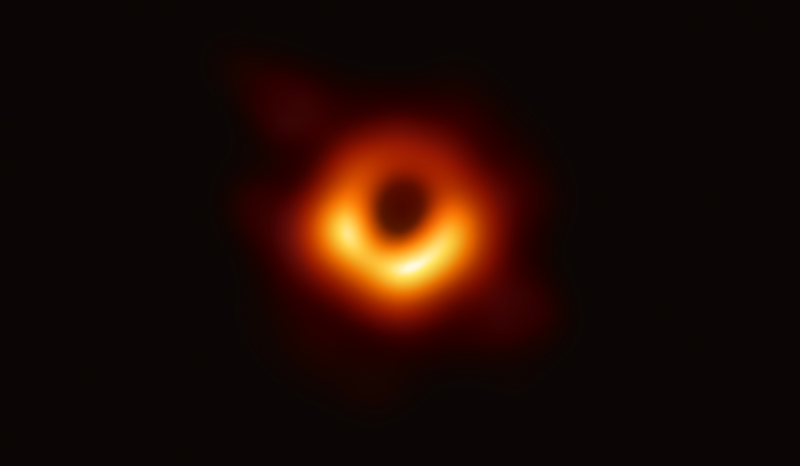}
			\caption{\label{fig:BH_img}The first image of a black hole which is located at the centre of the M87 galaxy, captured by Event Horizon Telescope. (\footnotesize{Photo credit: Event Horizon Telescope Collaboration})}
		\end{figure}
		Figure~\ref{fig:BH_img} is the first ever image of the black hole, Which was captured by the Event Horizon Telescope Collaboration (2019). Although the captured image of black hole, shown in  Figure~\ref{fig:BH_img}, is a supermassive black hole, the possible mass range of black holes is rather wider. A primordial black hole (PBH) can be significantly lighter than stellar masses black holes (see Figure~\ref{fig:dm_scl}). The idea of PBH was first introduced by two soviet scientists Y. B. Zel’dovich, I. D. Novikov \cite{PBH_0}. Primordial black holes are believed to be formed as an outcome of the collapse of the overdensity region in the early Universe \cite{khlopov_1,khlopov_2,khlopov_3,juan}. The limit of the such overdensity can be characterized by the Jeans length $R_j$ given by, $R_j = \sqrt{\displaystyle\frac {1} {3G \rho}}$. In this context, the density fluctuation $\delta \rho$ need to satisfy the condition $\delta_{\rm min} \leq \delta \rho \leq \delta_{\rm max}$, where $\delta_{\rm max}$ and $\delta_{\rm min}$ denote the maximum and minimum values of the density contrasts. The density fluctuation $\delta \rho$ can be expressed as $\rho = \rho_c + \delta \rho$, where $\rho_c$ indicates the critical density for collapse. The gravitational fluctuation arises in the density during the inflationary epoch is the most reliable conjecture for formation of primordial black holes \cite{stpbh_1,stpbh_2,stpbh_3,stpbh_4,stpbh_5,stpbh_6,stpbh_7}. Besides the standard scenarios, several alternative conjectures are available, those addresses the formation of primordial black holes namely, collapse of domain walls and cosmic strings \cite{ccs_1,ccs_2,ccs_3}, fragmentation of scalar condensation \cite{fsc_1,fsc_2,fsc_3} etc.
		
		\section{\label{sec:hawking} Hawking Radiation}
			\begin{figure}[h]
				\centering{}
				\includegraphics[trim={10 40 10 40},clip,width=\linewidth]{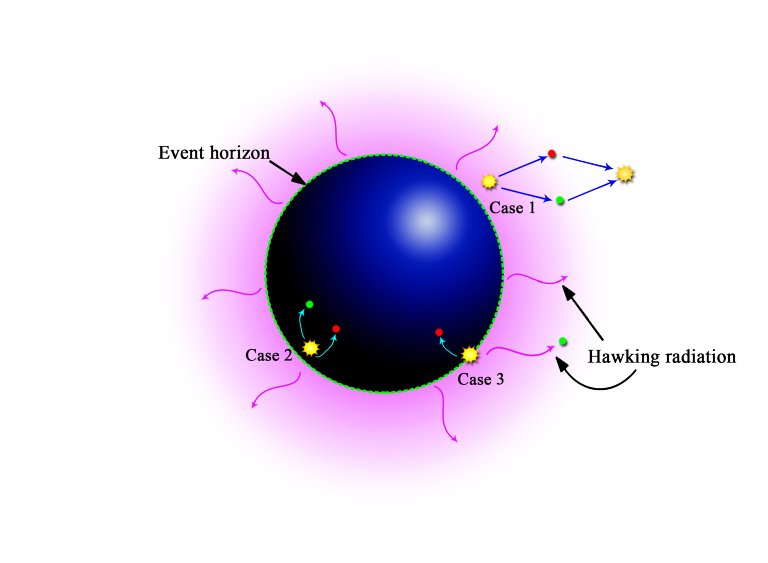}
				\caption{\label{fig:h_rad} Schematic diagram of Hawking radiation. In case 1, the virtual particle, antiparticle pair annihilate each other. In case 2, both virtual particle and antiparticle are pulled beyond the event horizon, as both of them are created inside the event horizon. However, in case 3, one of the particle-antiparticle pair falls inside the event horizon, while the other escapes in the form of Hawking radiation.}
			\end{figure}
			
			Black holes are assumed to be the final state of massive stars, which are so compact that, nothing can get rid of its tremendous gravitational field, not even light. As a result, a black hole should be entirely black. However, in practical, black holes are not absolutely black. Famous scientist Stephen W. Hawking predicted this fascinating phenomenon in the framework of Schwarzschild metric \cite{hawking}.
		
			According to the notion of quantum fluctuation, a particle-antiparticle is generated spontaneously in the empty space and they annihilate each other immediately after their formation. But when this quantum mechanical phenomenon produces virtual particle-antiparticle pair in the vicinity of the event horizon of a black hole, an interesting phenomenon manifests. In this particular case, one of them, which is closer to the event horizon, is unable to escape from the tremendous gravitational field of the black hole, while the other escapes the vicinity of the black hole in the form of radiation. This form of radiation is known as Hawking radiation (see Figure~\ref{fig:h_rad}).  
	
			In the case of an uncharged Schwarzschild black hole, the rate of particle emitted per degree of freedom within energy range $Q$ and $Q+{\rm d}Q$ is given by \cite{BH_F}, 
			\begin{equation}
			{\rm d}\dot{N}=\dfrac{\Gamma_s {\rm d}Q}{2\pi \hbar}\left[\exp\left(\dfrac{8\pi GQm}{\hbar c^3}\right)-(-1)^{2s}\right]^{-1},
			\end{equation}
			where $\Gamma_s$ is the dimensionless absorption probability of the species $s$, having mass $m$. The temperature of the black hole can be estimated as \cite{BH_F}
			\begin{equation}
			T_{\rm{BH}}=\dfrac{\hbar c^3}{8 \pi G M_{\rm BH}}\approx 1.05753 \times \left(\dfrac{10^{13} {\rm g}}{M_{\rm{BH}}}\right) {\rm GeV}.
			\label{eq:bh_temp}
			\end{equation}
			So, the change of entropy due to ${\rm d}Q$ amount of heat change is given by,
			\begin{equation}
			{\rm d}S = \dfrac{{\rm d}Q}{T_{\rm{BH}}}=8 \pi M_{\rm{BH}} {\rm d}Q.
			\end{equation}
			Consequently, the entropy of the entire black hole of surface $A$ takes the form
			\begin{equation}
			\mathcal{S}_{BH}=\dfrac{k_{\rm B} A}{4 l_{\rm P}^2}\label{eq:BHF},
			\end{equation}
			where $l_{\rm P}$ ($=\sqrt{G\hbar/c^3}$) is the Planck length. The Eq.~\ref{eq:BHF} is known as Bekenstein - Hawking formula. Now assuming pure photon emission in Hawking radiation, the luminosity of the black holes becomes \cite{SSBHT}
			\begin{equation}
			\mathcal{L}_{BH}=\dfrac{\hbar c^6}{15360 \pi G^2 M_{\rm{BH}}^2}\label{eq:BHL}.
			\end{equation}
			So, the life time of a black hole having mass $M_{\rm BH}$ can be estimated as \cite{SSBHT}
			\begin{eqnarray}
			t_{BH}=\dfrac{5120 \pi G^2 M_{\rm BH}^3}{\hbar c^4}&\approx& 2.1\times 10^{67} \left(\dfrac{M_{\rm BH}}{M_{\odot}}\right)^3 \,{\rm years}\label{eq:BHt}.
			\end{eqnarray}
			Now from the above equation (Eq.~\ref{eq:BHt}) one can conclude that a one solar mass black hole takes $\sim 2.1 \times 10^{67}$ years to evaporate completely, while the age of the Universe is only $\approx 14 \times 10^9$ years. However, the lifetime of a primordial black hole having mass $\sim 10^{14}$ g is more or less equal to the current age of the Universe.
	
	\section{Other Exotic stars}
		Besides the above mentioned highly compact stellar objects, there may exist several exotic compact stars, which are made of elementary particles other than protons, electrons, neutrons and muons \cite{Exotic_star_1,Exotic_star_2}. As a consequence, quark stars are also a candidate of this stellar category. A brief idea about quark stars is already drawn in the previous section (more in \Autoref{chp:chandra}). Preon star, Boson star, Planck star etc. are other notable candidates of exotic compact stars.
		
		\subsection{Preon Stars}
			Preons are assumed to be an elementary constituent of leptons and quarks \cite{preon}. This hypothetical particle comes in four flavours namely zero, anti-zero, plus and anti-plus. As preons are sub-elements of quark, a star made up of preons (Preon star) \cite{preonstr} is extremely dense in nature. According to the work of {\bf Hansson} \emph{et~al.} \cite{preonstr} the density may exceed $10^{23} \rm{g/cm^3}$. Such extremely dense and compact astrophysical objects are assumed to be generated due to the density fluctuation in the early Universe. In another possibility, the gravitational collapse inside highly massive stars may also produce Preon stars.
		
		\subsection{Boson Stars}
			The Boson stars are the most wired candidate of the category of compact astrophysical objects. Unlike others stars, these hypothetical compact stars are assumed to be transparent in nature \cite{transparentstr}. So, these stars are invisible, but their immense gravitational field bends the background light (they don't have event horizons like black holes) and produces gravitational lensing. The most possible constituent of the Boson stars is the transparent axions \cite{Exotic_star_1}. In another possibility, this kind of star can be made up of Helium-4 nuclei. There is no observational evidence for such hypothetical Boson stars till today. In 2016, {\bf Braaten} \emph{et~al.} proposed a different model of highly dense axion star \cite{Braaten:2015eeu}. According to that model, the self-gravity of the star is compensated by the mean-field pressure of the axion Bose-Einstein condensate. 
		
		\subsection{Planck Stars}
			Planck star is another extremely hypothetical astrophysical object, which is proposed by {\bf Carlo Rovell} and {\bf Francesca Vidotto} in 2014 \cite{planck_star}. In this case, the quantum-gravitational pressure counteracts the mass of the star. Planck stars may have both astrophysical and cosmological interests as they are able to generate a detectable signal of quantum gravitational origin which lies around the wavelength of $10^{-14}$ cm \cite{planck_star}.
		
		Such exotic celestial bodies are highly theoretical. It is extremely difficult to locate such novel celestial objects by analyzing conventional cosmic-ray signals. Hopefully, if such hypothetical celestial bodies really exist, future developments on space-based gravitational waves astronomy will take a significant role in exploring such exotic astrophysical bodies. 
		
\chapter{Chandrasekhar Limit for Rotating Quark Stars}\label{chp:chandra}
	\\
	The limiting mass of compact celestial bodies is a significant characteristic of such celestial bodies. In the case of highly dense quark stars, the maximum admissible mass essentially depends on the rotation frequency (for rotating stars), Bag constant and the fundamental constants. In this chapter, we address the maximum mass of rotating quark stars using a simple semi-analytic treatment, which mostly depends on the rotational frequency of the star apart from the bag constant and other fundamental parameters. Eventually, the limitation of the rotational frequency and compactness of the quark stars are also estimated and compared with relevant numerical results as well as several recent observations.
	
	\section{Introduction}
	The theoretical studies for quark stars have emerged as a worldwide enterprise after the first prediction of the highly-dense quark core at the centre of the core-collapsed neutron stars \cite{ivan_1}. In 1965, two soviet physicists Kurdgelaidze and Ivanenko  \cite{ivan_1,ivan_2} brought to light the concept of the quark star, just five years after the discovery of quarks by Gell-Mann \cite{Quark_Wikipedia}. Recently discovered comparatively cooler and compact stars (i.e. RXJ185635-3754, RXJ185635-3C58 \cite{Prakash:2002xx}, SWIFTJ1749.4-2807 \cite{hyu}) also support the conjecture of the quark stars. Almost all kinds of stars and compact celestial bodies are self-bounded by gravity. However, in case of quark stars, they are bounded by strong interaction rather than the gravitational field alone \cite{Witten,Jaikumar:2004zy}. In addition to that unique characteristic, the densities of such compact objects are significantly high (even denser than the nuclear density i.e. $2.4\times10^{14}\,\rm{g/cm^{3}}$ \cite{Bag_Model_2}), as a consequence, they earn immense significance as the natural laboratory for quark phase.
	
	Quark stars are assumed to be the very end phase of the stellar evolution. As the fuels of a massive neutron star tend to run out, the radiation pressure of such massive stars becomes too low to balance the self-gravity. Consequently, that star begins collapsing due to self-gravity at the core region of the star. However, in certain cases, the collapsed part of the star becomes substantially dense and as a result, a quark phase is generated spontaneously at the core part of the star \cite{key_40} (even without strangeness). Initially, the quark phase appears at the core region, the quark core grows over time and eventually captures the entire star by absorbing nearby free hadronic matters such as free neutrons from the vicinity of the quark core surface in absence of the Coulomb barrier. Besides the formation of quark star, an outer crust of comparatively lower density is also developed outside the quark star, that does not consist of any free neutrons \cite{Witten}. The density of such an outer crust is $\sim 1/1000$ of the average quark matter density of the star. As a result, the outer crust does not contribute remarkably to the estimation of the limiting mass of quark stars as well as the corresponding radius of the star. The newly generated quark star consists of two quark flavors (i.e. `u' quark and `d' quark), as this phase is developed by capturing the non-hyperonic baryons (the constituents of the parent neutron star). But later, the strangeness may appear in this quark state of two flavors via $ud \rightarrow us$ weak interaction process by taking in the free energy of the star (in excess of `d' quark) \cite{ns_to_qs, alcock} and thus turn into a strange quark star (SQS). The strange quark star is essentially composed of three quark flavors (SQS) confined in a hypothetical large bag \cite{Bag_Model_2,Bag_Model_1}, which is characterized by the parameter Bag Constant (SQS contains `s' quarks  (strange) along with the `u' and `d' flavours of quark). The matter with strange quarks (`s' quark) can be treated as the perfect ground state of the strongly interacting matter as indicated by {\bf Witten} \cite{Witten,Jaikumar:2004zy}. Moreover, there exist some alternative theories \cite{Strange_qs_1,Strange_qs_2,weber_structure} , those support the conjecture of strange quark stars (see Ref.~\cite{Strange_Q_Star}). However, the quark star may also be developed by the accumulation of ambient quarks by gravity like other ordinary stars. This formalism is possible only if there exist a huge fraction of strange quarks \cite{Witten} nearby. Such ``pure'' quark stars \cite{Pure_qs} might belong to hypothetical quark galaxies or might be generated by accumulating strangelets \cite{Quark_galaxy,qg1,qg2}. Primordial strange stars are another possible source of quark stars. According to this assumption, quark stars were generated as an outcome of the quark - hadron phase transition in the early epoch of the Universe. But in order to maintain the stability, those primordial quark stars may transform into strange quark stars, as discussed in the work of {\bf Witten} \cite{Witten}.
	
	In the case of an ordinary star, the almost entire mass of the star is coming from the contribution of its constituent atoms. But for a quark star without strangeness (only `u' and `d' quarks), the mass of the star comes from the effective mass of the quarks which are considered to be almost massless, this is the key difference between the quark stars and ordinary stars. As quark matters are expected to be generated from the hadron-quark transformation in extremely compact astrophysical bodies (in general neutron stars), the maximum allowed mass for such objects is expected to be similar to the limiting mass of neutron stars. As a consequence, it is possible to estimate the total mass of quark stars by solving the Tolman-Oppenheimer-Volkov (TOV) equations numerically \cite{tov,lm1,lm2} by incorporating the conjecture as described by {\bf Witten} \cite{Witten}. In several recent numerical analyses \cite{rotlm,rotlm_2}, the maximum allowed mass for fast rotating quark stars is also investigated using the similar formalism as used for the static quark stars. However, in the literature, there is no argument that addresses the contribution of fundamental constants in the estimation of limiting mass of quark stars, like ordinary compact objects \cite{shapiro}. 
	Moreover, unlike a gravitationally bound neutron star, quark stars are self-bounded by the strong interacting field (see Ref.~\cite{Jaikumar:2004zy,def_agnst}). As a result, it is not necessary to consider a radial distance-dependent density of the star. Although, quark stars are confined compact celestial objects, the density of such stars hover near the nuclear density. So, in this case, bag constant possibly covers appropriately the strong interaction section, particularly in view of the density-dependent modeling (QMD) of the quark stars. A significant feature of this formalism is that the quark state is almost independent of any specific EOS (equation of state) models and the entire dynamics is assumed to come from confinement mechanism for the density-dependent model.
	In the work of {\bf Banerjee} {\it et. al.} \cite{SBa}, an analytical computation was carried out for the estimation of the limiting mass of static quark stars. In this treatment, the simple energy balancing picture was adopted as proposed by Landau \cite{Landau} and the outcomes from the analysis are found to be dependent on the MIT bag constant \cite{Bag_Density,mmhs} and the fundamental constants. In the following work, we search for such theoretical limits (mass, radius etc.) for the rotating quark stars and also measure the maximum possible observed frequency of such compact rotating celestial bodies. The variation of the limiting mass and radius has also been studied with the rotational frequency and the bag constant of the star in the theoretical regime.
	
	As the quark stars are considered to be self-bound by strong interaction \cite{Jaikumar:2004zy,def_agnst} (strong interacting field is significantly stronger than the gravitational field), such stars can withstand high rotational frequency without getting significant deformation (the maximum possible ellipticity $\lesssim 10^{-4}-10^{-7}$ \cite{def_sphere_1,ushomirsky2000} even for millisecond stars (millisecond time period of rotation)). So, the modeled quark stars in the present analysis are considered to be spherical and rigid. In addition, quark stars are massive and highly dense, so the volume of such quark stars are no longer a Euclidean sphere (the volume of the star $\neq \frac{4}{3}\pi \rm{Radius}^3$) due to the space-time curvature. As a consequence, in the calculation of the volume of the star, the general relativistic correction is considered.
		
	\section{Bag Constant \label{sec_bag_qs}}
		In the study of hadronic structure, the MIT Bag model has gained immense success \cite{bag_sp1,bag_sp2,bag_sp3,TNPI}. According to this phenomenological model, point-like and massless quarks are assumed to be confined in a hypothetical bag and the corresponding state of the system is characterized by the parameter named bag constant $\mathcal{B}$.
		
		The bag constant basically measures the difference of the vacuum energy densities for the non-perturbative ground state to the same for the perturbative ground states of the quarks \cite{Bag_Density}. The bag constant $\mathcal{B}$ essentially depends on the number density (and temperature, in general) of the quarks. In the case of quark stars, the entire star is considered to be such a hypothetical bag, that contains all the quarks containing the star.
		
		
	\section{Fermi Energy \label{sec_ef_qs}}
		By adopting a simple picture of energy balancing, the maximum mass of a rotating QS can be estimated, as described by {\bf Landau} \cite{Landau}. According to this formalism, the total amount of energy contains per fermion (or quark) attains a minima at the limiting case, where the total energy per fermion is given by,
		\begin{equation}
			e=e_f+e_G+e_{rot},
		\end{equation}
		In the above equation, the terms $e_f$, $e_G$ and $e_{rot}$ are representing the Fermi energy per particle, gravitational energy per particle and the kinetic energy per fermions due to the spinning of the star respectively.  
		In a system of non-interacting fermions, the Fermi energy density indicates the maximum occupied energy per unit volume by the fermions at the ground state \cite{Fermi_Hyperphysics}. In the case of the quark star, a significant fraction of the total mass of the star comes from the contribution of the Fermi energy, which is mainly determined by the radius of the star ($R$) as well as the fermion number density ($n$). The Fermi energy density can be written as a function of the chemical potential ($\mu$) of the system, described as \cite{SBa}, 
		\begin{equation}
			\mathcal{E}_{f}=\frac{g}{8\pi^{2}}\mu^{4},\label{eq:1}
		\end{equation}
		where $g$ denotes the statistical degeneracy factor for the quark matter system, which contains $N$ number of fermions and $\mu$ is the chemical potential, given by,
		\begin{equation}
		\mu=\left(\frac{9\pi}{2g}\right)^{\frac{1}{3}}\frac{N^{\frac{1}{3}}}{R}.\label{eq:2}
		\end{equation}
		From Eqs.~\ref{eq:1} and \ref{eq:2}, the final expression of the Fermi energy per fermions can be obtained, given by,
		\begin{equation}
		e_{f}=\frac{\mathcal{E}_{f}}{n}=\frac{3}{4}\left(\frac{9\pi}{2g}\right)^{\frac{1}{3}}\frac{N^{\frac{1}{3}}}{R}\label{eq:3}.
		\end{equation}
		
	\section{Effective Mass Per Fermions \label{sec_emass}}
		The effective quark mass is a significant quantity, in the estimation the rotational kinetic energy and the gravitational potential energy and hence the total mass of the star. As we assumed the quark star as completely spherical and rigid, the effective mass of the entire quark star ($M$) can be expressed as,
		\begin{equation}
			M=e_{f}N+V\mathcal{B}.\label{eq:MM}
		\end{equation} 
		where $\mathcal{B}$ is the bag constant and $V$ represents the volume of the star.
		Now extremizing the total mass of the star $M$ with respect to the radius $R$, one can reduced the form of the bag constant ($\mathcal{B}$) by simplifying further.
		\begin{equation}
			\mathcal{B}=\frac{e_f N}{3 V},\label{bag}
		\end{equation}
		Now putting the above form of bag constant ($\mathcal{B}$) in the Eq.~\ref{eq:MM}, the total mass ($M$) of the star can be obtained in the most simplified form, given by
		\begin{equation}
			M=4V\mathcal{B},\label{eq:total mass}
		\end{equation}
		and hence the effective mass of each fermion (quark) inside the star (hypothetical bag in the bag model), as
		\begin{equation}
			m=\frac{M}{N}=\frac{4V\mathcal{B}}{N}.\label{eq:12}
		\end{equation}
		
		As the constituent of the quark star are fermions, the effective mass of the almost mass less quarks coincide with the chemical potential of the quark. Therefore, at the limit of vanishing quark density \cite{SBa,Fowler_Raha_Weiner,Fowler_Raha_Weiner_1} one can write, 
		\begin{equation}
			\begin{array}{cccc}
			&\mu&=&\mathcal{B}/n\\
			or,&n&=&\mathcal{B}/\mu.
			\end{array}
			\label{eq:13}
		\end{equation}
		Now applying Eq.~(\ref{eq:2}) and (\ref{eq:13}) in Eq.~(\ref{eq:12}), the desired expression for the effective mass per quark is estimated in terms of the bag constant and the fundamental constants.

	\section{Limiting Mass and Radius \label{sec_energy_qs}}
		In the present analysis, the limiting mass for the fast spinning quark star is studied, which is introduced here as ``Chandrasekhar limit for rotating quark stars". The rotational kinetic energy per quark at the $(R,\theta,\phi)$ position (in spherical polar coordinate system) with respect to the centre of mass of the star and the gravitational potential per particle, can be expressed in terms of effective mass per particle as,
		\begin{eqnarray}
			e_{G}&=&-G\dfrac{Mm}{R}\\
			e_{\rm{rot}}&=&mc^2(\gamma-1).
		\end{eqnarray}
		where $c$ is the speed of light in space and $\gamma=\dfrac{1}{\sqrt{1-(\frac{2\pi\omega R\:cos\:\theta}{c})^{2}}}$. From the above expressions it can be noticed that, unlike the expressions of the Fermi energy and the gravitational potential energy per particle, the kinetic energy per quarks are found to vary over the latitude of the star, $\theta$ for a fixed radial distance $R$. Therefore, in order to terminate the $\theta$-dependence from the kinetic energy term ($e_{\rm rot}$) in further computations, a new kinetic energy term ($\langle e_{\rm{rot}} \rangle$) is introduced, which is averaged over the latitude of the star $\theta$. Now the total energy per fermion becomes
		\begin{equation}
			e=e_{f}+e_{G}+\langle e_{\rm{rot}} \rangle.\label{eq:16}
		\end{equation}
		As described in the energy balance picture by Landau, the limiting mass ($M_{\text{max}}$) of the star and the corresponding limiting radius ($R_{\text{max}}$) can be estimated by extremizing the total energy per quarks ($e$) with respect to the number of quarks containing the star ($N$). Although the modeled quark star is assume to be spherical in our entire calculation, the volume of the star $V$ is not the same as one obtain for an Euclidean sphere $\left({\rm i.e.}\;\frac{4}{3}\pi R^3\right)$, because those highly compact stars are extremely massive and dense \cite{synge}. So, the effect of general relativistic in the stellar volume is not ignoreable for the current calculation. Therefore we use relativistic form of the volume in further calculation \cite{hobson}, given by
		\begin{equation}
			dV = \sqrt{|g_{\mu \nu}|}  dr d\theta d\phi,
		\end{equation}
		where ($r,~\theta,~\phi$) is the position of any point with respect to the centre of the star in spherical polar coordinate system and $g_{\mu \nu}$ denotes the $\mu \nu^{\rm{th}}$ component of the space-time metric.
		
		The limiting radius and the corresponding mass of quark stars, which are estimated by extremizing the total energy $e$ per particle, is independent to the statistical degeneracy factor of the system $g$ (see Ref.~\cite{SBa}, the results hold also for rotating stars). So, this treatment is viable for both the 2-flavored normal quark stars and the 3-flavored strange quark stars. It can be seen that the maximum allowed radius, mass and the corresponding total number of quarks containing the star $N$ are essentially depended on the rotational frequency as well as the bag constant.
		\begin{figure}
			\centering{}
			\includegraphics[trim=20 30 20 55, clip,width=0.7\columnwidth]{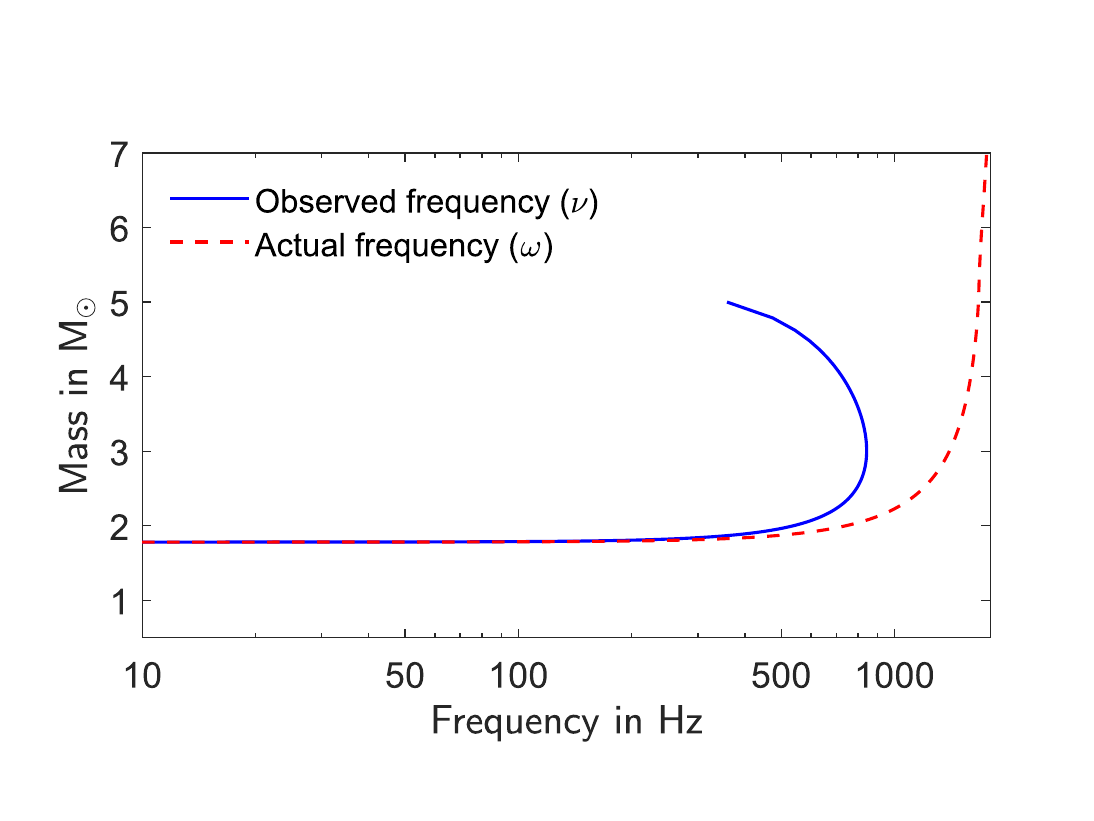}
			\caption{\label{fig:mmax vs freq2} Variation of the maximum allowed mass of the quark star ($M_{\text{max}}$) with the rotational frequency. The solid blue line denotes the variation of $M_{\text{max}}$ for the actual frequency of the star $\omega$ while the dashed red line represents the same for observed redshifted frequency ($\nu$).}
		\end{figure}
		The spinning effect is parameterized by the rotational frequency ($\omega$) of the quark star. But, the extreme compactness (radius to mass ratio of the star $\leq 2.2903$, see Figure~\ref{fig:max nu}c) and mass of such stars, the general relativistic effects turned out to be very prominently in the observed frequency ($\nu$), as observed by a far away observer \cite{gr_freq_2,gr_freq_1}. So, a far away observer would detect the redshifted form of the actual rotational frequency ($\omega$), which cen be expressed as
		\begin{equation}
			\nu=\omega \sqrt{1-\frac{2 G M}{c^2 R}},
		\end{equation}
		where, $M$ and $R$ are the mass of the star and its corresponding radius respectively. 
		As a consequence, a notable deviation in observed between the actual frequency $\omega$ and the observed frequency ($\nu$) for fast spinning stars, as described in Figure~\ref{fig:mmax vs freq2}. The variation of the limiting mass $M_{\rm{max}}$ with the actual frequency ($\omega$) and the observed redshifted frequency ($\nu$) are plotted in Figure~\ref{fig:mmax vs freq2} for a fixed bag constant $\mathcal{B}=(145~\rm{MeV})^4$.
		
		\begin{table}
			\centering{}
			\begin{tabular}{l c l}
				\hline \hline
				Object & Mass in $\rm{M}_{\odot}$ & Ref.\\
				\hline
				PRS J1614-2230 & $1.928^{+0.017}_{-0.017}$ & \cite{J1614-2230, Demorest}\\
				PSR J0348+0432 & $2.01^{+0.04}_{-0.04}$ & \cite{J0348+0432}\\
				PSR J0740+6620 & $2.14^{+0.10}_{-0.09}$ & \cite{J0740+6620}\\
				PSR J1311-3430 & $2.15-2.7$ & \cite{J1311-3430}\\
				PSR B1957+20 & $2.40^{+0.12}_{-0.12}$ & \cite{B1957+20}\\
				PSR J1600-3053	& $2.30^{+0.70}_{-0.60}$ & \cite{J1600-3053_1,J1600-3053}\\
				PSR J2215+5135	& $2.27^{+0.17}_{-0.15}$ &	\cite{J2215+5135_1,J2215+5135}\\
				PSR J0751+1807 & $2.10^{+0.20}_{-0.20}$ & \cite{J0751+1807,J0751+1807_1}\\
				PSR B1516+02B & $1.94^{+0.17}_{-0.19}$ & \cite{B1516+02B}\\
				\hline \hline
			\end{tabular}
			\caption{\label{tab:stars} Newly discovered massive pulsars.}
		\end{table}
		
		\begin{figure*}
			\centering{}
			\includegraphics[trim=0 40 30 60, clip,width=\textwidth]{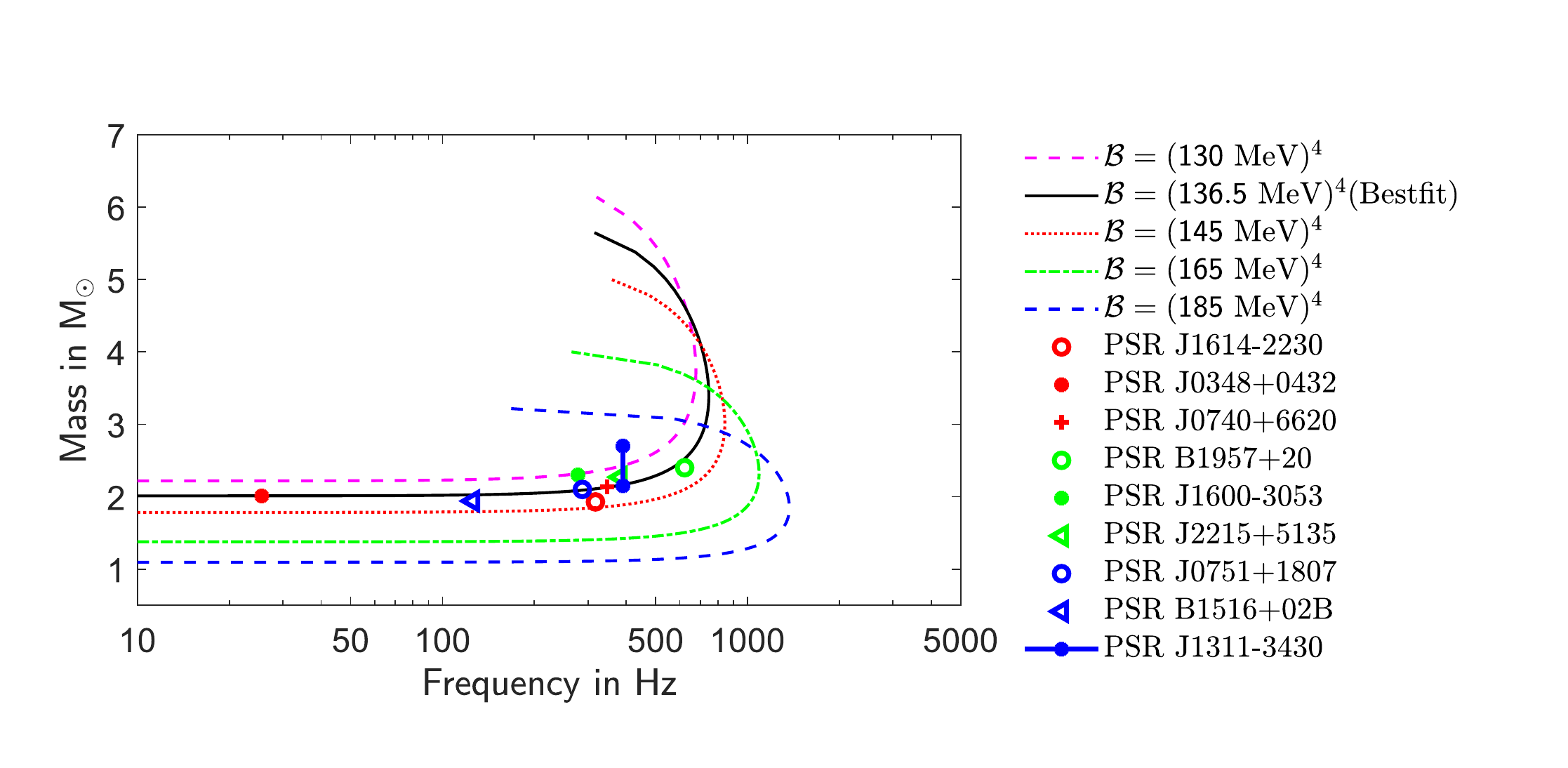}
			\caption{\label{fig:mmax nu} Variation of the limiting mass ($M_{\text{max}}$) of quark star with the observed rotational frequency ($\nu$) for different chosen values of ($\mathcal{B}$).}
		\end{figure*}

		\begin{figure}[h!]
			\centering{}
			\includegraphics[trim=0 40 30 60, clip,width=0.7\columnwidth]{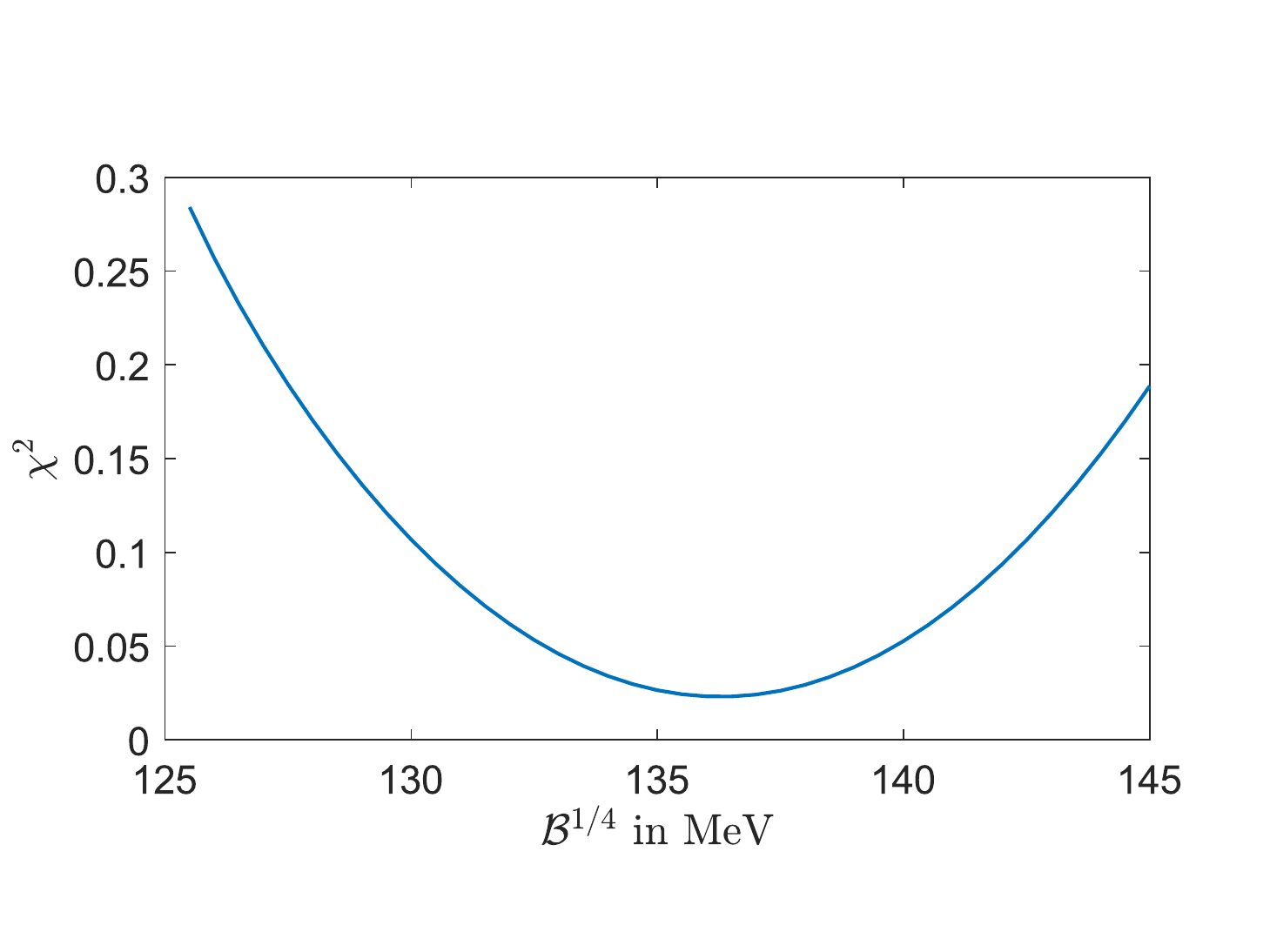}
			\caption{\label{fig:chi2} The $\chi^2$ analysis of bag constant $\mathcal{B}$ using recent the observational data of fast spinning stars.}
		\end{figure}
		
		\begin{figure*}[]
			\centering{}
			\begin{tabular}{c}
				\includegraphics[trim=15 30 20 50, clip,height=0.4\linewidth,width=0.65\columnwidth]{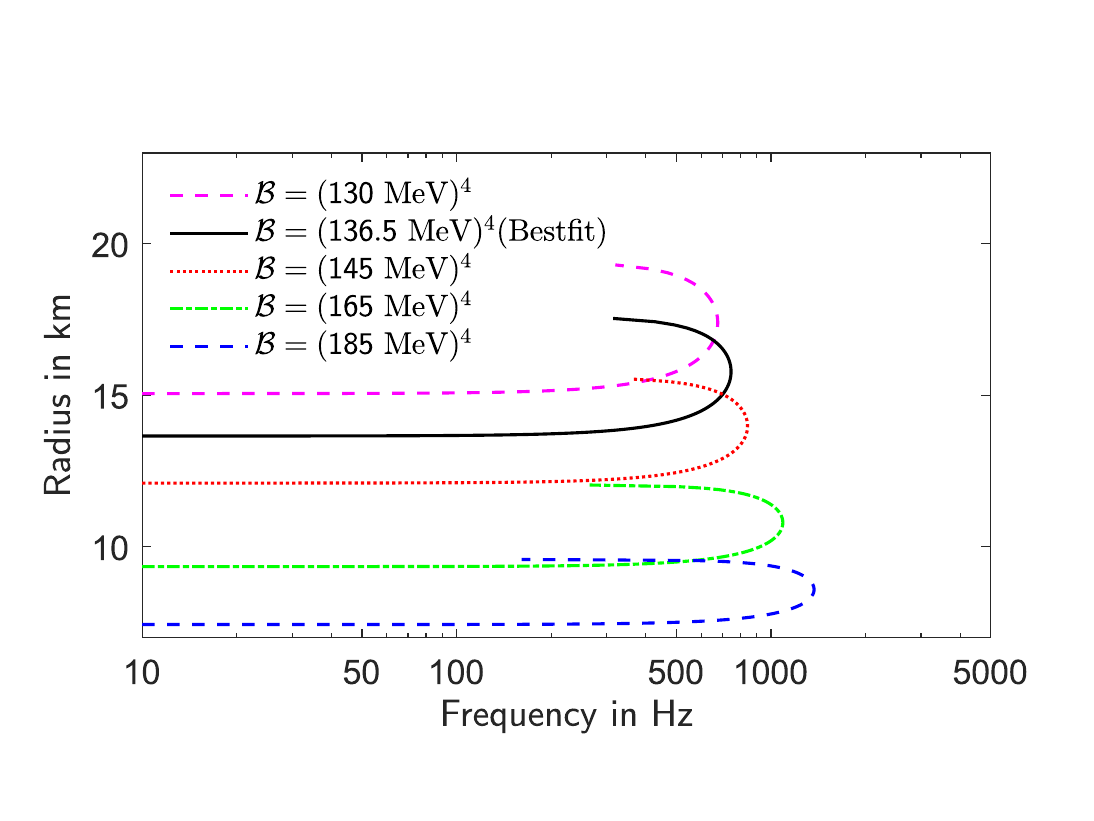} \\(a)\\
				\includegraphics[trim=20 30 20 50, clip,height=0.4\linewidth,width=0.65\columnwidth]{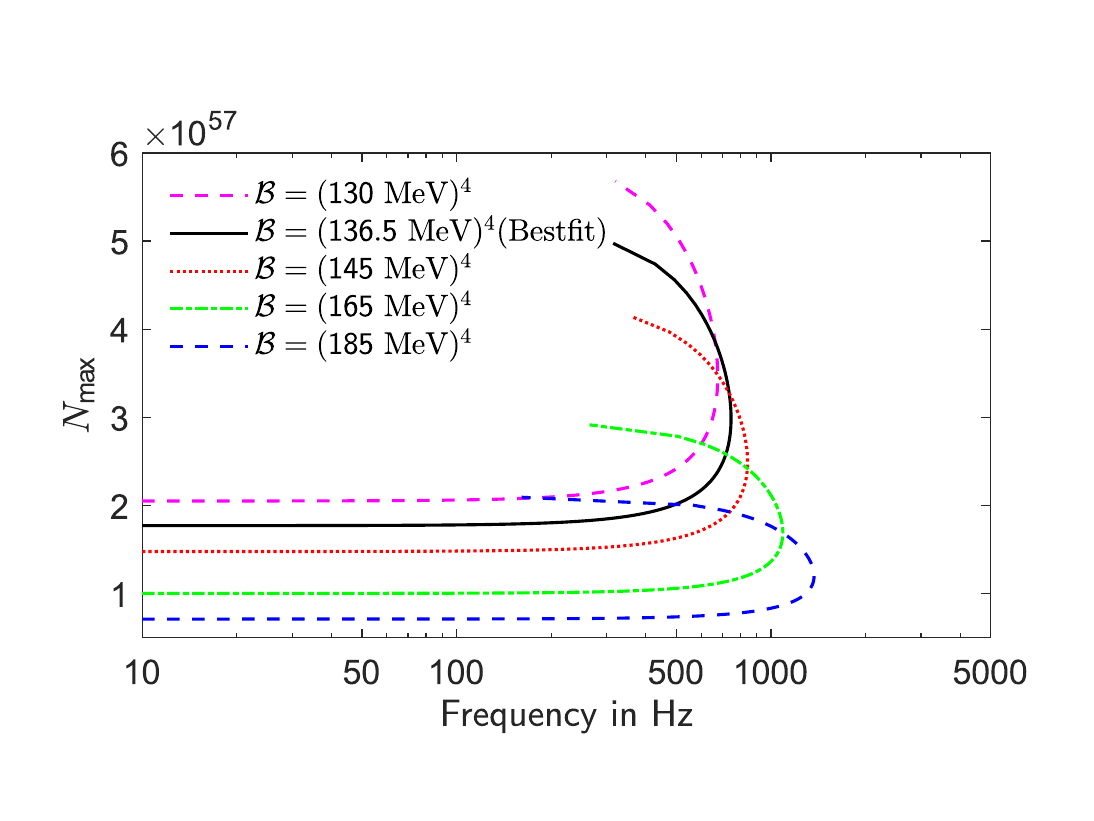}\\
				(b)\\
				\includegraphics[trim=0 30 20 50, clip,height=0.4\linewidth,width=0.65\columnwidth]{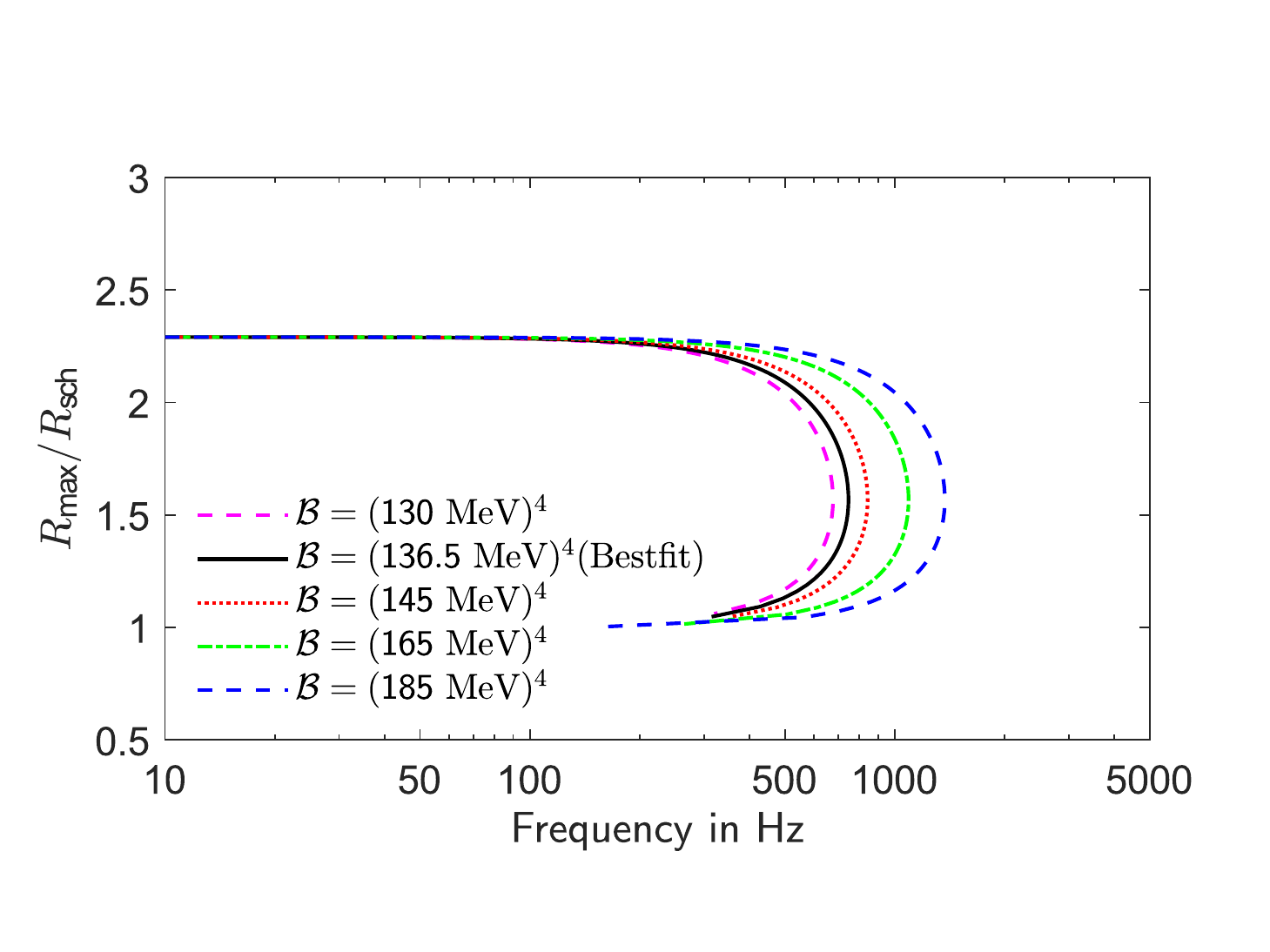}\\
				{(c)}\\
			\end{tabular}
			\caption{\label{fig:max nu} The variation of (a) the maximum allowed radius ($R_{\text{max}}$), (b) the number of quarks containing the quark star $N_{\text{max}}$ and (c) the radius to mass ratio with observed rotational frequency $\nu$ for four different chosen values of $\mathcal{B}$ and for the best-fit value of bag constant.}
		\end{figure*}
	
		From Figure~\ref{fig:mmax nu} it can be observed that, for individual chosen bag constant $\mathcal{B}$ values, the maximum allowed masses of the star and the corresponding radius of the star remains almost independent to the rotation for lower frequencies. However, as the observed rotational frequency ($\nu$) reaches $\sim 300$ Hz, the contribution of the kinetic energy turns out to be substantially high that, the kinetic energy becomes considerable in the estimation of the effective mass of quarks. As a consequence, the limiting mass $M_{\text{max}}$ starts raising gradually with frequency beyond 300 Hz. Above 600 Hz, the value of $M_{\text{max}}$ suffers a rapid increase. Beyond a certain value of $M_{\text{max}}$ the star becomes highly compact and massive, so that the observed frequency of the star ($\nu$) starts decreasing with increasing actual frequency $\omega$ as an outcome of general relativistic effect. From this phenomena, a maximum bound of the observed frequency $\nu$ (i.e. $\nu_{\text{max}}$) can be obtained for individual chosen values of bag constant (see Figure~\ref{fig:mmax nu}). The observed frequencies and masses of some recently discovered fast spinning massive compact stars (see Table~\ref{tab:stars}) are also plotted in the Figure~\ref{fig:mmax nu} and compared with the calculated observed frequencies and limiting masses for five chosen values of bag constant (one of which is obtained from the $\chi^2$ analysis). It is to be mentioned that, all the observed compact stars or pulsars, which are tabulated in Table~\ref{tab:stars}, are possibly belong to the neutron star category, having or without having a quark core, rather than a pure quark star. However as the estimated average densities of such compact celestial bodies are very close to the nuclear density, those stars are comparable to the calculated results from the present work. In the present analysis, the $\chi^2$ is defined as $\chi^2=\displaystyle\sum_{i}\left(\dfrac{M_{\rm{cal}}-M_{\rm{obs}}}{M_{\rm{cal}}}\right)^2$, where, $M_{\rm{obs}}$ is the observationally estimated mass of the newly discovered fast spinning stars, which are tabulated in Table~\ref{tab:stars}, while $M_{\rm {cal}}$ denotes the calculated mass from the present treatment. From this $\chi^2$ analysis the best-fitted value of bag constant $\mathcal{B}$ is found $\approx$(136.5 MeV)$^4$ (as shown in Figure~\ref{fig:chi2}). In this analysis, the contribution of PSR J1311-3430 is not considered, as the uncertainty in the mass of this star is substantially large. In Figure~\ref{fig:mmax nu}, the variation of limiting mass with the observed frequency for the best-fit value of $\mathcal{B}$, is plotted using the black solid line. The corresponding variation of $R_{\text{max}}$ and the total number of quark confined in the star ($M_{\rm{max}}$) are shown in  Figure~\ref{fig:max nu}a and Figure~\ref{fig:max nu}b respectively. In both figures, the maximum limiting frequencies ($\nu_{\text{max}}$) can be seen for all chosen values of bag constant $\mathcal{B}$.
	
	\section{Limiting Frequency \label{lim_freq}}
\begin{figure}[h!]
			\centering{}
			\includegraphics[trim=0 40 30 60, clip,width=0.7\columnwidth]{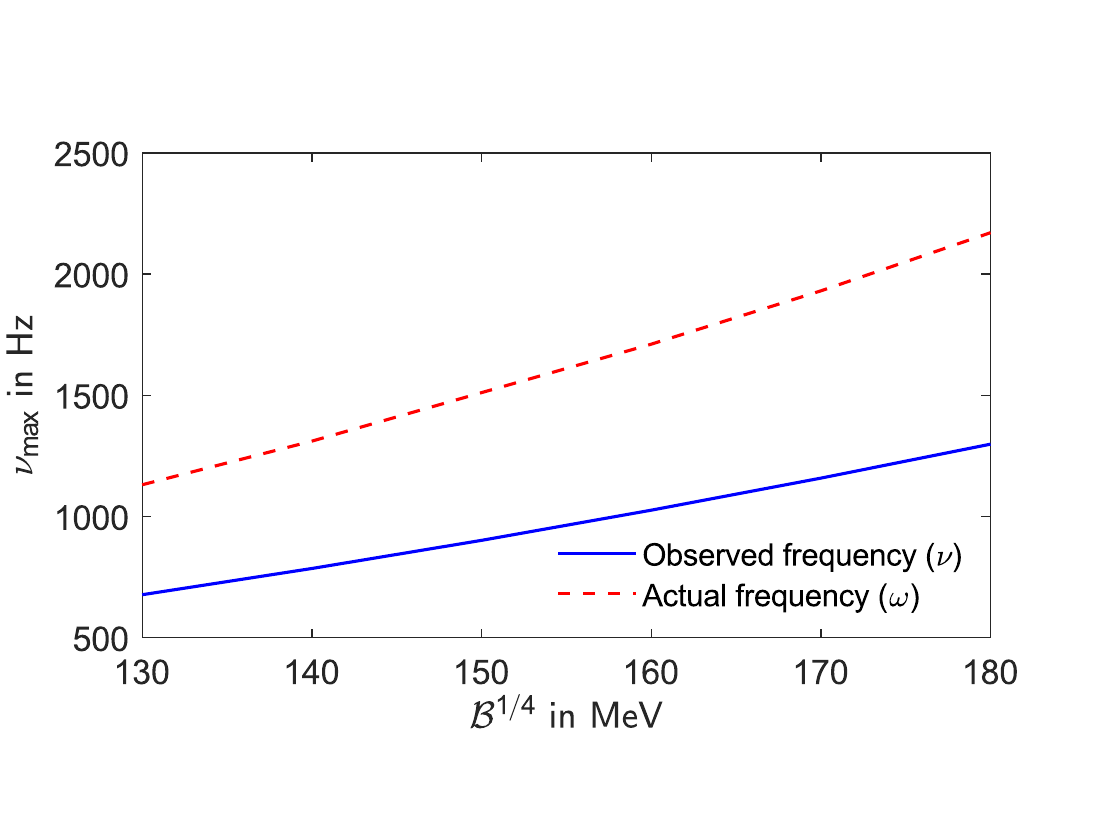}
			\caption{\label{fig:nu_max bag-1} The variation of the observed limiting frequency ($\nu_{\text{max}}$) and corresponding actual frequency $\omega$  of the star with $\mathcal{B}^{1/4}$.}
		\end{figure}

		The limiting frequencies ($\nu_{\text{max}}$) for quark stars are found to vary with different chosen values of $\mathcal{B}$. The variation of the limiting frequency $\nu_{\text{max}}$ with bag constant is described in Figure~\ref{fig:nu_max bag-1}, indicating that, the limiting observed frequency $\nu_{\text{max}}$ changes almost linearly with the bag constant $\mathcal{B}^{1/4}$.
		Recent numerical simulations \cite{bag_sp1,bag145,lim_bag} indicate that, in the case of stable system of quark matters, the numerical value of the bag constant lies within the range $\sim (130~\rm{MeV})^4-(162~\rm{MeV})^4$. This possible allowed range of $\mathcal{B}$ corresponds to the maximum observed frequency ($\nu$) within the range $677.4\sim 1052.6$ Hz (see Figure~\ref{fig:nu_max bag-1}). Therefore, this span of $677.4\sim 1052.6$ Hz can be considered as the possible upper-limit for the observed frequency ($\nu$) of rotating quark stars (millisecond order) \cite{atnf}. This maximum possible observed frequency ($\nu_{\max}$) agrees with recent observations \cite{J1614-2230,J0740+6620,J1311-3430,B1957+20,J1600-3053,J2215+5135,J0751+1807,atnf}. On the other hand, that range of bag constant ($(130~\rm{MeV})^4 \sim (162~\rm{MeV})^4$) corresponds to the actual frequency range ($\omega$) $1131 \sim 1755.0$ Hz, which fits well with the result of the numerical treatment by {\bf Gourgoulhon} \emph{et~al.} \cite{upperlim_fr}. For example, the best fitted values (from our $\chi^2$ analysis) of the bag constant ($\mathcal{B}=(136.5~\rm{MeV})^4$) corresponds to the limiting observed frequency ($\nu_{\text{max}}$) of quark star is $\sim 747.7346$ Hz. This limit agrees with the observational results for high frequency pulsars) and the corresponding value of the actual frequency ($\omega_{\text{max}}$) is $\sim 1248$ Hz.
		
		In the current analysis, a significant quantity has been investigated given by the quantity $\dfrac{R_{\text{max}}}{R_{\rm{sch}}}$ (radius to Schwarzschild radius ratio), i.e. inverse of the compactness of the star. The variation of the term $\left(\dfrac{R_{\text{max}}}{R_{\rm{sch}}}\right)$ with the observed rotational frequency ($\nu$) have been investigated for different chosen values of the bag constants (see Figure~\ref{fig:max nu}c). At the low-frequency region ($\omega<200$ Hz), the value of this ratio becomes independent to the variation of bag constant $\mathcal{B}$ and remains almost constant ($\sim 2.2903$ in the geometric unit system, in the work of {\bf Banerjee} \emph{et~al.} \cite{SBa}, the estimated ratio was 8/3, where the relativistic effects were not taken into account) with frequency (Figure~\ref{fig:max nu}c). However it starts decreasing gradually as the observed frequency of the star reaches $300$ Hz. In the case of $\nu \geq 300$, a very small dependence on $\mathcal{B}$ in the limiting mass can be observed. At further higher frequencies, all the $\dfrac{R_{\text{max}}}{R_{\rm{sch}}}$ vs frequency $\nu$ curves suffer a rapid fall toward unity for different sets of bag constants, but never become unity. From the present analysis, it can be noticed that the quark stars are extremely compact, however they can never collapse into a black hole, even for millisecond stars. Thus such compact celestial bodies (strange quark stars) and the black holes could co-exist as candidates of cold dark matter, where their phenomenological signatures may not be sufficient to distinguish between them. But their different signatures in the gravitational wave scenario may be interesting to study \cite{ssrmrp}.
		

\chapter{Bounds on the Abundance of PBH and Dark Matter from 21-cm Signal} \label{chp:21_feb}
\\
	Primordial black hole (PBH) is another hypothetical candidate of compact astrophysical objects, which has been the centre of interest for several decades in the field of cosmology and astrophysics. As PBHs are assumed to be substantially smaller in size, the effects of the evaporation (Hawking radiation \cite{hawking}) of such black holes (BHs) are very prominent. So, the phenomenon of PBH evaporation can be a promising tool in the estimation of the abundance of this hypothetical candidate of the black hole. The emitted particles in the form of Hawking radiation interact with the medium and hence modify the temperature evolution of the IGM (intergalactic medium). Those modifications can be estimated by looking over several astrophysical phenomena, e.g. the global 21-cm hydrogen absorption spectrum.
	
	
	The redshifted signature of the 21-cm radio signal has turned out to be a significant probe in the investigation of the dynamics of the Universe, essentially in the cosmic Dark Age, which started immediately after the recombination and continues till the reionization. The combined effects of the evaporation of primordial black holes (PBH) and the cooling of baryonic matter due to baryon - dark matter (DM) collisions fluid perturb the redshifted 21-cm absorption spectrum remarkably. As a consequence, the variation of the brightness temperature ($T_b$) also shows significant dependence on the mass of the DM particles ($m_{\chi}$) and the baryon - DM scattering cross-section ($\sigma_0$) besides PBH mass as well as its initial mass fraction ($\beta_{\rm BH}$). In this context, the bounds (both upper bound and lower bound) on the primordial black hole initial mass fraction are estimated for different values of dark matter and primordial black hole masses. A similar investigation has also been carried out in the $m_{\chi}$ - $\sigma_0$ parameter space and addressed its variations with different PBH masses and initial mass fractions ($\beta_{\rm BH}$).
	
	

	\section{\label{sec:result_21_2} Global 21-cm signal}
		The dynamic of the Universe in the Dark Age is still unexplored due to the lack of luminous sources. The redshifted signature of the 21-cm ($\sim 1.42$ GHz) hydrogen absorption spectrum is a promising probe in understanding the dynamics of the early Universe, in particular during this unexplored dark age. The 21-cm hydrogen absorption spectrum opens up a new window to understand the process of reionization and the factors in the early Universe influencing the same. Thus, the study of the 21-cm line in reionization era helps in understanding several cosmological and astrophysical aspects regarding the primordial black holes \cite{BH_21cm_0,BH_21cm_1,BH_21cm_2,BH_21cm_4,BH_21cm_5,legal_1,Domingo1,Domingo2,Zhou:2021ygz} as well as the interaction between baryons and dark matters, dark matter - dark energy interaction \cite{Li,upala} and neutrino physics \cite{21cm_nu_1,21cm_nu_2} in the high redshifted epoch.
		
		The ``Experiment to Detect the Global Epoch of Reionization Signature'' (EDGES) \cite{edges} reported a prominent footprint of the 21-cm line of hydrogen at cosmic dawn\footnote{Cosmic dawn: when the first star ignite at the end of the cosmic dark age.} with $99\%$ confidence level (C.L.). As reported by the EDGES's experiment, the brightness temperature of the 21-cm hydrogen absorption spectrum at $14<z<20$ is $-500^{+200}_{-500}$ mK, here $z$ denotes the cosmological redshift. But, according to the standard cosmology, the brightness temperature of the global 21-cm signal is estimated only $\approxeq-200$ mK. Therefore, the observed additional cooling can be explained either by lowering the temperature of the baryonic fluid, which becomes almost equal to the neutral hydrogen spin temperature during the epoch of $14<z<20$, or by enhancing the cosmic microwave background temperature $T_{\gamma}$. The interaction between baryon - dark matter fluid is such a possible source that may cool down the baryonic fluid, resulting in the lower background temperature. The evaporation of primordial black holes (PBHs), dark matter decay/annihilation, even dark matter - dark energy interaction are also some notable cosmological and astrophysical phenomena, that may modify significantly the separation between $T_{\gamma}$ and $T_s$ \cite{rdi,legal_2,atri} during this epoch ($14<z<20$).
		
		\begin{figure}
			\centering{}
			\includegraphics[width=\linewidth]{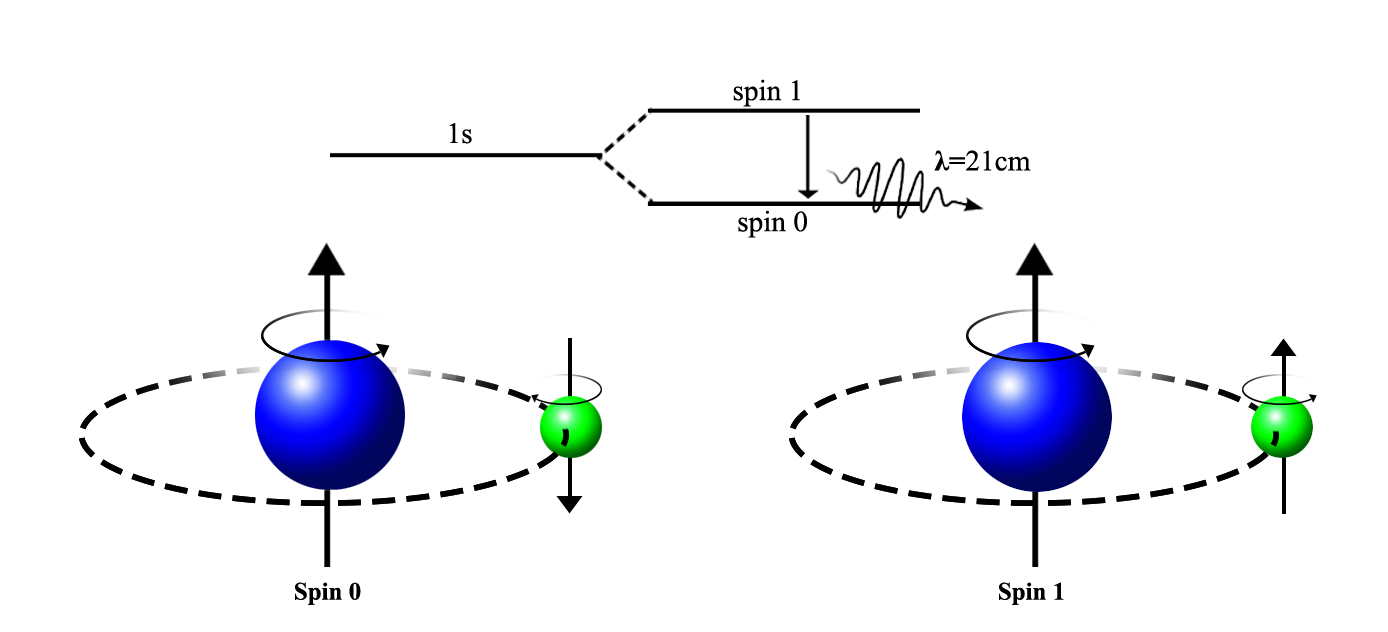}
			\caption{\label{fig:h_spin} The 21-cm transition of hydrogen atom.}
		\end{figure}
		
		Hydrogen is the most abundant baryonic component of the Universe, which occupies $\sim 75\%$ of the entire baryonic mass. The 21-cm spectrum appears as a result of the transition between two hyperfine spin-state of neutral hydrogen atoms, namely a singlet spin 0 state and a triplet spin 1 state. The expression of the brightness temperature of the 21-cm neutral hydrogen spectrum at different $z$ can be written as,
		\begin{equation}
		T_{21}=\dfrac{T_s-T_{\gamma}}{1+z}\left(1-e^{-\tau(z)}\right).
		\label{eq:t21}
		\end{equation}
		In the above equation, $T_s$ is the spin temperature of the baryonic medium at redshift $z$ and $\tau(z)$ denotes the optical depth of the medium, described as \cite{munoz},
		\begin{equation}
		\tau(z) = \dfrac{3}{32 \pi}\dfrac{T_{\star}}{T_s}n_{\rm HI} \lambda_{21}^3\dfrac{A_{10}}{H(z)+(1+z)\delta_r v_r},
		\label{eq:tau}
		\end{equation}
		where $A_{10}$ is the Einstein coefficient \cite{yacine,hyrec11} given by $A_{10}=2.85\times 10^{-15}\,{\rm s^{-1}}$, $T_{\star}=hc/k_B \lambda_{21}=0.068$ K is the temperature that corresponds to the wavelength of 21-cm spectrum, $\lambda_{21}\approx 21$ cm and $\delta_r v_r$ represents the radial gradient of the peculiar velocity.
		
		$T_s$ of the medium can be described as, 
		\begin{equation}
		\dfrac{n_1}{n_0}=3 e^{-T_{\star}/T_s},
		\end{equation}
		where $n_0$ and $n_1$ are the number densities of the neutral hydrogen atoms at the singlet spin-0 and the triplet spin-1 states respectively. The resonant scattering of Lyman$\alpha$ (Lyman-$\alpha$) photons and the background photons perturbed the spin temperature. In addition, Hawking radiation from PBHs and baryon-DM scattering also modify $T_s$ remarkably. The expression of $T_s$ as function of cosmic microwave background (CMB) temperature $T_{\gamma}$ ($=2.725\times (1+z)$ K) and baryon temperature $T_b$ is given by,
		\begin{equation}
		T_s = \dfrac{T_{\gamma}+y_c T_b+y_{\rm Ly\alpha} T_{\rm Ly\alpha}}
		{1+y_c+y_{\rm Ly\alpha}}.
		\label{eq:tspin}
		\end{equation}
		In the above expression, $y_{\rm Ly\alpha}$ describes the Wouthuysen-Field effect. $T_c$ and $T_{\rm Ly\alpha}$ are the collisional coupling parameter and the Lyman-$\alpha$ background temperature respectively \cite{BH_21cm_1}. The coefficients $y_{\rm Ly\alpha}$ and $y_c$ are given by \cite{BH_21cm_2,Yuan_2010,Kuhlen_2006,Yang_2019}
		\begin{equation}
		y_{\rm Ly\alpha}=\dfrac{P_{10}T_{\star}}{A_{10} T_{\rm Ly\alpha}}e^{0.3 (1+z)^{1/2} T_b^{-2/3} \left(1+\dfrac{0.4}{T_b}\right)^{-1}}
		\end{equation}
		and 
		\begin{equation}
		y_c=\dfrac{C_{10}T_{\star}}{A_{10} T_b}.
		\end{equation}
		Here, $S_{\alpha}$ is the spectral distortion factor \cite{salpha}, $P_{10}\approx1.3\times 10^{-21}S_{\alpha}J_{-21}\,{\rm s^{-1}}$ which denotes the rate of deexcitation due to the Lyman-$\alpha$ photons, $C_{10}$ is the collision deexcitation rate \cite{c10} and $J_{-21}$ denotes the background intensity of the Lyman-$\alpha$ photons \cite{jalpha}. 
		

	\section{\label{sec:PBH_21_2} Effects of Hawking Radiation in Baryon Temperature}
		The Primordial Black Holes (PBHs) are considered to be a hypothetical candidate of black hole, which are assumed to be generated subsequently after the Big Bang (see \Autoref{sec:bhs} of \Autoref{chp:compact_obj}). The mass of PBHs can be significantly lower than stellar mass as they are not produced from gravitational collapse of compact stellar objects. A lower mass black hole is a possible source, that may heat up the medium before the reionization. The radiation of steady $\gamma$ and $e^{\pm}$ in the form of Hawking radiation \cite{BH_F} may affect significantly the brightness temperature of the global 21-cm signal. Several numerical simulations and theoretical works \cite{BH_21cm_0,BH_21cm_1,BH_21cm_2,BH_21cm_4,BH_21cm_5,legal_1,BH_21cm_3,amar21,10.1093/mnras/stt1493} verify the effect of the primordial black holes in this context. It has been shown by {\bf Clark} \emph{et~al.} \cite{BH_21cm_1} that, in 21-cm scenario, the Hawking radiation is equally significant as that of the dark matter decay.
		
		The mass evaporation rate due to Hawking radiation can be expressed as \cite{hawking,BH_21cm_1} (see \Autoref{sec:hawking} of \Autoref{chp:compact_obj})
		\begin{equation}
			\dfrac{{\rm d}M_{\rm{BH}}}{{\rm d}t} \approx -5.34\times10^{25} \left(\sum_{i} \mathcal{F}_i\right) \left(\dfrac{M_{\rm{BH}}}{\rm g}\right)^{-2} \,\,\rm{g/sec}
			\label{eq:PBH}
		\end{equation}
		where, $M_{\rm BH}$ is the black hole mass at any instant $t$ and the coefficient $\sum\limits_{i}\mathcal{F}_i$ denotes the sum over all fraction of evaporation of black hole in the form of $i^{\rm th}$ particle which depends on the black hole temperature $T_{BH}$ as \cite{BH_F},
		\begin{eqnarray}
			\sum_{i} \mathcal{F}_i&=&1.569+0.569 \exp 
			\left(-\frac{0.0234}{T_{\rm{BH}}} \right)+
			3.414\exp \left(-\frac{0.066}{T_{\rm{BH}}} \right)\nonumber\\
			&&+1.707\exp \left(-\frac{0.413}{T_{\rm{BH}}} \right)+ 
			1.707\exp \left(-\frac{1.17}{T_{\rm{BH}}} \right)\nonumber\\
			&&+ 1.707\exp \left(-\frac{0.11}{T_{\rm{BH}}} \right)+ 
			0.569\exp \left(-\frac{0.394}{T_{\rm{BH}}} \right)\nonumber\\
			&&+1.707\exp \left(-\frac{22}{T_{\rm{BH}}} \right)+
			0.963\exp \left(-\frac{0.1}{T_{\rm{BH}}}\right).
		\end{eqnarray}
		The black hole temperature $T_{\rm BH}$ can be approximated as $T_{\rm BH}\approx 1.05753 \times \left(M_{\rm BH}/10^{13}{\rm g}\right)^{-1}$ GeV (see Eq.~\ref{eq:bh_temp}). In the case of massive BHs, only the photon and electron channels are significant. However, as the PBHs of masses, $10^{14}\sim 10^{16}$ g are considered in the present case, the black hole temperatures ($T_{\rm BH}$) are substantially high to radiate in the form of pions, muons, quarks, gluons etc. \cite{BH_21cm_2,PhysRevD.41.3052,PhysRevD.94.044029}. As a consequence, alongside the photon and electron channels, other channels also contribute to the IGM heating by producing photons, electrons and positrons via subsequent cascade decay \cite{BH_21cm_2,PhysRevD.41.3052,PhysRevD.94.044029,chen}. The rate of energy injected into the medium in the form of Hawking radiation is given by \cite{BH_21cm_2},
		\begin{equation}
			\left.\dfrac{{\rm d} E}{{\rm d}V {\rm d}t}\right|_{\rm{BH}}=-\dfrac{{\rm d} M_{\rm{BH}}}{{\rm d} t} n_{\rm BH}(z)
		\end{equation}
		where, $n_{\rm{BH}}(z)$ denotes the number density of the primordial black holes at redshift $z$. The black hole number density $n_{\rm BH}(z)$ essentially depends on the cosmological redshift ($z$) and initial mass fraction of the PBHs ($\beta_{\rm BH}$) as \cite{BH_21cm_2},
		\begin{eqnarray}
			n_{\rm{BH}}(z)&=&\beta_{\rm BH}\left(\dfrac{1+z}{1+z_{\rm eq}}\right)^3 \dfrac{\rho_{\rm c,eq}}{\mathcal{M}_{{\rm BH}}} \left(\dfrac{\mathcal{M}_{\rm H,eq}}{\mathcal{M}_{\rm H}}\right)^{1/2} \left(\dfrac{g^i_{\star}}{g^{\rm eq}_{\star}}\right)^{1/12}\nonumber\\
			&\approx&1.46 \times 10^{-4}\beta_{\rm BH} \left(1+z\right)^3 \left(\dfrac{\mathcal{M}_{{\rm BH}}}{\rm g}\right)^{-3/2} {\rm cm^{-3}}
		\end{eqnarray}
		In this expression, $\mathcal{M_{\rm H}}$ denotes the horizon mass and $\mathcal{M_{\rm BH}}$ is the initial mass of the primordial black hole \cite{BH_21cm_2,BH_21cm_3,betabh,PhysRevD.81.104019}.

	\section{\label{sec:T_evol_21_2} Temperature Evaluation of Intergalactic Medium (IGM)}
			
		The temperature evolution of the charge-neutral Universe can be estimated by evolving the temperatures of the DM fluid ($T_{\chi}$) and the baryonic fluid ($T_b$) with redshift. Now including the effects of the heating / cooling of baryonic and DM fluid, due to the dark matter scattering with the baryons and PBH evaporation, the evolution equations become  \cite{BH_21cm_1,BH_21cm_2,munoz,amar21,corr_equs},
		\begin{equation}
			(1+z)\frac{{\rm d} T_\chi}{{\rm d} z} = 2 T_\chi - \frac{2 \dot{Q}_\chi}{3 H(z)}, 
			\label{eq:T_chi}
		\end{equation}
		\begin{equation}
			(1+z)\frac{{\rm d} T_b}{{\rm d} z} = 2 T_b + \frac{\Gamma_c}{H(z)}
			(T_b - T_{\gamma})-\frac{2 \dot{Q}_b}{ 3 H(z)}-\frac{2}{3 k_B H(z)} \frac{K_{\rm BH}}{1+f_{\rm He}+x_e}.
			\label{eq:T_b}
		\end{equation}
		In the above equation (Eq.~\ref{eq:T_b}), the last term 
		denotes the contribution of the Hawking radiation from the evaporating PBHs, in energy deposition into the baryonic medium \cite{BH_21cm_1,BH_21cm_2,amar21}. In this expression (Eq.~\ref{eq:T_b}), $\Gamma_c$ represents the effect of the Compton scattering given by,
		\begin{equation}
			\Gamma_c=\frac{8\sigma_T a_r T^4_{\gamma}x_e}{3(1+f_{\rm He}+x_e)m_e c},
			\label{eq:gamma_c}
		\end{equation}
		where $\sigma_T$ is the cross-section for the Thomson scattering and  $a_r$ denotes the radiation constant. $f_{\rm He}$ denotes the Helium fraction of the baryonic medium and $x_e$ is the ionization fraction, given by $x_e=n_e/n_H$, where $n_e$ denotes the free electron number density and $n_H$ is the same for hydrogen atoms. In the present analysis, the heating rates of the dark matter fluid $\dot{Q}_{\chi}$ and the baryonic fluid $\dot{Q}_b$ as an outcome of the baryon - DM scattering are calculated as described in {\bf Mu\~{n}oz \emph{et~al.}} \cite{munoz} as,
		\begin{eqnarray}
			\dot{Q}_{\chi} =& \dfrac{{\rm d}Q_{\chi}}{{\rm d}t}=&\dfrac{1}{n_{\chi}}\left(\dfrac{\rho_{\chi}\rho_b}{\rho_m}D\left(V_{\chi b}\right)V_{\chi b}-n_b\dot{Q}_b\right),\label{eq:qchi}\\
			\dot{Q}_b =& \dfrac{{\rm d}Q_b}{{\rm d}t}=&\dfrac{2 m_b \rho_{\chi} \sigma_0 e^{-\frac{r^2}{2}}(T_{\chi}-T_b)}{\left(m_{\chi}+m_b\right)^2 \sqrt{2 \pi}u_{\rm th}^3} + \dfrac{\rho_{\chi} m_{\chi} m_b}{\rho_m(m_{\chi}+m_b)} V_{\chi b}D\left(V_{\chi b}\right),\label{eq:qb}
		\end{eqnarray}
		where $m_{\chi}$ ($m_b$) and $\rho_{\chi}$ ($\rho_b$) are respectively the mass and density of the DM (baryons) particles. In the above equations (Eqs.~\ref{eq:qb} and \ref{eq:qchi}), $u_{\rm th}^2=T_{\chi}/m_{\chi}+T_b/m_b$ measures the variance of the relative thermal motion and $r=V_{\chi b}/u_{\rm th}$. 
		
		The general form of the velocity dependent cross-section is $\bar{\sigma}=\sigma_0 (v/c)^n$, where the index $n$ depends on different physical dark matter processes and $c$ denotes the velocity of light in space (in natural unit $\bar{\sigma}=\sigma_0 v^n$). In the case of DM with magnetic and/or electric dipole moment $n=+2,-2$ are considered. $n=2,1,0,-1$ are applicable for scattering in presence of Yukawa potential \cite{yukawa}, $n=-4$ is attributed for millicharged dark matters \cite{mcharge1,mcharge2}. In Ref.~\cite{dvorkin2020cosmology}, the nature of the baryon - dark matter scattering is discussed for a wide mass range of dark matter. Similar investigations are also carried out in Ref.~\cite{Nadler_2019,Bhoonah_2018,Kovetz_2018}. In the present analysis, the baryon - dark matter scattering cross-section ($\bar{\sigma}$) is parameterized as $\bar{\sigma}=\sigma_0 v^{-4}$ \citep{upala,munoz,rennan_3GeV}. The term $\sigma_0$ is the DM scalar scattering cross-section with baryons (of the type $\alpha_q \bar{\chi}\chi \bar{q}q$ for dark dark matter particle $\chi$ with coupling $\alpha_q$). It may be mentioned in some earlier works \cite{Bhoonah_2018,Kovetz_2018}, millicharged dark matter is considered. But here, we assume a dark matter candidate, which is independent of any particle dark matter model. (\cite{Cheng:2002ej,servant_tait,Hooper:2007gi,Majumdar:2003dj,IDM_1,IDM_2,wimpfimp}) and adopt value of $\sigma_0 \sim 10^{-41} \rm{cm^2}$ consistent with the scalar cross-section bound obtained from ongoing experiments of direct dark matter search in the mass range discussed in this work. So, we parameterize $\sigma_0$ as a dimensionless quantity as $\sigma_{41}=\dfrac{\sigma_0}{10^{-41} {\rm cm^2}}$.
		Several recent investigations on EDGES 21-cm signal also suggest the similar velocity dependence ($n=-4$) of the cross-section \cite{munoz,rennan_3GeV,U66}. Moreover, $n=-4$ is chosen in many dark matter related cases namely hadronically interacting DM, millicharge DM, the Baryon Acoustic Oscillations (BAO) signal etc.
		
		The baryon and dark matter heating rate due to the baryon - DM interaction ($\dot{Q}_b$ and $\dot{Q}_{\chi}$) depend on the drag term $V_{\chi b}$, which is given by $V_{\chi b}\equiv V_{\chi}-V_{b}$\footnote{$V_{b}$ and $V_{\chi}$ are the relative velocities for the baryon fluid and dark matter fluid respectively}. The evolution of $V_{\chi b}$ can be expressed according to the work of {\bf Mu\~{n}oz} \emph{et~al.} \cite{munoz} as,
		\begin{equation}
			\frac{{\rm d} V_{\chi b}}{{\rm d} z} = \frac{V_{\chi b}}{1+z}+
			\frac{D(V_{\chi b})}{(1+z) H(z)}, \label{eq:V_chib}
		\end{equation}
		where $D(V_{\chi b})$ is given by, 
		\begin{equation}
			D(V_{\chi b})=\dfrac{{\rm d}(V_{\chi b})}{{\rm d}t}=\dfrac{\rho_m 
				\sigma_0}{m_b + m_{\chi}} \dfrac{1}{V^2_{\chi b}} F(r).
			\label{eq:dvchib}
		\end{equation}
		In the above expression of the drag term, $F(r)={\rm erf}\left(r/\sqrt{2}\right)-\sqrt{2/\pi}\,r e^{-r^2/2}$ where, $\rm erf$ denotes the Gauss error function.
		
		In addition to $T_{\chi}$ and $T_b$, the ionization fraction of the IGM ($x_e$) is also modified remarkably as an outcome of the energy deposition due to PBH evaporation. The evolution of the ionization fraction $x_e$ essentially depends on baryon temperature ($T_b$) and background temperature ($T_{\gamma}$) simultaneously as \cite{BH_21cm_2,munoz,amar21},	
		\begin{equation}
			\frac{{\rm d} x_e}{{\rm d} z} = \frac{1}{(1+z)\,H(z)}\left[I_{\rm Re}(z)-
			I_{\rm Ion}(z)-I_{\rm BH}(z)\right],
			\label{eq:xe}
		\end{equation}
		where $I_{\rm Re}(z)$ and $I_{\rm Ion}(z)$ describes the standard recombination rate and the ionization rate of the medium respectively, which can be written as \cite{munoz,yacine,hyrec11},
		\begin{equation}
			I_{\rm Re}(z)-I_{\rm Ion}(z) = C_P\left(n_H \alpha_B x_e^2-4(1-x_e)\beta_B 
			e^{-\frac{3 E_0}{4 k_B T_{\gamma}}}\right).
			\label{eq:xe_comp}
		\end{equation}
		In the above expression, $\alpha_B$ is the case-B recombination coefficient and the term $\beta_B$ denotes the photoionization coefficient. $C_P$ represents the Peebles C factor \cite{hyrec11,peeble}. 
		
		The case-B recombination coefficient in the unit of $\rm{m^3}s^{-1}$ can be calculated by fitting the data obtained from the work of {\bf Pequignot} \emph{et~al.} \cite{pequignot} as \cite{BH_21cm_5,pequignot},
		\begin{equation}
			\alpha_B=10^{-19} F\left(\frac{a t^b}{1+c t^d}\right), \label{alphaB}
		\end{equation}
		where $a=4.309$, $b=-0.6166$, $c=0.6703$, $d=0.5300$, $F=1.14$ \cite{BH_21cm_5,pequignot} are the fitted parameters. In this expression, $t$ denotes the temperature in $10^4$K \cite{pequignot,hummer,seager}. The photoionization rate ($\beta_B$) can be expressed as \cite{BH_21cm_5,seager},
		\begin{equation}
			\beta_B=\alpha_B \left(\frac{2 \pi \mu_e k_B T_{\gamma}}{h^2}\right)^{3/2} 
			\exp\left(-\frac{h \nu_{2s}}{k_B T_{\gamma}}\right), \label{betaB}
		\end{equation}
		where, $\mu_e$ denotes the reduces mass of the electron proton system (neutral hydrogen atom) and $\nu_{2s}$ is the frequency of the photon, which are produced due to the $2s\rightarrow 1s$ transition. The Peebles C factor can be expressed as a function of the escape rate of the ${\rm Ly\alpha}$ photons as \cite{hyrec11,peeble},
		\begin{equation}
			C_P=\dfrac{\frac{3}{4}R_{\rm Ly\alpha}+\frac{1}{4}\Lambda_{2s1s}}{\beta_B+
				\frac{3}{4}R_{\rm Ly\alpha}+\frac{1}{4}\Lambda_{2s,1s}}, \label{peeblec}
		\end{equation}		
		where $\Lambda_{2s,1s}\approx 8.22 \, \rm{s^{-1}}$ \cite{hyrec11} and the escape rate ($R_{\rm Ly\alpha}$) is defined as,
		\begin{equation}
			R_{\rm Ly\alpha}=\dfrac{8\pi H}{\left(3 n_H (1-x_e)\lambda_{\rm Ly\alpha}^3\right)}.
		\end{equation} 
		
		The term $I_{\rm BH}$ in Eq.~\ref{eq:xe} i.e. the evolution equation of ionization fraction due to PBH evaporation, is given by
		\begin{equation}
			I_{\rm BH}=\chi_i f(z) \frac{1}{n_b} \frac{1}{E_0}\times \left.
			\dfrac{{\rm d} E}{{\rm d}V {\rm d}t}\right|_{\rm{BH}}, \label{IBH}
		\end{equation}
		where $E_0=13.6$ eV is the absolute value of the ground state energy of hydrogen atom. In the equation for $T_b$ evolution (Eq.~\ref{eq:T_b}), the term $K_{\rm BH}$ describes the contribution of PBH evaporation in the baryon heating, given by,
		\begin{equation}
			K_{\rm BH}=\chi_h f(z) \frac{1}{n_b} \times \left.\dfrac{{\rm d} E}{{\rm d}V {\rm d}t}\right|_{\rm{BH}}. \label{KBH}
		\end{equation}
		In the above two equations (Eqs.~\ref{IBH}, \ref{KBH}), $\chi_i=(1-x_e)/3$ denotes the fraction of deposited energy helps to ionize the gas of the medium while $\chi_h=(1+2x_e)/3$ is the fractions of the deposited energy, which contribute in the heating of baryon fluid \cite{BH_21cm_2,PhysRevD.76.061301,chen,BH_21cm_4,Furlanetto:2006wp}. In Eqs.~\ref{IBH} and \ref{KBH}, the parameter $f(z)$ measures the ratio of the total amount of deposited energy to the energy injected to the medium due to PBH evaporation \cite{corr_equs,fcz001,fcz002,fcz003,fcz004}.
		
		It is to be noted that the effects of the Lyman-$\alpha$ photons from the first star have a significant contribution to the evolution of spin temperature for $z \lesssim 25$. In the epoch of the cosmic dark age, the background photons contribute to flip the spin state of hydrogen atoms. As a consequence, the spin temperature ($T_s$) and the background temperature ($T_{\gamma}$) become almost equal. However, later ($z \lessapprox 25$), the Lyman-$\alpha$ photons from the newborn stars lead to a quick transition of the spin temperature $T_s=T_b$. As a result, $T_s$ becomes almost equal to the $T_b$ during the age of cosmic dawn \cite{tstb2,tstb1} (Figure~\ref{fig:tspin}). This phenomenon of the flipping of spin temperature is known as the Wouthuysen-Field effect. The strength of the Wouthuysen-Field effect is essentially defined by the rate of scattering of the Ly$\alpha$ photons in the intergalactic medium \cite{lya001}.

	\section{\label{sec:result_21_2} Variation of $T_{21}$ with different parameters}
		In order to study the changes of baryon temperature $T_b$ and hence the spin temperature ($T_s$) and the brightness temperature ($T_{21}$) due to the Hawking radiation and baryon - DM interaction, five mutually coupled parameters are solved simultaneously (Eqs.~\ref{eq:PBH}, \ref{eq:T_chi}, \ref{eq:T_b}, \ref{eq:xe} and \ref{eq:V_chib}) with redshift.	
		\begin{figure}
			\centering{}
			\includegraphics[width=0.7\linewidth]{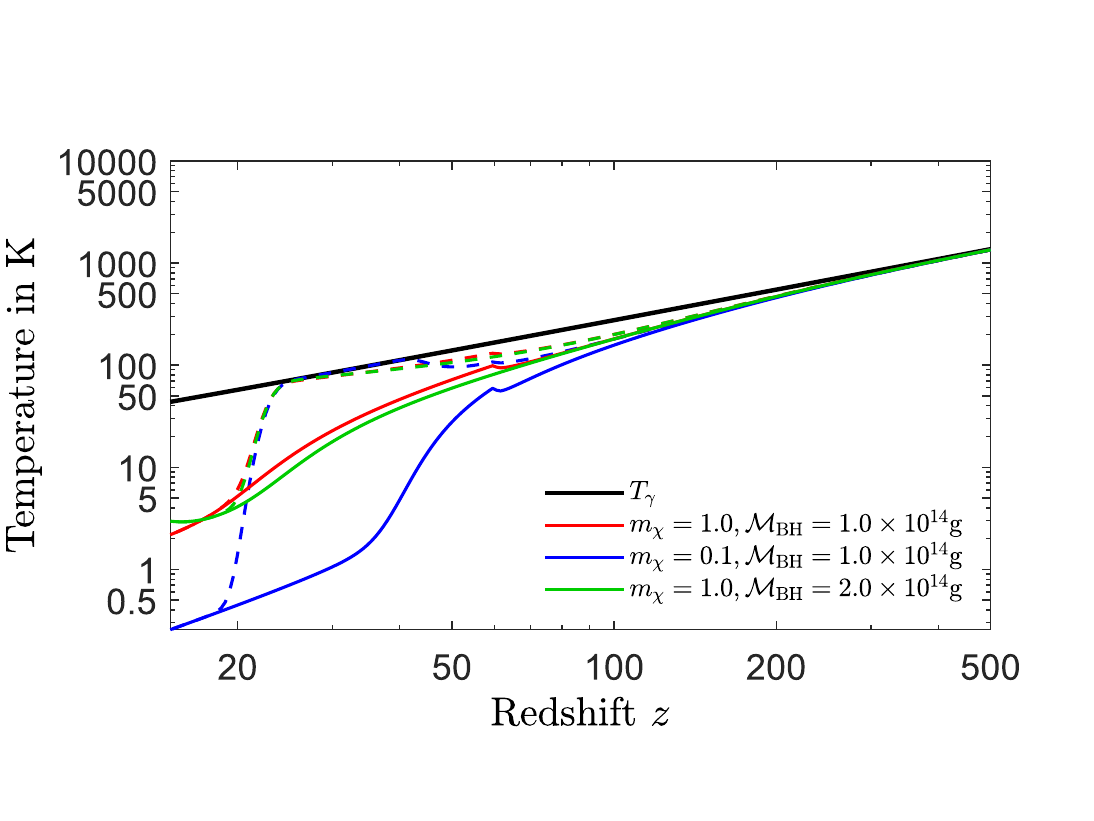}
			\caption{\label{fig:tspin} The evolution of $T_b$ and corresponding spin temperature $T_s$ with redshift $z$ (using the solid lines and dashed lines respectively) for different sets of dark matter mass and primordial black hole mass ($m_{\chi}$ and $\mathcal{M_{\rm BH}}$ respectively). In each case, $\beta_{\rm BH}=10^{-29}$ and $\sigma_{41}=1$ are considered. In this figure, the CMD temperature $T_{\gamma}$ is shown by the black solid line.}
		\end{figure}
		In Figure~\ref{fig:tspin}, the variation of the $T_b$ and the corresponding spin temperature $T_s$ with redshift $z$ are plotted for different masses of primordial black holes and dark matters. In Figure~\ref{fig:tspin}, the plotted red solid line describes the evolution of $T_b$ for the chosen mass of dark matter $m_{\chi}=1$~GeV and PBH mass $\mathcal{M}_{\rm BH}=10^{14}$ g. The corresponding evolution of spin temperature $T_s$ is described by the red dashed line (corresponds to the red solid line). In Figure~\ref{fig:tspin}, the solid green and solid blue lines are representing the evolution of $T_b$ for the DM and PBH masses $m_{\chi}=1$~GeV, $\mathcal{M}_{\rm BH}=2.0 \times 10^{14}$ g and $m_{\chi}=0.1$~GeV, $\mathcal{M}_{\rm BH}=1.0 \times 10^{14}$ g respectively. The green and blue dashed lines are indicating the variation of $T_s$ with redshift $z$ for the corresponding cases. In Figure~\ref{fig:tspin}, we choose $\beta_{\rm BH}=10^{-29}$ and $\sigma_{41}=1$ for each of the cases.
			
		\begin{figure*}
			\centering{}
			\begin{tabular}{c}
				\includegraphics[width=0.7\textwidth]{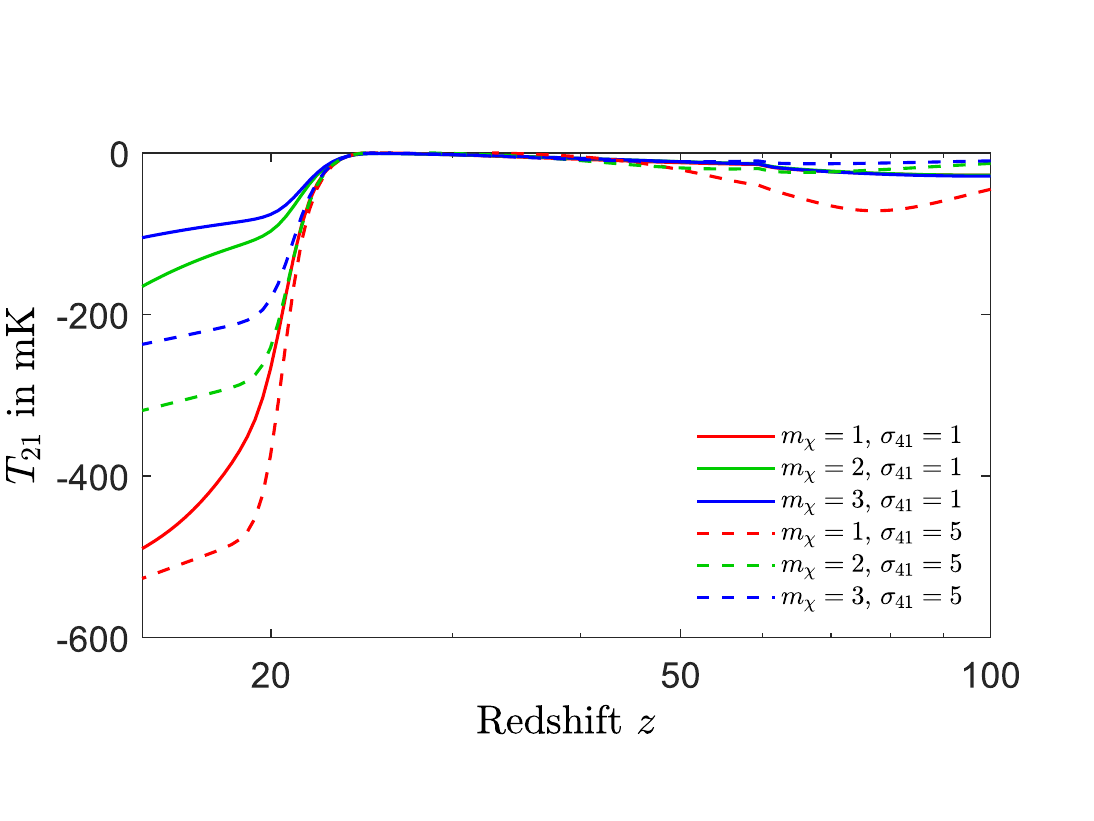}\\(a)\\
				\\
				\includegraphics[width=0.7\textwidth]{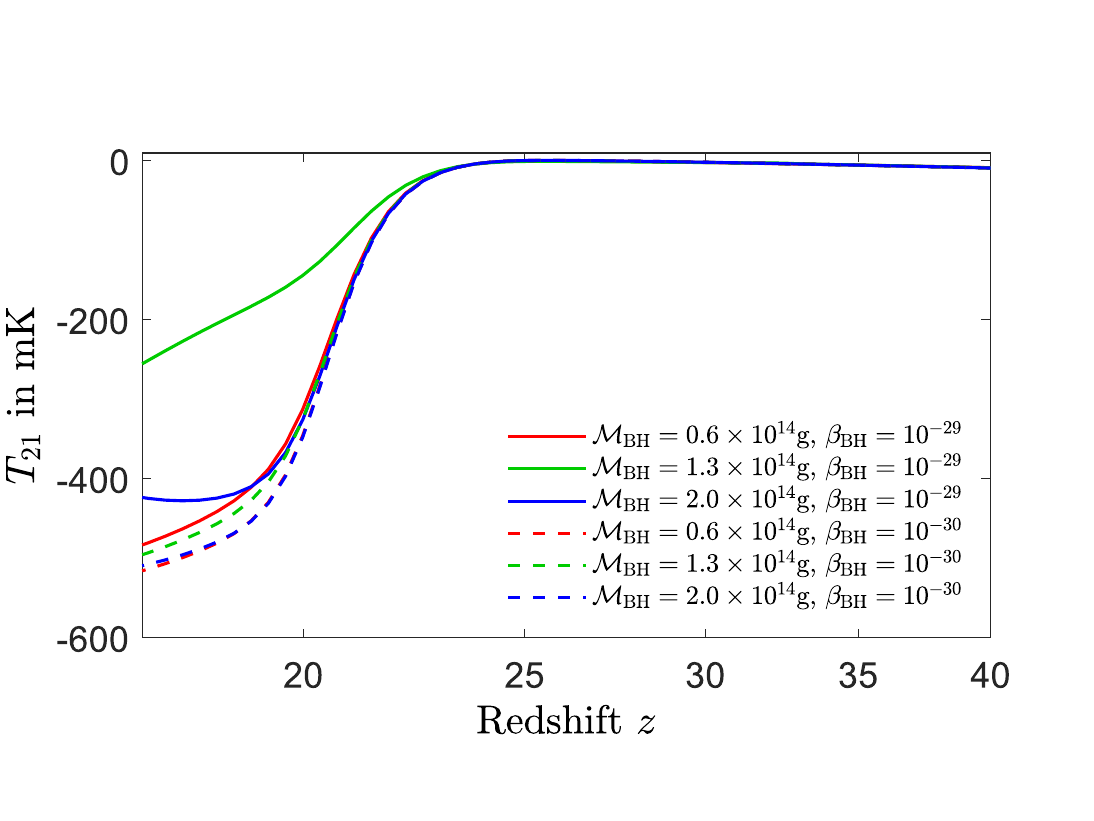}\\(b)
			\end{tabular}
			\caption{\label{fig:mchivar} The variation of brightness temperature $T_{21}$ (a) for different values of $\sigma_{41}$ and dark matter masses($m_{\chi}$) keeping $\beta_{\rm BH}=10^{-29}$ and $\mathcal{M_{\rm BH}}=10^{14}$ g, (b) for different initial mass fraction $\beta_{\rm BH}$ and $\mathcal{M_{\rm BH}}$ for $\sigma_{41}=1$ and $m_{\chi}=1$ GeV.}
		\end{figure*}
		The combined effect of Hawking radiation from PBHs and the baryon - dark matter scattering modifies the global 21-cm signature remarkably. The effects of the baryon - dark matter interaction parameters (dark matter mass $m_{\chi}$ and baryon - DM interaction cross-section $\sigma_{41}$) and the PBH parameters (i.e. PBH mass $\mathcal{M_{\rm BH}}$ and initial mass fraction $\beta_{\rm BH}$) in the brightness temperature of the 21-cm absorption spectrum are described graphically in Figure~\ref{fig:mchivar}. In Figure~\ref{fig:mchivar}a the evolution of the brightness temperatures ($T_{21}$) are plotted for three different DM masses and two different baryon - dark matter interaction cross-section parameter $\sigma_{41}$ (for fixed values of $\beta_{\rm BH}=10^{-29}$ and $\mathcal{M}_{\rm BH}=10^{14}$ g). On the other hand, the contribution of the different masses of PBHs and corresponding initial mass fraction $\beta_{\rm BH}$ in the 21-cm brightness temperature are demonstrated in Figure~\ref{fig:mchivar}b. In this particular case, the baryon - dark matter interaction cross-section and the dark matter mass are kept constant at $\sigma_{41}=1$ and $m_{\chi}=1$ GeV respectively.
		
		\begin{figure*}[h!]
			\centering
			\begin{tabular}{c}
				\includegraphics[width=0.7\textwidth]{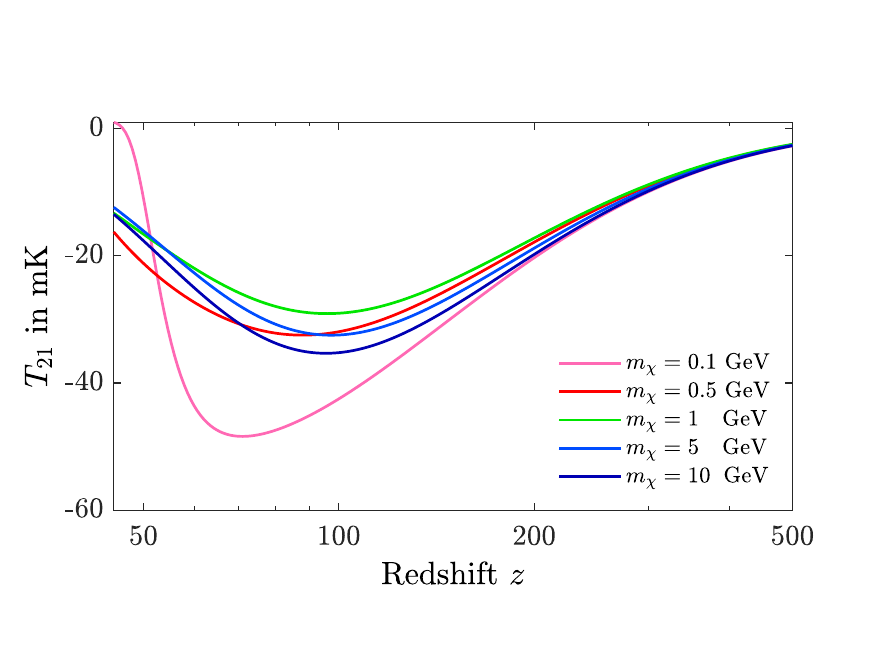}\\(a)\\
				\\
				\includegraphics[width=0.7\textwidth]{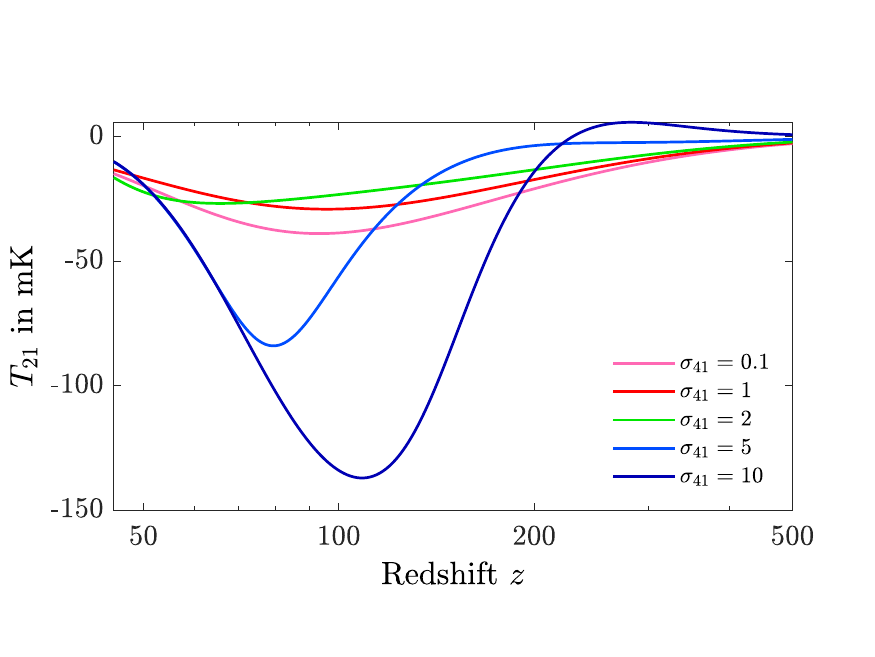}\\
				(b)
			\end{tabular}
			\caption{\label{fig:mchivar1} Variations of brightness temperature ($T_{21}$) with redshift ($z$) (a) for five chosen values of $m_{\chi}$ (0.1, 0.5, 1, 5 and 10 GeV) in presence of PBH having mass $\mathcal{M_{\rm BH}}=1.5 \times 10^{14}$ g and $\sigma_{41}=1$. Fig.~\ref{fig:mchivar1}b (right panel) describes the variations of $T_{21}$ for different values of $\sigma_{41}$ ($\sigma_{41}=$0.1, 1, 2, 5, 10) when the PBH mass $\mathcal{M_{\rm BH}}=10^{14}$ g and the dark matter mass $m_{\chi}=1$ GeV are chosen.}
		\end{figure*}
		In Fig.~\ref{fig:mchivar1}, a similar variations of the 21-cm brightness temperature $T_{21}$ are addressed for higher redshifted epoch (redshift $z\gtrapprox 50$). The dependence of the DM mass ($m_{\chi}$) in $T_{21}$ is very prominent in the epoch of $50<z<200$ as shown in Fig.~\ref{fig:mchivar1}a. Similarly, the effect of the baryon - DM cross-section is also described in Fig.~\ref{fig:mchivar1}b for $50<z<200$. From Figure~\ref{fig:mchivar} (a and b) one can conclude that, during the dark age, the fluctuation of $T_{21}$ is minimized for $m_{\chi}\approx 1$ GeV and $\sigma_{41}\approx 1$. It is to be noted that, Figure~\ref{fig:mchivar} deals with the higher redshifted epoch, which is irrelevant for ground based 21-cm observations in view of ionospheric opacity. However, future space-based radio observation may be capable to explore this salient feature in the global 21-cm signal \cite{Burns_2017}.

	\section{\label{sec:bounds} Bounds on The PBH Abundance and $m_{\chi}$}
		\begin{figure*}
			\centering{}
			\begin{tabular}{cc}
				\includegraphics[trim={0 0 75 0},clip,width=0.5\textwidth]{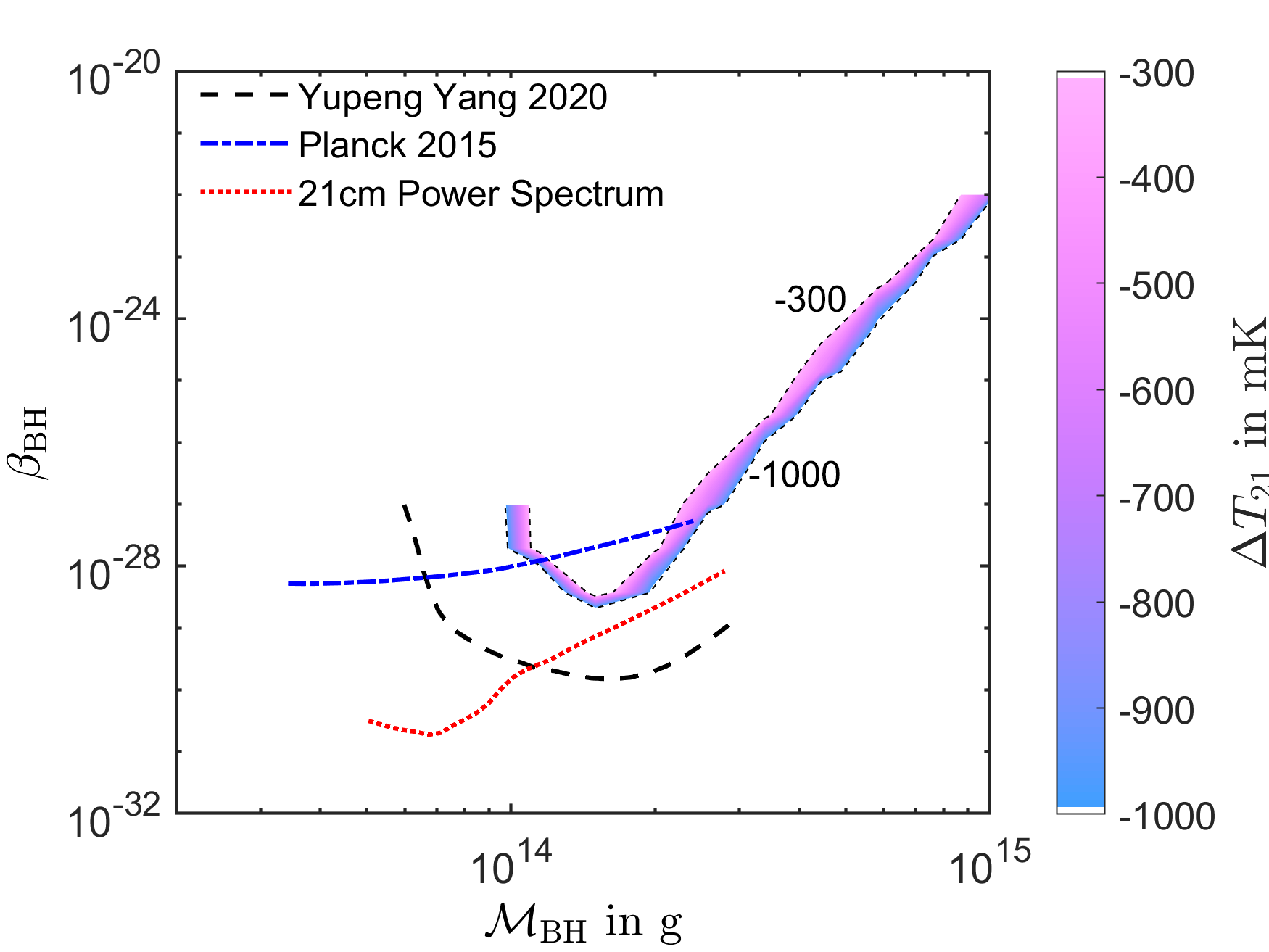}&
				\includegraphics[trim={0 0 75 0},clip,width=0.5\textwidth]{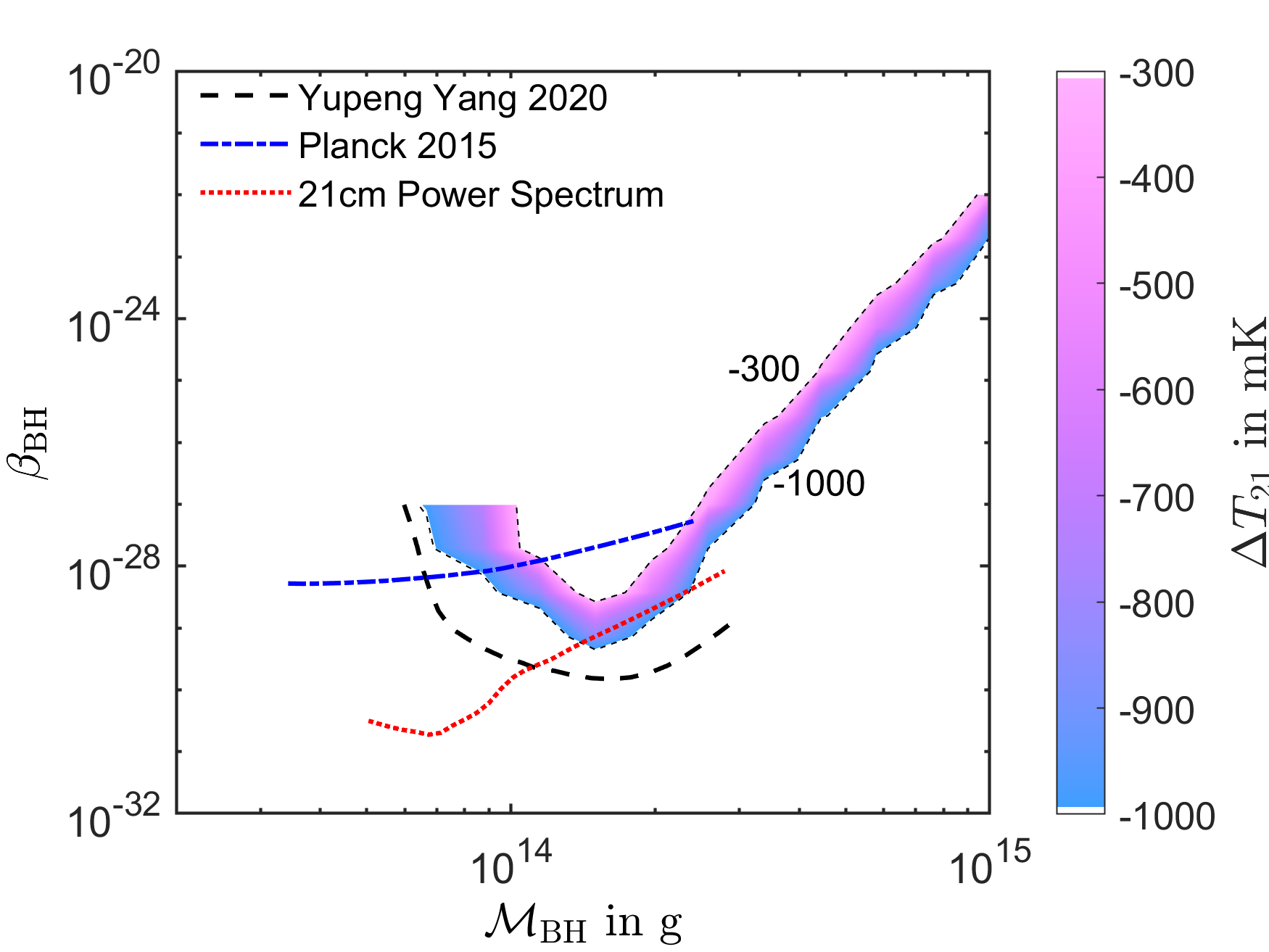}\\
				(a)&(b)\\
				\includegraphics[trim={0 0 75 0},clip,width=0.5\textwidth]{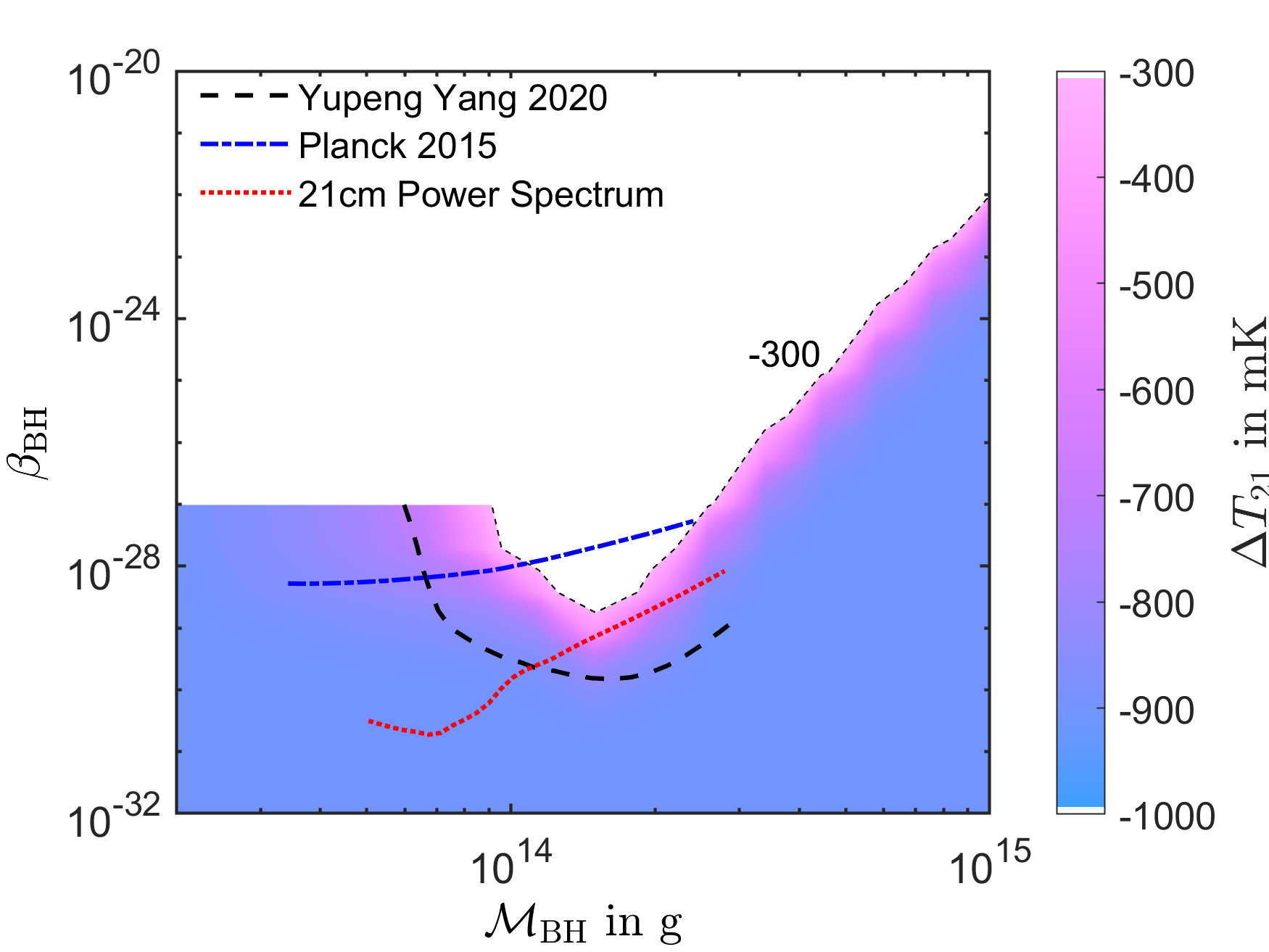}&
				\includegraphics[trim={0 0 75 0},clip,width=0.5\textwidth]{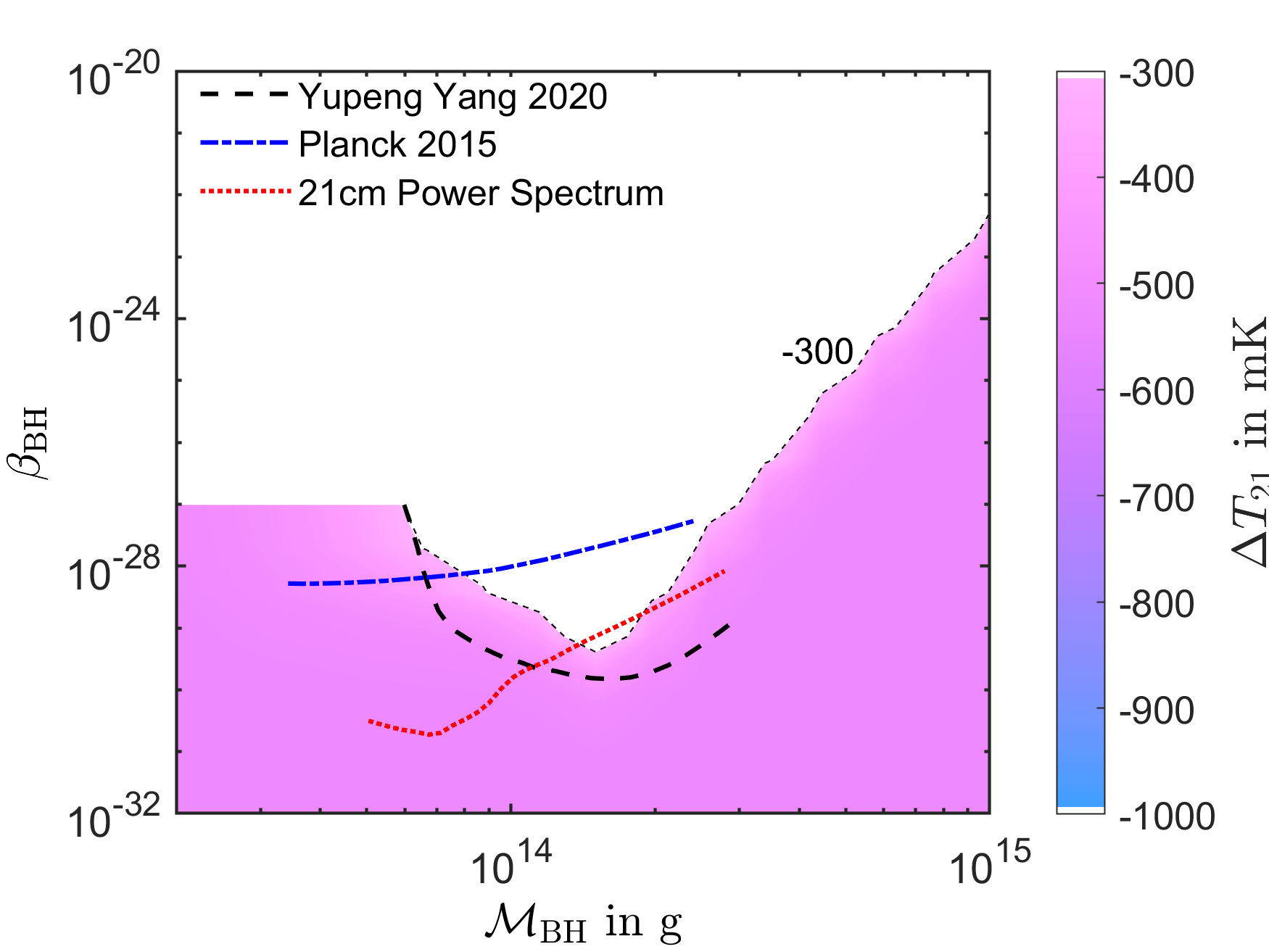}\\
				(c)&(d)\\
			\end{tabular}
			\begin{tabular}{c}
				\includegraphics[width=0.8\textwidth]{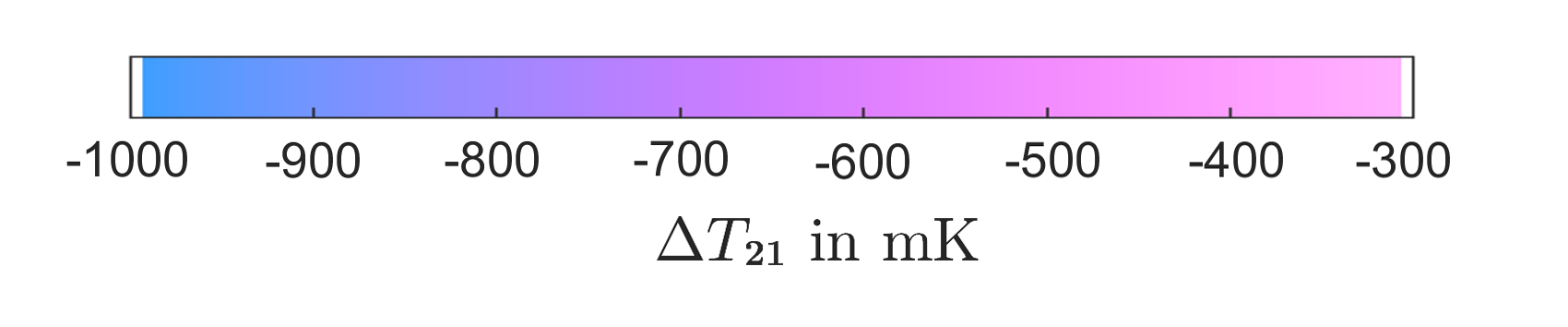}
			\end{tabular}
			\caption{\label{fig:c_mx} The bounds and allowed zones in the $\beta_{\rm BH}$ - $\mathcal{M_{\rm BH}}$ parameter space for different dark matter masses ((a) $m_{\chi}=0.1$ GeV, (b) $m_{\chi}=0.3$ GeV, (c) $m_{\chi}=0.5$ GeV and (d) $m_{\chi}=1.0$ GeV) for a fixed value $\sigma_{41}=1$. The allowed regions satisfy the limit ($-500^{+200}_{-500}$) of the 21-cm brightness temperature as obtained from the EDGES's result. Three other bounds are also shown in all four sub-figures. The dashed black line represents the upper bound (corresponds to the $\Delta T_{21}=-300$ mK) of $\beta_{\rm BH}$, as estimated by {\rm Yang} \cite{BH_21cm_2}. The dash-dotted blue line represents the same as computed from the Planck experiment data (2015) and the dotted red line describes the similar limit, which is evaluated from the power spectrum of 21-cm absorption line \cite{BH_21cm_2}.}
		\end{figure*}
\begin{figure*}
			\centering{}
			\begin{tabular}{cc}
				\includegraphics[trim={0 0 75 0},clip,width=0.48\textwidth]{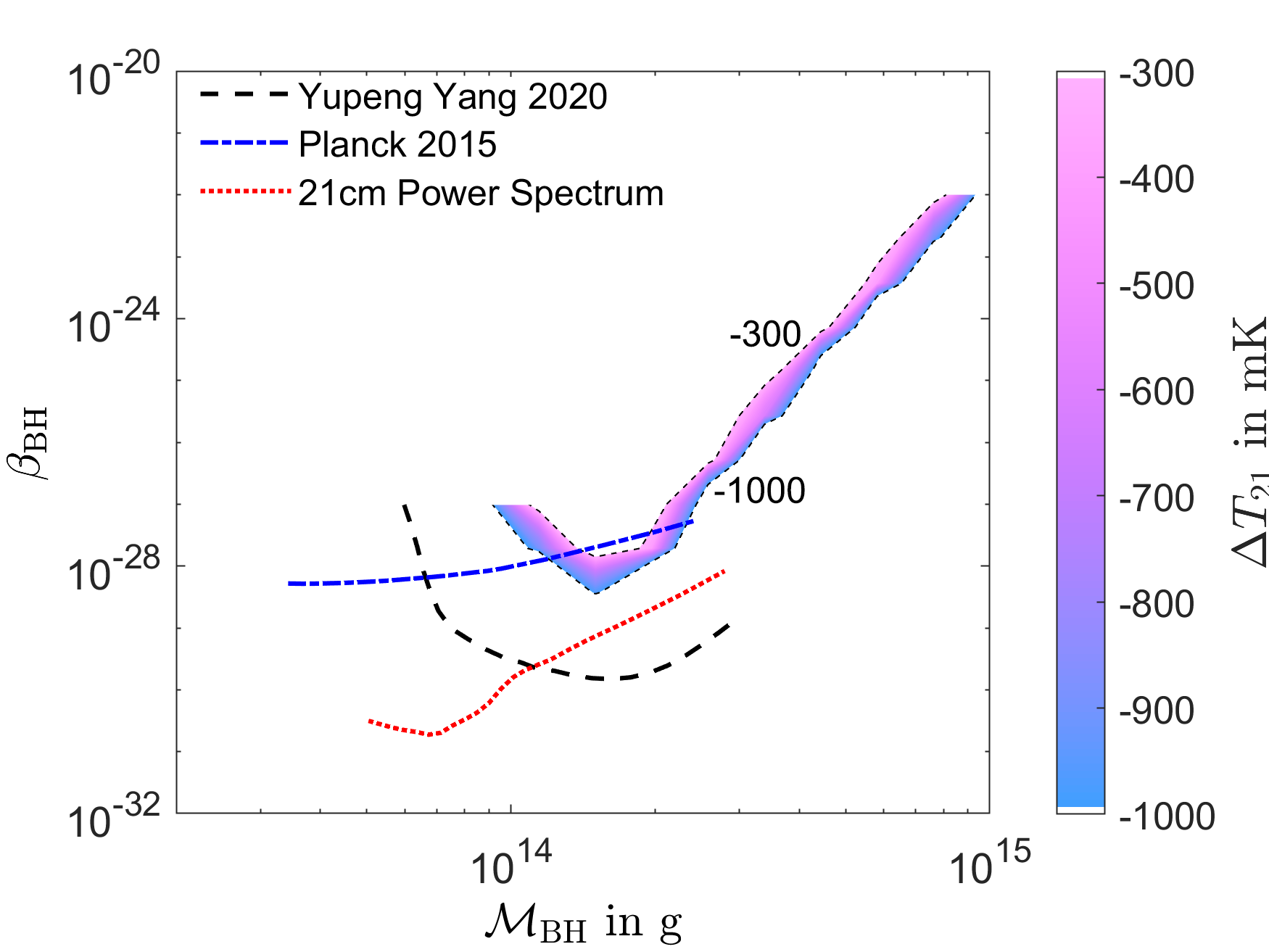}&
				\includegraphics[trim={0 0 75 0},clip,width=0.48\textwidth]{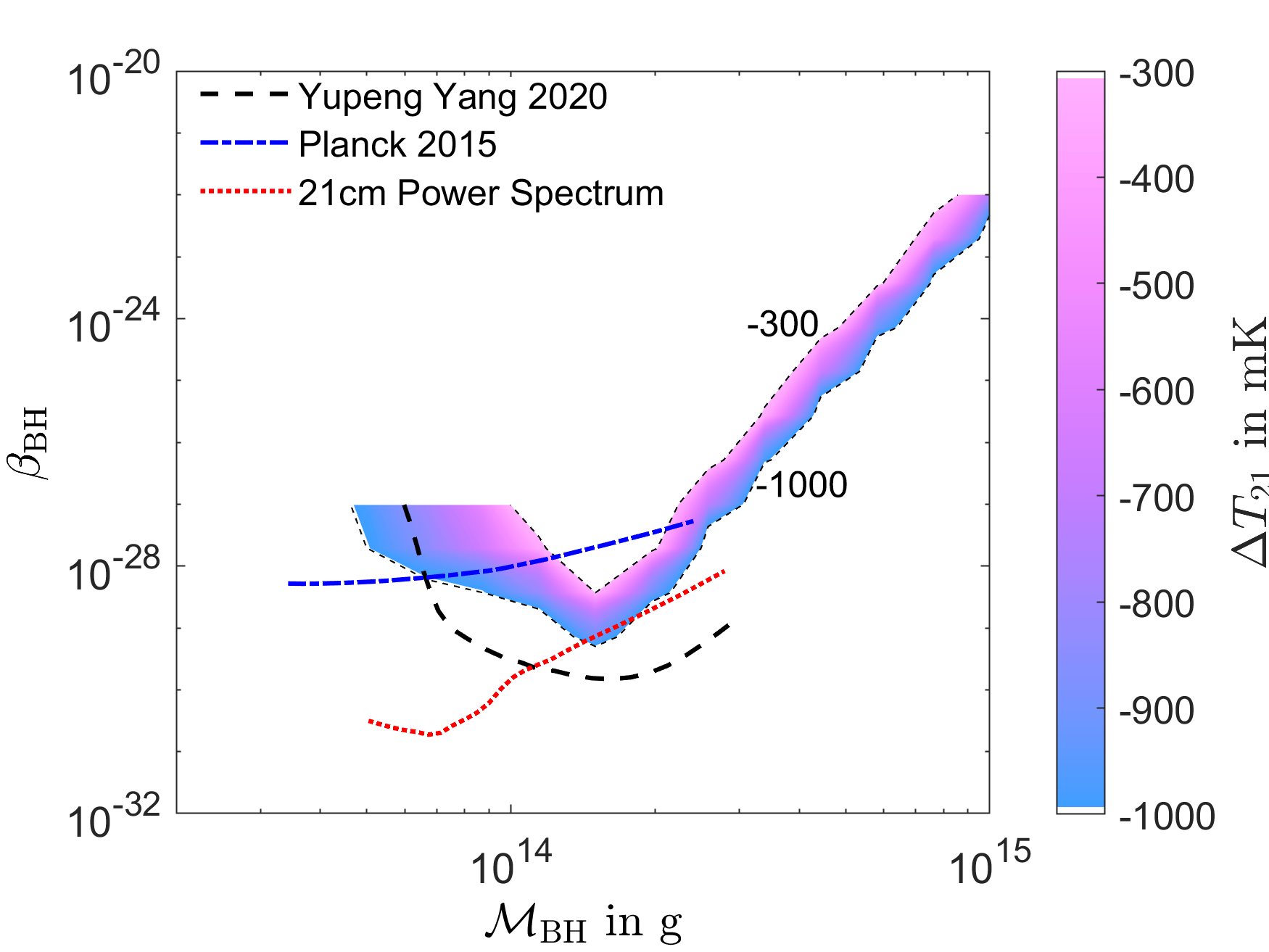}\\
				(a)&(b)\\
				\includegraphics[trim={0 0 75 0},clip,width=0.48\textwidth]{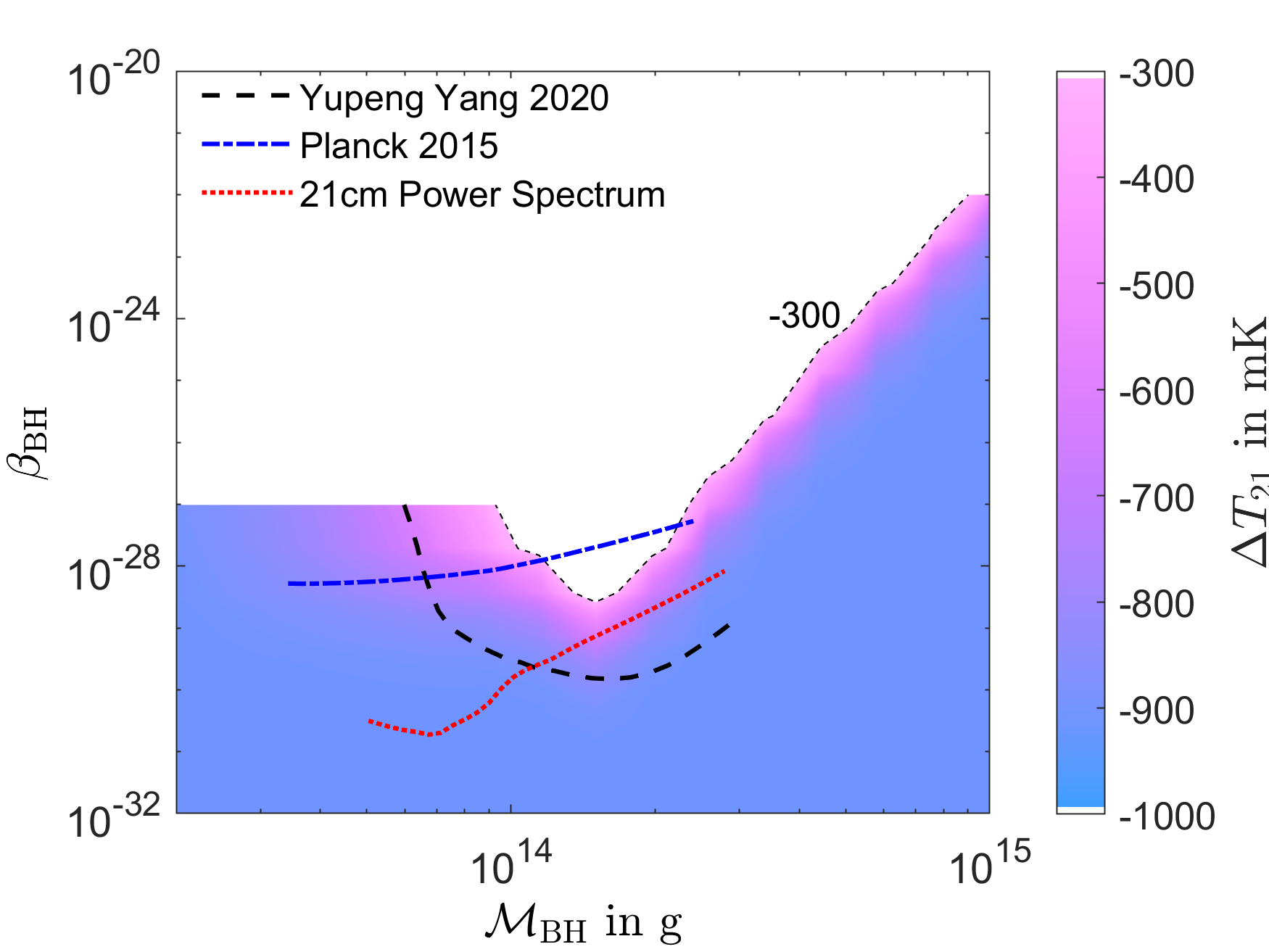}&
				\includegraphics[trim={0 0 75 0},clip,width=0.48\textwidth]{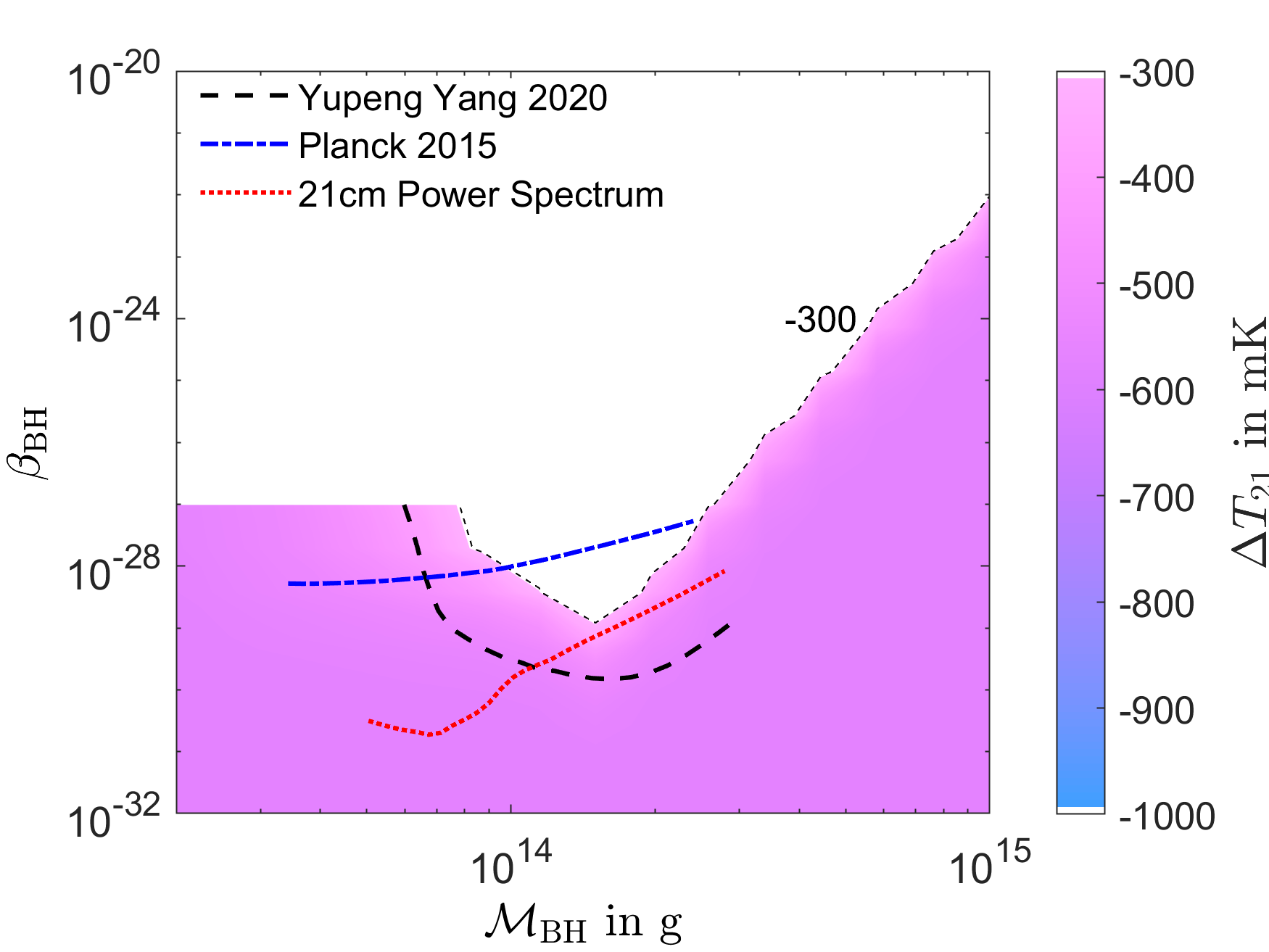}\\
				(c)&(d)\\
			\end{tabular}
			\begin{tabular}{c}
				\includegraphics[width=0.8\textwidth]{figure/21cm_2/scale}
			\end{tabular}
			\caption{\label{fig:c_mx_sigma5}Same as Figure~\ref{fig:c_mx} for $\sigma_{41}=5$.}
		\end{figure*}

		In the present work, we essentially tried to estimate of the upper and lower bounds for the initial mass fraction ($\beta_{\rm BH}$) for a wide range of primordial black hole mass $\mathcal{M}_{\rm BH}$ ({$6\times 10^{13}g\,\leq\mathcal{M}_{\rm BH}\leq 10^{15}$ g}). The variations of the allowed region (region between the upper bound and lower bound) with the dark matter mass $m_{\chi}$ and the cross-section parameters $\sigma_{41}$ are demonstrated in the graphs of Figure~\ref{fig:c_mx} and Figure~\ref{fig:c_mx_sigma5}. As the estimated bounds are put forwarded using the observational excess of the EDGES experiment (at at $z=17.2$, the brightness temperature $T_{21} = -500^{+200}_{-500}$), only the brightness temperature at $z=17.2$ is significant in this calculation. As a consequence, a new parameter is introduced i.e. $\Delta T_{21}$, which denotes the brightness temperature of the 21-cm  spectrum $T_{21}$ at redshift $z=17.2$. In Figure~\ref{fig:c_mx}, the furnish plots are representing the allowed region in the $\beta_{\rm BH}$-$\mathcal{M_{\rm BH}}$ parameter space for four chosen dark matter masses $m_{\chi}$. All four plots of Figure~\ref{fig:c_mx} are plotted for fixed baryon - DM cross-section $\sigma_{41}=1$. Figure~\ref{fig:c_mx}a describes both upper and lower bounds of the $\beta_{\rm BH}$ (i.e. initial mass fraction of PBHs) in the $\beta_{\rm BH}$ - $\mathcal{M_{\rm BH}}$ plane, for a fixed value of DM mass $m_{\chi}=0.1$ GeV. Here the lower bound (corresponds to $\Delta T_{21}=-1000$ mK) and the upper limit (corresponds to $\Delta T_{21}=-300$ mK) are estimated from the observational excess of the EDGES (i.e. $-1000 \leq \Delta T_{21}< -300$ mK). Figure~\ref{fig:c_mx}b, Figure~\ref{fig:c_mx}c and Figure~\ref{fig:c_mx}d are describing the same for $m_{\chi}=0.3$, 0.5 and 1.0 GeV respectively. The upper bounds estimated from the current treatment are also compared with the other three limits which are addressed by {\bf Yang} in his paper Ref.~\cite{BH_21cm_2}. From the Figure~\ref{fig:c_mx}, it can be seen that, for lower values of $m_{\chi}$ ($0.1\,{\rm GeV}<m_{\chi}<0.3$ GeV), the allowed region (the region bounded by the lower and the upper bounds) is very narrow. However, as the larger mass of the dark matter particle $m_{\chi}$ is considered, the lower limit falls rapidly keeping the upper limit remains almost same (the upper limit actually slightly decreases with increasing $m_{\chi}$).
		
		The previous analysis is also repeated for another value of baryon - DM cross-section ($\sigma_{41}=5$) as described in Figure~\ref{fig:c_mx_sigma5}. In this particular case, the region between the upper and lower bounds in the $\beta_{\rm BH}$ - $\mathcal{M_{\rm BH}}$ parameter space are slightly wider in comparison to the same for $\sigma_{41}=1$   for individual DM masses (that can be clearly noticed for case of DM mass $=0.1$ GeV and 0.3 GeV). The obtained upper bound of the $\beta_{\rm BH}$ is similar to the limits as addressed by {\bf Yang} \cite{BH_21cm_2}. In the case of higher DM mass ($m_{\chi}$), the trajectory of the upper bound for $\beta_{\rm BH}$ (which corresponds to $\Delta T_{21}=-300$ mK) is comparable (but higher in amplitude) to the same as proposed by {\bf Yang} \cite{BH_21cm_2}. It is to be noted that, in the work of {\bf Yang} \cite{BH_21cm_2}, the effects of baryon - DM scattering are not taken into account. On the other hand, for $m_{\chi}=0.1$ GeV and 0.3 GeV, the bounds are analogous to the same as estimated using the Planck experiment data for PBH mass $\mathcal{M_{\rm BH}}>1 \times 10^{14}$ g (especially for $\sigma_{41}=1$). In addition, the bounds obtained from the 21-cm power spectrum fits well for the PBH masses $\mathcal{M_{\rm BH}}>1 \times 10^{14}$ g at higher values of DM masses (0.5 GeV and 1 GeV).
		
		\begin{figure}
			\centering{}
			\includegraphics[width=0.8\linewidth]{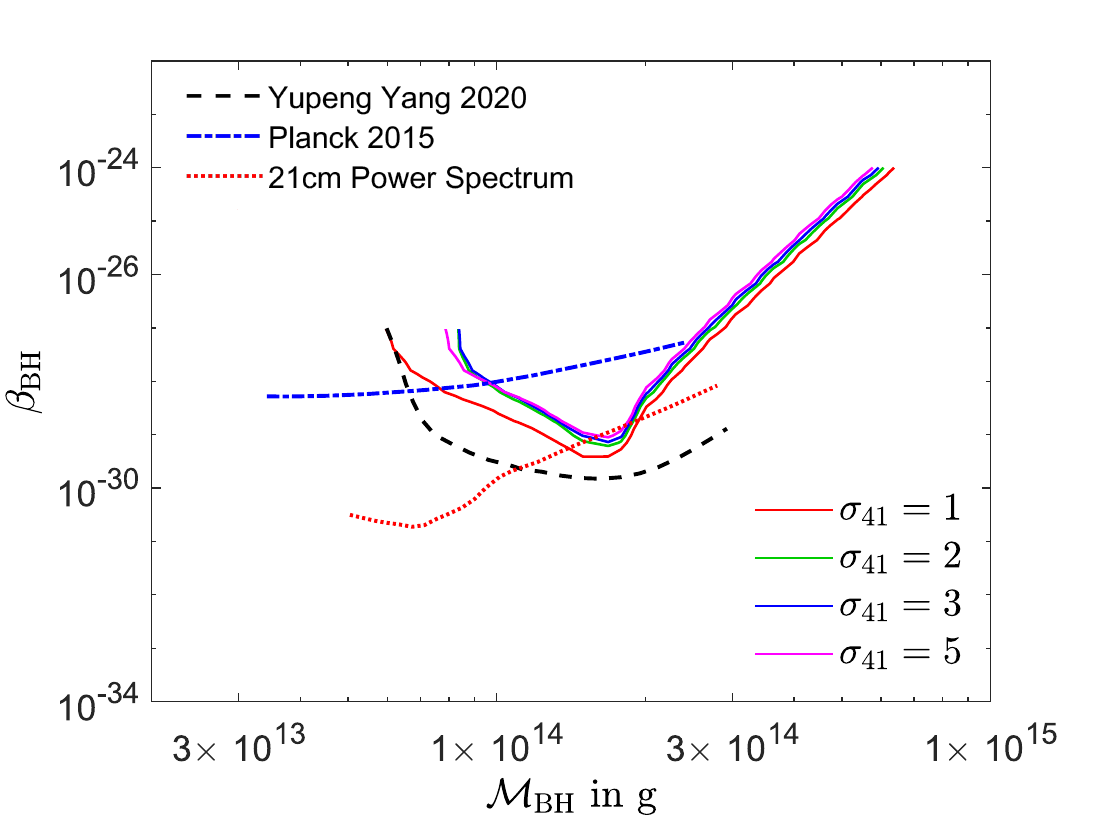}
			\caption{\label{fig:sigma_PBH} The variation of upper bound of $\beta_{\rm BH}$ in $\beta_{\rm BH}$-$\mathcal{M_{\rm BH}}$ parameter space with different values of $\sigma_{41}$ for a fixed value of the dark matter mass $m_{\chi}=1$ GeV.}
		\end{figure}
		The variation of the allowed region in the $\mathcal{M_{\rm BH}}$ - $\beta_{\rm BH}$ plane is also addressed for different values of baryon - dark matter interaction cross-section ($\sigma_{41}$) while $m_{\chi}$ is kept fixed at 1 GeV. In this particular case, only the upper limits are plotted in Figure~\ref{fig:sigma_PBH} with the proposed limits by other recent works. From Figure~\ref{fig:sigma_PBH}, one can notice that, for a fixed value of DM mass $m_{\chi}$, the upper limit decreases with the cross-section parameter $\sigma_{41}$. The bounds for all other chosen values of $\sigma_{41}$ (within $1<\sigma_{41}<5$) agree with the same as computed from the 21-cm power spectrum (for $\mathcal{M_{\rm BH}}>1.2\times 10^{14}$ g) and the investigation of {\bf Yang} \cite{BH_21cm_2} (for $\mathcal{M_{\rm BH}}<1.2\times 10^{14}$ g).
		
		\begin{figure}
			\centering{}
			\includegraphics[width=0.7\linewidth]{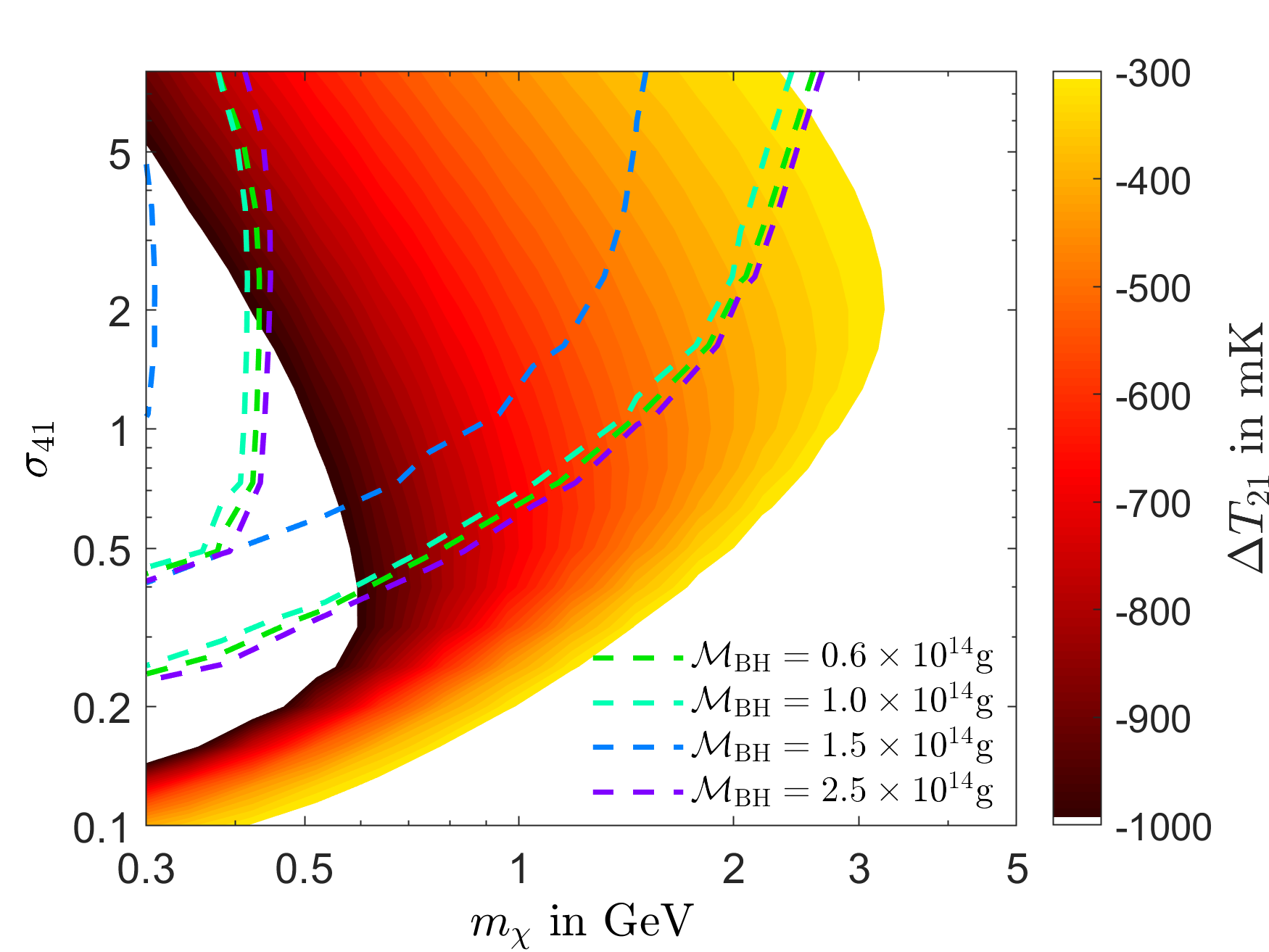}
			\caption{\label{fig:mchi_sigma} The allowed zone for the dark matter mass $m_{\chi}$ and baryon-dark matter scattering cross-section parameter $\sigma_{41}$. The colour area denotes the allowed zone for the cases, where the contribution of black hole evaporation is not incorporated. The dashed lines of different colours are representing the boundaries of the allowed region where the contribution of the Hawking radiation from different masses of PBHs is considered. In all four cases the initial mass fraction of PBH is chosen at $\beta_{\rm BH}=5\times 10^{-30}$.}
			\vskip 0.5cm
		\end{figure}
		
		Eventually, a similar bounds in $m_{\chi}$ - $\sigma_{41}$ plane are also addressed for different chosen values of black hole masses ($\mathcal{M_{\rm BH}}$). In Figure~\ref{fig:mchi_sigma}, the bound for the DM mass $m_{\chi}$ and the baryon - dark matter  scattering cross-section parameter $\sigma_{41}$ using the coloured region, where the contribution of the Hawking radiation is not considered. In this particular case, the values of $\Delta T_{21}$ at individual points are described using the different colour codes as described in the colourbar of Figure~\ref{fig:mchi_sigma}. However, as the heating due to PBH evaporation in the form of Hawking radiation is taken into account, the allowed region gets modified. In Figure~\ref{fig:mchi_sigma} similar allowed regions are plotted where different chosen values of PBH masses ($\mathcal{M_{\rm BH}}$) are considered in the system for a fixed value of initial mass fraction $\beta_{\rm BH}=5.0\times 10^{-30}$. In Figure~\ref{fig:mchi_sigma}, it can be observed that as the contribution of baryon heating due to PBH heating becomes prominent, the allowed region shifts toward lower values of dark matter mass ($m_{\chi}$) in the $m_{\chi}$-$\sigma_{41}$ parameter space. In addition, the allowed zone also found to be shifted toward higher $\sigma_{41}$ as the heating due to the evaporation of primordial black holes is incorporated into the system. It is to be noted that, the estimated bounds of the $m_{\chi}$ agree with the result of {\bf R.~Barkana} \cite{rennan_3GeV} for the individual cases (i.e. $m_{\chi} \leq 3$ GeV).

\chapter{Influence of DM-DE Interaction in 21-cm Signal} \label{chp:21_jan}
\\
	Besides compact celestial bodies, there are several unexplored mysteries in our Universe. Dark matter (DM) and dark energy (DE) are such candidates of the Universe, which are invisible to every form of electromagnetic radiation. The current cosmological paradigm tells that a very tiny percentage of the entire energy budget of our Universe belongs to ordinary baryonic matter i.e only 4.9\%. The remaining unseen part is essentially distributed among the two dark sector components of the Universe, namely dark energy and dark matter. The natures of those dark components are still a mystery.
	
	Till date, the evidence of dark matter - dark energy (DM-DE) interaction is still unexplored. So, one can assume a non-minimal coupling between these two dark sector elements in lieu of treated them separately. The interaction between dark matter and dark energy \citep{idem0,idem1,idem2,idem3,idem4,Kumar_2017,Kumar_2019} may have significant effects in the dynamics of the Universe. In this work, the effect of dark matter - dark energy interactions in 21-cm EDGES signal is principally addressed while taking into consideration other important effects arising out of dark matter scattering on baryons and also PBH evaporation that can possibly influence the observed temperature of the 21-cm signal during reionization epoch. The PBHs can inject energy into the system through their evaporation in the form of Hawking radiation. Also, the interactions between dark matter and dark energy can influence the global 21-cm signal as discussed in \Autoref{chp:21_feb}. To this end, three such interacting dark energy (IDE) models are adopted where the non-minimal coupling of two dark sectors namely dark matter and dark energy is adopted and the energy transfer due to the IDE and dark matter-baryon scattering are properly incorporated in the relevant evolution equations for baryon temperature, dark matter temperature etc.

\section{Introduction}

	The 21-cm cosmology is turning out to be a promising tool in understanding the dynamics of the early Universe. In the present analysis, we investigate the combined effects of the dark matter - dark energy interaction, primordial black holes (PBH) and cooling off of the baryonic matter due to dark matter (DM) - baryon collisions on the 21-cm brightness temperature. The variation of the brightness temperature shows significant dependence on the masses of the DM particles ($m_{\chi}$) as well as the baryon - dark matter scattering cross-section ($\sigma_{41}$) as discussed in the previous chapter. Bounds in $m_{\chi}$ - $\sigma_0$ parameter space are remarkably modified for different interacting dark energy models (IDE) in additional to the effect of PBH evaporation. These bounds are estimated based on the observed excess (21-cm brightness temperature at $=-500^{+200}_{-500} \; \rm{mK}$) of EDGES experiment. 
	It is to be mentioned that when the interaction between DM and DE is considered, the DM density $\rho_\chi$ and the DE density $\rho_{\rm DE}$ don't evolve as $\sim (1+z)^3$ and $(1+z)^{3(1+\omega)}$ respectively as suggested by standard cosmology. Therefore the evolution of the Universe ($H(z)$) is modified as an outcome of the dark matter - dark energy interactions which in turn affect the spin temperature $T_s$ of the baryonic medium and the optical depth. The evolution of the Hubble parameter is also computed in detail in the present work for all the three IDE models adopted. 
	The influence of DM - DE interaction on 21-cm signal has been discussed earlier in \cite{Li,upala}. Moreover, DM-DE interaction is discussed in several context such as in addressing the cosmological coincidence problem \cite{Wang_2016}, Hubble tension \cite{PhysRevD.96.043503}, Large Scale Structure formation \cite{Farrar_2004} etc.

\section{Dark Matter - Dark Energy Interaction} \label{sec:DMDE} 
	The Dark Matter - Dark Energy interaction may have a profound effect on the universal dynamics and hence on the spin temperature and the optical depth of the medium. According to the notion of the standard cosmological, the dark matter and dark energy density parameters ($\Omega_{\chi}$ and $\Omega_{\rm de}$ respectively) are expected to be evolved as $\Omega_{\chi,0}(1+z)^3$ and $\Omega_{\rm de,0}(1+z)^{3(1+\omega)}$ where, $\Omega_{\chi,0}$ and $\Omega_{\rm de,0}$ are the density parameters at $z=0$ and $\omega$ denotes the dark energy equation of state (EOS) parameter. However, if the DM - DE interaction is considered, the evolution of the dark matter density and the dark energy density takes the forms \cite{Li},
	\begin{equation}
		(1+z) H(z) \dfrac{{\rm d} \rho_{\chi}}{{\rm d} z}-3H(z)\rho_{\chi}  = -\mathcal{Q} 
		\label{eq:rho_chi}
	\end{equation}
	\begin{equation}
		(1+z) H(z) \dfrac{{\rm d} \rho_{\rm de}}{{\rm d} z} - 3 H(z) (1+ \omega) \rho_{de} = \mathcal{Q}
		\label{eq:rho_de}
	\end{equation}
	where $\mathcal{Q}$ represents the IDE model dependent energy transfer function, which measures the amount of energy transfer per unit time, per unit volume between dark energy and dark matter during the interaction. In the current analysis, three benchmark models are considered to investigate the effects of the dark matter - dark energy interaction in the brightness temperature. The energy transfer expressions of those benchmark models are described below \cite{model1,model2,model3,model4}.
	\begin{center}
		\begin{tabular}{ll}
			Model-I \hspace{5mm} & $\mathcal{Q}=3 \lambda H(z) \rho_{\rm de}$\\
			Model-II & $\mathcal{Q}=3 \lambda H(z) \rho_{\chi}$\\
			Model-III &$\mathcal{Q}=3 \lambda H(z) (\rho_{\rm de} +\rho_{\chi})$
		\end{tabular}
	\end{center}
	Here, $\lambda$ denotes the IDE coupling parameter, which determine the strength of the dark matter - dark energy interaction. The stability conditions for each of the chosen IDE models are described in Table~\ref{tab:stability}. Several phenomenological studies have been carried out with observational data of PLANCK, Supernova Ia (SNIa) Baryon Acoustic Oscillation (BAO) \cite{model3, model4, model_benchmark1, model_benchmark2, model_benchmark3, 	model_benchmark4, model_benchmark5, model_benchmark6} yielding the constraints for different models (in Table~\ref{tab:constraints}). It is to be mentioned that, all the IDE models discussed in this section are independent to the dark matter - baryon interaction.
	\begin{table}
		\centering
		\begin{tabular}{lccr}
			\hline
			Model & $\mathcal{Q}$ & EOS of dark energy & Constraints\\
			\hline
			I & 3 $\lambda H(z) \rho_{\rm de} $ & $\omega<-1$ & $\lambda<- 2 \omega \Omega_{\chi}$\\
			II & 3 $\lambda H(z) \rho_{\chi} $ & $\omega<-1$ & $0<\lambda<-\omega/4$\\ 
			III & 3 $\lambda H(z) (\rho_{\rm de} + \rho_{\chi}) $ & $\omega<-1$ & $0<\lambda<-\omega/4$\\
			\hline
		\end{tabular}
		\caption{\label{tab:stability} Stability conditions of the model parameters for different IDE models}
	\end{table}
	
	\begin{table}
		\centering
		\begin{tabular}{lccc}
			\hline
			Model & $\omega$ & $\lambda$ & $H_0$\\
			\hline
			
			$3 \lambda H \rho_{\rm de}$ & $-1.088^{+0.0651}_{-0.0448}$ & $0.05219^{+0.0349}_{-0.0355}$ & $68.35^{+1.47}_{-1.46}$\\
			
			$3 \lambda H \rho_{\chi}$ & $-1.1041^{+0.0467}_{-0.0292}$ & $0.0007127^{+0.000256}_{-0.000633}$ & $68.91^{+0.875}_{-0.997}$\\
			
			$3 \lambda H (\rho_{\rm de}+\rho_{\chi})$ & $-1.105^{+0.0468}_{-0.0288}$ & $0.000735^{+0.000254}_{-0.000679}$ & $68.88^{+0.854}_{-0.97}$\\
			\hline
		\end{tabular}
		\caption{\label{tab:constraints} Constraints of the different IDE models}
	\end{table}
	
	\begin{figure*}
		\centering
		\begin{tabular}{cc}
			\includegraphics[trim=0 40 0 55, clip, width=0.5\textwidth,height=0.3\textwidth]
			{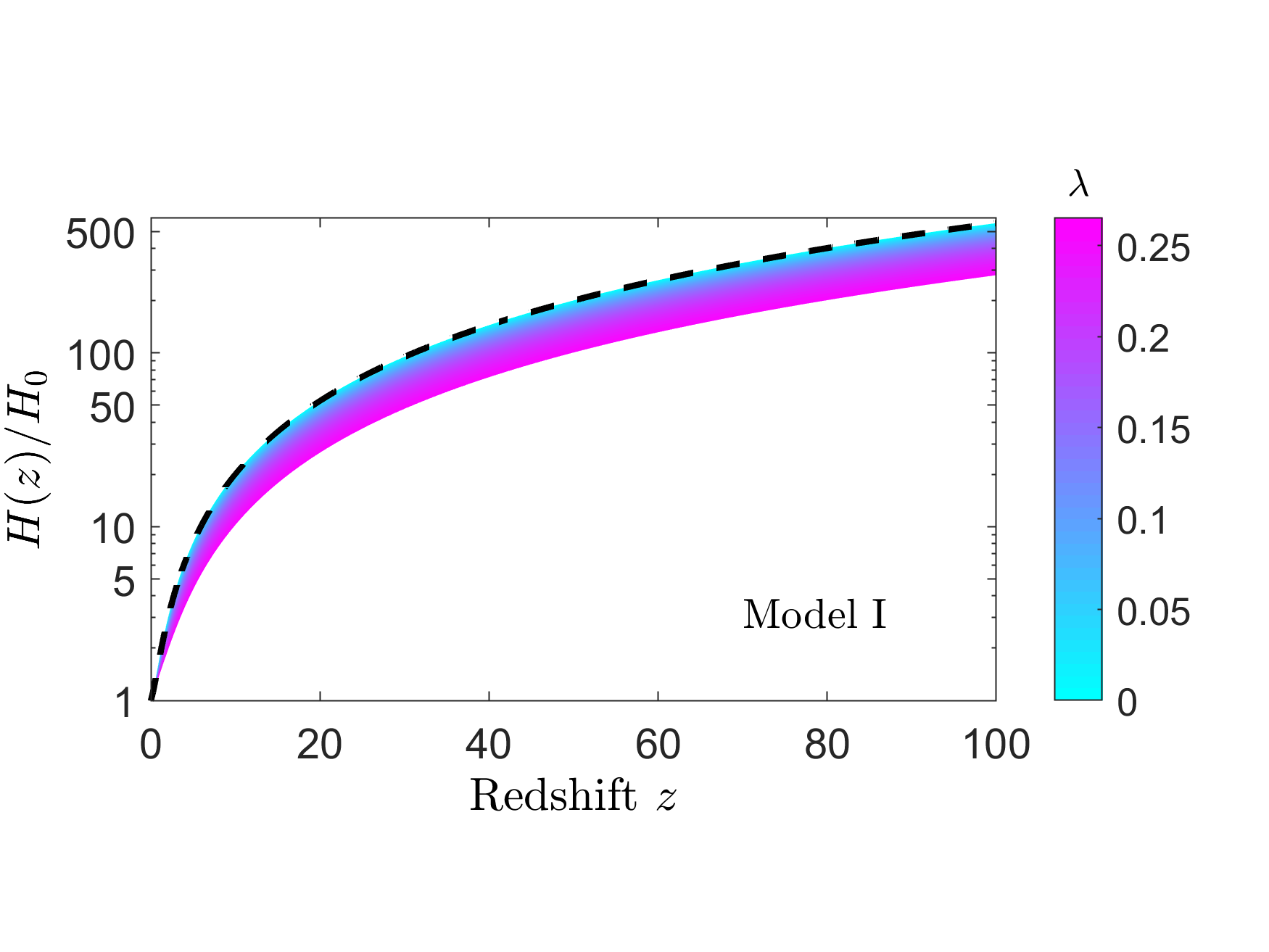}&
			\includegraphics[trim=0 40 10 55, clip, width=0.5\textwidth,height=0.3\textwidth]
			{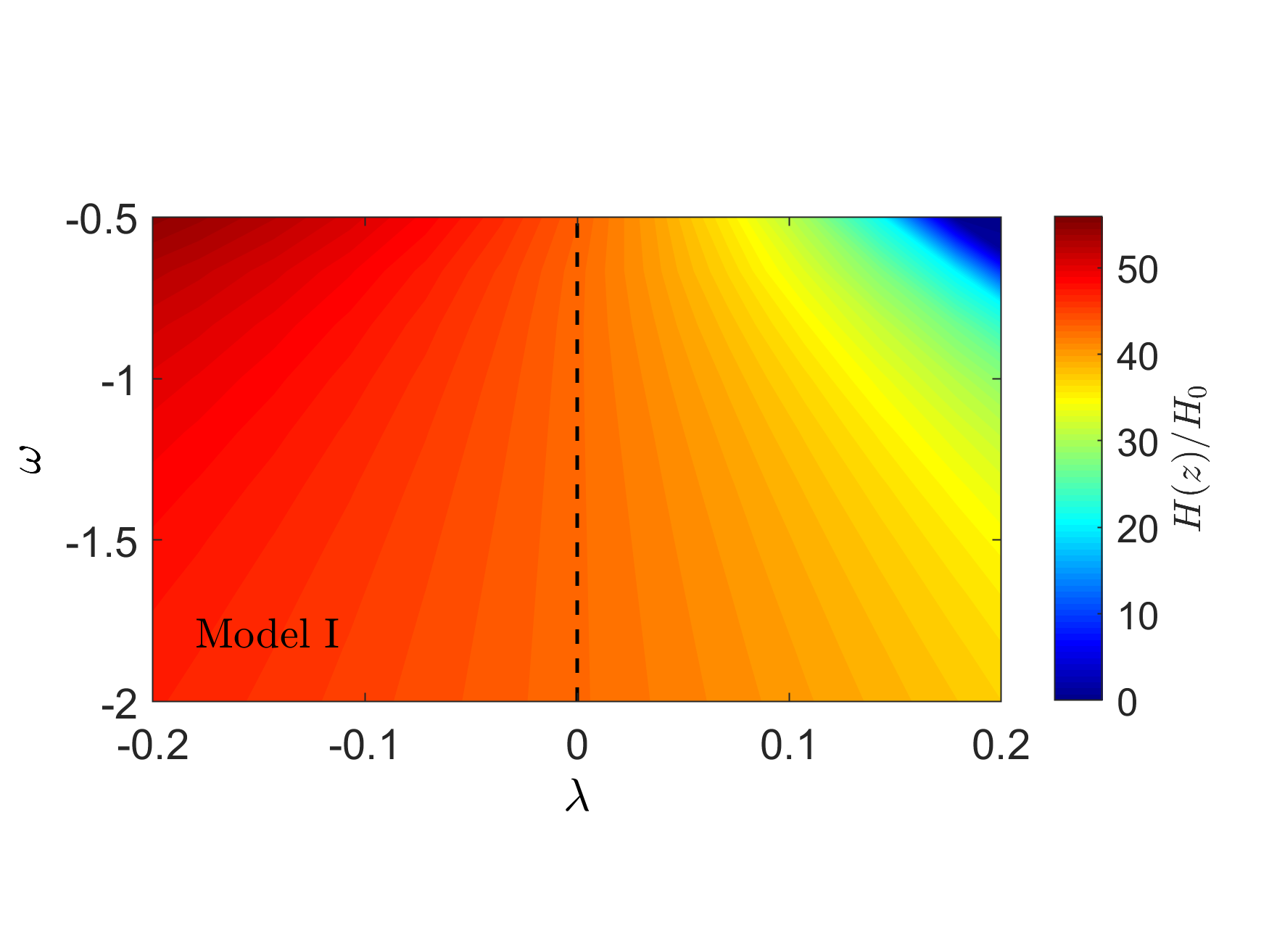}\\
			(a)&(b)\\
			\\
			\includegraphics[trim=0 40 0 55, clip, width=0.5\textwidth,height=0.3\textwidth]
			{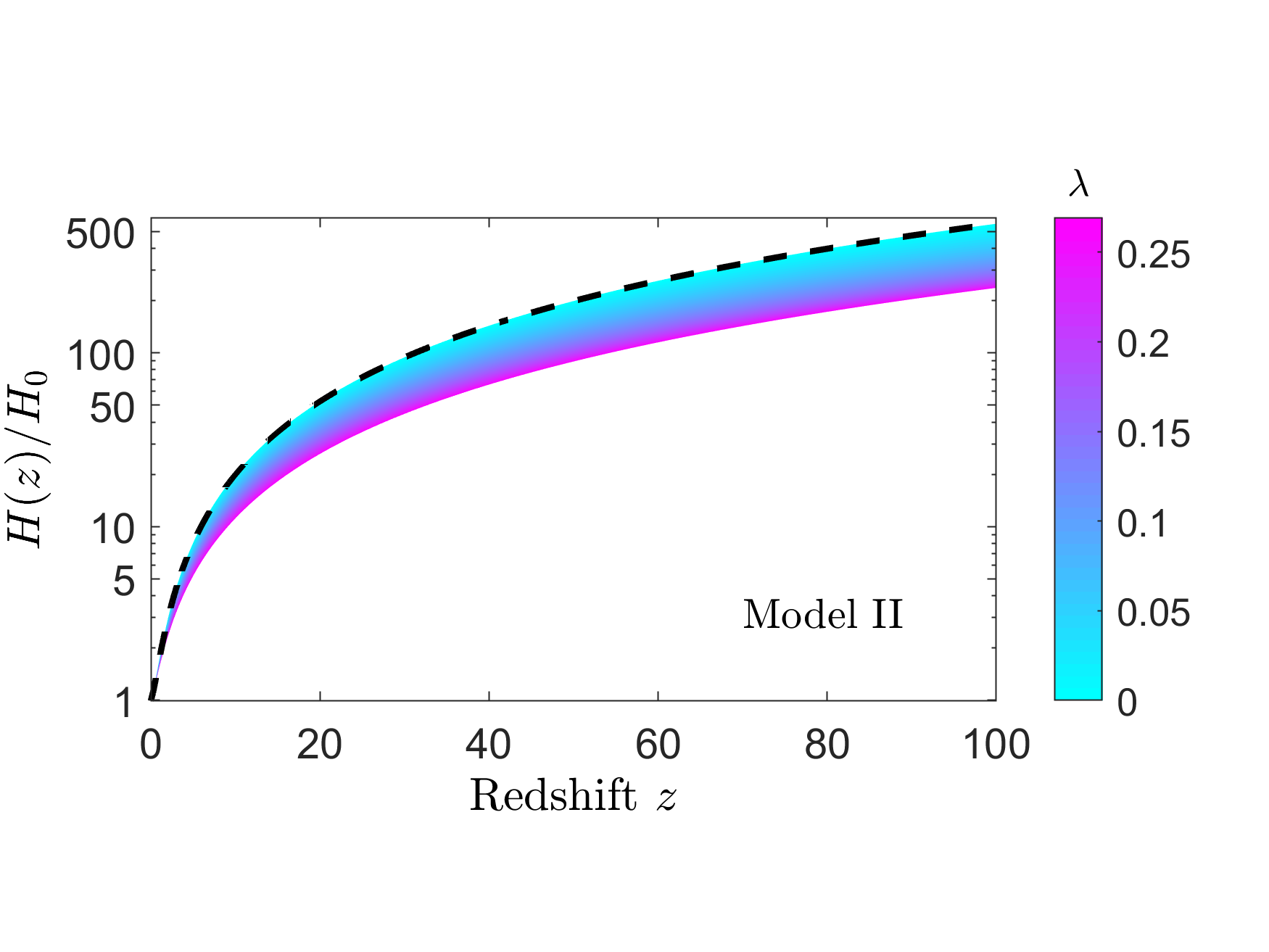}&
			\includegraphics[trim=0 40 10 55, clip, width=0.5\textwidth,height=0.3\textwidth]
			{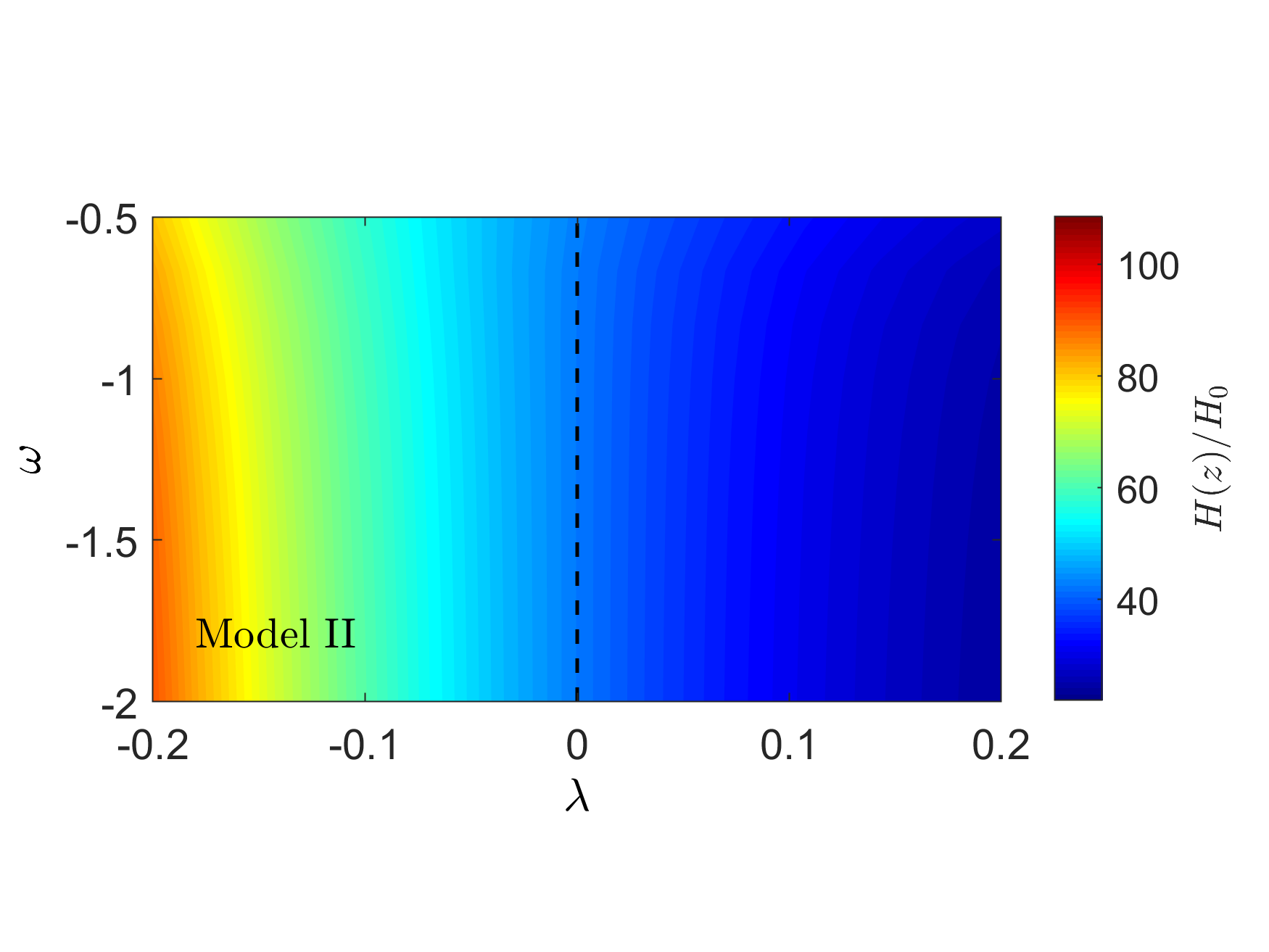}\\
			(c)&(d)\\
			\\
			\includegraphics[trim=0 40 0 55, clip, width=0.5\textwidth,height=0.3\textwidth]
			{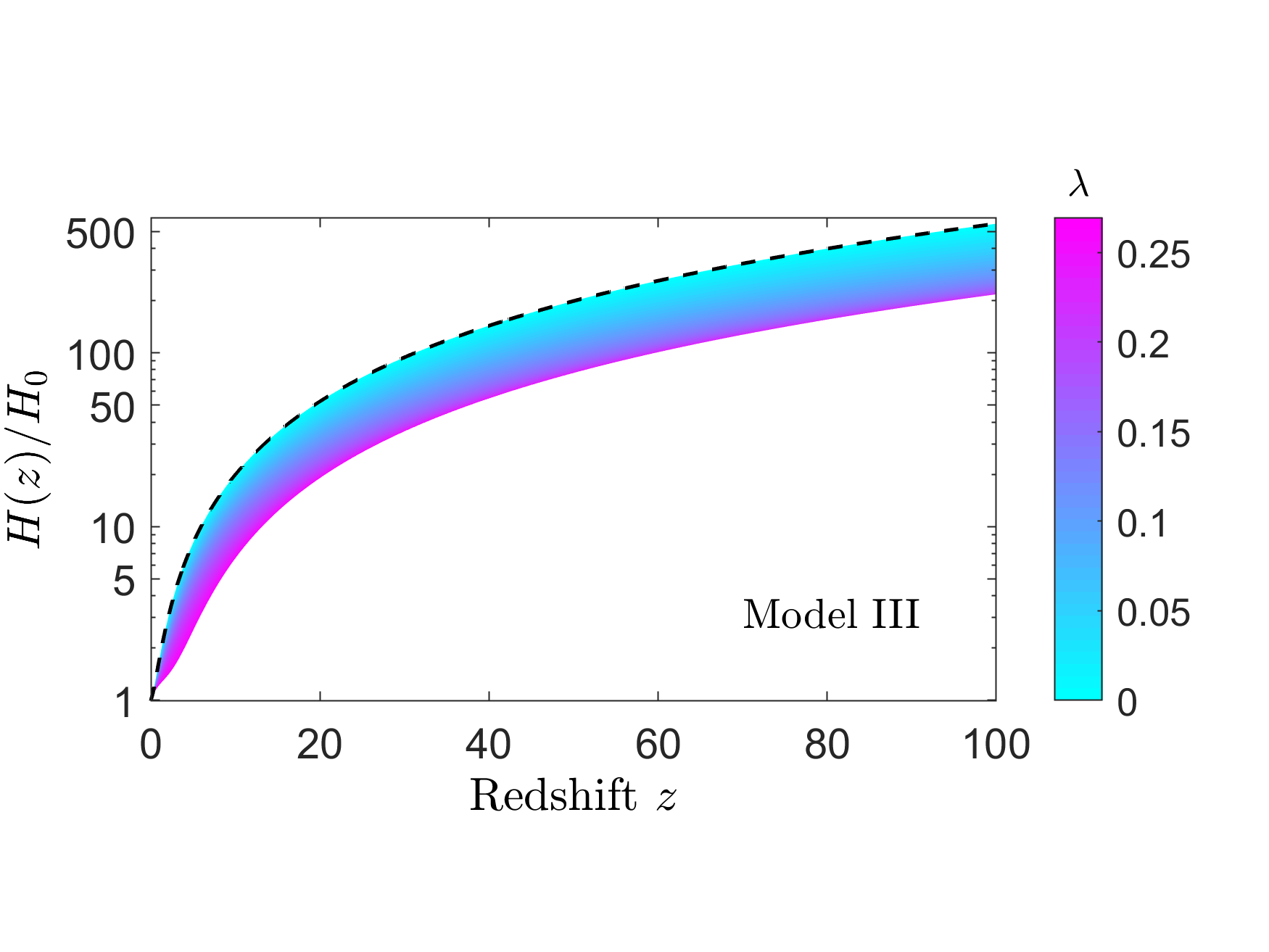}&
			\includegraphics[trim=0 40 10 55, clip, width=0.5\textwidth,height=0.3\textwidth]{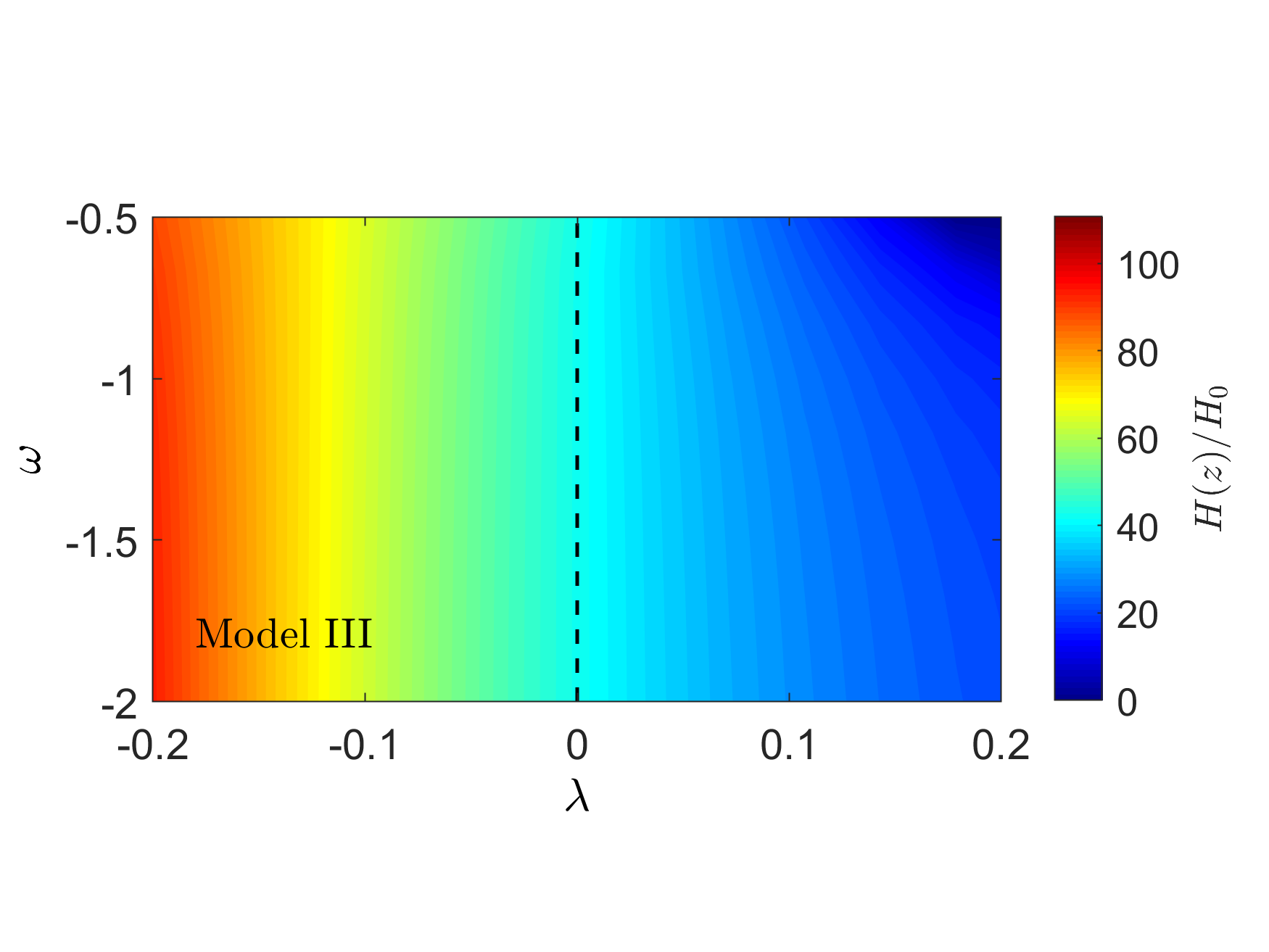}\\
			(e)&(f)\\
		\end{tabular}
		\caption{\label{fig:hubble} Evolution of Hubble parameter ($H(z)$) for IDE Model I, Model II and Model III with different values of coupling parameters $\lambda$. The plots on the left panel (Figure~\ref{fig:hubble}a, \ref{fig:hubble}c and \ref{fig:hubble}e) give the $H(z)$ evolution using different numerical values of $\lambda$ where the other parameters ($\omega$ and $H_0$) are chosen from Table~\ref{tab:constraints}. The plots on right panel (Figure~\ref{fig:hubble}b, Figure~\ref{fig:hubble}d and Figure~\ref{fig:hubble}f show the variation of $H(z)$ with simultaneous variations of the parameters $\lambda$ and $\omega$ (keeping $H_0$ fixed) at $z=17.2$. The dashed black line represents the variations of $H(z)$ for the $\Lambda$CDM case in each figure.}
	\end{figure*}
	
	In presence of the dark matter - dark energy interaction, the energy density parameters of the dark matter and dark energy evolve non-linearly with redshift $z$. As a consequence, remarkable fluctuation can be observed in the evolution of the Hubble parameter $H(z)$ at higher redshift $z$ (see Figure~\ref{fig:hubble}). In Figure~\ref{fig:hubble}, the variation of Hubble parameter as a dimensionless quantity $H(z)/H_0$ (here $H_0$ denotes the current value of the Hubble parameter) is shown with redshift $z$. The left panel of Figure~\ref{fig:hubble} (Figure~\ref{fig:hubble}a, \ref{fig:hubble}c and \ref{fig:hubble}e) shows the variation of the Hubble parameter with redshift for different chosen values of IDE coupling parameter $\lambda$, where the different values of $\lambda$ are represented by different colours (cyan to magenta, (see colour-bars)). On the other hand, Figures.~\ref{fig:hubble}b, \ref{fig:hubble}d and \ref{fig:hubble}f (right hand plots of Figure~\ref{fig:hubble}) show the variations of $H(z)/H_0$ with $\lambda$ and the equation of state parameter ($\omega$) simultaneously. The plots of Figure~\ref{fig:hubble}a, \ref{fig:hubble}c and \ref{fig:hubble}e (left panel plots of Figure~\ref{fig:hubble}) correspond to Model I, II and III (Table~\ref{tab:stability} and Table~\ref{tab:constraints}) respectively of DM - DE interaction. In the plots of Figure~\ref{fig:hubble}b, \ref{fig:hubble}d and \ref{fig:hubble}f also correspond to Model I, II and III respectively. In all three plots of the left panel of Figure~\ref{fig:hubble} (for three IDE Models), the values of $\omega$ for the corresponding models are adopted from Table~\ref{tab:constraints}. In each of the plots of Figure~\ref{fig:hubble}, the dashed black line represents the $\Lambda$CDM case ($\lambda = 0$). Analysing all the plots in Figure~\ref{fig:hubble}, one can conclude that, the evolution of Hubble parameter is almost identical for Model II and Model III where the variations mostly depend on $\lambda$ (see Figure~\ref{fig:hubble}d and \ref{fig:hubble}e, the dependence on $\omega$ is minimal) and the variation is almost linear in nature (see Figure~\ref{fig:hubble}c and \ref{fig:hubble}d). Nonetheless, very small difference can be observed between Model II and Model III in Figure~\ref{fig:hubble}d and \ref{fig:hubble}e (also comparing Figure~\ref{fig:hubble}c and \ref{fig:hubble}d). In contrast, for IDE Model I, both the parameters, $\lambda$ and $\omega$, are equally significant in the Hubble parameter evolution (see Figure~\ref{fig:hubble}b). From Figure~\ref{fig:hubble}a, it can be observed that, initially $H(z)/H_0$ decreases gradually with increasing coupling parameter $\lambda$. But later, (for higher values of $\lambda$) $H(z)/H_0$ suffers rapid fall with increasing $\lambda$. Therefore, although the allowed range of $\lambda$ for Model I is $\lambda<-2\omega \Omega_{\chi}$ (see Table~\ref{tab:stability}), we limit this range to $0<\lambda<0.25$ for the current work.

\section{Temperature Evolution} \label{sec:T_evol} 

	As already mentioned earlier, we considered the DM-DE interaction along with dark matter - baryon scattering and evaporation of PBH in the temperature evolution of baryonic fluid and the dark matter fluid and finally computed the corresponding effects in the brightness temperature of the 21-cm signal ($T_{21}$). The effect of the baryon - DM scattering has already been discussed in \Autoref{chp:21_jan}, in the context of the global 21-cm signature. In the present work, the effects of DM-DE interaction are also included along with the PBH evaporation in the form of Hawking radiation and the baryon - dark matter scattering in the thermal evolution of the Universe. In the conjecture of IDE models, the evolution equations (with redshift $z$) for dark matter ($T_{\chi}$) and the baryon temperature ($T_b$) take the form
	
	\begin{equation}
		(1+z)\frac{{\rm d} T_\chi}{{\rm d} z} = 2 T_\chi - \frac{2 \dot{Q}_\chi}{3 H(z)}-\frac{1}{n_\chi}\frac{2 \mathcal{Q}}{3 H(z)}, 
		\label{eq:T_chi_DE}
	\end{equation}
	\begin{equation}
		(1+z)\frac{{\rm d} T_b}{{\rm d} z} = 2 T_b + \frac{\Gamma_c}{H(z)}
		(T_b - T_\gamma)-\frac{2 \dot{Q}_b}{ 3 H(z)}-\mathcal{J}_{\rm BH}. 
		\label{eq:T_b_DE}
	\end{equation}
	where, the last term of Eq.~\ref{eq:T_chi_DE} indicates the effects of dark matter - dark energy interaction and the last term of Eq.~\ref{eq:T_b_DE} represents the contribution of PBHs in the form of Hawking radiation \cite{BH_21cm_2} given by,
	\begin{equation}
		\mathcal{J}_{\rm BH}=\frac{2}{3 k_B H(z)}
		\frac{K_{\rm BH}}{1+f_{\rm He}+x_e}.
		\label{JBH_DE}
	\end{equation}

	In this case, the expression of the evolution equation for ionization fraction $x_e$ is similar to that as described in the previous chapter (\Autoref{chp:21_feb}) given by,	
	\begin{equation}
		\frac{{\rm d} x_e}{{\rm d} z} = \frac{1}{(1+z)\,H(z)}\left[I_{\rm Re}(z)-
		I_{\rm Ion}(z)-I_{\rm BH}(z)\right]. 
		\label{eq:xe_DE}
	\end{equation}
	But, the evolution of $H(z)$ in the above equation (Eq.~\ref{eq:xe_DE}) is different from the standard cosmological approach, as described in \Autoref{chp:21_feb}. In this particular case, the evolution of $H(z)$ is evaluated by solving the IDE model equations (Eqs.~\ref{eq:rho_chi} and \ref{eq:rho_de}) simultaneously.

	In this work, we explore the 21-cm anomalous hydrogen absorption spectrum in the reionization era by considering the possible simultaneous effects of Hawking radiation from PBHs along with the baryon - dark matter scattering. Incorporating the energy exchange between dark matter and dark energy, and hence the modifications of the density parameters as well as the Hubble parameter due to the dark matter - dark energy interaction, the thermal evolution of the Universe is addressed where the effect of the evaporation of primordial black holes and the baryon - dark matter interaction (scattering) are considered. In order to estimate the thermal evolution, we numerically solve seven coupled equations, (Eqs.~\ref{eq:rho_chi}, \ref{eq:rho_de}, \ref{eq:PBH}, \ref{eq:T_chi_DE}, \ref{eq:T_b_DE}, \ref{eq:xe_DE} and \ref{eq:V_chib}) simultaneously. 
	\begin{figure}
		\centering
		\begin{center}
			\includegraphics[width=0.8\columnwidth]{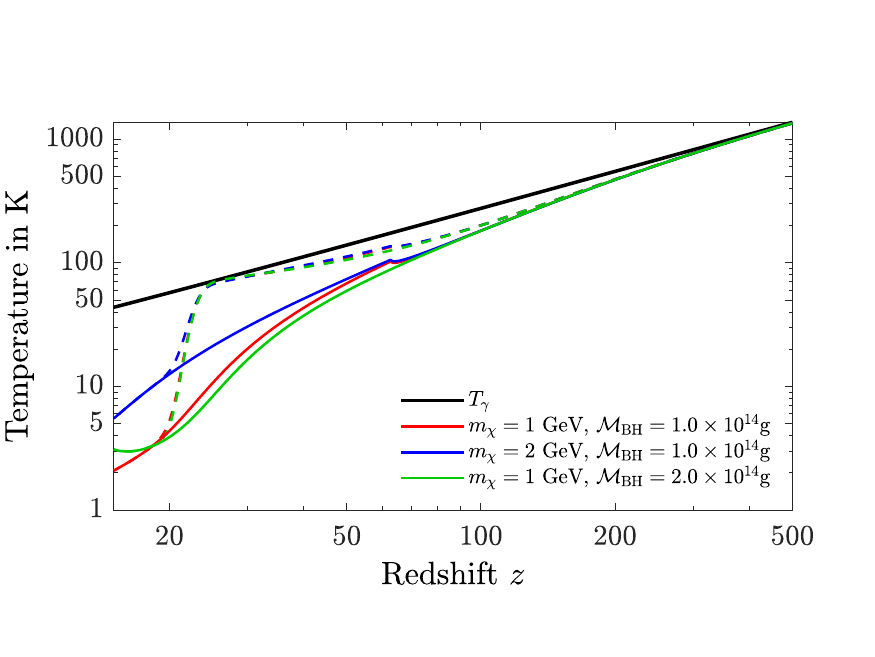}
		\end{center}
		\caption{\label{fig:tspin_DE} The variation of baryon temperature $T_b$, background temperature $T_{\gamma}$, spin temperature $T_s$ with redshift $z$. The black solid line represents the variation of $T_{\gamma}$ and the coloured solid lines and coloured dashed lines are representing the variations of baryon temperature $T_b$ and corresponding spin temperature $T_s$ respectively with redshift $z$, for different chosen sets of dark matter masses and PBH masses. Note that for spin temperature $T_s$, the plots for all three sets almost coincide at higher redshifted era. For both $T_b$ and $T_s$, the computations are made with Model I (Table~\ref{tab:stability} and \ref{tab:constraints}) only.}
	\end{figure}
	In Figure~\ref{fig:tspin_DE}, the evolution of baryon temperature ($T_b$), background temperature ($T_{\gamma}$) and corresponding spin temperature ($T_s$) are plotted with the cosmological redshift $z$. The solid red line, shown in Figure~\ref{fig:tspin_DE} describes the baryon temperature ($T_b$). Spin temperature ($T_s$, red dashed line) variations in presence of dark matter mass $m_{\chi}=1$ GeV and PBHs of mass $\mathcal{M}_{\rm{BH}}=1.0 \times 10^{14}$ g, with the IDE Model I (model parameters are chosen from Table.~\ref{tab:constraints}) is also shown. The green and blue solid and dashed lines are for the same with $m_{\chi}=1$~GeV, $\mathcal{M}_{\rm{BH}}=2.0\times 10^{14}$ g and $m_{\chi}=2$~GeV, $\mathcal{M}_{\rm{BH}}=1.0\times 10^{14}$ g respectively. In all the cases however Model I for DM - DE interaction is used. 
	
\begin{figure}
		\centering
		\includegraphics[width=0.8\columnwidth]{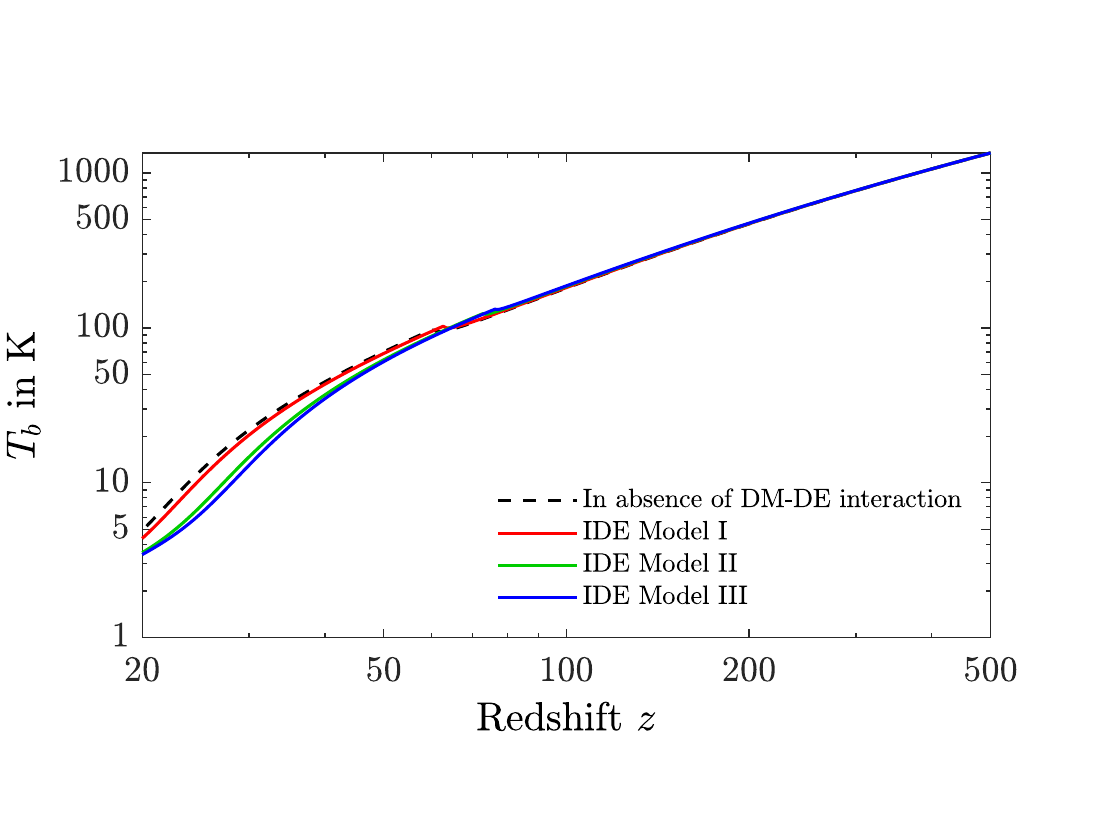}
		\caption{\label{fig:tb} Variation of baryon temperature ($T_b$) with redshift $z$ for different IDE models. For all these plots we consider $\lambda = 0.05$, $\mathcal{M_{\rm BH}}=10^{14}$g, $\beta_{\rm BH} = 10^{-29}$, $m_{\chi}=1$ GeV, $\sigma_{41}=1$, while $\omega$ is taken from Table~\ref{tab:constraints} for different IDE models.}
	\end{figure}

	The contribution of the dark matter - dark energy interaction in the evolution of baryon temperature ($T_b$) is significant in the context of global 21-cm signature. In Figure~\ref{fig:tb}, the variations of $T_b$ with $z$ are shown for different IDE models and compared with the same in absence of the interaction between two dark sector components (i.e. dark matter and dark energy). All the plots in Figure~\ref{fig:tb} are computed for $\lambda = 0.05$, $\mathcal{M_{\rm BH}}=10^{14}$g, $m_{\chi}=1$ GeV, $\beta_{\rm BH} = 10^{-29}$, $\sigma_{41}=1$, the equation of state parameter $\omega$ for different IDE models are taken from Table~\ref{tab:constraints}. It can also be seen that, the variation in the evolution of $T_b$ are barely observed except very mildly around $z\sim 20$, while for different choices of $m_{\chi}$ and $\mathcal{M_{\rm BH}}$ the variation is observable for $z\leq 100$.
	
	\begin{figure}
		\centering
		\includegraphics[width=0.8\columnwidth]{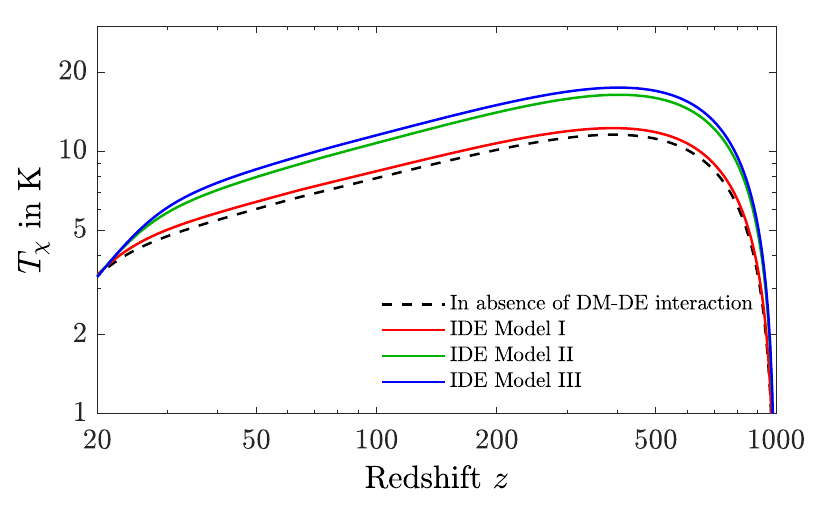}
		\caption{\label{fig:tchi} Same as Figure~\ref{fig:tb} for  dark matter temperature ($T_{\chi}$).}
	\end{figure}
	The effect of DM-DE interaction is also manifested in the dark matter temperature ($T_{\chi}$). From Figure~\ref{fig:tchi} it can be seen that the $T_{\chi}$ increases as an outcome of the interaction between two dark sector components (dark energy and dark matter). This phenomenon indicates that a significant amount of energy transfers from dark energy to dark matter, as an outcome of DM - DE interaction. It is also noticed from both Figure~\ref{fig:tb} and Figure~\ref{fig:tchi}, that the amount of energy transfer is remarkably high for the case of IDE Model II and III in comparison to the IDE Model I.

\section{\label{sec:result_feb}Influences of PBH and IDE models in $T_{21}$}
	As mentioned earlier, the 21-cm line is originated as an outcome of the transition of electrons between two hyperfiine states of hydrogen atom namely, a triplet spin 1 state and a singlet spin 0 states. The intensity of the 21-cm absorption line is represented by the brightness temperature ($T_{21}$), that is essentially depended on the optical depth and hence the Hubble parameter ($H(z)$) (see Eq.~\ref{eq:tau}). The variation of the 21-cm brightness temperature with redshift ($z$) can be expressed as,
	\begin{equation}
		T_{21}=\dfrac{T_s-T_{\gamma}}{1+z}\left(1-e^{-\tau}\right)
		\label{eq:t21_BH}
	\end{equation}
 	As the evolution of Hubble parameter depends on the DM - DE interaction, the optical depth of the medium and the brightness temperature ($T_{21}$) are also significantly modified by the IDE models.

	\begin{figure}
		\centering
		\includegraphics[width=0.8\columnwidth]{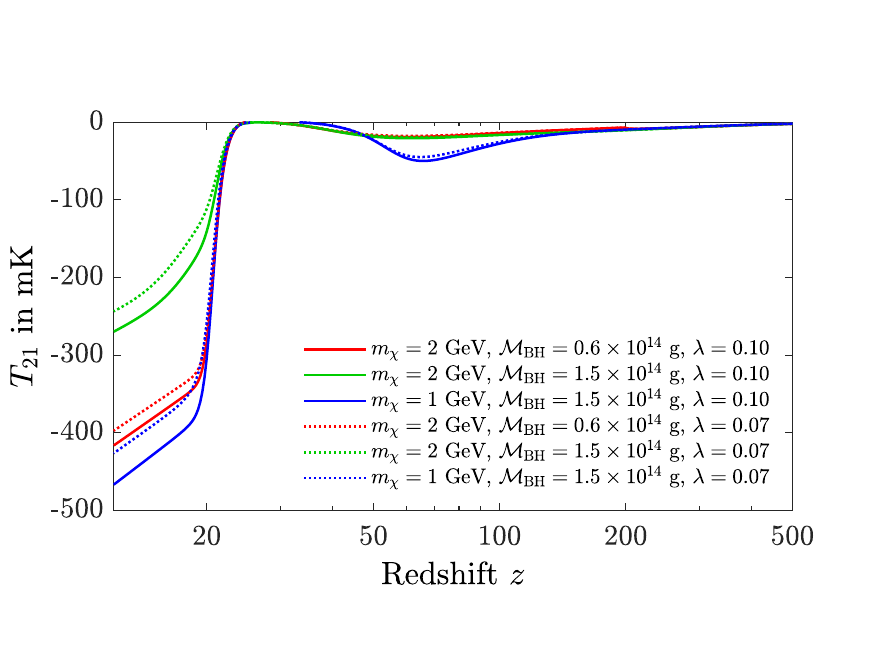}
		\caption{\label{fig:t21_DE} Brightness temperature ($T_{21}$) vs redshift ($z$) graph for different values of dark matter masses ($m_{\chi}$), PBH masses ($\mathcal{M_{\rm BH}}$) and  IDE coupling parameter $\lambda$ in the case IDE Model I.}
	\end{figure}
\begin{figure*}
	\centering
	\begin{tabular}{c}
		\includegraphics[width=0.7\linewidth]{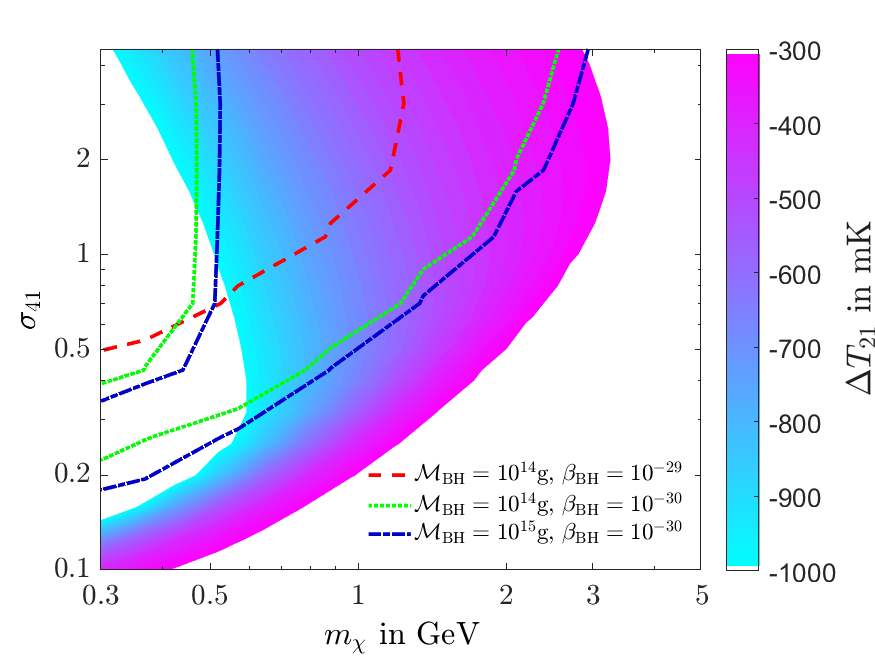}\\(a)\\
		\\
		\includegraphics[width=0.7\linewidth]{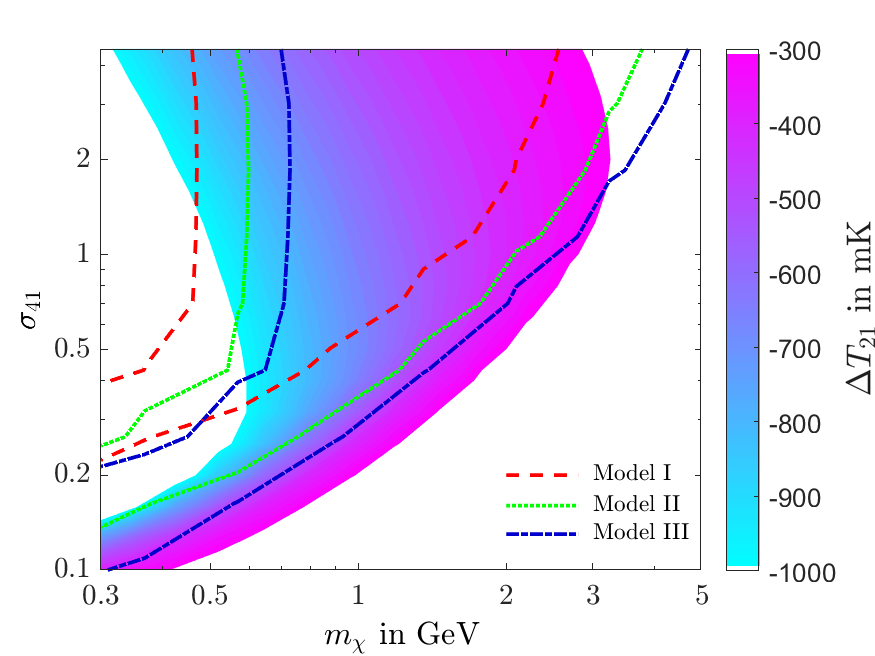}\\
		(b)\\
	\end{tabular}
	\caption{\label{fig:mchi_sigma_DE} The computed allowed region in the $m_{\chi}$ - $\sigma_{41}$ space using the EDGES's limit of $\delta T_{21}$. The coloured shaded region describes the allowed region for the $\Lambda$CDM model in both cases. In this coloured region, DM - DE interaction has not been considered. The coloured lines represent the similar bounds for different values of PBH parameters (Figure~\ref{fig:mchi_sigma_DE}a), for the case of IDE Model I whereas, in Figure~\ref{fig:mchi_sigma_DE}b, the estimated bounds are shown for different IDE models (with $\mathcal{M_{\rm BH}}=10^{14}$g, $\beta_{\rm BH}=10^{-30}$ and $\lambda=0.1$)}
\end{figure*}

	In Figure~\ref{fig:t21_DE}, we plot the 21-cm brightness temperatures ($T_{21}$) for different chosen values of IDE coupling parameter ($\lambda=$0.07 and 0.10) and the dark matter masses $m_{\chi} = 2$ GeV and 1 GeV and PBH masses $\mathcal{M}_{\rm BH} = 0.6 \times 10^{14}$ g and $0.6 \times 10^{14}$ g while keeping $\beta_{\rm BH}$ and $\sigma_{41}$ fixed at $10^{-29}$ and 3 respectively in the case of IDE Model I. From this figure (Figure~\ref{fig:t21_DE}), it can be noticed that, besides the DM mass and PBH mass, the IDE coupling parameter $\lambda$ also modifies the brightness temperature remarkably. A detail study for the variation due to $\sigma_{41}$ has been carried out in Figure~\ref{fig:mchi_sigma_DE}. 

	The allowed ranges of the dark matter mass $m_{\chi}$ and the baryon - dark matter scattering cross-section $\sigma_{41}$ are also calculated by incorporating the observational bound of the global 21-cm signal as provided by the EDGES experiment. The allowed regions in $m_{\chi} - \sigma_{41}$ plane are described in Figure~\ref{fig:mchi_sigma_DE} for different PBH masses (Figure~\ref{fig:mchi_sigma_DE}a) and also for three different IDE models (Figure~\ref{fig:mchi_sigma_DE}b). The contour plots are plotted for a fixed redshift $z=17.2$. In what follows the brightness temperature of 21-cm spectra at $z=17.2$ is represented by $\Delta T_{21}$ as mentioned in the previous chapter (\Autoref{chp:21_feb}). We estimate the limit for $m_{\chi}$ and $\sigma_{41}$ using the constraint which is obtained from the observational result of EDGES ($-300$ mK $\geq T_{21} \geq -1000$ mK). In both the graphs of Figure~\ref{fig:mchi_sigma_DE} ((a) and (b)), the calculated allowed zones in the $m_{\chi}$ - $\sigma_{41}$ plane are plotted for $\Lambda$CDM model using the colour code (coloured shaded region), where the individual colours within the coloured shaded region indicate the different values of $\Delta T_{21}$ (colour code is described in the corresponding colour-bar). It can be notices that the bounds on the DM masses $m_{\chi}$ obtained from these contour plots agree with a similar calculation given in \cite{rennan_3GeV} (i.e. $m_{\chi} \leq 3$ GeV). In both the plots, the coloured contour plots are generated by varying $m_{\chi}$ and $\sigma_{41}$ and computing $\Delta T_{21}$. For the coloured shaded regions (not the line contours) in both Figure~\ref{fig:mchi_sigma_DE}a and \ref{fig:mchi_sigma_DE}b, only the dark matter - baryon interaction is considered and the relevant coupled differential parameters are simultaneously solved numerically. The other contour plots are generated using all the effects considered here namely, the baryon - dark matter interaction, dark matter - dark energy interaction and PBH evaporations are shown by line contour plots (allowed region enclosed by lines) in both Figure~\ref{fig:mchi_sigma_DE}a and Figure~\ref{fig:mchi_sigma_DE}b. In these latter cases the allowed regions are bound by different pairs of lines in $m_{\chi}$ - $\sigma_{41}$ plane. In Figure~\ref{fig:mchi_sigma_DE}a, three different line contours represent three chosen sets of PBH parameters (PBH masses and $\beta_{\rm BH}$) where IDE Model I (Table~\ref{tab:stability} and \ref{tab:constraints}) is used to explain the DM - DE interaction for all the three allowed regions. In Figure~\ref{fig:mchi_sigma_DE}b, the three allowed contours (area enclosed by different lines) in $m_{\chi} - \sigma_{41}$ plane are generated with three IDE models (Model I, II and III, Table~\ref{tab:stability}, \ref{tab:constraints}) while the PBH mass and $\beta_{\rm BH}$ are kept fixed at values of $10^{14}$ g and $10^{-29}$ respectively. 

	From Figure~\ref{fig:mchi_sigma_DE}a it can be seen that the region enclosed by red dashed line corresponds to the $m_{\chi}$ - $\sigma_{41}$ allowed region when PBH mass $\mathcal{M_{\rm BH}} = 10^{14}$ g and $\beta_{\rm BH}=10^{-29}$ are chosen with Model I for DM - DE interaction. Similarly, the region enclosed by the green dotted lines specifies the allowed region when $\mathcal{M_{\rm BH}}=10^{14}$ g, $\beta_{\rm BH}=10^{-30}$ are considered. From Figure~\ref{fig:mchi_sigma_DE}a this can be observed that as the effect of Hawking radiation increases, the allowed region shifts towards higher $\sigma_{41}$ and lower $m_\chi$ region. A similar trend is also observed for the IDE Model I (Figure~\ref{fig:mchi_sigma_DE}b). On the other hand, for IDE Model II and III, the allowed region shifts toward higher values of $m_{\chi}$.


	\begin{figure*}
		\vspace{0.03\linewidth}
		\begin{center}
		\begin{tabular}{ccc}
			\includegraphics[trim=0 20 80 26, clip, height=0.27\textwidth]{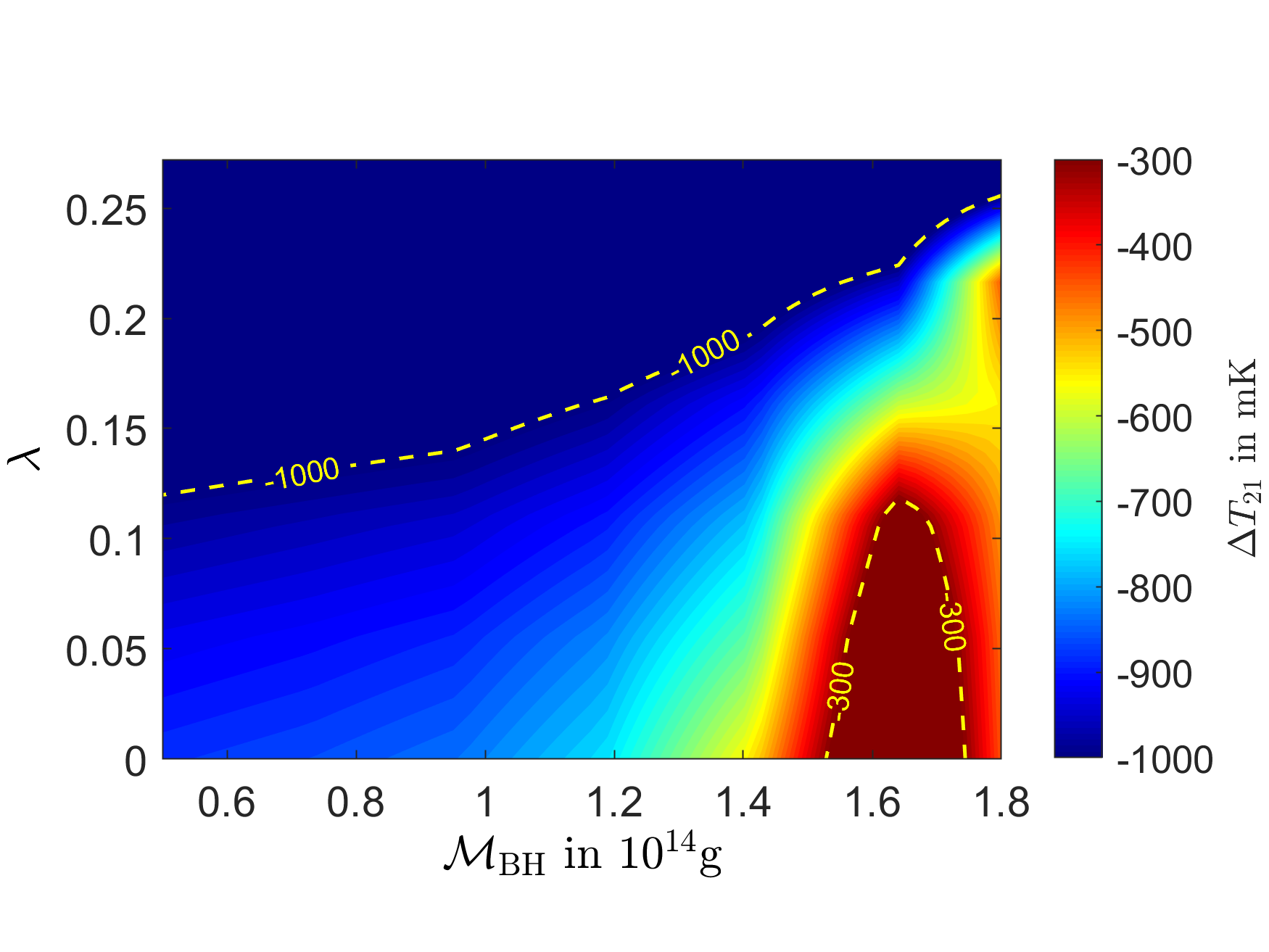}&
			\includegraphics[trim=51 20 80 26, clip, height=0.27\textwidth]{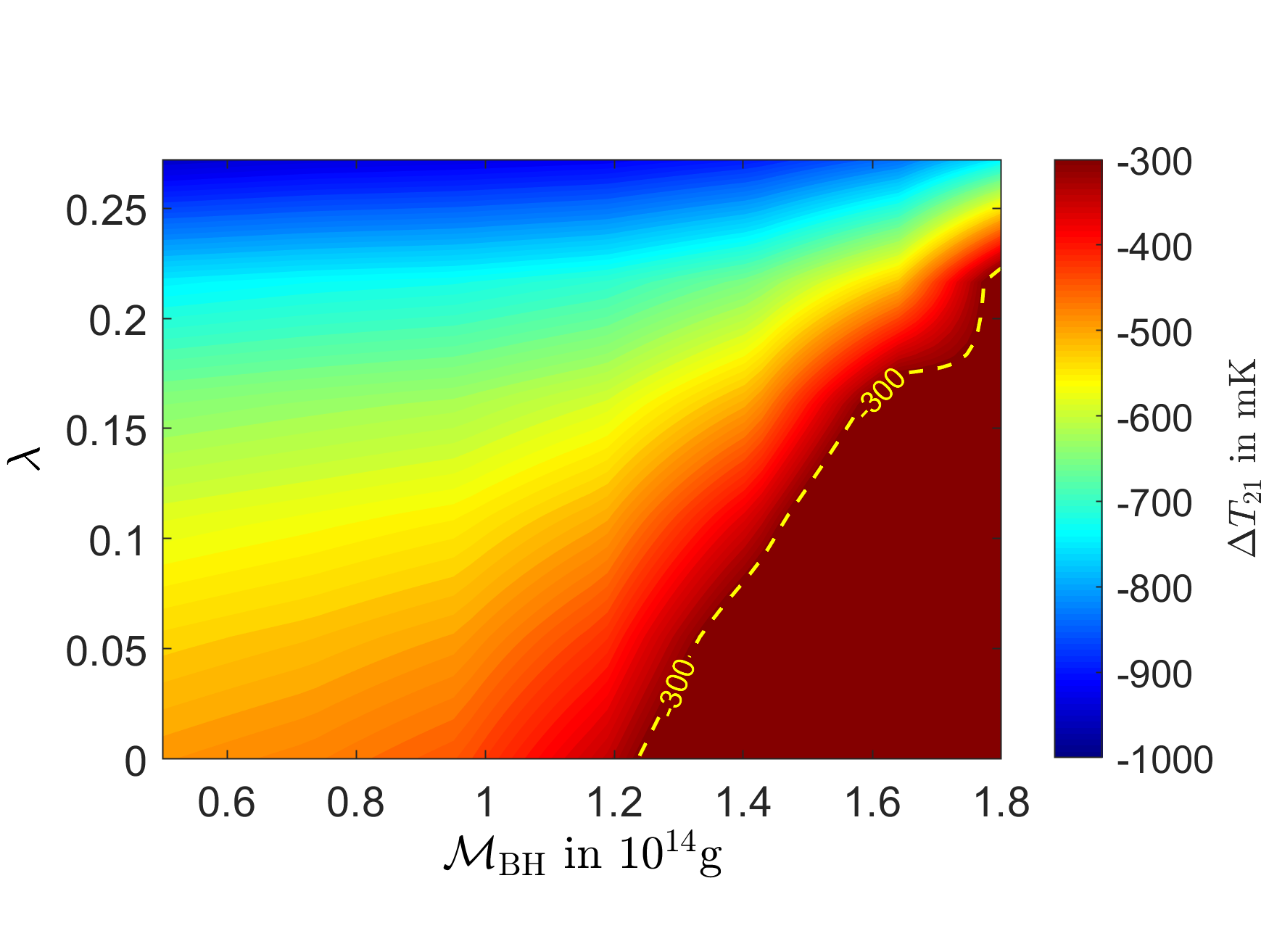}&
			\includegraphics[trim=51 20 80 26, clip, height=0.27\textwidth]{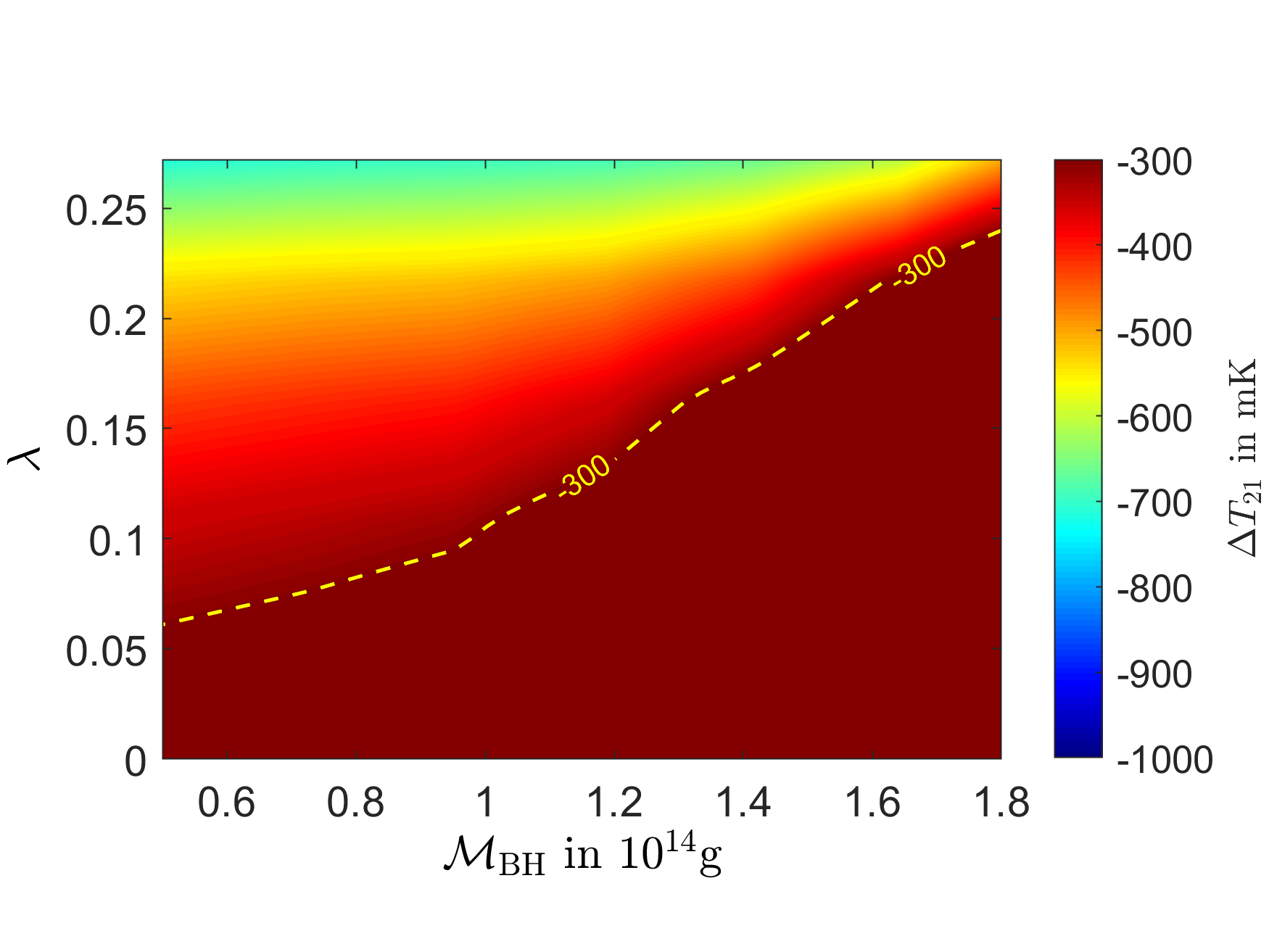}\\
			(a)&(b)&(c)\\
			\\
			\includegraphics[trim=0 20 80 26, clip, height=0.27\textwidth]{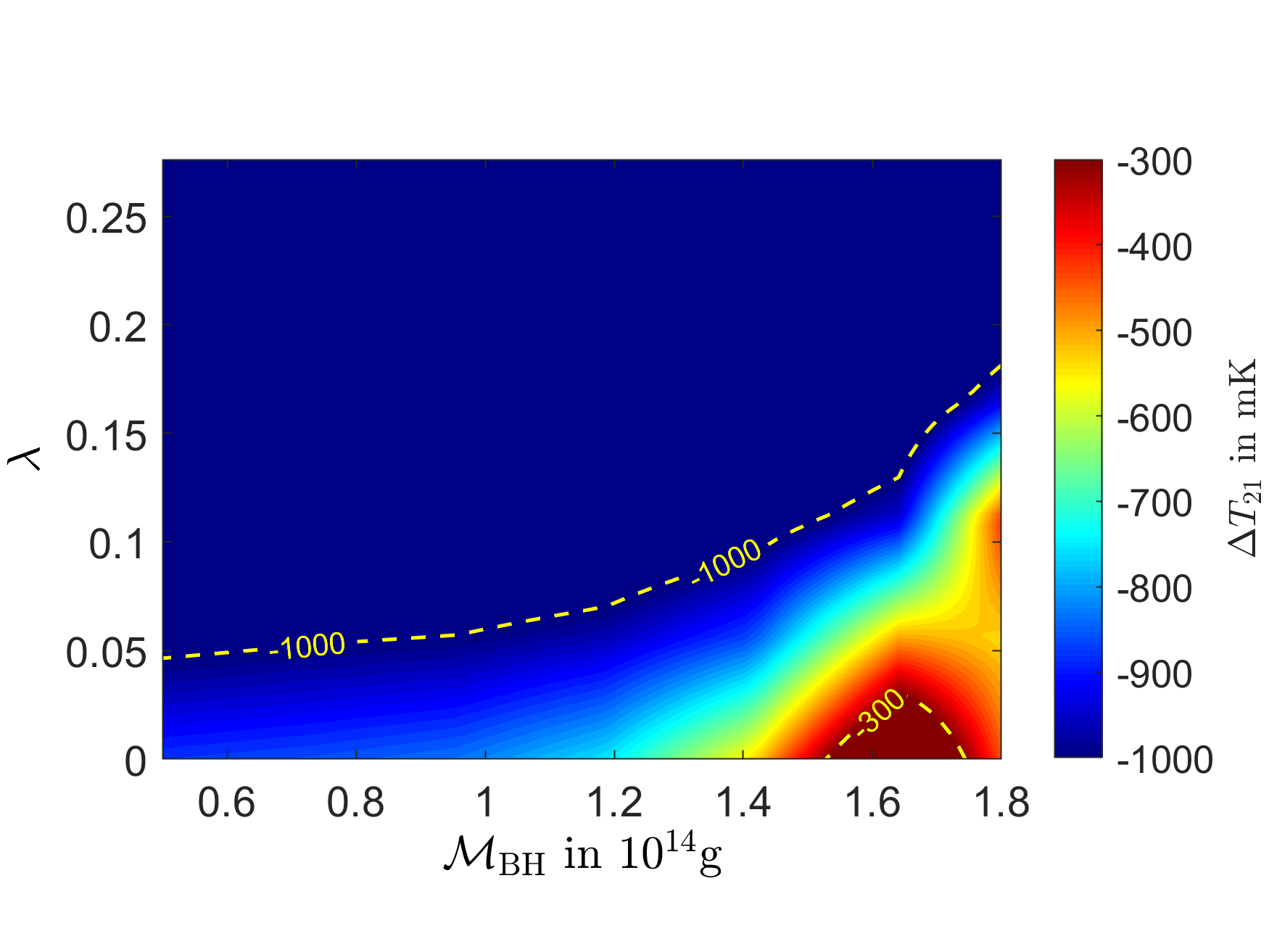}&
			\includegraphics[trim=51 20 80 26, clip, height=0.27\textwidth]{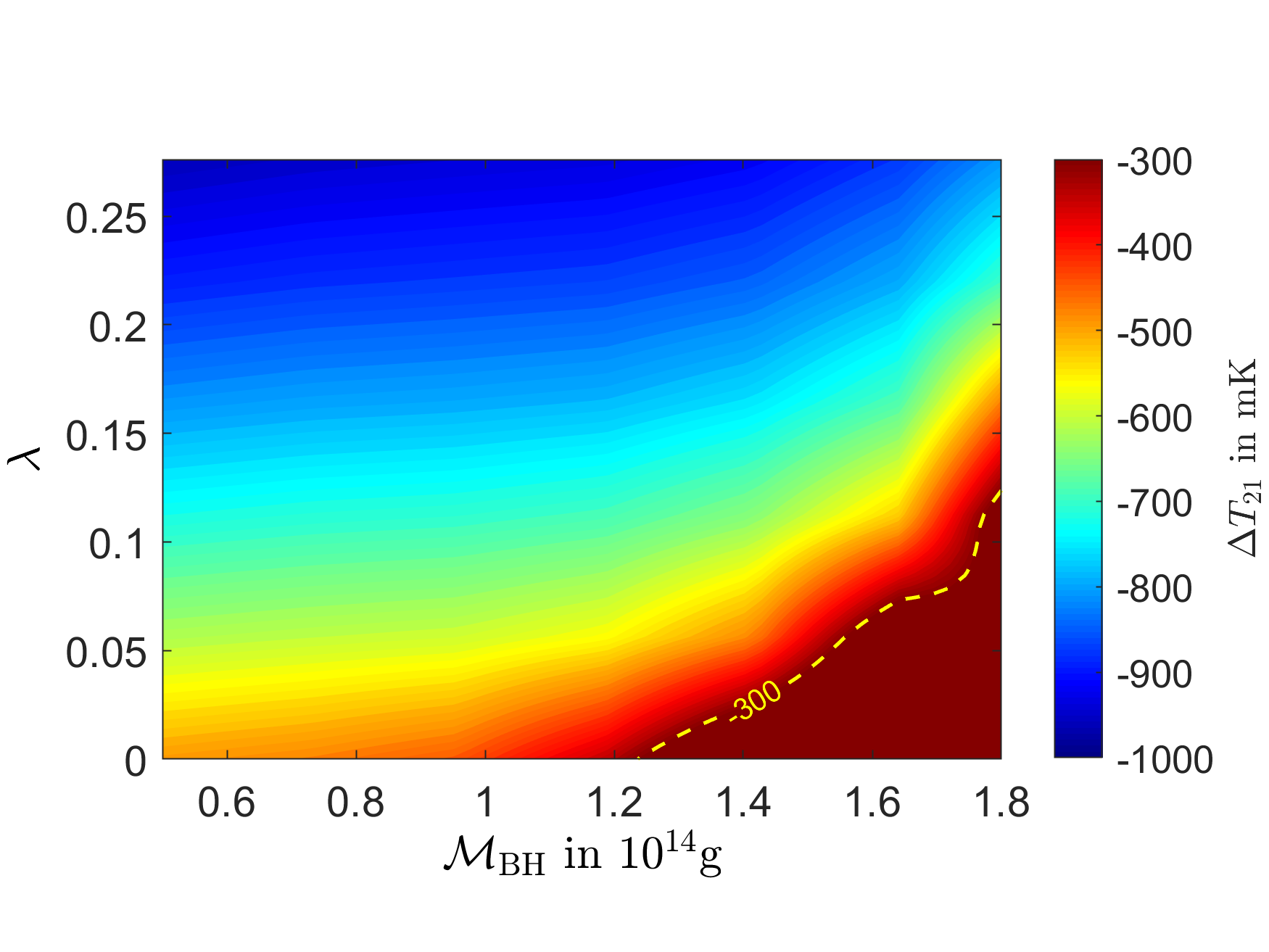}&
			\includegraphics[trim=51 20 80 26, clip, height=0.27\textwidth]{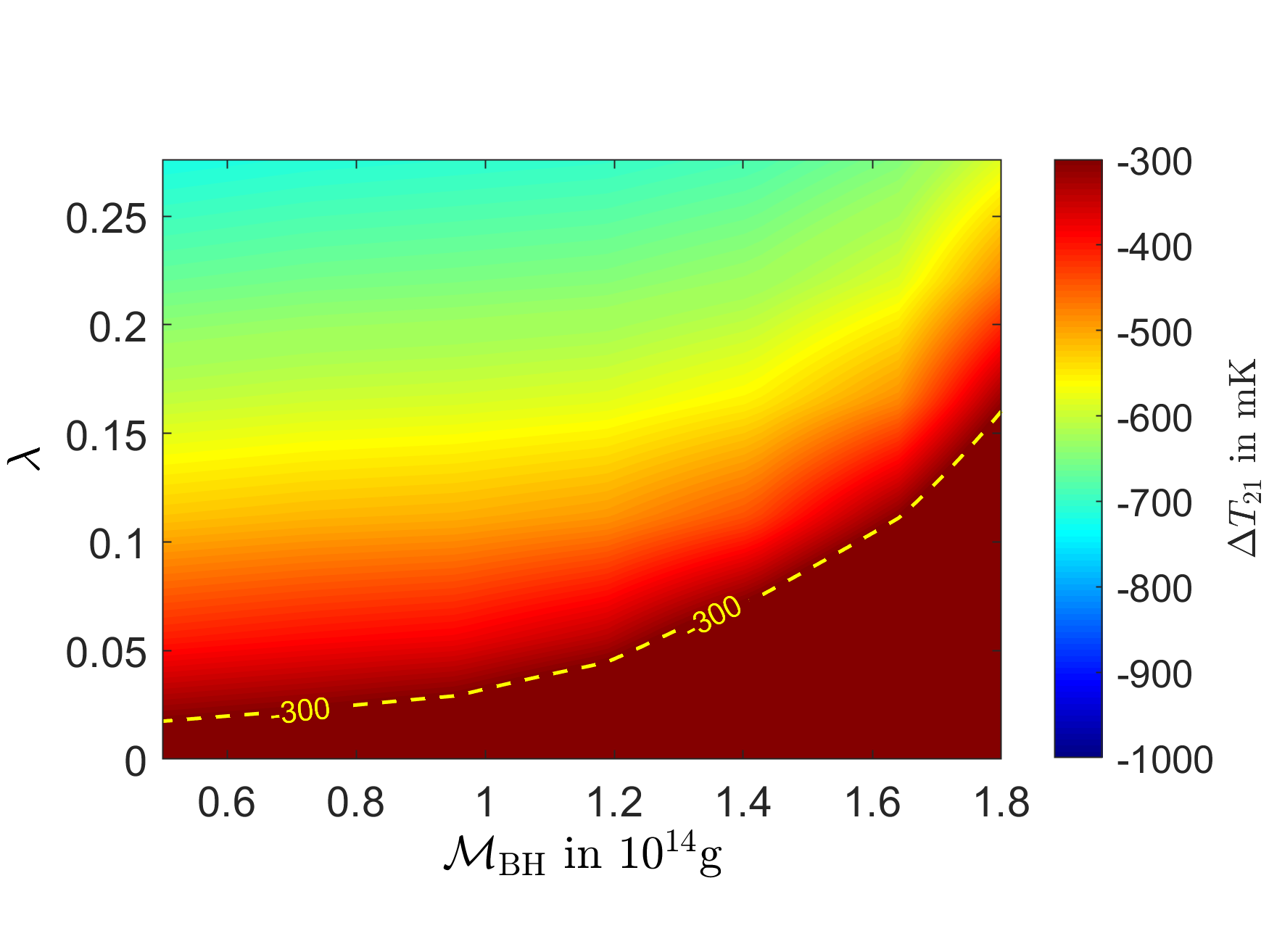}\\
			(d)&(e)&(f)\\
			\\
			\includegraphics[trim=0 20 80 26, clip, height=0.27\textwidth]{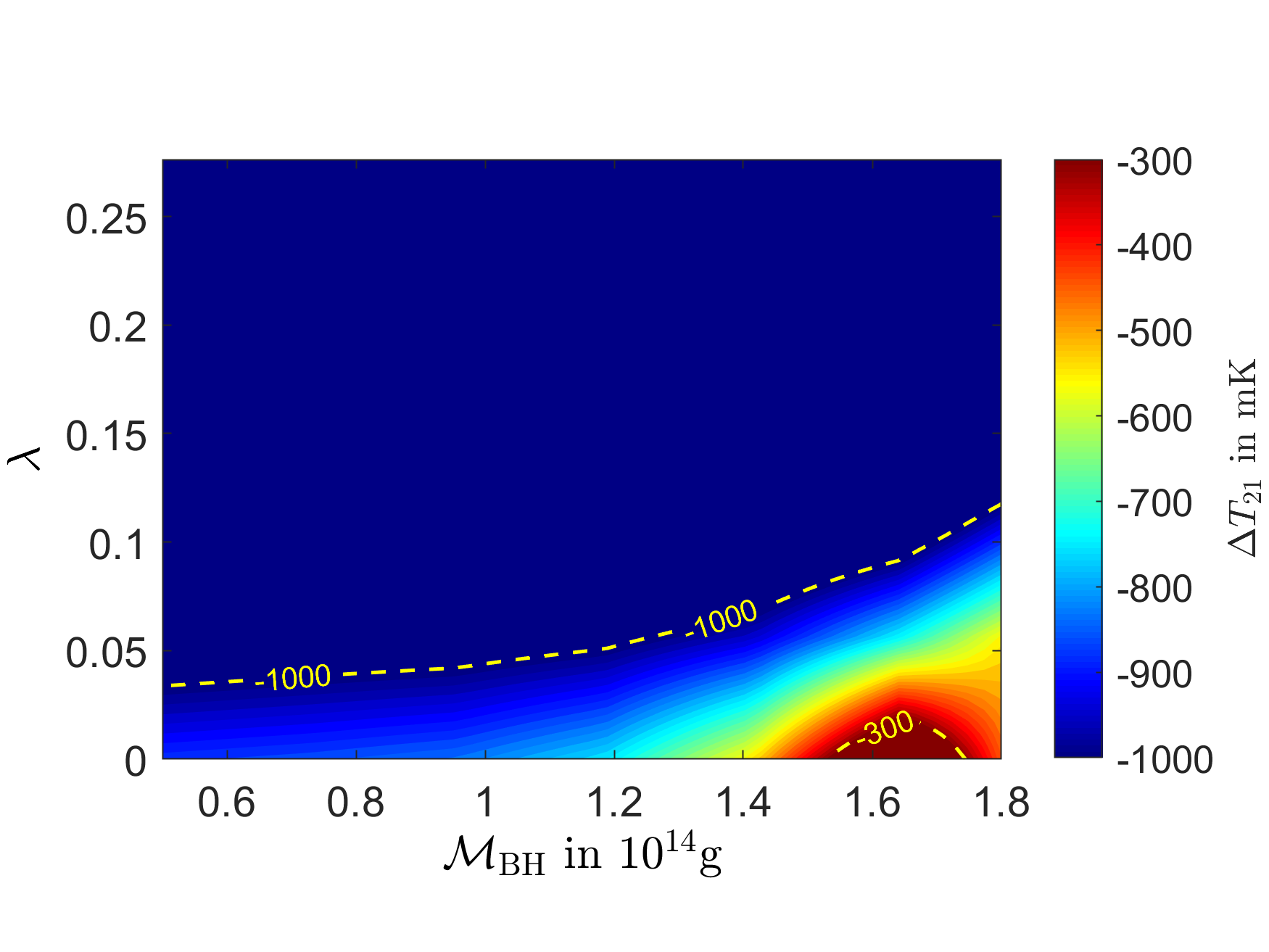}&
			\includegraphics[trim=51 20 80 26, clip, height=0.27\textwidth]{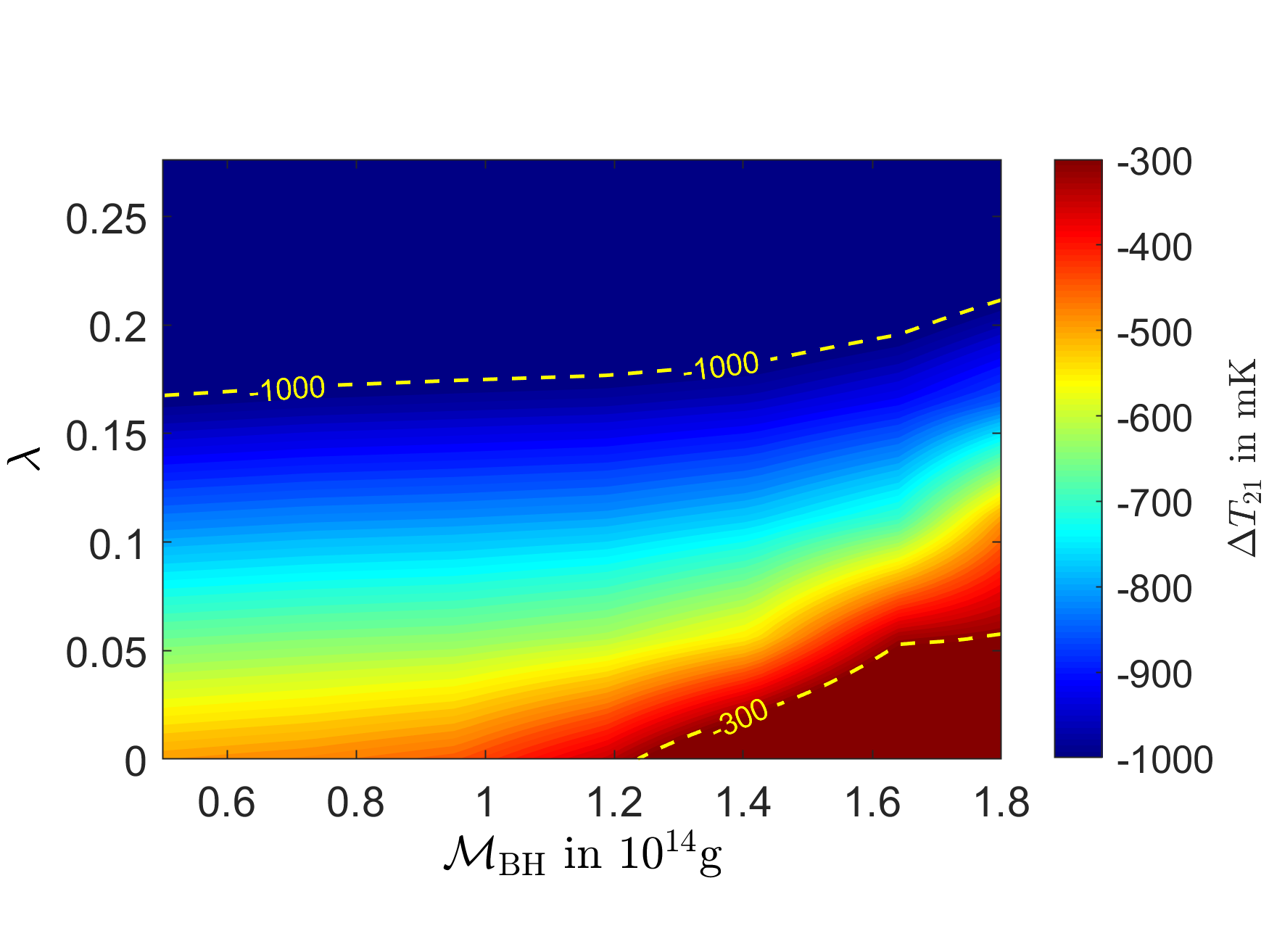}&
			\includegraphics[trim=51 20 80 26, clip, height=0.27\textwidth]{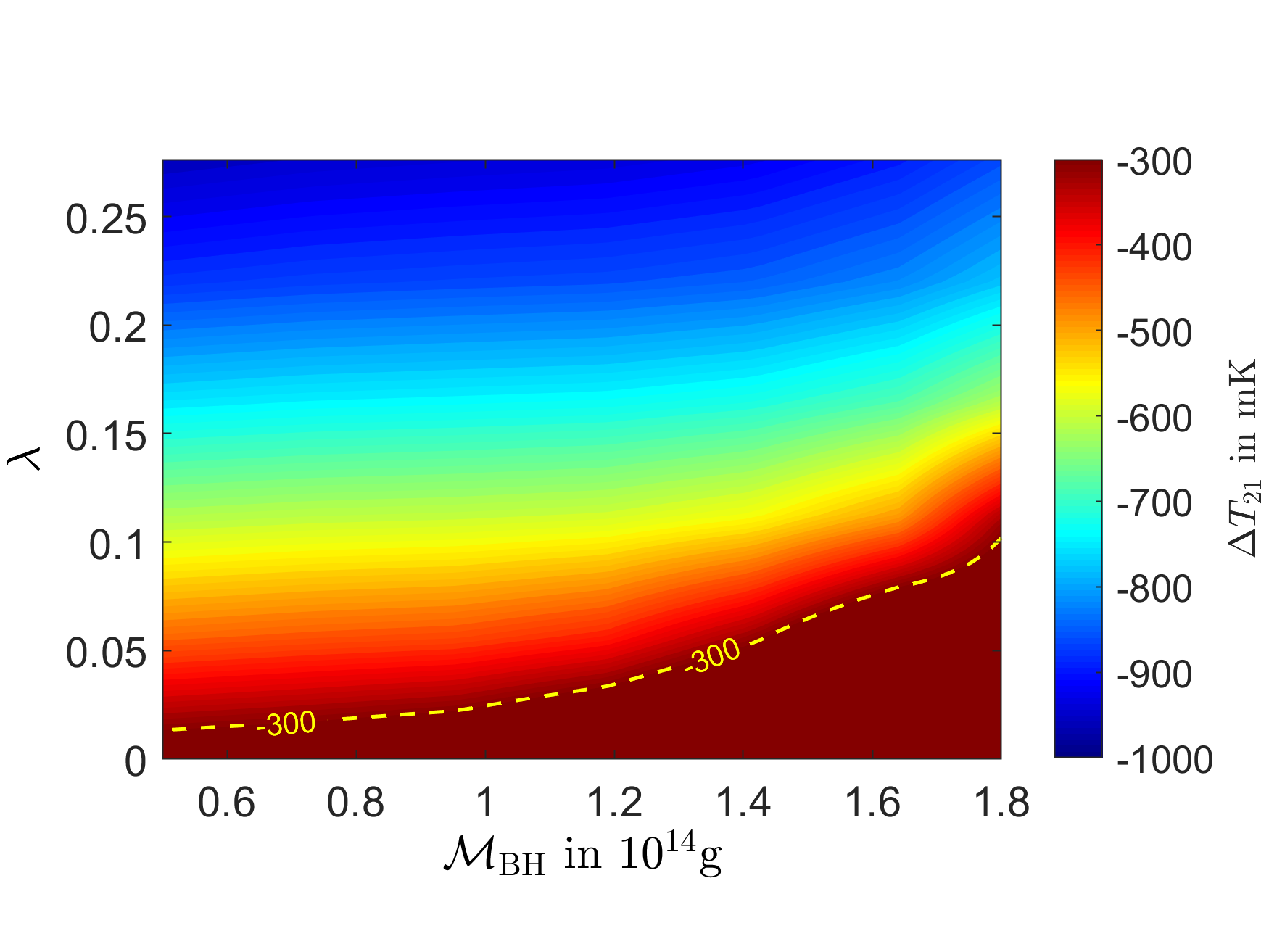}\\
			(g)&(h)&(i)\\
		\end{tabular}
		\vspace{0.03\linewidth}\\
		\begin{tabular}{c}
			\includegraphics[width=0.6\textwidth]{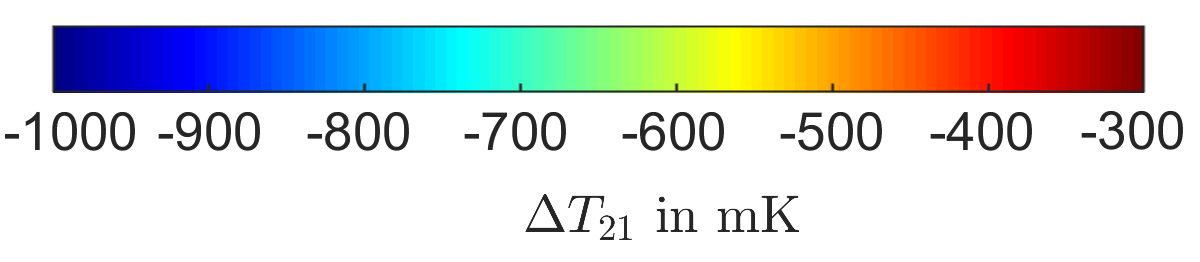}
		\end{tabular}
		\caption{\label{fig:mbh_lambda} Variation of $\Delta T_{21}$ with PBH mass and DE-DM coupling parameter $\lambda$ for different DM masses and IDE models (see text for details).}
		\vspace{0.03\linewidth}
		\end{center}
	\end{figure*}

	Finally, a detailed study has been carried out to explore similar bounds in the $\mathcal{M}_{\rm BH}$ - $\lambda$ plane for individual IDE models and its variation with $m_{\chi}$ as well. We use three different values of DM masses namely $m_{\chi}=$ 0.5 GeV, 1.0 GeV and 1.5 GeV. In each of these cases, the value of $\sigma_{41}$ is set to 1.0. These are shown in Figure~\ref{fig:mbh_lambda}. The first, second and third row correspond to the IDE Model I, II and III respectively while each of the columns 1, 2 and 3 are for three fixed dark masses, $m_\chi$ = 0.5 GeV, 1 GeV, 1.5 GeV respectively. For example, Figure~\ref{fig:mbh_lambda}a shows the fluctuation of $\Delta T_{21}$ in $\mathcal{M}_{\rm BH}$ - $\lambda$ space for $m_{\chi}=0.5$ GeV and the IDE Model I, where the value of $\Delta T_{21}$ at each point is described in the colour bar shown at the end of the figure. In all the plots of Figure~\ref{fig:mbh_lambda}, the yellow dashed lines represent the bounds from EDGES result. Figure~\ref{fig:mbh_lambda}b and Figure~\ref{fig:mbh_lambda}c describe the same for $m_{\chi}=$1.0 GeV and 1.5 GeV respectively while IDE Model I is considered. From these figures, one can see that, as $\mathcal{M}_{\rm BH}$ increases, $\Delta T_{21}$ increases for a fixed value of $\lambda$ (except a little distortion at $\mathcal{M_{\rm BH}}=1.7 \times 10^{14}$g). It can also be noticed (from Figure~\ref{fig:mbh_lambda}a, \ref{fig:mbh_lambda}b and \ref{fig:mbh_lambda}c) that as $m_{\chi}$ increases (from column 1 to column 3), $\Delta T_{21}$ also increases. As mentioned, the results for the case of Model II and Model III are furnished in plots~\ref{fig:mbh_lambda}d, \ref{fig:mbh_lambda}e, \ref{fig:mbh_lambda}f and in plots~\ref{fig:mbh_lambda}g, \ref{fig:mbh_lambda}h, \ref{fig:mbh_lambda}i respectively. From these figures (all plots of Figure~\ref{fig:mbh_lambda}), it can be noticed that, when $m_{\chi}=$ 0.5 GeV, $\lambda$ values for each of the three IDE model constraints (see benchmark points described in Table~\ref{tab:constraints}) satisfy the EDGES limits for $\mathcal{M_{\rm BH}}\lessapprox 1.5\times10^{14}$ g and $\mathcal{M_{\rm BH}}\gtrapprox1.8\times10^{14}$ g. Those constraints also agree with the EDGES limit for PBH masses $\mathcal{M_{\rm BH}}\lessapprox1.2\times10^{14}$ g, when $m_{\chi}=1.0$ is considered. But, for higher values of dark matter mass ($m_{\chi}=1.5$ GeV), the benchmark values of $\lambda$ does not satisfy the EDGES limit of $\Delta T_{21}$ (see Figure~\ref{fig:mchi_sigma}c, Figure~\ref{fig:mchi_sigma}f and Figure~\ref{fig:mchi_sigma}i). As a consequence, one can conclude that the model constraints described in Table~\ref{tab:constraints} satisfy the EDGES limit for relatively lower masses of dark matter ($\leq 1.0$ GeV).

	
\chapter{Multimessenger Signals from Heavy Dark Matter Decay} \label{chp:21_mar}
\\
	Keeping in view of the 21-cm scenario, we also explore the multimessenger signals from possible rare decay of fundamental heavy dark matter or superheavy dark matter (HDM) from dark ages leading to the epoch of cosmic reionization. The heavy dark matter candidates of mass as high as $10^8$ GeV or more could be created via gravitational production mechanism or non-linear quantum effects during the reheating or preheating stages after inflation. These dark matters are therefore produced non-thermally in the early Universe and are long-lived. Thus these heavy or super heavy dark matter, if exists, exhibits a rare decay process. Although these heavy dark matter candidates could undergo rare decays, the energy and decay products produced by such decay process may affect the global 21-cm signal, which are originated from the hyperfine transition between two energy states of hydrogen during the reionization. We have explored, possible multimessenger signals from the decay of such superheavy or heavy dark matter candidates. The ultra high energy neutrino (in the $\sim$ PeV energy regime) signals as detected at IceCube square-kilometer detector can be one of those signals \cite{PeV-1,PeV-2,PeV-3,PeV-4,IND-ice,esmaili2019,esmaili2017,esmaili2013,galacticPeV1,galacticPeV2,galacticPeV3,ICnuExcess,DM-nu,esmaili2014,esmaili2015} whereas the other signal attributes to the heating of the baryonic medium by exchanging of the energy, which is produced from the rare decay process of HDM candidate and the corresponding impact on the global 21-cm signal. The effects of the baryon scattering with the CDM-type (cold dark matter) light DM candidates and the Hawking radiation from primordial black holes are also considered in the formalism of the brightness temperature of the global 21-cm signal and corresponding effects are also addressed.
	
	\section{Introduction}
	Heavy dark matter has turned out to be a topic of interest in high-energy particle physics after the detection of high-energy neutrinos at the square-kilometer detector by IceCube Collaboration. Although one of those events coincides with the multimessenger $\gamma$-ray observations by the collaboration of MAGIC, Fermi-LAT, MAGIC, HESS, AGILE and HAWC, which is indicating the flairing blazer TXS0506+056 \cite{IceCube:2018dnn}, the sources of other observed events are still unknown. Therefore those unknown neutrinos could be originated from the rare decay of the HDM particles.
	
	The so-called HESE (high energy starting events, also known as the contained vertex events) events that are obtained from track and shower events with energy deposition $> 20$ TeV apparently fits a power-law spectrum $\sim E^{-2.9}$. But up-going muon neutron events in the energy range $\sim 120$ TeV or above in the PeV range, show a different power-law dependence. In this work, the IceCube 7.5 year muon neutrino event data \cite{IC_7.5yr} are considered. We calculate the neutrino flux that can be obtained from heavy dark matter decay of a given mass and then compares the theoretical events with the IceCube experimental data in the PeV regime. We may mention here the analysis of similar nature had been addressed earlier by previous authors \cite{nu_line,nu_broad,Murase:2015gea}.
	
	The decay process of primordial superheavy dark matter (HDM) may proceed via QCD cascades. The primary products $q\bar{q}$, on hadronization and decay produce leptons, $\gamma$ etc. as the end products. Ultra high energy (UHE) neutrinos are one of the possible final products of this rare decay process. We attribute these high-energy neutrinos from the decay of primordial heavy dark matter to be the source of UHE $\nu$ signal, as detected at IceCube neutrino detector in the $\sim$ PeV energy range. On the other hand, we also explore the energy associated with the HDM decay process and hence the effect of the consequent heating on the evolution of absorption temperature of 21-cm hydrogen spectra during the reionization era. We have also included other cooling/heating effects that could have been originated due to the collisions of cold DM particles with baryons besides the evaporation of primordial black holes or PBH.
	 
	
	
	In addition to the HDM decay, we also consider another thermal WIMP type cold dark matter (CDM) within the mass range of few hundred MeV to few GeV (This is consistent with {\bf Barkana} \emph{et~al}. \cite{rennan_3GeV}). But the contribution of the heavy dark matter to the total amount of dark matter content of the Universe is presumed to be small while the lighter cold dark matter (CDM) component accounts for the remaining amount of dark matter containing our Universe. Also, the collisional effect of the WIMP dark matter with the baryons modifies the baryon temperature ($T_b$) and consequently the brightness temperature of 21-cm neutral hydrogen spectra ($T_{21}$) during the reionization epoch.

	As mentioned earlier, energy injection is also possible in case of the decay of heavy primordial dark matter. Such heavy candidates of DM can be gravitationally produced in the early Universe or it can also be produced during the reheating phase after inflation. Such a heavy dark matter can undergo cascade decay processes through the hadronic channel or leptonic channel (or both) to produce UHE neutrinos (and other particles). The UHE neutrinos can be observed by Km$^2$ detectors such as the IceCube experiment. Such possibilities are earlier explored in Ref.~\cite{mpandey}.
	
	In this work, we study possible multimessenger signals of HDM decay whereby the two effects namely the neutrinos from HDM decay and the possible effects on 21-cm signal at reionization epoch due to the heat transfer following the decay process are addressed. In the present work, we consider two additional effects on the thermal evolution of the Universe, while computing the temperature of the 21-cm line at dark ages and reionization epoch. One is the heavy dark matter decay and the other is the evaporation of primordial black holes. In addition, the effects of the baryon - dark matter interactions in the evolution of $T_b$ are also included.

\section{\label{sec:HDM} Heavy Dark Matter Decay}
	The decay of superheavy or heavy dark matter (HDM) having mass significantly higher than the electroweak scale takes place via cascading of QCD partons \cite{kuz,bera,bera1,bera2}. We used only the hadronic decay channel ($\chi\rightarrow q\bar{q}$) in our calculation as the contribution of the leptonic decay channel ($\chi\rightarrow l\bar{l}$) is much smaller than the hadronic channel \cite{kuz,mpandey}. The total spectrum at scale $s=M_{\rm HDM}^2$ can be obtained by summing over all parton fragmentation function $D^{h} (x,s)$ from the decay of particle $h$. The fragmentation functions at higher values of $s$ can be estimated by evolving DGLAP (Dokshitzer - Gribov - Lipatov - Altarelli - Parisi) equation \cite{bera1,addicaler,PhysRevD.94.063535}, given by,
	\begin{equation}
		\dfrac{\partial D_i^h(x,M_{\rm HDM}^2)}{\partial \ln M_{\rm HDM}^2}=\dfrac{\alpha_s(M_{\rm HDM}^2)}{2 \pi} \sum_j \int_{\chi}^1 \dfrac{dz}{z} P_{ij}(z,\alpha_s(M_{\rm HDM}^2))\times D_j^h\left(\dfrac{x}{z},M_{\rm HDM}^2\right),
		\label{eq:dglap}
	\end{equation}
	where $i$ denotes the parton of different types (quark ($q$), gluon ($g$)). In fact, one has to evolve coupled evolution equation for the case of quarks and gluons \cite{bera1,PhysRevD.94.063535}. 
	
	Here $x=\dfrac{2E}{M_{\rm HDM}}$ is the dimensionless energy. $i$ denotes the different types of parton (quark ($q$) and gluon ($g$)) and $P_{ij}$ represents the probability that the `$i$' parton produces the `$j$' parton having dimensionless energy $x$. This ($P_{ij}$) is also known as the splitting function for parton branching. In our current analysis, only muon decay is considered in this calculation as the total contribution of the other mesons is negligible ($<10\%$) \cite{kuz,bera1}. The DGLAP equation is evaluated numerically by considering the conventional standard model QCD splitting function \cite{Jones:1983eh}. For this purpose, we follow the prescription given in Ref.~\cite{PhysRevLett.86.3224}.
	
	The spectra of neutrinos ($\nu$), photons ($\gamma$) and electrons ($e$) are written as \cite{kuz},
	\begin{equation}
		\frac {dN_{\nu}}{dx}=2R \int_{xR}^{1} \frac{dy}{y} D^{\pi^{\pm}}(y)+2\int_{x}^{1}\frac{dz}{z}f_{\nu_i}\left(\frac{y}{z}\right)D^{\pi^{\pm}}(z),
		\label{form1_hdm}
	\end{equation}
	\begin{equation}
		\frac {dN_{\gamma}} {dx} = 2 \int_{x}^{1} \frac{dz}{z} D^{\pi^0}(z),
		\label{eq:gamma_hdm}
	\end{equation}
	\begin{equation}
		\frac {dN_{e}}{dx}=2R \int_{x}^{1} \frac{dy}{y}\left(\frac{5}{3}+3y^2+\frac{4}{3}y^3\right)\int_{x/y}^{x/(ry)}\frac{dz}{z}D^{\pi^{\pm}}(z).
		\label{eq:electron_hdm}
	\end{equation}
	In the above equations (Eqs.~\ref{form1_hdm}, \ref{eq:gamma_hdm} and \ref{eq:electron_hdm}), we can define $D^{\pi} (x,s)$ as $D^{\pi} \equiv [D_{q}^{\pi} (x,s) + 
	D_{g}^{\pi} (x,s)]$, where $x=\dfrac{2E}{M_{\rm HDM}}$ and $R = \displaystyle\frac {1} {1-r}$, $r = (m_\mu/m_\pi)^2 \approx 0.573$. In Eq.~\ref{form1_hdm}, the term $f_{\nu_i} (x)$ is adopted from {\bf Kelner} \emph{et~al.} \cite{kelner}
	\begin{eqnarray}
		f_{\nu_i} (x) &=& g_{\nu_i} (x) \Theta (x-r) +(h_{\nu_i}^{(1)} (x) + 
		h_{\nu_i}^{(2)} (x))\Theta(r-x) \,\, , \nonumber\\
		g_{\nu_\mu} (x) &=& \displaystyle\frac {3-2r} {9(1-r)^2} (9x^2 - 6\ln{x} -4x^3 -5), \nonumber\\
		h_{\nu_\mu}^{(1)} (x) &=& \displaystyle\frac {3-2r} {9(1-r)^2} (9r^2 - 6\ln{r} -
		4r^3 -5), \nonumber\\
		h_{\nu_\mu}^{(2)} (x) &=& \displaystyle\frac {(1+2r)(r-x)} {9r^2} [9(r+x) - 
		4(r^2+rx+x^2)], \nonumber\\
		g_{\nu_e} (x) &=& \displaystyle\frac {2} {3(1-r)^2} [(1-x) (6(1-x)^2 + 
		r(5 + 5x - 4x^2)) + 6r\ln{x}],\nonumber\\
		h_{\nu_e}^{(1)} (x) &=& \displaystyle\frac {2} {3(1-r)^2} [(1-r)
		(6-7r+11r^2-4r^3) + 6r \ln{r}], \nonumber\\
		h_{\nu_e}^{(2)} (x) &=& \displaystyle\frac {2(r-x)} {3r^2} (7r^2 - 4r^3 +7xr 
		-4xr^2 - 2x^2 - 4x^2r).
		\label{form2_hdm}
	\end{eqnarray}
	
	In this work, the neutrino spectra are obtained by computing Eqs.~\ref{form1_hdm} and \ref{form2_hdm} for different values of $M_{\rm HDM}$. The photon and electron spectra also can be compute using Eqs.~\ref{eq:gamma_hdm} -- \ref{form2_hdm}. The neutrino spectrum $\frac{dN_{\nu}}{dx}$ in Eq.~\ref{form1_hdm} represents the combination of neutrinos and antineutrinos spectrum for three flavours ($\nu_e$, $\nu_{\mu}$ and $\nu_{\tau}$) where the neutrinos are considered to be originated via pion decay at the ratio 1:2:0 at the source.
	
	The extragalctic counterpart of	muon flux from the rare decay of HDM particles is given by \cite{kuz}, 
	\begin{equation}
		\frac{d\Phi_{\rm EG}}{dE} (E_\nu) = \frac{\mathcal{K}}{4\pi M_{\rm HDM}} \int_{0}^{\infty}\frac{\rho_0 c /H_0}{\sqrt{\Omega_m (1+z^3) + (1-\Omega_m)}} \frac{dN_{\nu_{\mu}}}{dE} [E(1+z)] dz.
		\label{form3}
	\end{equation}
	In the above equation (Eq.~\ref{form3}), $\mathcal{K}=f_{\rm HDM}\Gamma$, where $f_{\rm HDM}$ is the fraction of the heavy dark matter and $\Gamma$ is the decay width of the HDM candidate. The Hubble radius (or proper radius of the Hubble sphere) is given as $c/H_0 = 1.37 \times 10^{28}$ cm. The average dark matter density at $z=0$ (the present epoch) is denoted as $\rho_0 = 1.15 \times 10^{-6}$ GeV cm$^{-3}$) and $\Omega_m = 0.316$ where $\Omega_m$ represents the density parameter of the dark matter. The neutrinos undergo flavour oscillate during their propagation from the source to the Earth.
	But given the astronomical distance they traverse before reaching the Earth, the neutrinos originated with the flavour ratio $\nu_e:\nu_{\mu}:\nu_{\tau}=1:2:0$ and reaches the Earth (as the oscillation part is averaged out) with a flavour ratio $1:1:1$. Thus the muon neutrino spectrum $\frac{dN_{\nu_{\mu}}}{dE}=\frac{1}{3}\frac{dN_{\nu}}{dE}$ on reaching the Earth.
	
	In addition to the extragalactic part, the galactic $\nu_{\mu}$ flux from the HDM decay is given by \cite{kuz}
	\begin{equation}
		\frac{d\Phi_{\rm G}}{dE} (E_\nu) = \frac{\mathcal{K}}{4\pi M_{\rm HDM}} \int_{V} \frac{\rho_\chi (R[r])}{4\pi r^2} \frac{dN_{\nu_{\mu}}}{dE}(E,l,b) dV,
		\label{form4}
	\end{equation}
	where $(l,\, b)$ denotes position of any point in the galactic coordinate system. The density of DM at a distance $R$ from the centre of the galactic halo distribution is denoted as $\rho_{\chi} (R[r])$ and $r$ is the distance to the earth (observer) from the galactic centre. In our calculation, we have taken into account the Navarro-Frenk-White (NFW) density profile of dark matter halo and the density profile is integrated over the entire galactic halo of the Milky way galaxy for which $R_{\rm max}$ is considered to be 260 Kpc. 
	
	The total neutrino flux, which is also referred to as the total theoretical flux ($\phi^{\rm Th}$), at energy $E_{\nu}$ can be obtained by considering the contribution of both the extragalactic and the galactic $\nu_{\mu}$ flux as
	\begin{equation}
		\phi^{\rm Th} (E_\nu) = \frac{d\Phi_{\rm EG}}{dE} (E_{\nu}) + \frac{d\Phi_{\rm G}}{dE} (E_\nu).
		\label{form5}
	\end{equation}
	The quantities $\mathcal{K}$ and $M_{\rm HDM}$ are two parameters to be obtained by comparing and fitting the IceCube experimental data with $\phi^{\rm Th} (E_\nu)$.

\section{\label{sec:T_evol} Heavy Dark Matter Decay and Baryon Temperature Evolution in Dark Ages}
	The energy injection rate per unit time per unit volume due to the decay of HDM is given by
	\begin{equation}
		\left.\dfrac{{\rm d} E}{{\rm d}V {\rm d}t}\right|_{\rm{HDM}}=\rho_{\chi} f_{\rm HDM} \Gamma,
		\label{eq:hdm_inj_0}
	\end{equation}
	where, as mentioned in Section~\ref{sec:HDM}, $\Gamma$ is the decay width, $f_{\rm HDM}$ denotes the fraction of DM in the form of heavy dark matter and $\rho_{\chi}$ is the total dark matter density given by $\rho_{\chi}=\rho_{\chi,0} (1+z)^3$ where $\rho_{\chi,0}$ is the total dark matter density at the current epoch. With $\mathcal{K}=f_{\rm HDM} \Gamma$, the above equation is reduced to  
	\begin{equation}
		\left.\dfrac{{\rm d} E}{{\rm d}V {\rm d}t}\right|_{\rm{HDM}}=\rho_{\chi}\mathcal{K}.
		\label{eq:hdm_inj}
	\end{equation}

	In the present analysis, the temperature evolution of the Universe is studied by evolving the baryon temperature ($T_b$) and the dark matter temperature ($T_{\chi}$) with cosmological redshift $z$, where the Universe is considered to be charge-neutral ($n_p = n_e$, $n_p$ is the abundance of protons and $n_e$ is the same for electrons). After incorporating the energy injected into the system due to rare decay of HDM particles as also due to the baryon-dark matter scattering and primordial black hole evaporation, the evolution equation of the baryon temperature $T_b$ takes the form \cite{BH_21cm_1,BH_21cm_2,munoz,amar21,corr_equs},
	\begin{equation}
		(1+z)\frac{{\rm d} T_b}{{\rm d} z} = 2 T_b + \frac{\Gamma_c}{H(z)}
		(T_b - T_{\gamma})-\frac{2 \dot{Q}_b}{ 3 H(z)}-\frac{2}{3 k_B H(z)} \frac{K_{\rm HDM}+K_{\rm BH}}{1+f_{\rm He}+x_e}.
		\label{eq:T_b_HDM}
	\end{equation}
	However, the evolution equation for dark matter temperature $T_{\chi}$ remains the same as described in \Autoref{chp:21_feb}
	\begin{equation}
		(1+z)\frac{{\rm d} T_\chi}{{\rm d} z} = 2 T_\chi - \frac{2 \dot{Q}_\chi}{3 H(z)}.
		\label{eq:T_chi_HDM}
	\end{equation}
	In Eq.~\ref{eq:T_b_HDM}, $\Gamma_c$ represents the effect of Compton interaction, given by
	\begin{equation}
		\Gamma_c=\dfrac{8\sigma_T a_r T^4_{\gamma}x_e}{3(1+f_{\rm He}+x_e)m_e c},
	\end{equation} 
	where $\sigma_T$ is the Thomson scattering cross-section and $a_r$ is the radiation constant. $x_e$ denotes the ionization fraction and $f_{\rm He}$ is the fractional abundance of He. The last term of the Eq.~\ref{eq:T_b_HDM} represents additional contributions to the evolution of baryon temperature $T_b$ due to heavy dark matter (HDM) decay \cite{BH_21cm_1,Mitridate:2018iag} and PBH evaporation \cite{BH_21cm_1,BH_21cm_2,amar21}. The quantities $K_{\rm HDM}$ and $K_{\rm BH}$ are defined as
	\begin{equation}
	 	K_{\rm HDM}=\chi_h f(z)_{\rm HDM} \frac{1}{n_b} \times \left.\dfrac{{\rm d} E}{{\rm d}V {\rm d}t}\right|_{\rm{HDM}}, \label{KBH_HDM}
	\end{equation}
	\begin{equation}
		K_{\rm BH}=\chi_h f(z)_{\rm BH} \frac{1}{n_b} \times \left.\dfrac{{\rm d} E}{{\rm d}V {\rm d}t}\right|_{\rm{BH}}. \label{KBH_HDM_1}
	\end{equation}
 	Here $\chi_h=(1+2x_e)/3$ is the fraction of the deposited energy into the system in the form of baryon heating \cite{BH_21cm_2,BH_21cm_4,chen,PhysRevD.76.061301,Furlanetto:2006wp}, $f(z)_{\rm HDM(BH)}$ denotes the ratio of the total energy deposited into the system, to the total amount of energy injected by HDM decay (PBH evaporation) \cite{corr_equs,fcz001,fcz002,fcz003,fcz004}. $n_b$ is the baryon number density due to HDM decay (PBH evaporation).
	
	The ionization fraction $x_e$ essentially depends on the total amount of energy deposited in the medium. The combined effect of heavy dark matter decay and PBH evaporation transform the evolution equation of the ionization fraction $x_e$ with redshift $z$ given by \cite{BH_21cm_2,munoz,amar21},	
	\begin{equation}
		\frac{{\rm d} x_e}{{\rm d} z} = \frac{1}{(1+z)\,H(z)}\left[I_{\rm Re}(z)-
		I_{\rm Ion}(z)-(I_{\rm HDM}(z)+I_{\rm BH}(z))\right], 
		\label{eq:xe_HDM}
	\end{equation}
	where, $I_{\rm Re}(z)$ is the standard recombination and $I_{\rm Ion}(z)$ is the ionization rate respectively \cite{munoz,yacine,hyrec11,amar21,amar21_1}. As both terms ($I_{\rm Re}(z)$ and $I_{\rm Ion}(z)$) are functions of $T_b$ and $T_{\gamma}$ \cite{BH_21cm_5,hyrec11,amar21,peeble,pequignot,hummer,seager,amar21_1}, the term $x_e$ depends on baryon temperature and DM temperature simultaneously.
	
	The terms $I_{\rm HDM}$ and $I_{\rm BH}$ appear in the expression of the evolution of ionization fraction (Eq.~\ref{eq:xe_HDM}) are described as
	\begin{equation}
		I_{\rm HDM}=\chi_i f(z)_{\rm HDM} \frac{1}{n_b} \frac{1}{E_0}\times \left.
		\dfrac{{\rm d} E}{{\rm d}V {\rm d}t}\right|_{\rm HDM}. \label{IBH_HDM_1}
	\end{equation}
	\begin{equation}
		 I_{\rm BH}=\chi_i f(z)_{\rm BH} \frac{1}{n_b} \frac{1}{E_0}\times \left.
		 \dfrac{{\rm d} E}{{\rm d}V {\rm d}t}\right|_{\rm BH}. \label{IBH_HDM}
	\end{equation}
	where $E_0$ is the amplitude of the ground state energy of hydrogen atom, $\chi_i=(1-x_e)/3$ is the fractions of total deposited energy in the form of ionization \cite{BH_21cm_2,PhysRevD.76.061301,chen,BH_21cm_4,Furlanetto:2006wp}. In the case of HDM decay, the term $f(z)_{\rm HDM}$ is estimated by incorporating the photon and electron spectra as obtained by the Eqs.~\ref{eq:gamma_hdm} and \ref{eq:electron_hdm} (see section~\ref{sec:HDM}) ~\cite{corr_equs,fcz001,fcz002,fcz003,fcz004}. 	
	At higher redshift and energy, almost entire flux of electrons transfer to photons by the process of inverse Compton scatter (ICS) and above the threshold of pair-production, the photons efficiently produce $e^{-}e^{+}$ pair. Those electrons and positrons will rapidly reduce to low energy $e^{-}$ and $e^{+}$ via ICS and cascade. As a consequence, in the calculation of $f(z)$, the transfer function does not change significantly with increasing DM mass at higher values. Moreover, it can be noticed that, the available grid data of energy transfer function \cite{fcz001,fcz002} is almost constant for $\gtrapprox 10^3$ GeV for fixed values of input and output redshift \cite{corr_equs,fcz001,fcz002,fcz003,fcz004}. So, in the case of HDM with different masses, we use the transfer function for $10^{12.75}$ eV in our calculations (the electron and photon spectrum are estimated from Eqs.~\ref{eq:gamma_hdm} and \ref{eq:electron_hdm} of Section~\ref{sec:HDM}).
	
	The baryon-DM fluid ($\dot{Q}_b$ and $\dot{Q}_{\chi}$) interaction is principally dependent by the drag term $V_{\chi b}$ \cite{munoz}. The evolution of $V_{\chi b}$ remains unchanged due to the incorporation of an additional dark matter candidate, given by \cite{munoz},
	\begin{equation}
		\frac{{\rm d} V_{\chi b}}{{\rm d} z} = \frac{V_{\chi b}}{1+z}+
		\frac{1}{(1+z) H(z)}\dfrac{\rho_m \sigma_0}{m_b + m_{\chi}} \dfrac{1}{V^2_{\chi b}} F(r), \label{eq:V_chib_HDM}
	\end{equation}
	where, $F(r)={\rm erf}\left(r/\sqrt{2}\right)-\sqrt{2/\pi}r e^{-r^2/2}$ ($\rm erf$ denotes the error function), $r=V_{\chi b}/u_{\rm th}$ and $\sigma_{41}=\frac{\sigma_0}{10^{-41} {\rm cm^2}}$. The term $u_{\rm th}^2$ denotes the variance of thermal relative motion.
	
	It is to be noted that two different DM candidates are considered in the present work. One is the heavy dark matter produced in the early epoch of the Universe. Decay of such DM candidates inject energy into the system and produce UHE neutrinos, while the other is the usual WIMP dark matter that scatters with baryons and exchange heat. Both the processes influence the temperature of the 21-cm line.

\section{\label{sec:result} Calculations and Results}
	\begin{figure}
		\centering{}
		\includegraphics[trim=0 20 0 57, clip, width=0.7\linewidth]{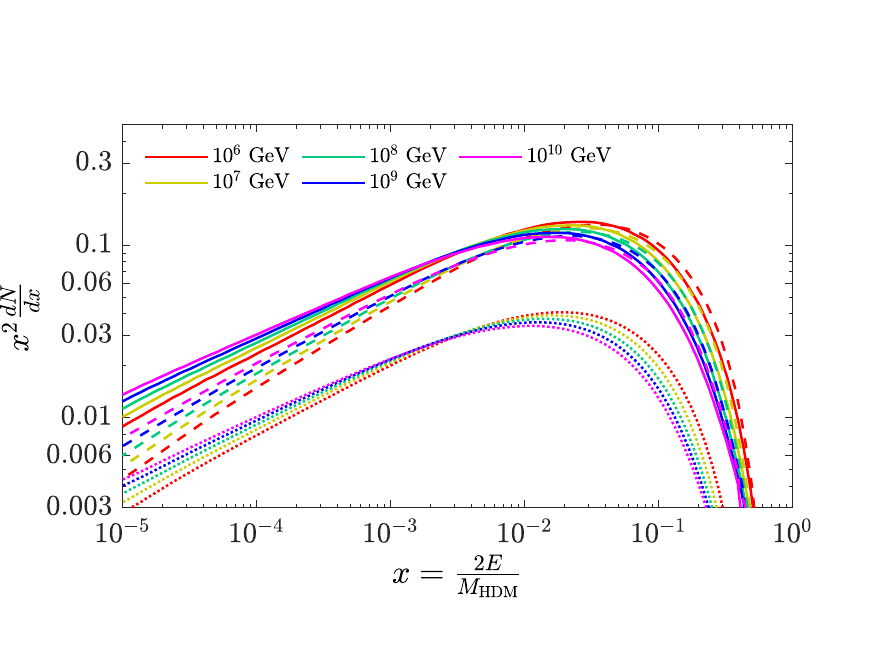}
		\caption{\label{fig:spectrum} Spectra of the final products from HDM decay of different HDM masses $M_{\rm HDM}$. The solid lines represent the total neutrino spectra ($\nu_e+\nu_{\mu}+\nu_{\tau}$, $\nu+\bar{\nu}$) for different $M_{\rm HDM}$. The corresponding $\gamma$ and $e$ spectra are described by dashed and dotted lines respectively.}
	\end{figure}
	
	
	The computation of the neutrino spectrum is performed by numerically evaluating Eqs.~\ref{form1_hdm}--\ref{form2_hdm}. The calculations involve Monte Carlo simulation of numerical evolution of DGLAP equations for obtaining QCD spectrum. These computed neutrino spectra are then used to obtain neutrino flux using Eqs.~\ref{form3}--\ref{form5}. These are furnished in Fig.\ref{fig:spectrum}. In Figure~\ref{fig:spectrum}, the total neutrino flux obtained from the above computation is shown for five chosen values of $M_{\rm HDM}$, the mass of heavy dark matter. These are shown in Figure~\ref{fig:spectrum} using solid lines of different colours for different chosen $M_{\rm HDM}$ values. Spectra for $\gamma$ and $e^-$ (from HDM decay) are also shown in Figure~\ref{fig:spectrum} for reference using coloured dashed lines.
	
	The experimentally observed data are taken from the IceCube 7.5 yr data corresponding to muon through going events. These events beyond the energy $\sim$120 TeV appear to describe a power law of High Energy Starting Events or HESE that includes the shower, as well as track events and are assumed to be obtained from diffuse extragalactic UHE neutrino flux \cite{Kopper:2017Df}. 
	
	\begin{figure}
		\centering{}
		\includegraphics[width=0.7\linewidth]{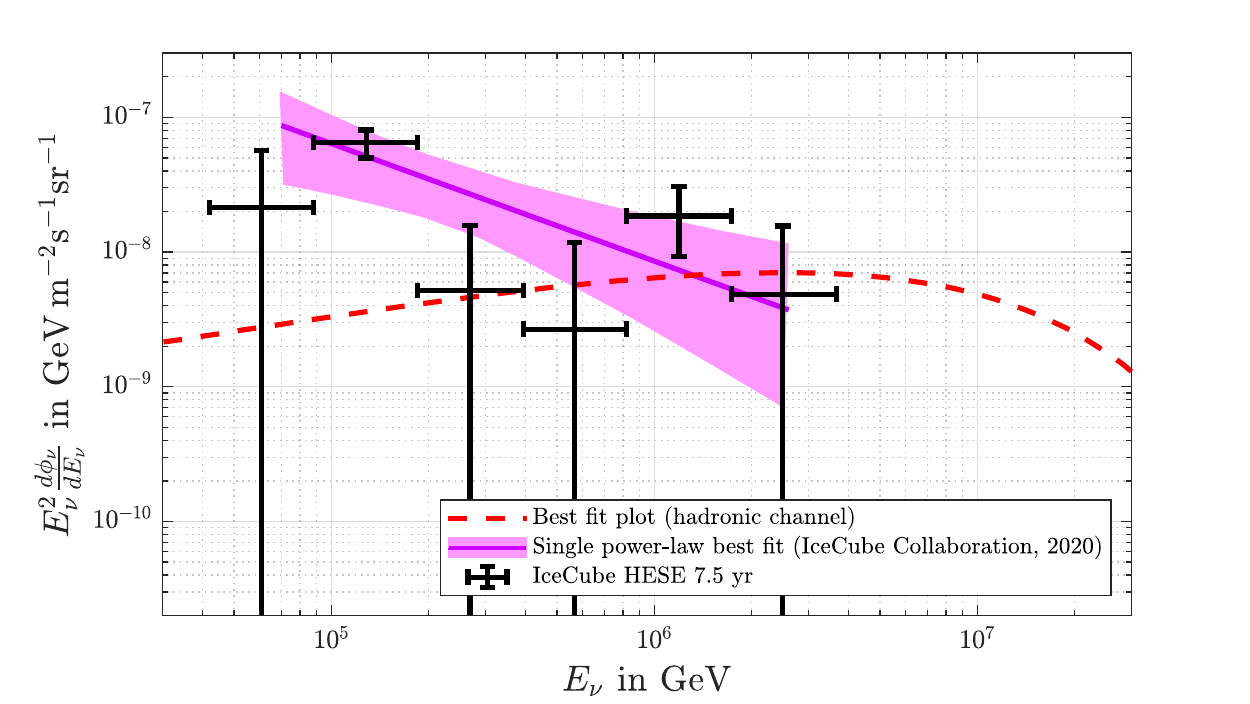}
		\caption{\label{fig:flx} Bestfit $\nu_{\mu}$ spectrum from heavy dark matter decay.}
	\end{figure}
	In Figure~\ref{fig:flx}, the through-going best-fit power-law (different from HESE single power-law fit of $E^{-\gamma}$ with $\gamma \sim 2.9$) is shown along with its statistical fluctuation is shown by the pink band. The actually observed data points for through-going muon events (from 7.5 yr data) are shown in Figure~\ref{fig:flx} with error bars. We have chosen four experimentally observed neutrino signal within the energy range $2\times 10^5$ GeV -- $4 \times 10^6$ GeV. The other two data points in the lower energy regime in Figure~\ref{fig:flx} belong to Astrophysical Neutrino events \cite{Chianese_2016} and hence are not considered in the present analysis. These four experimental data points are referred to as $\phi^{\rm Ex}$.
	
	The parameters $\mathcal{K}$ and $M_{\rm HDM}$ are obtained by making a $\chi^2$ fit ($\chi^2$ minimization) of data $\phi^{\rm Ex}$ with $\phi^{\rm Th}$. The $\chi^2$ is defined as,
	\begin{equation}
	\chi^2 = \dfrac{1}{n}\sum_{i=1}^{n} \left(\frac{E_i^2 \phi_i^{\rm Th}-E_i^2 \phi_i^{\rm Ex}} {(\rm err)_i} \right)^2,
	\label{cal1}
	\end{equation}
	where $n(=4)$ is the number of chosen points and $E_i$ represents the energy corresponding to data point $i$. In Eq.~\ref{cal1}, $({\rm err})_i$ represents the error for the $i^{\rm th}$ number of chosen point. For this two-parameter $\chi^2$ fit, only the hadronic decay channel of HDM is considered, as the contribution due to the leptonic channel in the energy range for the chosen data points is negligible \cite{mpandey}.
	From this treatment, the obtained best-fitted values of the HDM mass ($M_{\rm HDM}$) and $\mathcal{K}$ are $M_{\rm HDM}=2.75\times 10^8$ GeV and $\mathcal{K}=2.56\times 10^{-29}\,{\rm sec^{-1}}$ respectively. The red dashed line in Figure~\ref{fig:flx} indicates the calculated $\nu_{\mu}$ flux for the best-fit values of $M_{\rm HDM}$ and $\mathcal{K}$. Moreover, best-fit values of $\mathcal{K}$ for the cases of different HDM masses are also estimated (see Figure~\ref{fig:Mx_mathK}).
	
	The other multimessenger effect considered here is related to exploring the influence of heavy dark matter (HDM) decay in the context of global 21-cm signature, where the contribution of Hawking radiation from evaporation primordial black holes and the baryon cooling due to the baryon-DM collision have also been included. Two categories of dark matter are considered here. One is the possible decay of heavy dark matter that produce neutrinos in ultra-high energy region and their detection by IceCube, while the other is a WIMP type cold dark matter (CDM) interaction strength in the weak interaction regime. We presume that the fraction of dark matter in the form of HDM is negligibly small in comparison to the lighter CDM-type dark matter candidate. 
	From Eqs.~\ref{eq:T_b}, \ref{KBH}, \ref{eq:hdm_inj}, it can be noticed that the decay of HDM candidate (and the PBH evaporation) contributes to the baryon temperature $T_b$ evolution and hence in the brightness temperature of the 21-cm absorption spectrum $T_{21}$. Also, the heavy dark matter cascading decay can produce UHE neutrinos that could be tracked out by km$^2$ detector like IceCube. Here in this work, the best-fit values for heavy dark matter decay width (in fact $\mathcal{K}=\Gamma f_{\rm HDM}$) is further constrained for different possible HDM masses $M_{\rm HDM}$ in case of different $m_{\chi}$ values by the EDGES result for 21-cm absorption line. While computing $T_{21}$, in addition to the effects of heavy dark matter decay, the scattering effects of baryons with CDM (that is assumed to account for almost all DM containing the Universe), the evaporation of primary black holes etc. By performing this analysis we also attempted to explore the contribution of a very small fraction of heavy dark matter through its decay and the contribution of the overwhelming CDM type lighter dark matter through their collisional effects with baryon while also including PBH evaporation contributions to 21-cm Hydrogen absorption line during reionization era. As mentioned, various constraints are estimated using the experimental results of the EDGES experiment ($T_{21} = -500^{+200}_{-500}$ at $z=17.2$). The brightness temperature at redshift $z=17.2$ is an important quantity in this analysis, represented by $\Delta T_{21}$.
	
	\begin{figure*}
		\centering{}
		\includegraphics[width=0.8\textwidth]{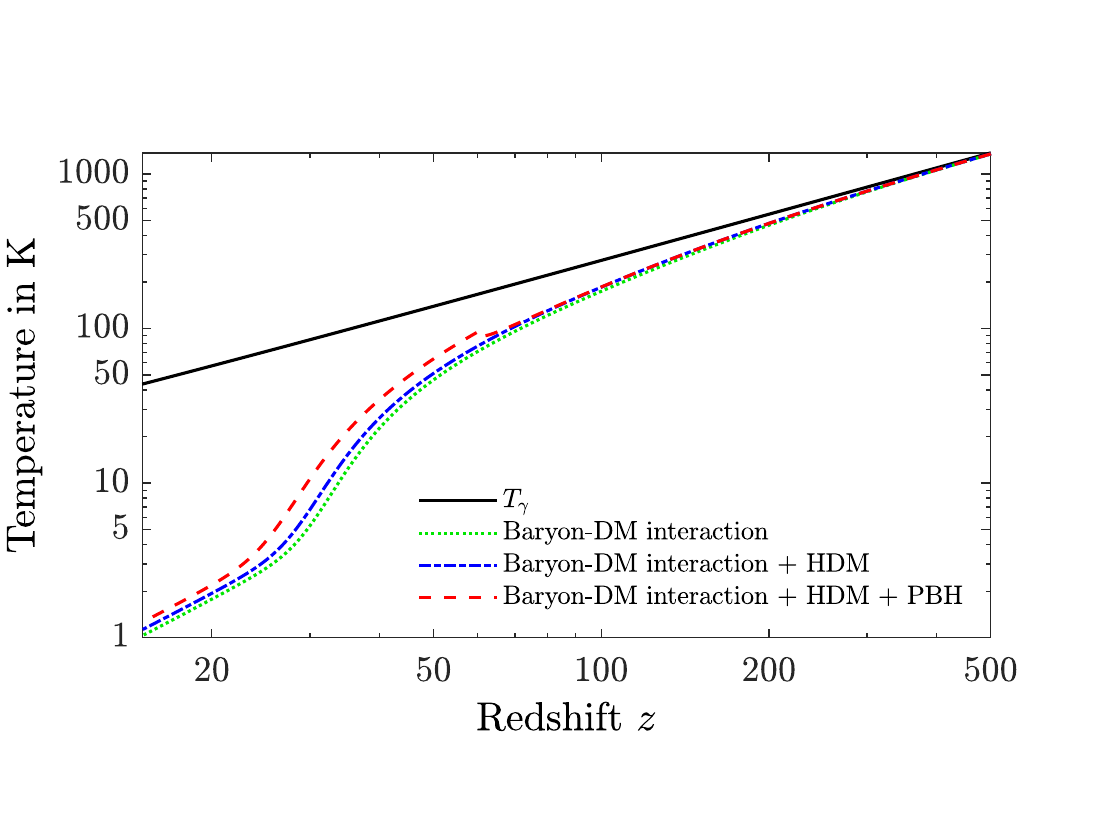}
		\caption{\label{fig:tbs_21cm_mar} The evolution of baryon temperature ($T_b$) in presence of baryon-DM collisions, PBH evaporation and HDM decay is plotted using the dashed red line. On the other hand, the blue dash-dotted line is for the same where the effects of baryon - DM interaction and the HDM decay are only incorporated. When only the effect of baryon - DM interaction is considered, the evolution of $T_b$ is denoted by the dotted green line, while the solid black curve denotes the variation of background temperature with redshift $z$. See text for detail.}
	\end{figure*}
	
	\begin{figure*}
		\centering{}
		\begin{tabular}{cc}
			\includegraphics[trim={0 0 75 0},clip,width=0.48\textwidth]{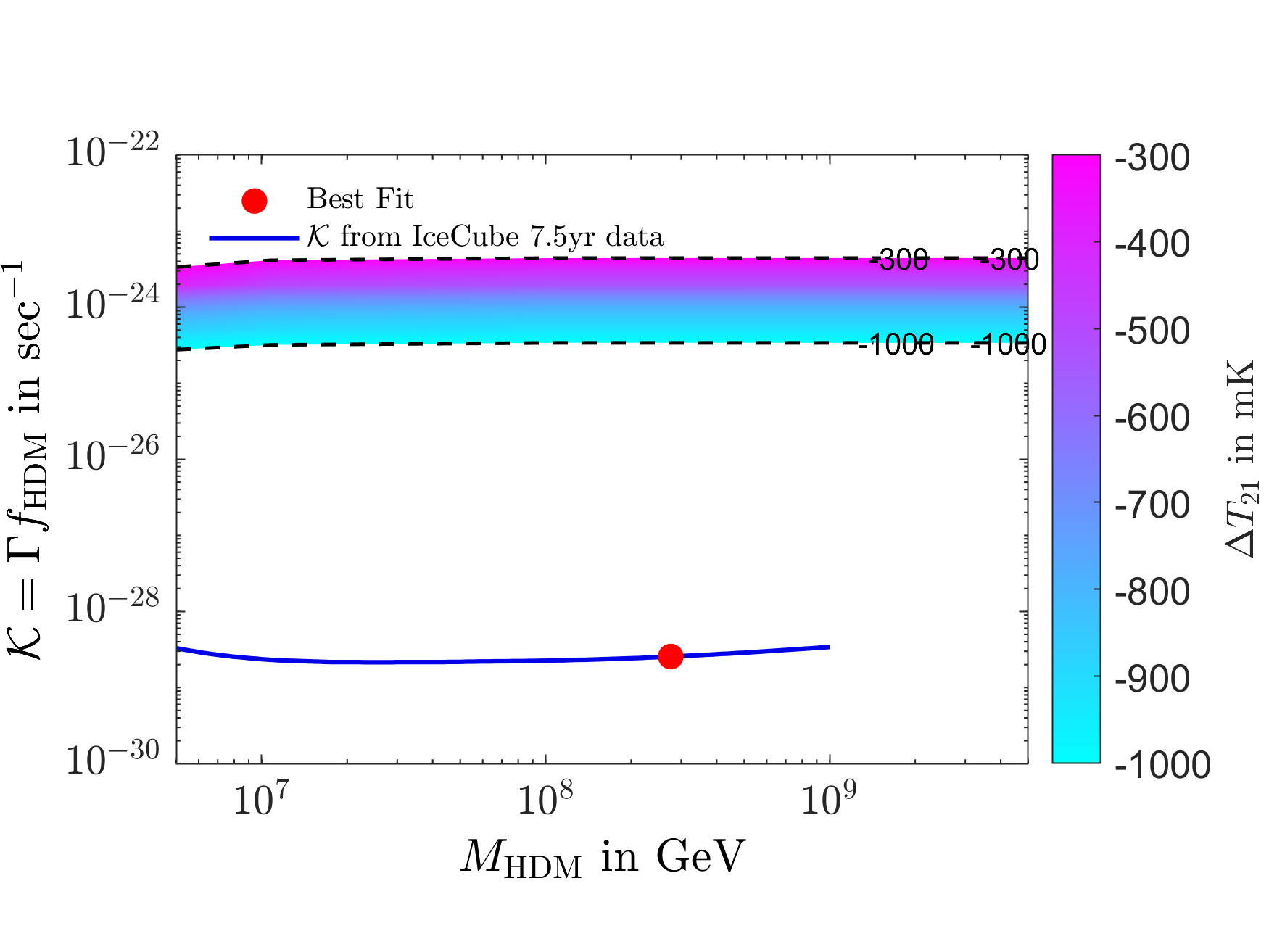}&
			\includegraphics[trim={0 0 75 0},clip,width=0.48\textwidth]{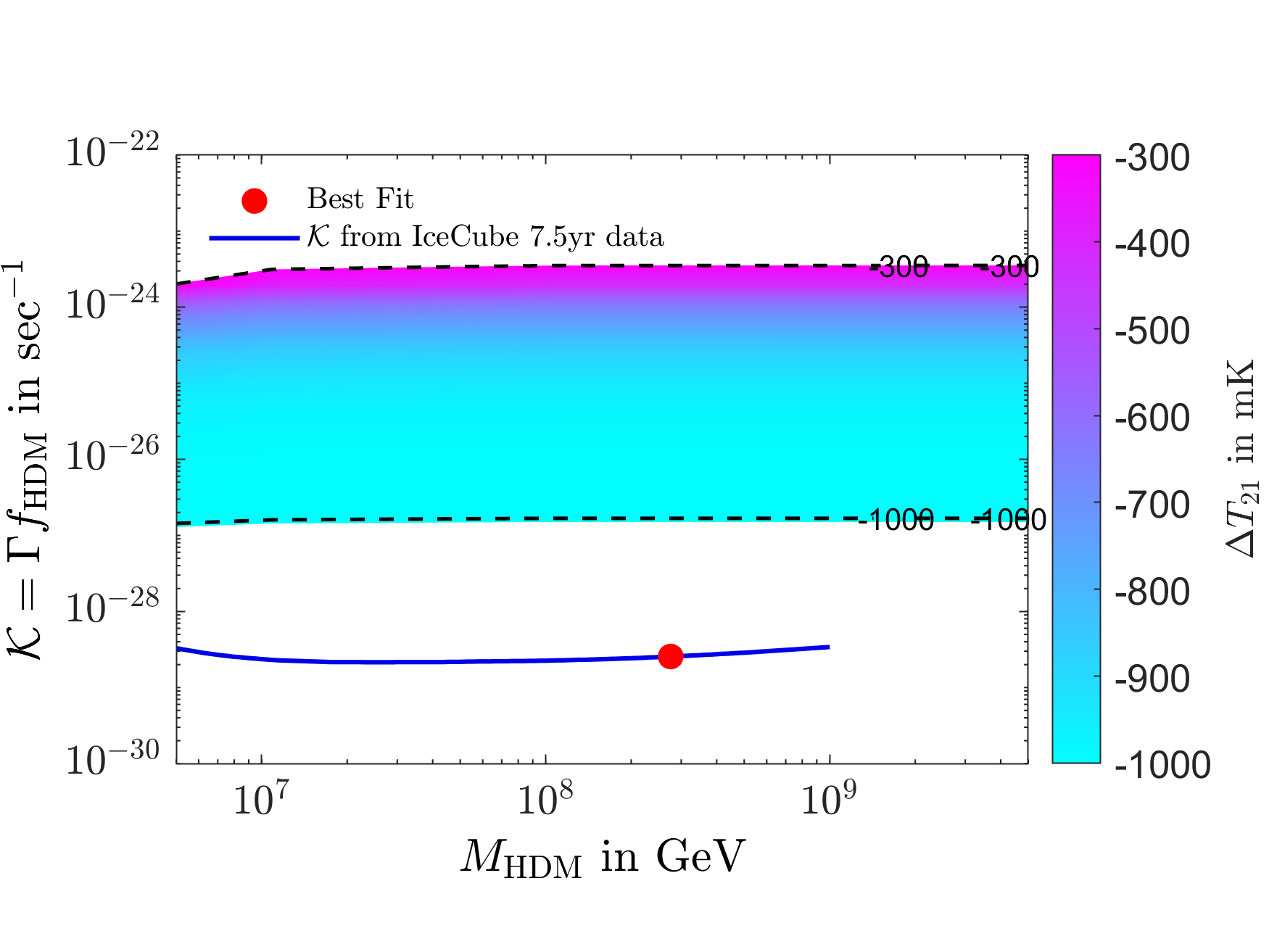}\\
			(a)&(b)\\
			\includegraphics[trim={0 0 75 0},clip,width=0.48\textwidth]{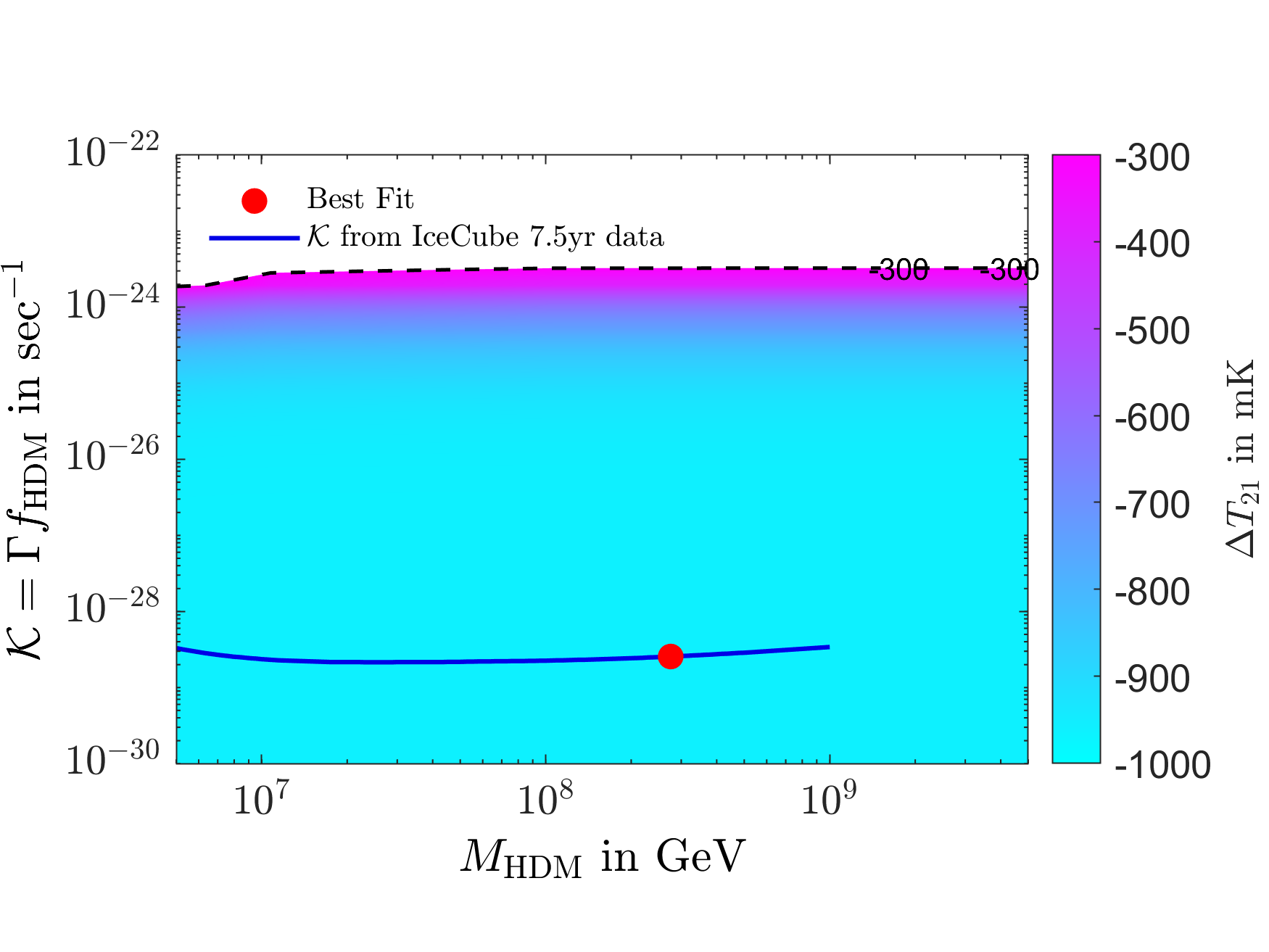}&
			\includegraphics[trim={0 0 75 0},clip,width=0.48\textwidth]{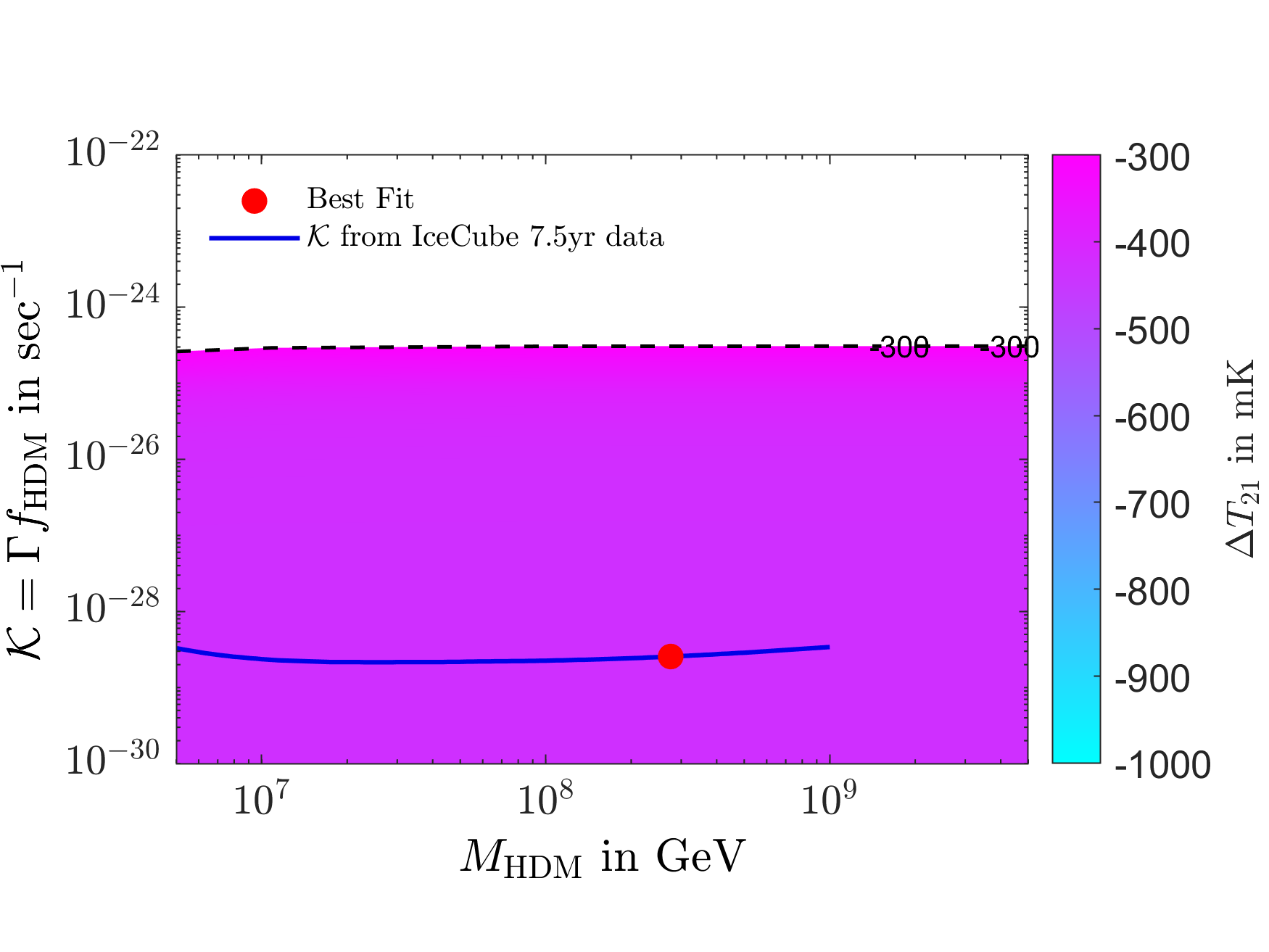}\\
			(c)&(d)\\
		\end{tabular}
		\begin{tabular}{c}
			\includegraphics[width=0.8\textwidth]{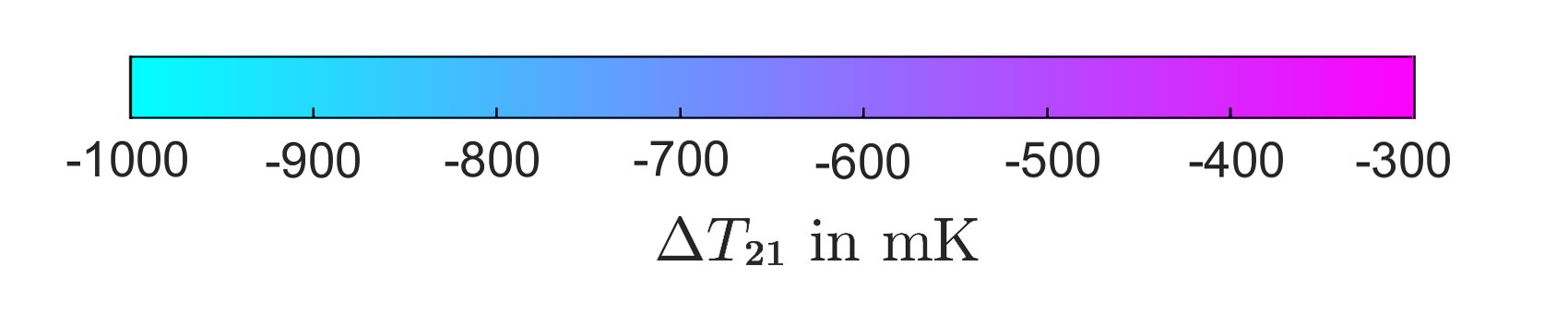}
		\end{tabular}
		\caption{\label{fig:Mx_mathK} The allowed region in the $\mathcal{K}$-$M_{\rm HDM}$ parameter space for (a) $m_{\chi}=0.3$ GeV, (b) $m_{\chi}=0.38$ GeV, (c) $m_{\chi}=0.4$ GeV and (d) $m_{\chi}=1.0$ GeV, where other variables are kept fixed at $\sigma_{41}=1$, $\mathcal{M_{\rm BH}}=10^{14}$g and $\beta_{\rm BH}=10^{29}$. The upper limits and lower limits for individual plots are related to the $\Delta T_{21}=-300$ mK and $-1000$ mK respectively. The solid blue line indicates best-fitted values of $\mathcal{K}$ for different values of $M_{\rm HDM}$ (from IceCube 7.5 year up-going event data), while the red dot symbol specifies the bestfit $\mathcal{K}$ for bestfit $M_{\rm HDM}$.}
	\end{figure*}
	We simultaneously solve computationally Eqs.~\ref{eq:hdm_inj} -- \ref{eq:V_chib_HDM} and obtain the baryon temperature. The effect of heavy dark matter decay, primordial black hole evaporation and baryon-DM scattering in the evolution of $T_b$ are graphically demonstrated in Figure~\ref{fig:tbs_21cm_mar}. In this figure (Figure~\ref{fig:tbs_21cm_mar}), the dashed red line represents the evolution of the baryon temperature ($T_b$) in presence of baryon-dark matter interaction, PBH evaporation and HDM decay (for this particular plot we choose $\beta=10^{-29}$ and $\mathcal{M}_{\rm BH}=10^{14}$ g, $\mathcal{K}=10^{-25}$ s$^{-1}$, $M_{\rm HDM}=10^9$ GeV, $m_{\chi}=0.5$ GeV and $\sigma_{41}=1$). On the other hand, the blue dash-dotted line indicates the same where the effects of baryon - DM interaction and the HDM decay are only incorporated. The separation between these two plotted lines is denoting the effect of Hawking radiation from the black bole evaporation in the evolution of baryon temperature. In the case of the only baryon - dark matter interaction, the evolution of $T_b$ is given by the dotted green line, while the solid black line indicates the background temperature evolution with redshift $z$.

	The spin temperature can be computer from the Eq.~\ref{eq:tspin} where the effect of Ly$\alpha$ forest is also included. Finally $T_{21}$ is calculated using Eqs.~\ref{eq:t21}, \ref{eq:tau}. In Figure~\ref{fig:Mx_mathK}, the allowed regions in the $\mathcal{K}$-$M_{\rm HDM}$ parameter plane are shown for different chosen values of dark matter masses ($m_{\chi}$). The four demonstrative plots in Figure~\ref{fig:Mx_mathK} correspond to (a) $m_{\chi}=0.3$ GeV, (b) $m_{\chi}=0.38$ GeV, (c) $m_{\chi}=0.4$ GeV and (d) $m_{\chi}=1.0$ GeV respectively. The allowed region is estimated using the experimental excess of EDGES. The uppermost lines of the shaded region (upper dashed black line) correspond to $\Delta T_{21}=-300$ mK, while the lower dashed lines (can be seen only in Figure~\ref{fig:Mx_mathK}a and Figure~\ref{fig:Mx_mathK}b) are related to $\Delta T_{21}=-1000$ mK. The values of $\Delta T_{21}$ at any points between the upper and lower limits are indicated by the corresponding colour codes (see colour bar). The best-fitted values of $\mathcal{K}$ and $M_{\rm HDM}$ as estimated from the $\chi^2$ minimization analysis of the IceCube 7.5 yr through going data, which are plotted in all the four plots of Figure~\ref{fig:Mx_mathK}. The best-fit values of $\mathcal{K}$ for every chosen value of $M_{\rm HDM}$ in the region $\sim 10^6$--$\sim 10^9$ GeV (as obtained from $\chi^2$ minimization of IceCube 7.5 yr through going data) are plotted using the solid blue line in all the plots of Figure~\ref{fig:Mx_mathK}.
	For dark matter mass $m_{\chi}=0.3$ GeV, the allowed zone in $\mathcal{K}$-$M_{\rm HDM}$ plane lies far above the $\mathcal{K}$-best-fit line. However, comparing the plots of Figure~\ref{fig:Mx_mathK}(a, b, c, d), it can be noticed that, as $m_{\chi}$ increases, the lower bound in the $\mathcal{K}$-$M_{\rm HDM}$ region falls rapidly, while the upper bound of the allowed region decreases gradually. It can be seen from Figure~\ref{fig:Mx_mathK} that for $m_{\chi} \gtrapprox 0.4$ GeV, the best-fit line lies within the $\mathcal{K}$-$M_{\rm HDM}$ allowed region. For all the cases, the values of PBH parameters are chosen to be $\mathcal{M_{\rm BH}}=10^{14}$g and $\beta_{\rm BH}=10^{29}$, while $\sigma_{41}$ is fixed at 1. 
	
	\begin{figure}
		\centering{}
		\includegraphics[width=0.7\linewidth]{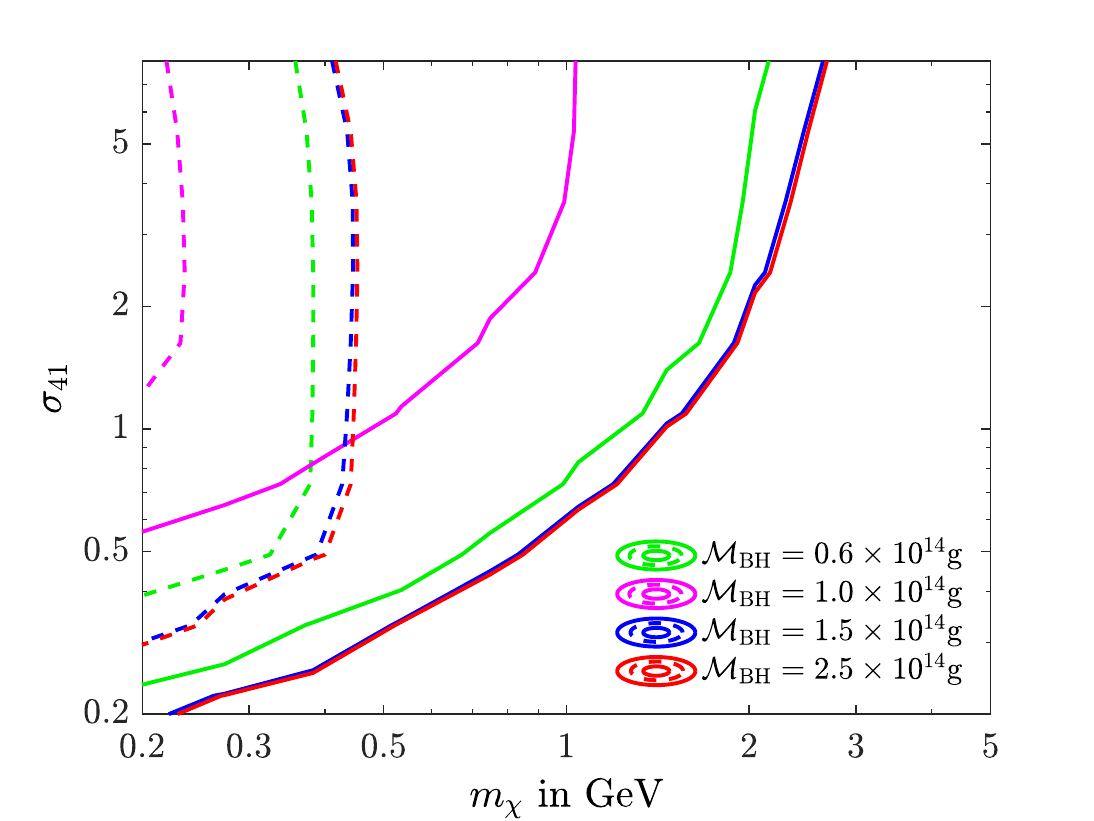}
		\caption{\label{fig:mchi_sigma_mar} The allowed region in the $\sigma_{41}-m_{\chi}$ parameter space for different values of black hole masses where, the chosen values of $M_{\rm HDM}$ and $\mathcal{K}$ are the best-fit values, obtained from the $\chi^2$ analysis using the IceCube 7.5 year data \cite{IC_7.5yr}. The different coloured solid lines are addressing the bound corresponds to $\Delta T_{21}=-300$ mK for different PBH masses while the dashed lines are representing the same for $\Delta T_{21}=-1000$ mK. Consequently, in each case, the region between the solid and the corresponding dashed line (of individual colour) indicates the allowed zone in the $m_{\chi}$-$\sigma_{41}$ plane.}
	\end{figure}
	We also address the bounds in the $m_{\chi}$-$\sigma_{41}$ plane for different masses of primordial black holes ($\mathcal{M_{\rm BH}}$), where the values of HDM mass $M_{\rm HDM}$ and corresponding $\mathcal{K}$ are taken into account from the $\chi^2$ fit using the IceCube 7.5 yr data. The results are graphically described in Figure~\ref{fig:mchi_sigma_mar}. In Figure~\ref{fig:mchi_sigma_mar}, the area between the dashed and solid lines (of the same colour) represents the allowed range in the $m_{\chi}$-$\sigma_{41}$ plane for particular chosen masses of PBH $\mathcal{M_{\rm BH}}$. It can be seen that as the value of $\mathcal{M}_{\rm BH}$ decreases upto around $10^{14}$g, the allowed zone shifts toward lower dark matter mass (WIMP type dark matter) $m_{\chi}$ and higher values of $\sigma_{41}$ (for $\mathcal{M}_{\rm BH}\lessapprox10^{14}$g). But in the case when $\mathcal{M}_{\rm BH}=0.6\times10^{14}$g, the $\sigma_{41}$-$m_{\chi}$ allowed region appears to shift to higher $m_{\chi}$ - lower $\sigma_{41}$ domain. It is shown that this trend in the shift of $m_{\chi}$-$\sigma_{41}$ allowed zone toward the opposite direction sets in for $\mathcal{M}_{\rm BH} \lessapprox 1 \times 10^{14}$ g. For PBH mass $\sim 0.6 \times 10^{14}$ g (\ref{fig:mchi_sigma_mar}), the black holes evaporate very early. Therefore such PBHs are unable to heat up the baryonic fluid significantly. As a consequence, in this particular case, the allowed region is close to those for higher values of $\mathcal{M}_{\rm BH}$. In all four cases, the $\beta_{\rm BH}$ is chosen to be $10^{-29}$.
	It is to be noted that, for the best-fitted values of HDM mass ($M_{\rm HDM}$) and $\mathcal{K}$, the maximum possible value of the $m_{\chi}$ is $2\sim3$ GeV, when the energy injection by PBH is comparatively low and agree with the results of {\bf Barkana} \cite{rennan_3GeV} (i.e. $m_{\chi} \leq 3$ GeV).

\chapter{Estimation of Baryon Asymmetry from HDM Decay} \label{chp:IC}
\\


\section{Introduction}

	Besides several significant aspects of dark sector components, the shortage of antimatter in the Universe is another open question of modern physics and cosmology. According to the notion of particle physics, every particle has its own antiparticle having identical mass but of opposite quantum numbers. If matters and corresponding antimatters are created simultaneously, there must exist significant amounts of antimatters in our Universe. However, antimatter is extremely rare on the earth and even in our solar system. The baryon asymmetry of the Universe is usually expressed by the baryon-to-entropy ratio $\eta_B$ ($=\dfrac{n_B}{s}$, $n_B$ and $s$ are the baryon number density the entropy of the Universe respectively). According to result of high precision Planck \cite{planck18} observation, $\eta_B=\left(8.61 \pm 0.09 \right) \times 10^{-11}$.

	Based on the investigations regarding CMB and CP-violation in the neutral kaon system \cite{PhysRevLett.13.138}, famous soviet physicist Andrei Sakharov proposed (1967) \cite{sakharov} that, there are three necessary conditions, that must satisfy in order to produce matter and corresponding antimatter in unequal rate, which are enumerated as\\
	1. Violation of baryon number.\\
	2. Violation of both charge conjugation and charge-parity (CP)-symmetry.\\
	3. Interactions out of thermal equilibrium.\\
	
	In the present analysis, we estimate the amount of baryon asymmetry present in the Universe due to the decay of heavy dark matter candidates. Such estimation can be carried out from the decay time of dark matter candidate and its mass as described in the work of {\bf Ziaeepour} \cite{wimpzilla_decay}. Here the rare decay of HDM particles are considered and generate high energy neutrinos (PeV range neutrinos) as an end product. Such highly energetic neutrinos can be detected by the square-kilometer neutrino observatory of IceCube. The possible decay lifetime of such heavy dark matter particles and the corresponding mass can be estimated by comparing the estimated muon neutrino flux with the observed PeV range flux by IceCube experiment \cite{ice-2010}. These are then incorporated into the calculation to measure the amount of baryon asymmetry of the Universe.

	The process of rare decay of heavy dark matter (HDM) particles is already discussed in the previous chapter (\Autoref{chp:21_mar}). As shown in the work of {\bf Berezinsky} \emph{et~al.} \cite{bera}, such HDM particles undergo rare decay to produce known particle-antiparticle pairs of Standard Model candidates \cite{kuz,bera} following the QCD cascade through hadronic and leptonic channel using the Altarelli-Parisi formalism. The flux of the HDM decay final products contains all three possible flavours of neutrinos. Unlike the previous analysis (described in \Autoref{chp:21_mar}), in the present treatment, both hadronic and leptonic channels of HDM decay are taken into account to estimate the ultra high energy flux of diffused muon neutrinos. Such highly energetic neutrinos can be produced from the decay of HDM particles from both extra-galactic and galactic sources. The contribution of astrophysical neutrinos flux \cite{Chianese_2016,ICastroph-nu,cosmic-nu} is also included in the calculation of total flux. Now fitting the obtained muon neutrino flux with observed PeV range flux by IceCube, the most probable values of HDM mass and corresponding decay lifetime can be estimated. In this case, several additional data points are also considered along with the 7.5 yr and 6 yr IceCube datasets. The adopted additional data points chosen from the pink colour band of Figure~2 from the paper of {\bf Kopper (2017)} \cite{Kopper:2017Df}. We also repeat the entire treatment neglecting the leptonic decay contribution and compare the result with the previous case.

\section{Neutrino Flux}
	As mentioned earlier (\Autoref{chp:21_mar}), heavy dark matters are possibly created via gravitational production mechanism or non-linear quantum effects during the reheating or preheating stages after inflation. Such heavy dark matter particles produce highly energetic particle-antiparticle pairs of Standard Model candidates and photons via rare decay. Later, from these particle-antiparticle pairs electrons, muons, tauons their antiparticles and their corresponding neutrinos and antineutrinos are produced along with photons. These final products are known as secondary products. The produced secondary neutrinos of different flavours undergo further flavour oscillation during the propagating toward the detector. 
	
	Although the observed neutrino flux are essentially extra-galactic \footnote{For example, the rare decay of HDM particles from galactic origin, having mass 100 PeV produced neutrino flux $\sim10^{-26}\rm{GeV\,cm^{-2}s^{-1}sr^{-1}}$, while for the extragalactic sources of identical HDM decay, the obtained neutrino flux is $\sim10^{-7}\rm{GeV\,cm^{-2}s^{-1}sr^{-1}}$}, from recent investigations by {\bf Neronov} \emph{et~al.} \cite{neronov2018} and {\bf Kachelrie\ss{}} \emph{et~al.} \cite{kuz} it is evident that the ultra high energy neutrino flux (UHE), as detected at the IceCubes's square-kilometer detector, has a significant contribution from the galactic sources. Therefore, in the present calculation, both galactic and extra-galactic contributions are considered.


	Several celestial sources may also produce ultra-high energy neutrinos via the proton acceleration mechanism. In such cases, high-energy protons undergo self-interaction or sometimes interact with photons ($\gamma$) to produce highly energetic neutrinos. Active Galactic Nuclei (AGN) \cite{AGN-nu, AGN1, AGN2, AGN3}, Gamma-Ray Bursts (GRBs) \cite{grb-nu, GRB}, Extra-galactic Supernova Remnants (SNR) \cite{SNR} etc. are some possible astrophysical sources of UHE neutrinos. As an outcome of such particle acceleration process, highly energetic shockwaves are generated, which propagate in the form of fireballs of energy as high as $\sim 10^{53}$ ergs. For the case of astrophysical sources of neutrinos, the fluxes follow an Unbroken Power Law (UPL) given by,
	\begin{equation}
	E_{\nu}^2\dfrac{{\rm d}\phi_{\nu}^{\prime}}{{\rm d}E_{\nu}}=N \left(\dfrac{E_{\nu}}{100\,{\rm TeV}}\right)^{-\gamma}\: {\rm GeV\,cm^{\-2}\,s^{-1}\,sr^{-1}},
	\label{eq:IC_ast_ini}
	\end{equation}
	where $\gamma$ denotes the spectral index of the power spectrum which is normalized using the coefficient $N$. In the present work, the numerical values of $\gamma$ and $N$ are chosen as 1.0 and $1 \times 10^{-8}\,{\rm GeV\,cm^{\-2}\,/s/sr}$ respectively. Now assuming the initial flavor ratio of neutrinos is 1:2:0 and observed final ratio 1:1:1, the flux for individual neutrino flavor is given by, $\dfrac{{\rm d}\phi_{\nu}^{ast}}{{\rm d}E_{\nu}}=\dfrac{1}{3} \dfrac{{\rm d}\phi_{\nu}^{\prime}}{{\rm d}E_{\nu}}$.
	
	In the present treatment, such astrophysical neutrino flux is considered, which is essentially dominated within the energy range $\sim 60$ TeV $- 120$ TeV. Some recent works by the authors {\bf Chianese} \cite{Chianese_2016} and {\bf Yicong} and {\bf PS Bhopal} \cite{bhupal} have provided a mechanism for the estimation of neutrino flux from  astrophysical origins. Adopting the similar formalism, the contribution of the astrophysical neutrino flux are considered in further analysis. As a result, the total muon neutrino flux (per unit solid angle $\Omega$) consists of contribution of galactic, extra-galactic and the astrophysical flux given by,
	\begin{equation}
	\left( \dfrac{d\phi_{\nu}}{d\Omega dE_{\nu}} \right)_{\rm th} = \dfrac{d\phi_{\nu}^{\rm G}}{d\Omega dE_{\nu}} + \dfrac{d\phi_{\nu}^{\rm EG}}{d\Omega dE_{\nu}} + \dfrac{d\phi_{\nu}^{\rm ast}}{d\Omega dE_{\nu}}.
	\label{eq:IC_1}
	\end{equation}
	In the above expression, $\dfrac{d\phi_{\nu}^{G}}{d\Omega dE_{\nu}}$ and $\dfrac{d\phi_{\nu}^{EG}}{d\Omega dE_{\nu}}$ represent the galactic and extra-galactic flux respectively at energy $E_{\nu}$, which are obtained from rare decay of heavy dark matter cascade (from both hadronic and leptonic channels). The term $\dfrac{d\phi_{\nu}^{\rm ast}}{d\Omega dE_{\nu}}$ is the astrophysical flux as described in the work of \cite{Chianese_2016}. Thus, the Eq.~\ref{eq:IC_1} can be expressed as,
	\begin{eqnarray}
	\left( \dfrac{d\phi_{\nu}}{d\Omega dE_{\nu}} \right)_{\rm th} &=& \left( \dfrac{d\phi_{\nu}^{\rm G}}{d\Omega dE_{\nu}} \right)_{\rm had} + \left( \dfrac{d\phi_{\nu}^{\rm G}}{d\Omega dE_{\nu}} \right)_{\rm lep} + \nonumber \\ 
	&&\left( \dfrac{d\phi_{\nu}^{\rm EG}}{d\Omega dE_{\nu}} \right)_{\rm had} + \left( \dfrac{d\phi_{\nu}^{\rm EG}}{d\Omega dE_{\nu}} \right)_{\rm lep} + \nonumber \\
	&& \dfrac{d\phi_{\nu}^{\rm ast}}{d\Omega dE_{\nu}}.
	\end{eqnarray}
	
	In the case of galactic contribution, the expression for the differential galactic neutrino flux is given by \cite{diff-flux},
	\begin{equation}
	\dfrac{d\phi_{\nu}^{G}}{d\Omega dE_{\nu}} = \dfrac{1}{4{\pi}{\alpha}M_{\rm HDM}\tau} \int_{los} dl \dfrac{dN}{dE} \rho[r(l,\theta)].
	\label{eq:galactic_IC}
	\end{equation}
	Here, $M_{\rm HDM}$ denotes the mass of the HDM candidates while $\tau$ is the decay lifetime of HDM particles. $\alpha = 1$ is considered for the case of Majorana-type candidates. To estimate the ultra-high energy muons in the PeV region as observed in IceCube detectors, we choose HDM candidates having masses $M_{\rm HDM} \geq 10^6$ GeV. In the above expression, the integral $\displaystyle\int_{los} \rho{(r(l, \theta))}^2 dl$ represents the integration over the line of sight, where the term $\rho(r)$ is the density of the dark matter halo at a distance $r$ from the centre of the galactic dark matter halo distribution. In this context, the Navarro-Frenk-White (NFW) profile (discussed in \Autoref{chp:intro}) is used to estimate the DM halo density profile $\rho(r)$. The NFW density profile can be expressed as \cite{Navarro:1996gj,nfw1},
	\begin{equation}
	\rho_{\rm NFW}(r) = \rho_s \dfrac{r_s}{r}{ \left(1 + \dfrac {r_s}{r} \right)}^{-2}.
	\end{equation}
	where the distance scale $r_s$ = 20 kpc and the density scale $\rho_s$ = 0.259 $\rm{GeV/cm^{3}}$.
	
	The distance any point ($l,\,\theta$) from the centre of the galactic halo distribution can be expressed as,
	\begin{equation}
	r = \sqrt{r^2_\odot\ + l^2 - 2 r_\odot\ l \cos\theta},
	\end{equation}
	where $r_\odot$ represents the radial distance to the sun from the centre of the galactic halo distribution given by $r_\odot=8.5$ kpc (approximately equal to the distance to the observer placed on earth) and $l$ is the line of sight. In Eq.~\ref{eq:galactic_IC} the neutrino spectrum $\dfrac{dN}{dE}$ is calculated by adopting the numerical treatment as described in the works of {\bf Kachelrie\ss{}} \emph{et~al.} and {\bf Berezinsky} \emph{et~al.} \cite{kuz, bera} (see \Autoref{chp:21_mar}). It is to be mentioned that, the decay lifetime $\tau$ of the HDM candidates should be larger than the approximate age of Universe, i.e. $\sim 10^{17}$ seconds.
	
	In the present scenario, the expression for the extra-galactic differential neutrino flux is given as,
	\begin{equation}
	\dfrac{d\phi_{\nu}^{EG}}{d\Omega dE_{\nu}} = \dfrac{1}{4{\pi}{\alpha}M_{\rm HDM} \tau} \int_{0}^{\infty} \dfrac{\rho_0 c/H_0}{\sqrt{\Omega_m (1 + z)^3 + (1 - \Omega_m)}} \dfrac{dN}{dE_z} dz,
	\label{eq:eg_gal_IC}
	\end{equation}
	where the term $c/H_0 = 1.37 \times 10^{28}$ cm describes the Hubble radius, $H_0$ is the Hubble constant at the present epoch and $c$ is the speed of light. $\rho_0$ denotes the average density of dark matter at the current time, which is $\approxeq 1.15 \times 10^{-6}$ GeV/$ \rm {cm^3}$. The term $\Omega_m = 0.311$ \cite{planck18} is the cosmological density parameter of matter. In Eq.~\ref{eq:eg_gal_IC}, $\dfrac{dN}{dE_z}$ is the spectrum, which is produced from of the decay of DM particle, depending on the redshifted energy $E_z = E(z) = E\times (1 + z)$, given by
	\begin{eqnarray}
	&\dfrac{dN}{dE_z} &= \dfrac{dN}{dE} \dfrac{dE}{dE_z}\nonumber\\
	{\rm or,}&\dfrac{dN}{dE_z} &= \dfrac{dN}{dE} \dfrac{1}{(1 + z)}.
	\end{eqnarray}
	
	In the present work, HDM mass $M_{\rm HDM}$ and the decay width $\tau$ are considered as free parameters. The numerical values of $M_{\rm HDM}$ and $\tau$ are estimated from a $\chi^2$ analysis which is carried out with the observed data of IceCube 7.5 yr HESE data and IceCube 6 yr HESE data.

\section{Heavy Dark Matter Decay Spectra}

\begin{figure}
		\centering{}
		\begin{tabular}{c}
			\includegraphics[width=0.8\columnwidth]{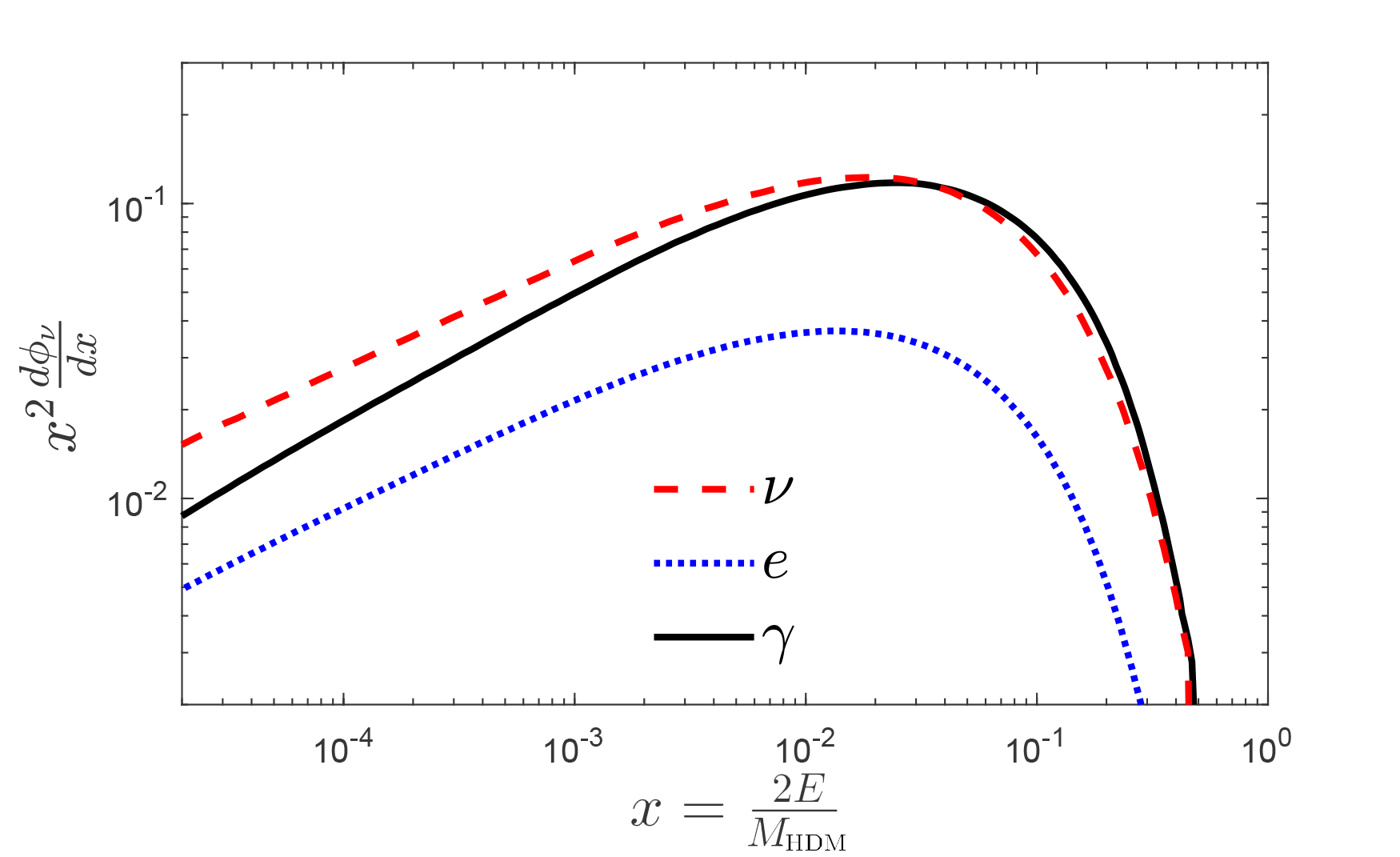}\\
			(a)\\
			\includegraphics[width=0.8\columnwidth]{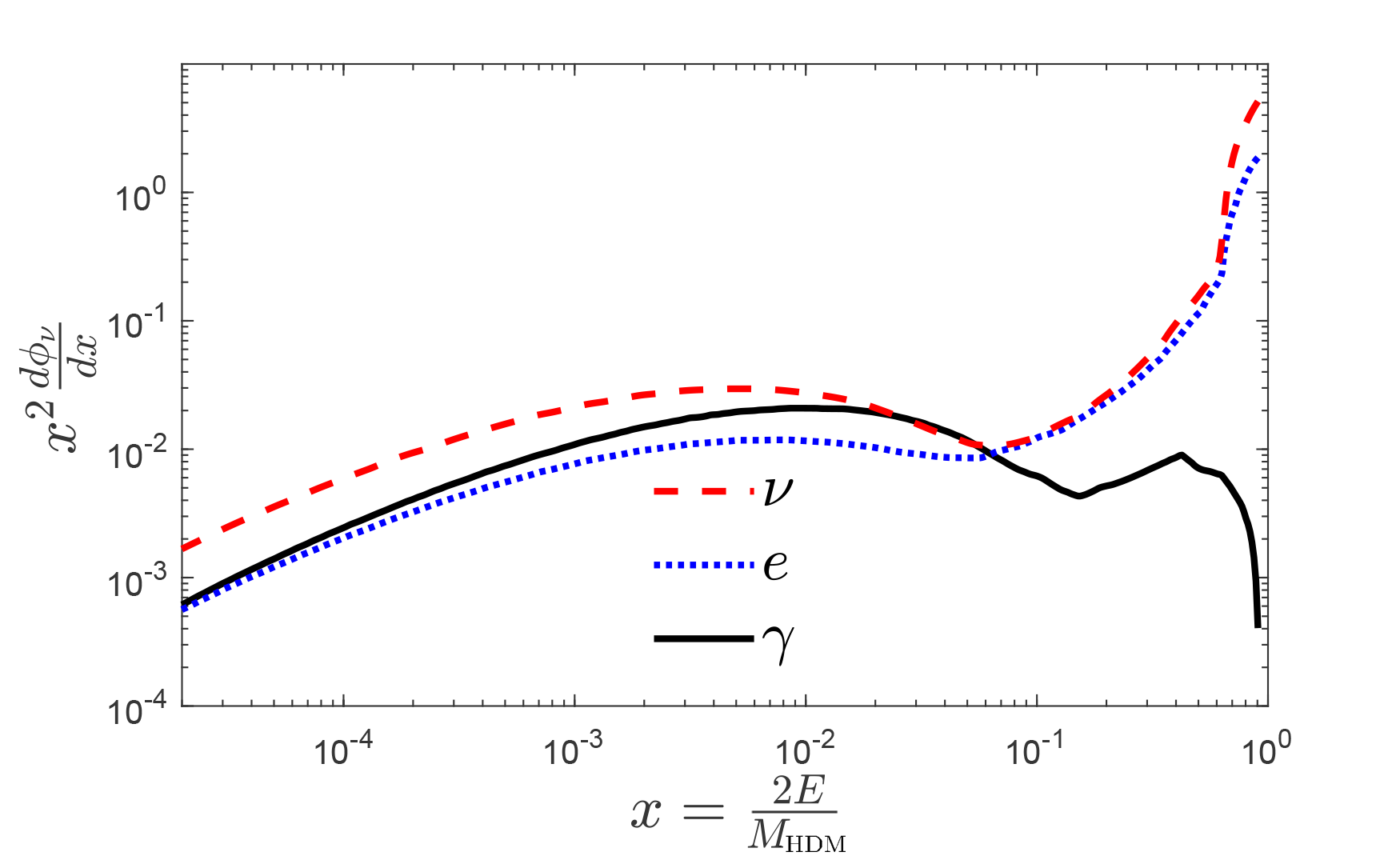}\\
			(b)\\
		\end{tabular}
		\caption{\label{Fig:3}Prompt spectrum due to the decay of HDM particles having mass $M_{\rm HDM}=1.9498\times10^8\:$ GeV via (a) hadronic, (b) leptonic decay channel.}
	\end{figure}

	In the present work, both hadronic and leptonic decay channels of the HDM candidate are considered simultaneously. The decay process of the hadronic channel is already discussed in \Autoref{chp:21_mar}. In the case of the leptonic channel, the decay of HDM ($M_{\rm HDM}\gg M_W$, where $M_W$ represents the mass of the $W$ boson) results in the electroweak cascade \cite{bera}. For leptonic decay $\chi \rightarrow l \bar{l}$ ($l$ denotes the lepton) as $M_{\rm HDM}\gg M_W$, the quantity  $\ln{\left(\dfrac{M_{\rm HDM}^2}{M_W^2}\right)}$ becomes substantially large and thus compensate the electroweak coupling. In addition, the perturbation theory also breaks down and produces the cascade. In order to evaluate the decay spectrum from the leptonic decay channel, one needs to evolve evolution equations, which are identical to the DGLAP equation (discussed in \Autoref{chp:21_mar}). The contribution of the leptonic channel only marginally affect the total calculated flux, which contains both leptonic and hadronic contribution. We adopt the formalism as described in {\bf Berezinsky} \emph{et~al.} \cite{ref53} and compute the DGLAP-equivalent evolution equation for the electroweak cascade, as described in the Eq.~6-7 of {\bf Berezinsky} \emph{et~al., (1998)} \cite{ref53} and obtained the corresponding spectra of neutrinos. In the present analysis, the considered mass of the HDM candidate is remarkably higher in comparison to the electroweak scale. As a consequence, during the decay process of HDM an electroweak cascade \cite{bera} is manifested besides the QCD cascade \cite{kuz,Berezinsky:1997sb,ref50}. We have also performed a simulation using the electroweak and QCD sectors as addressed in several works \cite{bera,bera2}. The calculated spectrum of the final products $\nu$, $e$ and $\gamma$ from the heavy dark matter decay via leptonic channel are furnished in Figure~\ref{Fig:3}b for $M_{\rm HDM}\simeq2 \times 10^8$ GeV, while the same ($\nu$, $e$, $\gamma$ spectra) from the hadronic channels are shown in Figure~\ref{Fig:3}a.
	
	
\section{Matter-antimatter Asymmetry}
	The consequence of late time decay of such superheavy or heavy candidate of dark matter has a significant contribution in the context of the matter-antimatter asymmetry, as it further take a part in the production of lepton asymmetry and baryon asymmetry, that satisfies the necessery conditions for the baryogenesis as proposed by Sakharov \cite{wimpzilla, UHDMdecay&baryonAsymmetry, Kolb_EarlyUniverse,CMBbaryogenesis}. In order to estimate the asymmetry appears as an outcome of HDM decay, one need to solve a Boltzmann-like equation in order to estimate the rate of baryon number density per dark matter decay, which is expressed as \cite{wimpzilla}
	\begin{equation}
	\dfrac{d(n_{b} - n_{\bar{b}})}{dt} + 3 \dfrac{\dot{a}(t)}{a(t)} (n_{b} - n_{\bar{b}}) = \dfrac{\epsilon n_{\chi} (t)}{\tau},
	\label{eq:dnbdt}
	\end{equation}
	where, $n_b$, $n_{\bar{b}}$ represent the baryon and antibaryon number densities respectively, $n_{\chi}$ denotes the same for the decaying heavy dark matter particles (having decay lifetime $\tau$) and the quantity $a$ represents the cosmological scale factor. In the above expression, $\epsilon$ denotes the number violation parameter of baryon, which is related to the dark matter decay. According to the standard cosmological scenario of scaling, the average matter number density is given by (with scale factor $a$ as $\propto a^{-3}$),
	\begin{eqnarray}
	n_{\chi} &\propto& (z + 1)^3, \nonumber\\
	n_{\chi} (t) &=& n_{\chi} (t_0) \dfrac{(z + 1)^3}{(z_0 + 1)^3},
	\label{eq:nchi}
	\end{eqnarray}
	where, $t_0$ and $z_0$ are representing the initial time after the Big Bang and corresponding redshift respectively. In the case of photon, we also have,
	\begin{eqnarray}
	n_{\gamma} &\propto& (z + 1)^3, \nonumber\\
	n_{\gamma} (t) &=& n_{\gamma} (t_0) \dfrac{(z + 1)^3}{(z_0 + 1)^3}.
	\end{eqnarray}
	In the above expressions, we introduce $n_{\gamma} (t_0)$ and $n_{\chi} (t_0)$ which represent the background photon (CMB) number density and the same for DM particles respectively at the time $t_0$ after the Big Bang (corresponding redshift $z_0$).
	
	Solving the Eq.~\ref{eq:dnbdt} within the time limit $t_0$ to $t$ ($z_0$ and $z$ are the corresponds values of redshift), the difference between the number densities of baryon and antibaryon $(n_b - n_{\bar{b}})$ can be estimated as (see \hyperref[APP:IC]{Appendix}),
	\begin{equation}
	(n_{b} - n_{\bar{b}}) = \dfrac{\epsilon n_{\chi} (t_0)}{3} \left [ 1 - \exp \left (-\dfrac{t - t_0}{\tau} \right ) \right ]\dfrac{{(1 + z)}^3}{{(1 + z_{0})}^3}.
	\label{eq:nbnbbar}
	\end{equation}
	Now introducing the expression of baryon asymmetry as $\Delta B = \dfrac{(n_{b} - n_{\bar{b}})}{2 g_{*} n_{\gamma} (t)}$, the above equation becomes (Eq.~\ref{eq:nbnbbar}),
	\begin{equation}
	\Delta B = \dfrac{\epsilon n_{\chi}(t_{0})}{6 g_{*} n_{\gamma}(t_{0})} \left[ 1 - \exp \left(-\dfrac{t - t_{0}}{\tau} \right) \right].
	\label{eq:delB}
	\end{equation}
	
	In the comoving frame, $t_0 = t_{\rm dec}$ is chosen, where $t_{\rm dec}$ represents the time when the CMB photons were decoupled, $z_{\rm dec} \simeq 1100$ is the corresponding value of redshift. Thus, Eq.~\ref{eq:delB} takes the form
	\begin{equation}
	\Delta B = \dfrac{\epsilon n_{\chi}(t_{\rm dec})}{6 g_{*} n_{\gamma}(t_{\rm dec})} \left[ 1 - \exp \left(-\dfrac{t - t_{\rm dec}}{\tau} \right) \right].
	\label{eq:deltaB}
	\end{equation}
	At the time of recombination, the dark matter number density $n_{\chi} (t_{\rm dec})$ can be expressed as,
	\begin{eqnarray}
	n_{\chi} (t_{\rm dec}) &=& n_{\chi,0} \left(1 + z_{\rm dec}\right)^3,\nonumber \\
	&=& \dfrac{\rho_{c}\left[\Omega_{m} - \Omega_{\rm hot} - \Omega_{b} \left(1 + \dfrac{m_{e}}{m_{p}}\right)\right]}{M_{\rm HDM}} {(1 + z_{\rm dec})}^3.
	\label{eq:nchitdec}
	\end{eqnarray}
	
	Here, $n_{\chi,0}$ represents the present value of the DM number density (i.e. at $z = 0$) and $\rho_{c}$ denotes the critical density of the Universe described by,
	\begin{equation}
	\rho_{c} = \dfrac{3 {H_{0}}^2}{8 \pi G}\,\, .
	\end{equation}
	In the above expression, $H_0$ is the Hubble constant at the current epoch ($H_{0} = 67.27 \pm 0.60$ ${\rm km/sec/Mpc}$) \cite{planck18} and the term $\Omega_{\rm hot}$ can be expressed as,
	\begin{equation}
	\Omega_{\rm hot} = \dfrac{{\pi}^2}{30} g_{*} {T_{\rm CMB}}^4,
	\end{equation}
	where $T_{\rm CMB} \sim 2.73$K is the current value of the CMB temperature and $g_{*}$ denotes the effective degrees of freedom. Now applying $n_{\gamma,0} = 410 {\rm \,cm^{-3}}$, i.e. the present value of the number density of the background photons, the amount of baryon asymmetry $\Delta B$ can be calculated by incorporating the chosen values of $M_{\rm HDM}$ and $\tau$ from Eqs.~\ref{eq:dnbdt}-\ref{eq:nchitdec}. The used best-fitted values of $M_{\rm HDM}$ and $\tau$ are estimated from the $\chi^2$ analysis using IceCube HESE data as mention earlier.
	
	\section{Calculations and Results}
	In the present calculation the recent 7.5 yr IceCube HESE (High-Energy Starting Events) flux is considered within the neutrino energy range $\sim$ 60 TeV to 5 PeV as well. In this particular case, the fluxes of UHE neutrino data are extracted from the demonstrative plot mentioned by C. Wiebusch in the presentation at SSP – Aachen, June 2018 \cite{aachen}. The UHE neutrino signals for the energy $E_{\nu} >$ 20 TeV are considered as only the High Energy Starting Events (HESE) by the IceCube Collaboration are taken into account. In the above-mentioned plot, the best-fit line is observed at the centre of a pink band and the pink band denotes the corresponding 1$\sigma$ uncertainty around the best-fitted line. In this analysis, five actually observed data points are taken into account along with twelve other data points, which are suitably chosen from the 1$\sigma$ pink band region (all 12+5 data points are tabulated in \Autoref{table:7.5yr}). For the case of 12 adopted best-fit points, the width of the pink band at corresponding energy is considered to be the error in neutrino flux. Beyond the energy range of $\sim 5 \times 10^6$ GeV, some upper bounds are can be noticed in this figure. As these bounds are only the predicted signal, no experimental data points are taken into the calculation from that region. As already mentioned earlier, the ultra-high energy neutrino events from possible astrophysical origins are also taken into account for the present treatment. Such astrophysical contribution is denoted by the first two observed data points by IceCube within the range of neutrino energy $\sim$ 60 TeV - 120 TeV, plotted in the same figure. In the present analysis, a power-law spectrum of $\sim E^{-2.9}$ is incorporated in order to describe the astrophysical neutrino flux. 
	
	\begin{table}
		\begin{center}
			\begin{tabular}{ccc}
				\hline
				Energy  & Neutrino Flux $\left(E_{\nu}^2 \dfrac {{\rm d}\Phi_{\nu}}{{\rm d}\Omega dE_{\nu}}\right)$  & {Error}
				\\
				(GeV) & (GeV cm$^{-2}$ s$^{-1}$ sr$^{-1}$) & \\ \hline
				\hline
				
				60837.7& 2.3831$\times 10^{-8}$& 1.99171$\times 10^{-8}$\\
				128208&	2.14501$\times 10^{-8}$& 1.15915$\times 10^{-8}$\\
				268693&	4.3081$\times 10^{-9}$&	8.02824$\times 10^{-9}$\\
				1193270& 4.14136$\times 10^{-9}$& 6.96905$\times 10^{-9}$\\
				2500810& 4.36516$\times 10^{-9}$& 7.96376$\times 10^{-9}$\\
				\hline
				3548130& 5.25248$\times 10^{-9}$& 4.1258$\times 10^{-9}$\\
				2304090& 5.71267$\times 10^{-9}$& 4.1600$\times 10^{-9}$\\
				1528890& 6.21317$\times 10^{-9}$& 3.9882$\times 10^{-9}$\\
				1059250& 6.61712$\times 10^{-9}$& 3.7349$\times 10^{-9}$\\
				718208&	7.04733$\times 10^{-9}$& 3.9777$\times 10^{-9}$\\
				446684&	7.66476$\times 10^{-9}$& 3.6478$\times 10^{-9}$\\
				286954&	8.16308$\times 10^{-9}$& 4.1571$\times 10^{-9}$\\
				190409&	8.87827$\times 10^{-9}$& 6.2069$\times 10^{-9}$\\
				143818&	9.65612$\times 10^{-9}$& 6.8856$\times 10^{-9}$\\
				2511890& 4.16928$\times 10^{-9}$& 8.2726$\times 10^{-9}$\\
				1192790& 5.03649$\times 10^{-9}$& 7.5383$\times 10^{-9}$\\
				268960& 7.50551$\times 10^{-9}$& 8.1583$\times 10^{-9}$\\
				\hline
			\end{tabular}
		\end{center}
		\caption{\label{table:7.5yr} List of neutrino flux data from IceCube 7.5 yr dataset, which is used in the $\chi^2$ analysis to estimate the best-fitted HDM mass $M_{\rm HDM}$ and corresponding HDM decay lifetime $\tau$. It is to be noted that only the first five points are the actual observed data, obtained from IceCube 7.5 yr dataset. The remaining 12 points are suitably chosen from the 1-$\sigma$ pink band region.}
	\end{table}
	
\begin{figure}
		\centering
		\includegraphics[width=14cm]{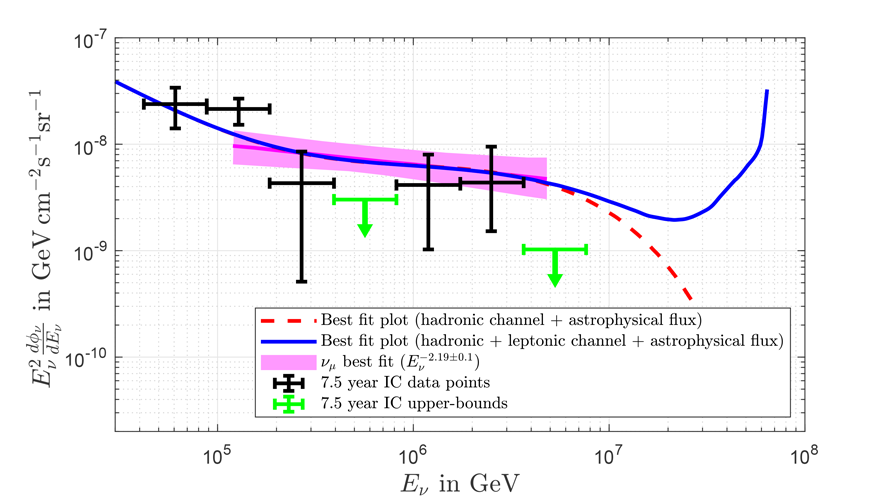}
		\caption{The calculated flux for the best-fitted dark matter mass $M_{\rm HDM}$ and the corresponding best-fit decay lifetime $\tau$ by considering all points from IceCube 7.5 yr data (within energy $\sim 60$ TeV - 5 PeV) where (i) the astrophysical flux and hadronic channel of HDM decay are considered (using red dashed line), (ii) astrophysical flux is considered along with both the hadronic and leptonic decay channels of HDM (using blue solid line).}
		\label{Fig:7.5yr}
	\end{figure}

	We have performed a $\chi^2$ analysis using the observed data points by IceCube of the entire range of energies in order to estimate the best-fit heavy dark matter mass ($M_{\rm HDM}$) and the corresponding best-fit lifetime ($\tau$). The $\chi^2$ is defined as $\chi^2 = \displaystyle\sum_{i = 1}^{n} \left (\dfrac {E_i^2 \phi_i^{\rm th} - E_i^2 \phi_i^{\rm Ex}} {({\rm err})_i} \right)^2$, where $\phi_i^{\rm th(Ex)}= \left(\dfrac{d\phi_{\nu}}{d\Omega dE_\nu} \right)_{\rm th(Ex)}$, denotes the theoretical (experimental) neutrino flux for energy $E_i$ having error (err)$_i$. The chosen data points of the experimental neutrino flux are tabulated in Table~\ref{table:7.5yr}. The first 5 data points out of total 17 tabulated points in Table~\ref{table:7.5yr}, are the actually observed events while the remaining 12 points are suitably taken from the pink band region (see Figure~\ref{Fig:7.5yr} and \ref{Fig:6yr}). In this analysis, the theoretical data points are computed, where HDM decay via the hadronic channel and the astrophysical flux are considered. We also repeat our analysis, where both the hadronic channel and leptonic channel of HDM decay are considered along with the contribution from the astrophysical flux and estimate the best-fitted mass ($M_{\rm HDM}$) of the decaying heavy dark matter and the corresponding value of the decay lifetime ($\tau$) for this particular case. These best-fit $M_{\rm HDM}$ and $\tau$ are tabulated in Table~\ref{table:3}. The best-fit neutrino fluxes along with the observed events and possible upper bounds are shown in Figure~\ref{Fig:7.5yr}. The pink band region is also mentioned in this plot, from which the last 12 points of Table~\ref{table:7.5yr} are adopted. The upper bounds of possible events beyond 5 PeV, as given in the figure of Ref.~\cite{aachen} are included in Figure~\ref{Fig:7.5yr} for reference (the upper bounds are shown in green in Figure~\ref{Fig:7.5yr} and Figure~\ref{Fig:6yr}).
	
\begin{figure}
		\centering{}
		\begin{tabular}{c}
			\includegraphics[width=0.8\columnwidth]{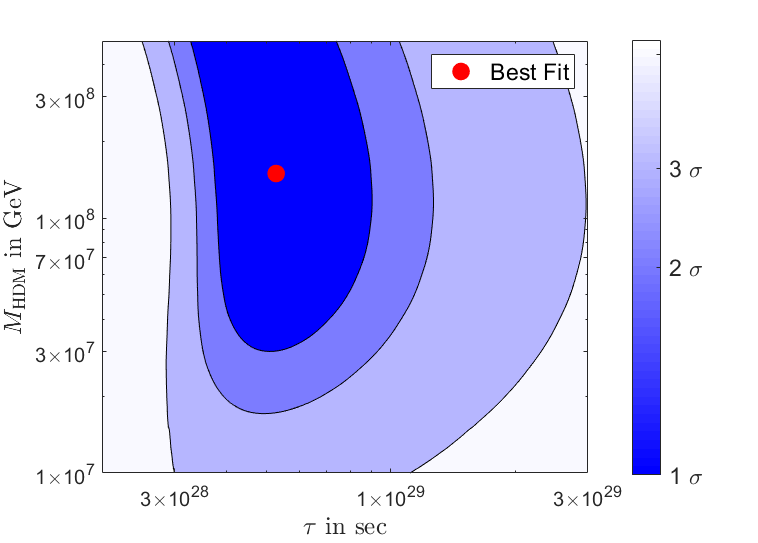}\\
			(a)\\
			\includegraphics[width=0.8\columnwidth]{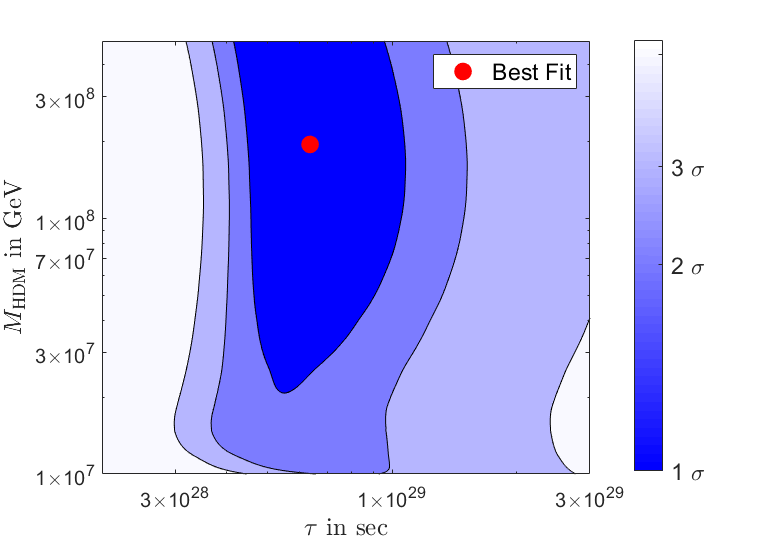}\\
			(b)\\
		\end{tabular}
		\caption{\label{Fig:contour}The contour-plot of the $\chi^2$ analysis in the $M_{\rm HDM} - \tau$ plane by considering all points of IceCube 7.5 yr data within the energy range of $\sim 60$ TeV - $\sim 5$ PeV, which are tabulated in Table~\ref{table:7.5yr}. In this analysis, we considered (a) only the hadronic decay channel along with the astrophysical neutrino flux, (b) both leptonic and hadronic decay channels along with the astrophysical neutrino flux.}
	\end{figure}

	The analysis is performed using \\
	{\bf Case: 1 } Only the hadronic channel for HDM decay with the contribution of astrophysical flux of muon neutrino.\\
	{\bf Case: 2 } Both the leptonic and hadronic channels along with astrophysical flux. \\
	The fitted fluxes for both cases (case1 and case2) are shown in Figure~\ref{Fig:7.5yr}. In Figure~\ref{Fig:7.5yr} one can observe that the fitted flux which corresponds to the hadronic channel and astrophysical flux, posses a downward nature for energy $>5$ PeV (beyond the pink band region). In contrast, due to the incorporation of the leptonic decay channel, the other fitted flux (blue solid line) displays an upward trend for $\gtrapprox 5$ PeV. Thus the effect of the leptonic channel manifests essentially beyond $\sim$5 PeV.
	In Figure~\ref{Fig:contour}a, the computed 1$\sigma$, 2$\sigma$, 3$\sigma$ regions in the $M_{\rm HDM} - \tau$ plane are described graphically, while the contribution of the hadronic channel and the astrophysical are considered. The best-fitted value of HDM mass $M_{\rm HDM}$ and decay time $\tau$ are also shown in the same contour representation. Figure~\ref{Fig:contour}b, demonstrates the same, where the contribution of the astrophysical contribution is considered along with hadronic and leptonic decay channels.

	\begin{figure}
		\centering{}
		\begin{tabular}{c}
			\includegraphics[width=0.8\columnwidth]{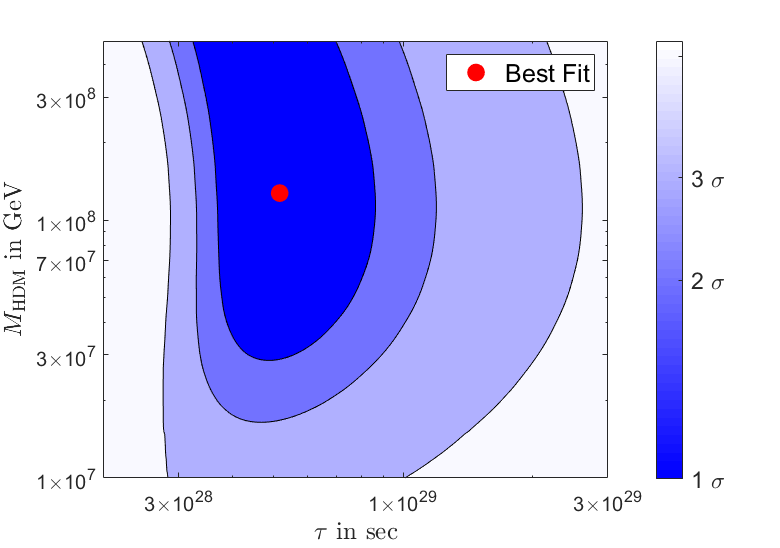}\\
			(a)\\
			\includegraphics[width=0.8\columnwidth]{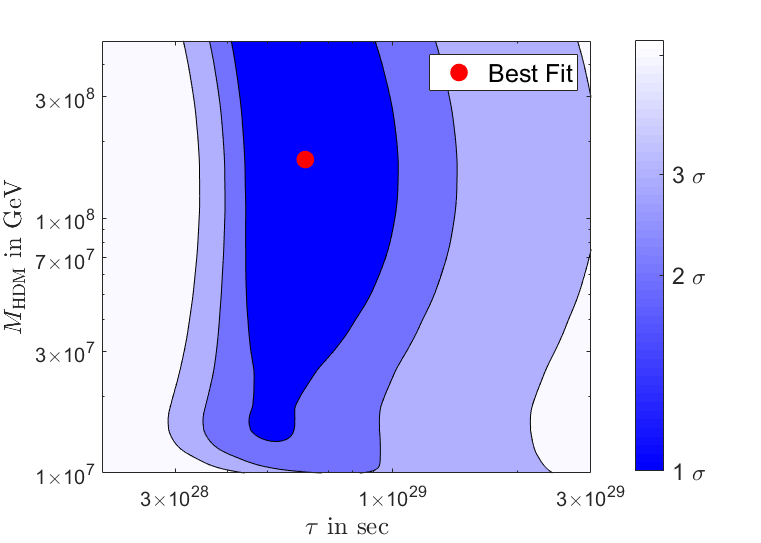}\\
			(b)\\
		\end{tabular}
		\caption{\label{Fig:contour6}Same as Figure~\ref{Fig:contour} for IceCube 6 yr data (see \autoref{table:6yr}).}
	\end{figure}


	
	\begin{figure}
		\centering
		\includegraphics[width=14cm]{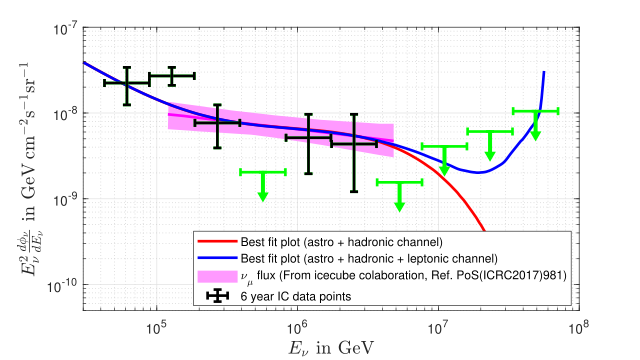}
		\caption{The calculated flux of neutrino with the best-fitted values of HDM mass $M_{\rm HDM}$ and corresponding best-fitted decay lifetime $\tau$. The best-fit points are estimated by performing a $\chi^2$ analysis using the IceCube 6 yr data (Tabulated in Table~\ref{table:6yr}) with both the hadronic and leptonic decay channels of HDM.}
		\label{Fig:6yr}
	\end{figure}
	
	\begin{table}
		\begin{center}
			\begin{tabular}{ccc}
				\hline
				Energy  & Neutrino Flux ($E_\nu^2 \displaystyle\dfrac {d\Phi_{\nu}}{d\Omega dE_\nu}$)  & {Error}
				\\
				(GeV) & (GeV/cm$^{2}$/s/sr) & (GeV/cm$^{2}$/s/sr) \\ \hline
				\hline
				6.13446$\times 10^4\, ^*$ & 2.23637$\times 10^{-8}$ & 2.16107$\times 10^{-8}$\\
				1.27832$\times 10^5\, ^*$ & 2.70154$\times 10^{-8}$ & 1.30356$\times 10^{-8}$\\
				2.69271$\times 10^5\, ^*$ & 7.66476$\times 10^{-9}$ & 8.5082$\times 10^{-9}$\\
				1.19479$\times 10^6\, ^*$ & 5.14335$\times 10^{-9}$ & 7.6982$\times 10^{-9}$\\
				2.51676$\times 10^6\, ^*$ & 4.34808$\times 10^{-9}$ & 8.4481$\times 10^{-9}$\\
				\hline
				3.54813$\times 10^6$ & 5.25248$\times 10^{-9}$ & 4.1258$\times 10^{-9}$\\
				2.30409$\times 10^6$ & 5.71267$\times 10^{-9}$ & 4.1600$\times 10^{-9}$\\
				1.52889$\times 10^6$ & 6.21317$\times 10^{-9}$ & 3.9882$\times 10^{-9}$\\
				1.05925$\times 10^6$ & 6.61712$\times 10^{-9}$ & 3.7349$\times 10^{-9}$\\
				7.18208$\times 10^5$ & 7.04733$\times 10^{-9}$ & 3.9777$\times 10^{-9}$\\
				4.46684$\times 10^5$ & 7.66476$\times 10^{-9}$ & 3.6478$\times 10^{-9}$\\
				2.86954$\times 10^5$ & 8.16308$\times 10^{-9}$ & 4.1571$\times 10^{-9}$\\
				1.90409$\times 10^5$ & 8.87827$\times 10^{-9}$ & 6.2069$\times 10^{-9}$\\
				1.43818$\times 10^5$ & 9.65612$\times 10^{-9}$ & 6.8856$\times 10^{-9}$\\
				2.51189$\times 10^6$ & 4.16928$\times 10^{-9}$ & 8.2726$\times 10^{-9}$\\
				1.19279$\times 10^6$ & 5.03649$\times 10^{-9}$ & 7.5383$\times 10^{-9}$\\
				2.68960$\times 10^5$ & 7.50551$\times 10^{-9}$ & 8.1583$\times 10^{-9}$\\
				\hline
			\end{tabular}
		\end{center}
		\caption{\label{table:6yr} The chosen data points of the UHE neutrino flux with errorbars for the $\chi^2$ analysis with the IceCube 6 yr data.}
	\end{table}

	We also repeat the entire treatment using 6 yr IceCube HESE data within the same energy range (Figure~2 of Ref.~\cite{Kopper:2017Df}). The chosen points from the pink band region are kept the same as adopted in the previous part of the analysis. The data set for this part of the analysis are tabulated in Table~\ref{table:6yr} (pink band points + read-out data points), where the first 5 data points are observed data and the remaining points are obtained from the pink band. As in the case of IceCube 7.5 yr data, here the analysis has also been performed by considering the total calculated neutrino flux as \\
	{\bf Case 1} The summation of the fluxes obtained from the hadronic channel and the astrophysical neutrino flux,\\ 
	{\bf Case 2} The summation of the fluxes obtained from both the hadronic channel and leptonic channel along with the astrophysical neutrino flux.\\
	The corresponding best-fit points along with the computed 1$\sigma$, 2$\sigma$, 3$\sigma$ confidence levels in the $M_{\rm HDM} - \tau$ parameter plane are described graphically in Figure~\ref{Fig:contour6}a and \ref{Fig:contour6}b respectively. The chosen data points for this analysis are tabulated in Table~\ref{table:6yr}. In Figure~\ref{Fig:6yr} we describe both fitted fluxes for IceCube 7.5 yr data along with the IceCube 6 yr HESE data. The best-fitted points for both cases (Case 1 and Case 2) are tabulated in Table~\ref{table:3}.
	
	In this context, it is to be mentioned that, in several recent treatments on similar topics \cite{Feldstein2013}, a comparatively much smaller value for heavy dark matter mass is suggested ($\sim 10^6$ GeV). However, in the present works, neutrino events $\gtrapprox 1$ PeV energy were considered and obtained value of the best-fitted HDM mass is $\sim 10^{8}$ GeV.
	
	Now using the best-fitted HDM mass $M_{\rm HDM}$ and decay lifetime $\tau$ are estimated from each of the two analyses with both datasets (IceCube 7.5 yr and 6 yr HESE data) the total amount of baryon asymmetry $\Delta B$ is estimated, which is manifested as an outcome of the HDM decay by adopting the formalism discussed in the previous section (see Appendix). The estimated value of $\Delta B$ for all four cases are tabulated in Table~\ref{table:3} besides the corresponding best-fit points. 
	Throughout the present analyses, the baryon number violation parameter $\epsilon=1$ is chosen in order to explain the estimated amount of baryon asymmetry from observational results i.e. $\sim 10^{-10}$. Such choice of baryon number violation parameter $\epsilon$ is evident in the proton decay model by Georgi-Glashow model \cite{GGProtonDecay}, which is applicable during the baryogenesis based on Grand Unified Theory (GUT).
	
	\begin{figure}
		\centering
		\includegraphics[width=11cm]{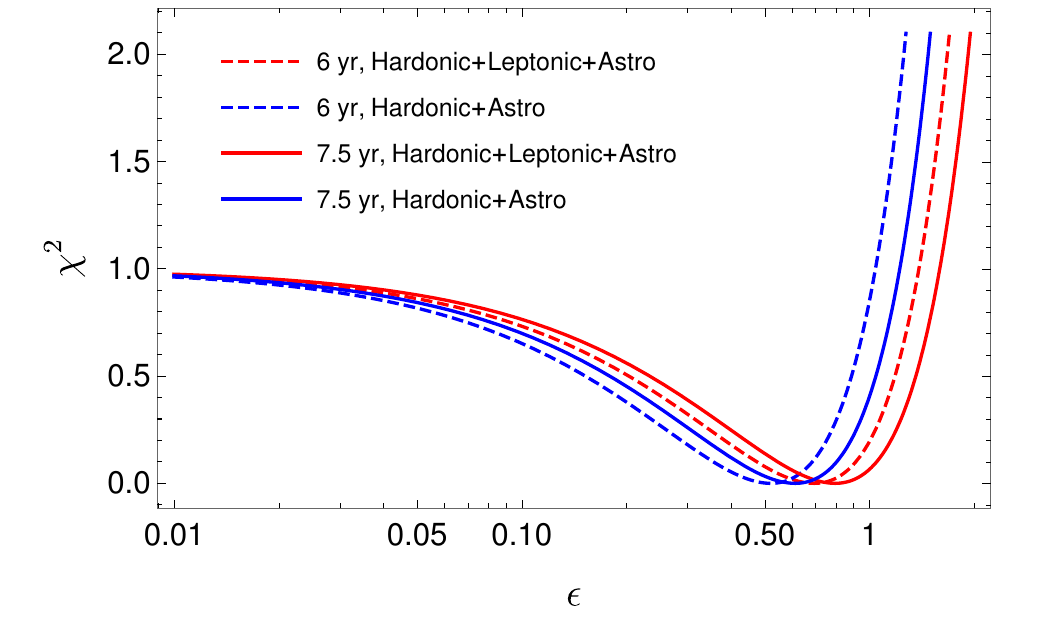}
		\caption{\label{fig:chi2_epsilon} Variation of $\chi^2$ with $\epsilon$ for all four best-fit $M_{\rm HDM}$, which are tabulated in Table~\ref{table:3}.}
	\end{figure}
	
\begin{figure}
		\centering
		\includegraphics[width=0.8\linewidth]{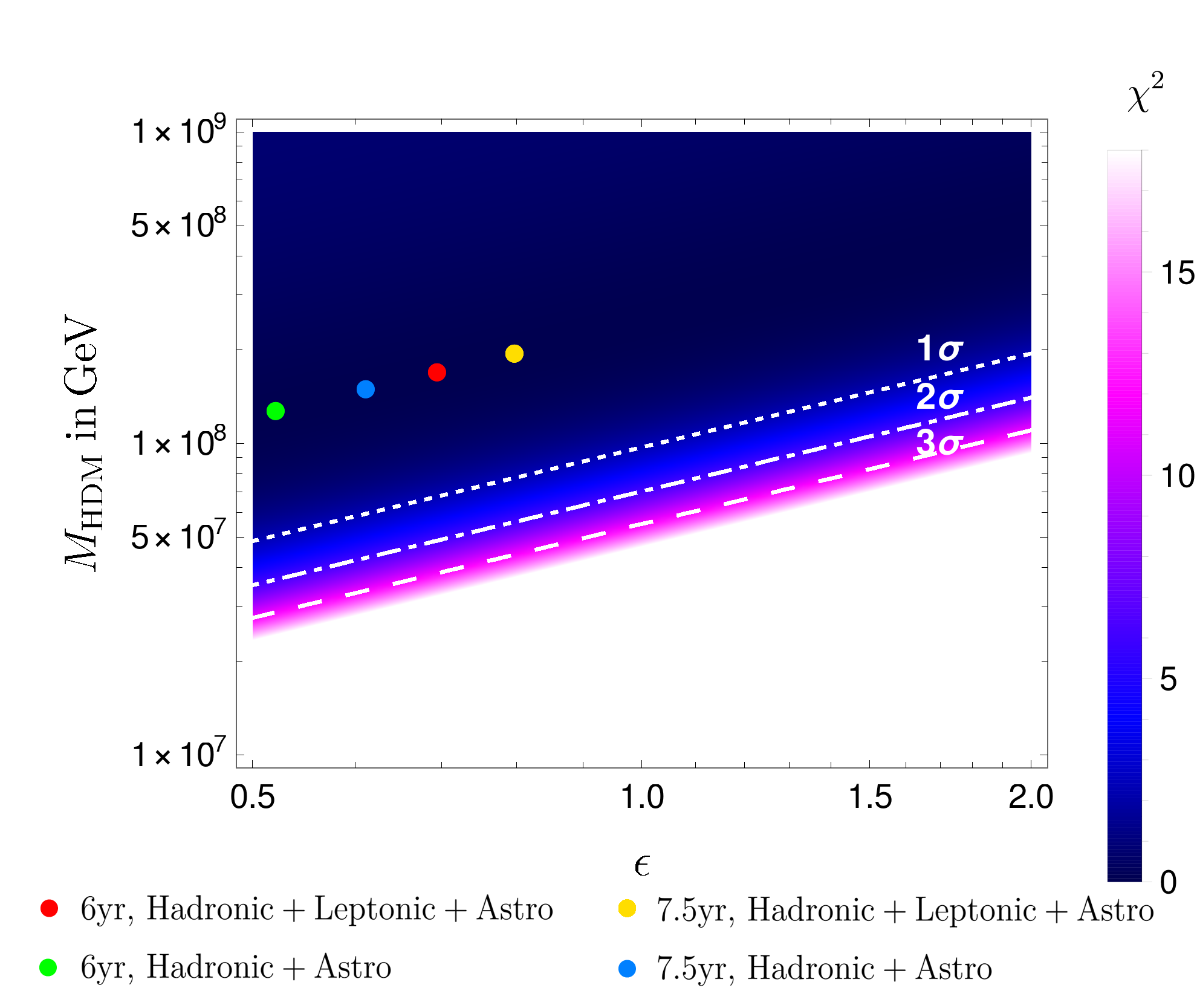}
		\caption{Contour representation of the $\chi^2$ value for baryon asymmetry $\Delta B$ in the $M_{\rm HDM}-\epsilon$ parameter space. Four best-fitted points are also shown in the same figure.}
		\label{Fig:MassEpsilon}
	\end{figure}

	The choice $\epsilon = 1$ is also justified in Figure~\ref{fig:chi2_epsilon}. Here another $\chi^2$ analysis is carried out, which is defined as $\chi^2 = \displaystyle\sum_{i = 1}^{n} \left (\dfrac {\Delta B_{\rm calculated} - \Delta B_{\rm observed}} {{\Delta B}_{\rm observed}} \right )^2$. In this expression, ${\Delta B}_{\rm observed}$ is the observed value of $\Delta B$ by PLANCK experiment \cite{planck18}. In Figure~\ref{fig:chi2_epsilon}, it can be noticed that, for each values of best-fit $M_{\rm HDM}$ and the corresponding values of decay lifetime $\tau$, the best-fitted (minimum $\chi^2$) values of the parameter $\epsilon$ is $\sim 1$. It is to be mentioned that, when the only hadronic channel along with the astrophysical flux are considered (for both IceCube 6 yr data and IceCube 7.5 yr data), the obtained values of the best-fit $\epsilon$s are slightly smaller ($\sim0.5-0.7$). In contrast, as both leptonic and hadronic channels are considered (along with astrophysical flux), the best-fit values of $\epsilon$ become almost unity ($\sim0.7- 1$). Besides the best-fit values of $M_{\rm HDM}$, a general $\chi^2$ analysis has also been performed in the $M_{\rm HDM}-\epsilon$ parameter plane for $10^7$ GeV $\leq M_{\rm HDM}\leq 10^9$ GeV and $0.5\leq \epsilon \leq2.0$ (see Figure~\ref{Fig:MassEpsilon}). In Figure~\ref{Fig:MassEpsilon}, the values of $\chi^2$ at different values of $M_{\rm HDM}$ and $\epsilon$ are represented using colour code, which is described in the colour bar of Figure~\ref{Fig:MassEpsilon}. The corresponding 1$\sigma$, 2$\sigma$ and 3$\sigma$ contour lines are addressed using the dotted, dash-dotted and the dashed line respectively. The best-fitted masses and the corresponding best-fitted $\epsilon$ are also plotted for individual cases (Hadronic, Leptonic and astrophysical contribution with IceCube 7.5 yr and Icecube 6 yr HESE data and the same with only hadronic and astrophysical contribution) in this figure. It can be observed that at the best-fitted values of $M_{\rm HDM}$ (with both 6 yr and 7.5 yr IceCube data) the minimum  $\chi^2$ are located at $\epsilon \sim 1$, which justifies the choice of $\epsilon = 1$ in the present work.
	
	\begin{table}
		\begin{center}
			\begin{tabular}{|c||c|c|c|c|c|}
				\hline
				&& $M_{\rm HDM}$ & $\tau$ & $\Delta B$ & $\Delta B$\\
				Set & Flux & in GeV & in sec & from  & for\\
				&& (best-fit) & (best-fit) & Eq.~\ref{eq:deltaB} & $\epsilon=1$\\
				\hline \hline 
				
				\multirow{5}{*}{6 yr}& Hadronic + &&&&\\
				& Leptonic +  & $1.6982 \times 10^8$ & $6.1660 \times 10^{28}$ & $1.26 \times 10^{-10} \epsilon$ & $1.26 \times 10^{-10}$\\
				& astro &&&&\\
				\cline{2-6}
				& Hadronic + & \multirow{2}{*}{$1.2735 \times 10^8$} & \multirow{2}{*}{$5.1582 \times 10^{28}$} & \multirow{2}{*}{$1.67 \times 10^{-10} \epsilon$} & \multirow{2}{*}{$1.67 \times 10^{-10}$}\\
				& astro &&&&\\
				\hline
				
				\multirow{5}{*}{7.5 yr}& Hadronic + &&&&\\
				& Leptonic +  & $1.9498 \times 10^8$ & $6.3460 \times 10^{28}$ & $1.09 \times 10^{-10} \epsilon$ & $1.09 \times 10^{-10}$\\
				& astro &&&&\\
				\cline{2-6}
				& Hadronic + & \multirow{2}{*}{$1.4962 \times 10^8$} & \multirow{2}{*}{$5.2784 \times 10^{28}$} & \multirow{2}{*}{$1.42 \times 10^{-10} \epsilon$} & \multirow{2}{*}{$1.42 \times 10^{-10}$}\\
				& astro &&&&\\
				\hline
			\end{tabular}\\
			\caption{\label{table:3} The best-fit values of HDM mass and corresponding best-fit decay lifetimes, which are estimated from $\chi^2$ analyses using IceCube 7.5 yr and IceCube 6 yr HESE data. The corresponding values of baryon asymmetry ($\Delta B$) for $\epsilon = 1$ are also tabulated.}
		\end{center}
	\end{table}

			
			
			
		
\begin{figure}
		\centering
		\includegraphics[width=\linewidth]{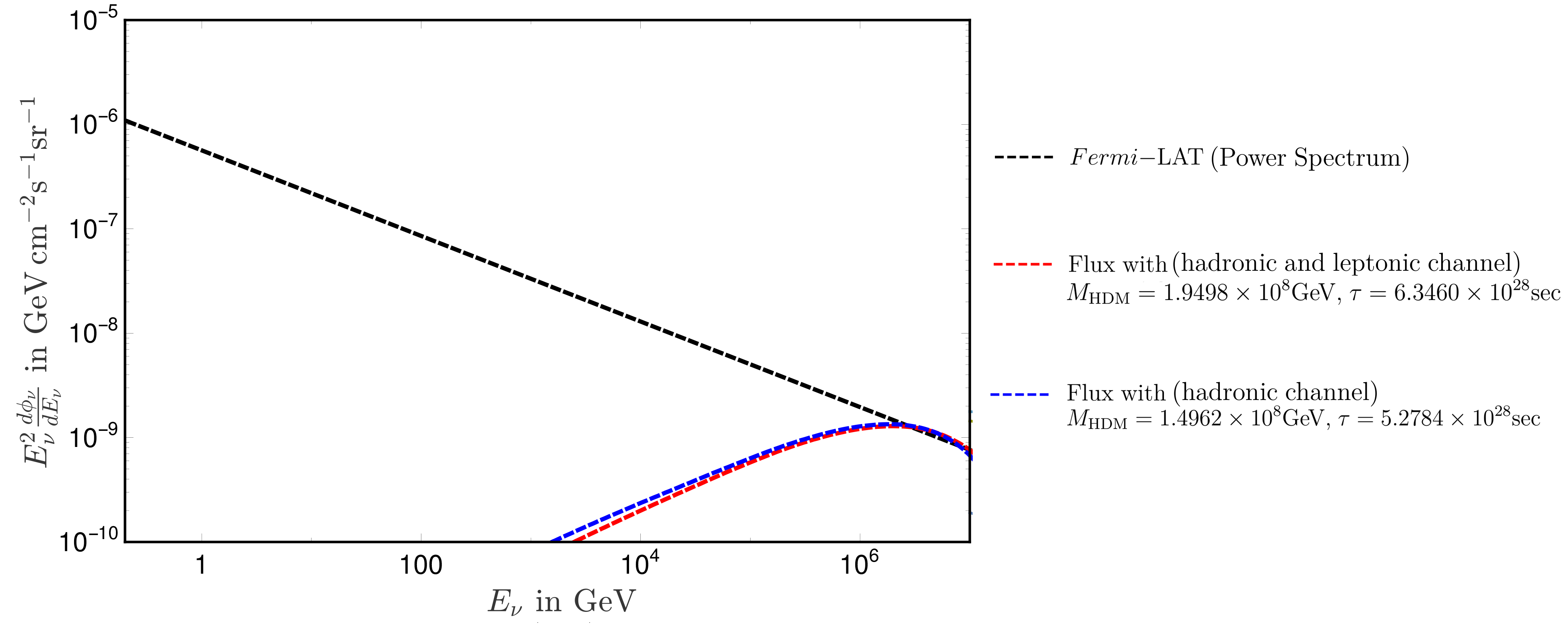}
		\caption{Comparing calculated extragalactic $\gamma$-ray flux, with the empirical relation of the extra-galactic $\gamma$-ray flux as observed by \emph{Fermi}-LAT. In the current analysis, the calculated $\gamma$-ray flux are estimated for best-fitted values of HDM mass $M_{\rm HDM}$ and corresponding decay lifetime $\tau$ for individual cases.}
		\label{Fig:Fermi_IC}
	\end{figure}

	We also check, whether the $\gamma$-rays (evaluated using Eq.~\ref{eq:gamma_hdm}) from the leptonic and hadronic channel of heavy dark matter decay for the best-fit masses and lifetimes of HDM (see Table~\ref{table:3}) satisfy the constraint proposed by Fermi. The extra-galactic $\gamma$ flux obtained from the \emph{Fermi}-LAT observations obeys a power-law behavior (see Eq.~2.2 of {\bf Murase} \emph{et~al.} \cite{Murase:2012xs}). In Figure~\ref{Fig:Fermi_IC}, that analytical form of the Fermi observation is plotted using the black dashed line. The $\gamma$-ray fluxes for the combined hadronic and leptonic channels (red dashed line), and the only hadronic channel are plotted in the same graph (blue dashed line). In both the cases, the values of $M_{\rm HDM}$ and $\tau$ are chosen from Table~\ref{table:3}, which are calculated from the $\chi^2$ analysis using IceCube 7.5 yr data. Comparing the empirical flux given by {\bf Murase} \emph{et~al.} \cite{Murase:2012xs} with the calculated extra-galactic $\gamma$-ray flux, it can be seen that for both cases addressed in Figure.~\ref{Fig:Fermi_IC} the calculated $\gamma$-ray fluxes lie below the extra-galactic flux of Fermi-LAT as obtained from derived from {\bf Murase} \emph{et~al.} \cite{Murase:2012xs}.
	
\chapter{Summary and Future Outlook} \label{chp:conclu}
\\
	\begin{changemargin}{40pt}{40pt}
		\centering
		\setstretch{1}
		{\myfont ``Dark matter and dark energy are two things we measure in the universe that are making things happen, and we have no idea what the cause is."}
		\vspace{-0.5cm}
		\begin{flushright}
			{\bf \color{qtcol} -Neil deGrasse Tyson}
		\end{flushright}
	\end{changemargin}

	Although the nature of dark matter and dark energy are yet to be ascertained, these dominant components of the Universe play a significant role in the formation and dynamics of galaxies, galaxy clusters and more importantly in shaping the evolution and fate of the Universe.
	
	The thesis is essentially based on few open questions regarding the unknown dark sector part of our Universe, which occupies $\sim 95\%$ of the total energy budget. The thesis addresses these and other aspects of astroparticle physics and investigates several aspects of compact astrophysical objects and the effects of heavy dark matter decay in the dynamics of the Universe. The inferences that emerged out from this thesis are summarized as follows.
	
	\Autoref{chp:intro} and \Autoref{chp:compact_obj} are introductory in nature where brief overviews are given for the topics addressed in the thesis. In \Autoref{chp:intro}, we motivate the physics of two dark sector components namely dark matter and dark energy that together constitute about $\sim 95\%$ of the total mass-energy budget of the Universe. To this end basics of dark matter and dark energy are reviewed in \Autoref{chp:intro}. These include the evidence of dark matter and dark energy, the types and nature of dark matter components, their detection, density profiles of the galactic dark matter halo etc. In \Autoref{chp:compact_obj}, we give an introductory account of the astrophysical compact objects such as neutron stars, quark stars, black holes and Hawking radiation and other exotic forms of compact stars, e.g. preon star, boson star etc.
	
	
	Our Universe seems to be full of objects which elude common interpretation. The quark star \cite{Witten} is such an exotic star of our universe which was proposed by two Soviet physicists {\bf D. F. Kurdgelaidze} and {\bf D. D. Ivanenko} (in 1965) just five years after the discovery of quark. In \Autoref{chp:chandra}, we studied the maximum allowed mass and corresponding limiting radius for rotating quark stars, which are introduced in the present work as ``Chandrasekhar limit for rotating quark stars" \cite{Qs_chandra}. Since a quark star is made up of fundamental constituents of matter, they are extremely dense, conceivably even higher than a pure neutron star. Unlike typical neutron stars, quark stars are self-bounded by the strong interaction. Consequently, the maximum admissible mass is essentially depending on fundamental constants and is addressed by employing the method of energy balancing as proposed by {\bf Landau} \cite{Landau}. Besides fundamental constants, angular frequency and MIT Bag constant also have a significant role in the estimation of the limiting mass of a star and its corresponding radius and compactness \cite{SBa}. The maximum possible observed frequency ($\nu_{\max}$) for such a stellar object is also evaluated which agrees with the recent observations (observed pulsars namely PSR J0740+6620 \cite{J0740+6620}, PSR J1614-2230 \cite{J1614-2230}, PSR B1957+20 \cite{B1957+20}, PSR J1600-3053 \cite{J1600-3053}, PSR J2215+5135 \cite{J2215+5135}, PSR J0751+1807 \cite{J0751+1807}, PSR J1311-3430 \cite{J1311-3430} etc.) as well as the numerical simulations based results \cite{upperlim_fr}. A relation between the maximum admissible mass (mass in terms of the Schwarzschild radius $R_{\rm{sch}}$) and the corresponding radius ($R_{\text{max}}$) for different chosen values of bag constant has also been put forward, for a range of rotational frequencies of the star, which yields the numerical value of $\dfrac{R_{\text{max}}}{R_{\rm{sch}}} \sim 2.2903$ for the non-rotating quark stars (similar result is obtained in {\bf Banerjee}~\emph{et~al.} \cite{SBa} (i.e. 2.6667), where the effects due to the relativistic corrections were ignored). However, the quantity $\dfrac{R_{\text{max}}}{R_{\rm{sch}}}$ tends toward unity as the rotational frequency increases toward an extreme case (i.e. $\omega \rightarrow R_{\text{max}} c$). In the case of a fast rotating compact celestial object, the Schwarzschild radius does not denote the event horizon \cite{hobson}. In this case, the Kerr spacetime needs to be considered in the estimation of the event horizon for such rotating objects, which essentially provides a smaller horizon in comparison to the same as calculated using the Schwarzschild metric system \cite{hobson}. As a consequence, one may therefore conjecture that, {\bf ``a quark star can never behave like a black hole''}.
	
	The primordial black hole is another candidate for compact astrophysical objects, which has been the center of interest for several decades in the field of cosmology and astrophysics. For the cases of primordial black holes, the masses of PBHs are assumed to be substantially low, the effects of the evaporation of such black holes (Hawking radiation \cite{hawking}) are very prominent. As a consequence, the phenomenon of PBH evaporation can be a promising tool in the estimation of the PBH abundance. The emitted particles in the form of Hawking radiation interact with the medium and hence modify the thermal evolution of IGM. Such modification can be estimated by looking over the global 21-cm signature.
	
	In \Autoref{chp:21_feb}, the combined effects of PBH evaporation and the cooling of the baryonic matter due to baryon - dark matter collision are addressed \cite{amar21_1}, which may substantially affect the brightness temperature of the global 21-cm signal. As a consequence, the variation of the 21-cm brightness temperature also shows significant dependence on dark matter mass $m_{\chi}$ and the baryon-dark matter interaction cross-section $\sigma_0$, besides PBH mass and its initial mass fraction ($\beta_{\rm BH}$). In this context, the bounds on the initial mass fraction of PBH are estimated for different masses of primordial black holes and dark matter particles (both upper and lower bounds).
	
	It can be observed that the estimated allowed regions (the regions between the upper and lower bound) for $\beta_{\rm BH}$ are extremely narrow when essentially lighter dark matter particles are considered in the dark matter - baryon collision. However, as the higher masses of DM ($m_{\chi}$) are taken into account, the lower limit falls rapidly and eventually vanishes for $m_{\chi}\gtrsim 0.5$ GeV. A similar variation is observed for higher values of $\sigma_{41}$ (for example $\sigma_{41} = 5$, $\sigma_{41}$ being the dark matter-baryon scattering cross-section expressed in the units of $10^{-41}$ cm$^2$). It is to be mentioned that, the calculated upper bounds are identical to the bounds as estimated from the observational data of the Planck-2015 experiment when a lower dark matter mass is considered ($m_{\chi}\approx 0.1$ GeV). In this case of the bound estimated from the Planck-2015 data, the limit is calculated using the \texttt{COSMOMC} code with the observational results of Planck-2015. On the other hand, for higher values of $m_{\chi}$, the obtained limits for initial mass fraction $\beta_{\rm BH}$ are comparable to the same as proposed by {\bf Yang} \cite{BH_21cm_2} \footnote{In the work of {\bf Yang} \cite{BH_21cm_2}, the effects of baryon - dark matter collision are not taken into the account}. In addition, the estimated bound using the 21-cm power spectra also fits well for the dark matter mass $m_{\chi}\approx 1$ GeV and the black hole mass $\mathcal{M}_{\rm BH}\geq 1.1 \times 10^{14}$ gm. We repeat our analysis for other values of $\sigma_{41}$, while the $m_{\chi}$ is kept fixed at 1 GeV. It can be observed that at a certain value of DM mass $m_{\chi}$, the upper limit for $\beta_{\rm BH}$ increases with increasing value of $\sigma_{41}$. The similar allowed zone in the $m_{\chi} - \sigma_{41}$ parameter space is also put forward in this chapter, from where one can notice that the maximum possible value of $m_{\chi}$ obtained from this analysis, agrees with the results of {\bf Barkana} \cite{rennan_3GeV} for the entire PBH mass range discussed in this work.
	
	In \Autoref{chp:21_jan}, the effects of the interaction between two dark sector components (dark matter - dark energy interactions) are principally addressed in the framework of 21-cm EDGES signal, while taking into consideration other important effects arising out of dark matter scattering on baryons and also PBH evaporation that can possibly influence the observed temperature of the 21-cm signal during the reionization \cite{amar21}. The interactions between two dark sector components (dark matter and dark energy) can influence the 21-cm brightness temperature. To this end, three such interacting dark energy (IDE) models are adopted where the non-minimal coupling of two dark sectors (i.e. dark matter and dark energy) is adopted and the amount of energy transfer due to IDE and dark matter-baryon scattering are properly incorporated in the relevant evolution equations for baryon temperature, dark matter temperature etc. 
	
	It can also be noted that when the dark matter - dark energy interactions are considered, the DM density $\rho_\chi$ and the DE density $\rho_{\rm DE}$ do not evolve as $\sim (1+z)^3$ and $(1+z)^{3(1+\omega)}$ respectively, as suggested by standard cosmology. Therefore the evolution of the Universe ($H(z)$) is modified as an outcome of the interaction between two dark sector components (DM and DE) which in turn affects the spin temperature $T_s$ and the optical depth of the medium. The evolution of the Hubble parameter is also computed in detail in the present work for all the three IDE models adopted. 
	
	The dark matter-dark energy interaction parameter $\lambda$ has been probed in this work along with the effects of PBH and DM-baryon interaction. The upper and lower limits of the IDE parameter $\lambda$ for three IDE models (adopted in this work) have been investigated for different PBH masses (within the ball park of $10^{14}$ gm) for the range of $T_{21}$ given by the EDGES experiment at reionization era. This appears that the model constraints described in Table~\ref{tab:constraints} satisfy the EDGES limit for relatively lower masses of dark matter ($\leq 1.0$ GeV). Moreover, for the case of IDE Model II and III, the obtained upper bounds of the dark matter mass $m_{\chi}$ increases significantly, which lies $\sim 3$ GeV in the case $\Lambda$CDM model \cite{rennan_3GeV}. Future results related to 21-cm physics from the early Universe would throw more light to all these issues and the thermal evolution and dynamics of the early Universe.
	
	Keeping in view of the 21-cm scenario, we also explore the multimessenger signals from possible rare decay of fundamental heavy dark matter or superheavy dark matter (HDM) from dark ages leading to the reionization epoch \cite{21cm_mar}. Heavy dark matter of mass as high as $10^8$ GeV or more could be created via gravitational production mechanism or non-linear quantum effects during the reheating or preheating stages after inflation. These dark matters are therefore produced non-thermally in the high redshifted epoch of the Universe and are long-lived. Thus these heavy or super heavy dark matter, if exists, exhibits a rare decay process.
	
	In \Autoref{chp:21_mar}, two possible multimessenger signals from rare decay of heavy or superheavy dark matter are discussed. One of them is the neutrino signals (of $\sim$ PeV energies) from rare decays of such heavy dark matter at the IceCube neutrino experiment and the other is the influence of this decay process on the absorption temperature of 21-cm Hydrogen signal. The decay process of primordial superheavy dark matter may proceed via QCD cascades. The primary products $q\bar{q}$, on hadronization and decay, produce leptons, $\gamma$ etc. as the end products. The decay of heavy dark matter having mass significantly higher than the electroweak scale takes place via the cascading of QCD partons \cite{kuz,bera,bera1,bera2}. We used only the hadronic decay channel ($\chi\rightarrow q\bar{q}$) in our calculation as the contribution of the leptonic decay channel ($\chi\rightarrow l\bar{l}$) is much smaller than the hadronic channel \cite{kuz,mpandey}. The total spectrum at scale $s=M_{\rm HDM}^2$ can be obtained by summing over all Parton fragmentation functions $D^{h} (x,s)$ from the decay of particle $h$. Here $x=\frac{2E}{M_{\rm HDM}}$ is the dimensionless energy of the particles. In our current treatment, only the muon decay is considered in the calculation as the total contribution of the other mesons is negligible ($<10\%$) \cite{kuz,bera1}. 
	
	Ultra-high energy neutrinos could be a possible final product of this decay process. We attribute these ultra high energy (UHE) neutrinos from the decay of primordial HDM to be the source of UHE $\nu$ signal at IceCube neutrino detector in the $\sim$ PeV energy regime. The so called HESE (high energy starting events, also known as contained vertex events) events that are obtained from 82 track and shower events with energy deposition $> 20$ TeV apparently fit a power-law spectrum $\sim E^{-2.9}$. But the up-going muon neutron events in the energy range $\sim 120$ TeV to few PeV show a different power-law dependence. For this purpose, 7.5 years of IceCube up-going muon data are chosen in the neutrino energy window of $2\times 10^5$ GeV to $4\times10^6$ GeV \cite{IC_7.5yr}. It contains four data points. The calculated neutrino spectrum is then fitted with the IceCube results and the best fit value of mass and $\mathcal{K}$ ($\mathcal{K}$ is the product of decay width and the density fraction) of the HDM are obtained.
	
	The other possible multimessenger signal from the heavy dark matter decay is addressed by including the effect of this decay in the formalism for computing 21-cm brightness temperature ($T_{21}$). To this end, the effect of HDM decay has been included in the evolution of baryon temperature. In addition, we also consider another thermal WIMP type cold dark matter (CDM) with mass lie within few hundred MeV to few GeV (This is consistent with {\bf Barkana et~al.} \cite{rennan_3GeV}). But the abundance of the HDM to the total dark matter containing the Universe is considered to be small, while the lighter cold type dark matter (CDM) component accounts for the remaining part of the DM content of the Universe. This lighter candidate of dark matter takes a part in IGM heating/cooling effect that could have been caused by the collisions of CDM types DM candidates with baryons. We have also included other heating effects due to the evaporation of PBHs in the form of Hawking radiation.
	
	The EDGES observational results for $T_{21}$ are then used to constrain the $\mathcal{K} - M_{\rm HDM}$ plane. It is found that both the multimessenger signals considered for heavy dark matter decay signal, namely the IceCube PeV neutrinos and $T_{21}$ temperature are satisfied when the dark matter mass $m_{\chi}$ (with which baryon collides to influence $T_{21}$ signal in the early Universe) is $m_{\chi} \geq 0.4$ GeV and PBH mass $\mathcal{M}_{\rm BH} \approx 10^{14}$ GeV. In addition, it is also investigated how the mass of PBH affects the baryon - dark matter scattering cross-section ($\sigma_{41}$) with different chosen values of DM masses ($m_{\chi}$) of the order of GeV in presence of heavy dark matter decay contributions to $T_{21}$. To this end, we use $T_{21}$ results and constrain $\sigma_{41} - m_{\chi}$ plane. The heavy dark matter mass $M_{\rm HDM}$ and $\mathcal{K}$ are kept fixed at their best fit values ($M_{\rm HDM} = 2.75\times10^8$ GeV, $\mathcal{K} = 2.56\times10^{-29}$ sec$^{-1}$), which are obtained from the $\chi^2$ analysis using the IceCube 7.5 year HESE data. We find $m_{\chi} < 3$ GeV in order to satisfy the EDGES results for $\mathcal{M}_{\rm BH} \leq 2.5 \times 10^{14}$ gm which agrees with the work of {\bf Barkana} \cite{rennan_3GeV}. Thus this work is a detailed multimessenger study of the possible heavy dark matter decay signals using IceCube 7.5 year data and 21-cm absorption line temperature of hydrogen and EDGES results, while simultaneously exploring the effects of Ly$\alpha$ forest and PBH evaporation on 21-cm signal.

	The consequence of late time decay of such superheavy DMs may also have significant implications to the matter - antimatter asymmetry. In \Autoref{chp:IC}, we discussed about baryon asymmetry \cite{tista_IC}, which can be developed in the Universe due to the decay of such DM candidates. The matter-antimatter asymmetry may give rise to asymmetry in the baryon number and lepton number, which satisfies the Sakharov's conditions for baryogenesis \cite{wimpzilla, UHDMdecay&baryonAsymmetry, Kolb_EarlyUniverse}. In order to estimate the amount of asymmetry that arises as an outcome of the decay of heavy dark matter, a Boltzmann-like expression has to be evaluated in order to estimate the rate of baryon production number per DM decay. In this context, the baryon number violation parameter ($\epsilon$) plays a significant role, which is related to the heavy dark matter decay.
	
	In this particular case, 7.5 year IceCube neutrino events are considered in the present calculation, within the energy range $\sim$ 60 TeV - $\sim 8$ PeV. The data corresponding to those events are obtained from the plot shown by C. Wiebusch in his presentation at SSP – Aachen, June 2018 \cite{aachen}. We choose five data points and adopt twelve other points which are suitably chosen from the 1-$\sigma$ bestfit region (with power-law spectrum $E_{\nu}^{-2.9}$) within the energy range of 120 TeV $<$ $E_{\nu}$ $<$ 5 PeV. In the case of chosen points from the best-fit region, the width of the 1-$\sigma$ limits is considered as the error bars for the corresponding points.
	
	The heavy dark matter may undergo rare decay to produce ultra-high energy neutrinos as the final product by following two dark matter decay cascade channels - hadronic and leptonic. It can be seen that the incorporation of the leptonic decay channel in the computation modifies the fitted spectrum beyond the energy region of 5 PeV. The astrophysical power-law flux (with power-law behavior $\sim E^{-2.9}$) is also included in our $\chi^2$ analysis for the IceCube flux in the energy range $\sim 60 - 120$ TeV in this work. From the $\chi^2$ analysis with the combined flux from both leptonic and hadronic decay channels for HDM along with the astrophysical flux, the best fit values for the heavy dark matter mass ($M_{\rm HDM}$) and corresponding decay lifetime ($\tau$) are estimated. The entire analysis is performed for both 6 year and the recent 7.5 year IceCube data in two separate analyses.
	
	It is also investigated that, the $\gamma$-rays from the leptonic and hadronic channels of the HDM decay satisfy the FermiLAT constraint. The extra-galactic $\gamma$-ray spectrum can be obtained from the FermiLAT observations, which follows a power-law behavior as mentioned in the work of {\bf Murase} \emph{et~al.} \cite{Murase:2012xs}. In the present analysis, both hadronic and leptonic decay channels are considered in the calculation of the $\gamma$-ray spectrum for the best-fitted HDM mass and lifetime for the individual cases and then compared that with the empirical spectrum, as proposed in the work of {\bf Murase} \emph{et~al.} \cite{Murase:2012xs}.	
	

	The research work described in the thesis can be extended to explore new areas and to further understand these very interesting topics, these will be taken up in the future. They include the following future possibilities.
	\begin{enumerate}
		\item Due to the intense gravity of Compact stars (CS) (such as Neutron stars (NS), Quark stars etc.), they can capture and trap dark matter inside its core. If this dark matter is asymmetric dark matter and Bosonic in nature, then they may form Bose-Einstein (BE) Condensate at the compact star core leading ultimately to the formation of a black hole. On the other hand, if we consider the existence of the dark sector (along with the visible standard model sector) then we may conceive of the existence of U(1) dark photon. Contrary to the formation of BE condensate, these dark photons, if exist, may prevent any coagulation of the dark matter particles. Such dark photons may be created inside compact star core by the process of Compton scattering and mixing with standard model (SM) photons. They can even mix with a standard model photon and the mixed state may decay to lepton - antilepton pairs, which under the influence of compact star magnetic field can emit synchrotron radiation, which is detectable by terrestrial radio detectors (eg. SKA, GMRT etc.). It is worth exploring these issues that are important not only to understand the nature of dark matter but also the properties of compact stars.
		\item The 21-cm Hydrogen absorption line is an important probe for the cosmic dark ages in the absence of any significant signal from this era, the nature of the 21-cm line from this age may carry information of this era. The possible dependence of the nature of the 21-cm hydrogen line on various other factors such as DM interactions, DE interactions, PBH evaporations and also others may help understand better the physics and cosmological processes during the Dark age. Future lunar and space-based radio observation may be capable to explore several salient features of the Universal dynamics at the higher redshifted epoch using the global 21-cm signal \cite{Burns_2017,plice2017dare,21cm_moon}.
		\item The IceCube signals from high energy neutrino events may content enormous possibilities to understand the high energy phenomena in the Universe. A thorough and systematic analysis of such data may not only help understand the high energy phenomena like GRB, AGN, Blazar etc., these data might indicate other exotic cosmic phenomena, that could produce these ultra high energy neutrinos. Therefore the studies of these neutrinos along with multimessenger astronomy may help understand better not only these phenomena but also other cosmic processes which are still unknown. In addition, these IceCube neutrinos may through light on other properties of neutrinos still undiscovered (e.g. Lorentz invariance violation, CP violation, the origin of neutrino mass etc.).
	\end{enumerate}
\lhead{\ifthenelse{\isodd{\value{page}}}{}{\color{hcol} \emph{Appendix}}} \rhead{\ifthenelse{\isodd{\value{page}}}{\color{hcol} \emph{Appendix}}{}}
\chapter{Appendix: \\
	Solving Boltzmann-like Equation} \label{APP:IC}

	To estimate the baryon asymmetry appears as an outcome of heavy DM decay, one need to solve a Boltzmann-like equation, that measures the baryon number production rate per unit decay, given by \cite{wimpzilla}
	\begin{equation}
		\dfrac{{\rm d}(n_{b} - n_{\bar{b}})}{{\rm d}t} + 3 \dfrac{\dot{a}(t)}{a(t)} (n_{b} - n_{\bar{b}}) = \dfrac{\epsilon n_{\chi} (t)}{\tau},
		\label{eq:app_1}
	\end{equation}
	where, $n_b$, $n_{\bar{b}}$ are respectively the baryon and antibaryon number density of the Universe, $n_{\chi}$ denotes the same for the HDM particles at redshift $z$ while the decay lifetime of the heavy dark matter is represented by $\tau$. It can be easily recognized that the Eq.~\ref{eq:app_1} has the form of a linear homogeneous first-order differential equation,
	\begin{equation}
		\dfrac{{\rm d}y}{{\rm d}t} + P y = Q,
	\end{equation}
	where,
	\begin{eqnarray}
		y &=& (n_{b} - n_{\bar{b}}), \nonumber \\
		P &=& 3 \dfrac{\dot{a}(t)}{a(t)}, \nonumber \\
		Q &=& \dfrac{\epsilon n_{\chi} (t)}{\tau}.
	\end{eqnarray}
	
	The Eq.~\ref{eq:app_1} can be solved by introducing an integrating factor ($C_{\rm int}$) given by,
	\begin{equation}
		C_{\rm int} = \exp{\left(\displaystyle\int P {\rm d}t\right)}.
	\end{equation}
	So, the solution of the Eq.~\ref{eq:app_1} is expressed as,
	\begin{equation}
		y \times C_{\rm int} = Q \int C_{\rm int} dt.
		\label{eq:app_5}
	\end{equation}
	Now applying $P = 3 \dfrac{\dot{a}(t)}{a(t)}$ in Eq.~\ref{eq:app_1} we obtain,
	\begin{eqnarray}
		\exp{\left(\displaystyle\int P {\rm d}t\right)} &=& \exp{\left(\displaystyle\int 3 \dfrac{\dot{a}(t)}{a(t)} {\rm d}t\right)}\nonumber \\
		&=& a^3.
	\end{eqnarray}
	From Eq.~\ref{eq:app_5}, the solution takes the form,
	\begin{equation}
		y a^3 = \dfrac{\epsilon n_{\chi} (t)}{\tau} \int_{t_0}^t a^3 {\rm d}t,
		\label{eq:app_7}
	\end{equation}
	here, $t$ denotes the time since the Big Bang takes place and $z$ is the corresponding redshift. Here, $t_0$ and $z_0$ are the time and redshift, when the difference between baryon and antibaryon number density was developed.
	
	The baryon asymmetry is originated in the epoch of matter-dominated Universe. According to the well-known cosmological scaling relations in the matter dominated Universe, the scale factor ($a$) is related to the time $t$ as,
	\begin{eqnarray}
		a &\propto& t^{2/3}, \nonumber \\
		a^3 &=& a_0 t^2.
	\end{eqnarray}
	Now substituting the above expression of scale factor in Eq.~\ref{eq:app_7}, we obtain
	\begin{eqnarray}
		&y a_0 t^2 =& \dfrac{\epsilon n_{\chi} (t)}{\tau} a_0 \int_{t_0}^t t^2 {\rm d}t \nonumber \\
		or,&y a_0 t^2 =& \dfrac{a_0^2}{3} \dfrac{\epsilon n_{\chi} (t)}{\tau}  (t^3 - t_{0}^3) \nonumber \\
		or,&y =& \dfrac{1}{3} (t-t_0) \dfrac{\epsilon n_{\chi} (t)}{\tau} \dfrac{(t^2 + t t_0 + t_{0}^2)}{t^2} \nonumber \\
		or,&y =& \dfrac{(t-t_0)}{\tau} \dfrac{\epsilon n_{\chi} (t)}{3} \left[1 + \dfrac{t_0}{t} + \dfrac{t_{0}^2}{t^2}\right].
	\end{eqnarray}
	Now considering, $\dfrac{t_0}{t}$ and $\dfrac{t_{0}^2}{t^2} \ll 1 \simeq 0$, it can be can approximate as,
	\begin{equation}
		\dfrac{t - t_0}{\tau} = 1 - \left( 1 - \dfrac{t - t_0}{\tau} \right) \approx 1 - \exp \left( -\dfrac{t - t_0}{\tau} \right)
	\end{equation}
	Therefore, Eq.~\ref{eq:app_7} becomes,
	\begin{equation}
		y = \dfrac{\epsilon n_{\chi}}{3} \left[ 1 - e^{\left( -\dfrac{t - t_0}{\tau} \right)} \right].
		\label{eq_app_11}
	\end{equation}
	Applying Eq.~\ref{eq:nchi} in the above equation (Eq.~\ref{eq_app_11}), the final form is obtained, given by,
	\begin{equation}
		y = (n_{b} - n_{\bar{b}}) = \dfrac{\epsilon n_{\chi} (t_0)}{3} \left(\dfrac{z + 1}{z_0 + 1}\right)^3 \left[ 1 - e^{\left( -\dfrac{t - t_0}{\tau} \right)} \right].
	\end{equation}

\backmatter
\nocite{*}

\lhead{\ifthenelse{\isodd{\value{page}}}{}{\color{hcol} \emph{Bibliography}}} \rhead{\ifthenelse{\isodd{\value{page}}}{\color{hcol} \emph{Bibliography}}{}}

\bibliographystyle{thsbst}
\bibliography{bib/Project1.bib}

\appendix
\cleardoublepage
\newpage
\newcommand{\lop}{List of Publications}
\lhead{\ifthenelse{\isodd{\value{page}}}{}{\color{hcol} \emph{\lop}}} \rhead{\ifthenelse{\isodd{\value{page}}}{\color{hcol} \emph{\lop}}{}}
\chapter*{\lop}
\addcontentsline{toc}{chapter}{\lop}
\phantomsection


\subsection*{In Refereed Journals:}

\begin{enumerate}
\item{
	``Chandrasekhar limit for rotating quark stars'',
	{\bf Ashadul Halder}, Shibaji Banerjee, Sanjay K. Ghosh and Sibaji Raha, 
	\href{http://dx.doi.org/10.1103/PhysRevC.103.035806}{\color{jnl} \emph{\prc}, {\bf 103}, \emph{3}, \texttt{035806}} (2021), 
	[\href{https://arxiv.org/abs/2005.14567}{{\tt \color{arx} arXiv:2005.14567}}].}

\item{
	``Bounds on abundance of primordial black hole and dark matter from EDGES 21-cm signal'',
	{\bf Ashadul Halder} and Shibaji Banerjee,
	\href{https://doi.org/10.1103/PhysRevD.103.063044}{\color{jnl} \emph {\prd}, {\bf 103}, \emph{6}, \texttt{063044}} (2021),
	[\href{https://arxiv.org/abs/2102.00959}{{\tt \color{arx} arXiv:2102.00959}}].}
\item{
	``Probing the effects of primordial black holes on 21-cm EDGES signal along with interacting dark energy and dark matter - baryon scattering'',
	{\bf Ashadul Halder} and Madhurima Pandey,
	\href{https://doi.org/10.1093/mnras/stab2795}{\color{jnl} \emph{\mnras}, {\bf 508} \emph{3}, \texttt{3446}} (2021), [\href{https://arxiv.org/abs/2101.05228}{{\color{arx} \tt arXiv:2101.05228}}].}

\item{
	``Exploring multimessenger signals from heavy dark matter decay with EDGES 21-cm result and IceCube'',
	{\bf Ashadul Halder} Madhurima Pandey, Debasish Majumdar and Rupa Basu,
	\href{https://doi.org/10.1088/1475-7516/2021/10/033}{\color{jnl} \emph{\jcap}, {\bf 2021}, \emph{10}, \texttt{033}} (2021),
	[\href{https://arxiv.org/abs/2105.14356}{{\color{arx} \tt arXiv:2105.14356}}].}
   
\item{
	``Estimation of Baryon Asymmetry from Dark Matter Decaying into IceCube Neutrinos'',
	Tista Mukherjee, Madhurima Pandey, Debasish Majumdar and {\bf Ashadul Halder},
	\href{https://doi.org/10.1142/S0217751X21500780}{\color{jnl} \emph{\ijmp A}, {\bf 36}, \emph{13}, \texttt{2150078}} (2021), $\qquad$ [\href{https://arxiv.org/abs/1911.10148}{{\tt \color{arx} arXiv:1911.10148}}].} 

\end{enumerate}

\subsection*{Publications not included in the Thesis\\
	(in Refereed Journals):}

\begin{enumerate}
	
\item{
	``Addressing $\gamma$-ray emissions from dark matter annihilations in 45 milky way satellite galaxies and in extragalactic sources with particle dark matter models'',
	{\bf Ashadul Halder}, Shibaji Banerjee, Madhurima Pandey and Debasish Majumdar, 
	\href{http://dx.doi.org/10.1093/mnras/staa3481}{\color{jnl} \emph{\mnras}, {\bf 500}, \emph{4}, \texttt{5589}}, 
	[\href{https://arxiv.org/abs/1910.02322}{{\tt \color{arx} arXiv:1910.02322}}].}

\item{
	``Mass and Life Time of Heavy Dark Matter Decaying into IceCube PeV Neutrinos'',
	Madhurima Pandey, Debasish Majumdar, {\bf Ashadul Halder} and Shibaji Banerjee, 
	\href{http://dx.doi.org/10.1016/j.physletb.2019.134910}{\color{jnl} \emph{Phys. Lett. B}, {\bf 2019}, \emph{797}, \texttt{134910}}, 
	[\href{https://arxiv.org/abs/1905.08662}{{\color{arx} \tt arXiv:1905.08662}}].} 
\end{enumerate}

\subsection*{Preprints (not included in the Thesis):}
\begin{enumerate}
\item{
	``IceCube PeV Neutrino Events from the Decay of Superheavy Dark Matter; an Analysis'',
	    Madhurima Pandey, Debasish Majumdar and {\bf Ashadul Halder}, [\href{https://arxiv.org/abs/1909.06839}{{\color{arx} \tt arXiv:1909.06839}}].}
	
\item{
	``The Violation of Equivalence Principle and Four Neutrino Oscillations for Long Baseline Neutrinos'',
	    Madhurima Pandey, Debasish Majumdar, Amit Dutta Banik and {\bf Ashadul Halder}, [\href{https://arxiv.org/abs/2003.02102}{{\color{arx} \tt arXiv:2003.02102}}].}
    
\item{
	``Speeding up of Binary Merger Due to "Apparent" Gravitational Wave Emissions'',
	Shibaji Banerjee, {\bf Ashadul Halder}, Sanjay K. Ghosh, Sibaji Raha and Debasish Majumdar, [\href{https://arxiv.org/abs/1810.04477}{{\color{arx} \tt arXiv:1810.04477}}].}

\item{
	``Intensification of Gravitational Wave Field Near Compact Star'',
	{\bf Ashadul Halder}, Shibaji Banerjee and Debasish Majumdar, [\href{https://arxiv.org/abs/1902.06903}{{\color{arx} \tt arXiv:1902.06903}}].}
 
\end{enumerate}

\end{document}